\documentclass[10pt,journal]{IEEEtran}
\usepackage{cite}
\usepackage{amsmath,amssymb,amsfonts}
\usepackage{algorithmic}
\usepackage{graphicx}
\usepackage{textcomp}
\usepackage[table]{xcolor}
\usepackage{longtable}
\usepackage{array}
\usepackage{ragged2e}
\usepackage{wrapfig}
\usepackage{bm}
\definecolor{lightgray}{gray}{0.9}

\IEEEaftertitletext{%
	\begin{center}
		{\small\bfseries Preprint --- submitted to IEEE Access for consideration.\par}
		\smallskip
		{\footnotesize This work has been submitted to the IEEE for possible publication. Copyright may be transferred without notice, after which this version may no longer be accessible.\par}
	\end{center}
	\vspace{-0.4\baselineskip}
}

\begin{document}
	\title{EEG-Based Motor Imagery BCI Algorithms and Technologies: A Review}
	\author{Mohammad Hossein Koohi Ghamsari$^{1}$, Seyede Fatemeh Ghamkhari$^{2}$, Siavash Bayat$^{1}$, and Ahmed Hemani$^{2}$%
		\thanks{$^{1}$Department of Electrical Engineering, Sharif University of Technology, Azadi St., Tehran, 14588-89694, Iran.}%
		\thanks{$^{2}$School of Electrical Engineering and Computer Science, KTH Royal Institute of Technology, 100 44 Stockholm, Sweden.}%
		\thanks{Corresponding author: Siavash Bayat (e-mail: bayat@sharif.ir).}}
	\markboth{Koohi Ghamsari \MakeLowercase{\textit{et al.}}: EEG-Based Motor Imagery BCI Algorithms and Technologies: A Review}{Koohi Ghamsari \MakeLowercase{\textit{et al.}}: EEG-Based Motor Imagery BCI Algorithms and Technologies: A Review}
	\maketitle
	
	\begin{abstract}
		Brain-computer interfaces (BCIs) have emerged as transformative technologies that enable direct communication between the brain and external devices. Among various BCI paradigms, EEG-based motor imagery (MI) has gained prominence due to its simplicity, non-invasiveness, and potential to restore motor function and facilitate rehabilitation for patients with motor impairments. This paper presents a comprehensive review of the most practical processing algorithms developed over the past decade for decoding brain sensorimotor cortex signals. Specifically, this paper discusses the integration of artificial intelligence (AI)-based algorithms, particularly machine learning and deep learning techniques, and their contributions to improving the performance and efficiency of MI-BCI systems in detail. Furthermore, the paper reviews state-of-the-art hardware platforms and emerging converging technologies—including system-on-chip (SoC) architectures, application-specific integrated circuits (ASICs), field-programmable gate arrays (FPGAs), wearable devices, the Internet of Things (IoT), and augmented/virtual reality (AR/VR)—and discusses their integration with advanced signal processing algorithms to enable next-generation MI-BCI systems. By highlighting current achievements of EEG-based MI-BCI technology and predicting future research directions that could further enhance real-time capabilities, this paper aims to provide valuable insights for researchers and practitioners, fostering innovation in high-performance EEG-based MI-BCI systems.
	\end{abstract}
	
	
	\section{Introduction}
	\label{sec:Introduction}
	
	\IEEEPARstart{T}{he} history of brain-computer interfaces (BCIs) spans nearly a century, beginning with Hans Berger's pioneering work in 1924 when he first recorded human brain activity using electroencephalography (EEG), which provided the first evidence that neural activity associated with mental processes can be measured through electrical brain signals \cite{miller2020current,varbu2022past, kudale2026comprehensive}. BCIs represent a transformative communication paradigm that establishes a direct, neuromuscular-independent pathway between the central nervous system and external computational devices \cite{kawala2021summary,qin2004motor,tung2013motor, samal2024role}. By definition, a BCI is a hardware-software system that acquires neural activity, processes it in real-time, and translates it into discrete control commands for assistive tools, neuroprosthetics, or digital interfaces \cite{carmena2003learning,ifft2013brain,kim2006continuous}. Among the various brain-signal acquisition modalities, EEG records scalp electrical potentials arising predominantly from the summed postsynaptic activity of neuronal populations and remains one of the most widely used non-invasive approaches for BCI applications because of its high temporal resolution, portability, relatively low cost, and ease of use \cite{cohen2017does,michel2012towards,nicolas2012brain,feng2022efficient}.

\begin{figure}[!t]
		\centerline{\includegraphics[scale=0.5,trim=6cm 5cm 7cm 2.5cm,clip=true]{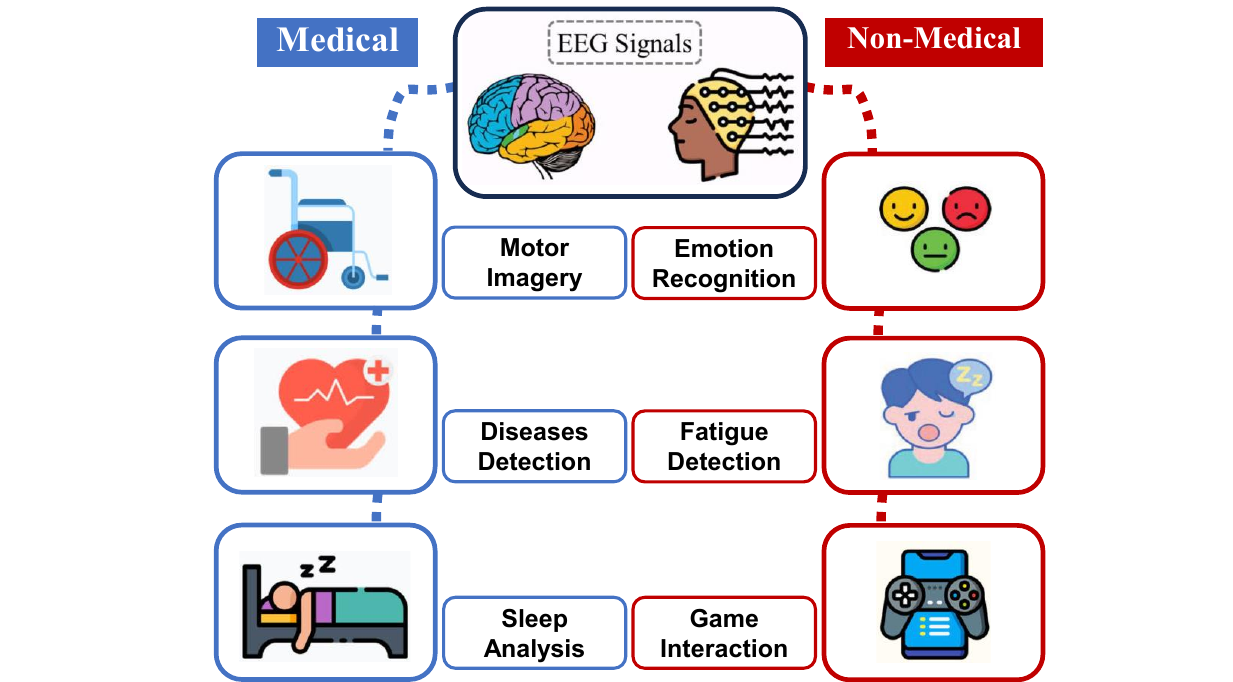}}
		\caption{Various medical and non-medical applications of EEG-based BCI modality.}
		\label{EEG_based_BCI_Applications}
	\end{figure}

	Early experiments demonstrating that animals could control devices through neural activity paved the way for significant developments in neuroprosthetics \cite{carmena2003learning,ifft2013brain,kim2006continuous}. These studies ultimately aimed to restore lost motor and communication functions. The mid-1990s saw the advent of the first neuroprosthetic devices for humans, demonstrating the potential of BCIs to enhance the quality of life for individuals with severe disabilities \cite{bird2019mental,bird2018study,levine2000direct,vidal1977real}. Today, as shown in Fig. \ref{EEG_based_BCI_Applications}, BCIs are recognized not only for their profound medical applications but also for their expanding non-medical utility and potential to augment human capabilities, making them a critical area of research spanning neuroscience and engineering \cite{allison2007brain,lebedev2017brain,wahalla2020cerebridge}.

	Within the medical domain, EEG-based BCIs are pivotal for motor imagery (MI) tasks, crucial for rehabilitation and control, as well as for disease detection and sleep analysis. Concurrently, non-medical applications are rapidly emerging, including emotion recognition, fatigue detection for safety and performance monitoring, and interactive gaming \cite{raza2025deep, gong2025multi}. As technology continues to evolve, the implications of BCIs extend into ethical and philosophical domains, challenging our understanding of consciousness and human-machine interaction \cite{boly2012brain,chatelle2012brain,edlinger2015many,gibson2014multiple}.

	EEG-based motor imagery BCIs (MI-BCIs) enable individuals to control external devices by imagining limb movements, without any physical action \cite{ cardoso2021effect, shin2024sparse}. This non-invasive approach offers significant advantages, including its portability, relatively low cost, and ease of use compared with invasive BCI approaches \cite{rao2026calibration, yao2026multiview}. These benefits make MI-BCIs highly accessible for widespread research and practical applications, particularly in rehabilitation and assistive technologies for individuals with motor impairments \cite{lazarou2018eeg, bang2021spatio}.

	This paper provides a comprehensive review of EEG-based MI-BCI systems, with a specific focus on system architecture and the complete signal-processing pipeline, including benchmark datasets, software and hardware implementations, and emerging enabling technologies for next-generation wearable and edge MI-BCI systems. Unlike broader BCI surveys or reviews that focus primarily on algorithms, rehabilitation, or individual technologies, this work jointly examines classical and AI-based signal-processing methods, hardware realization, and converging technologies within a unified MI-BCI framework. By providing a foundational roadmap for researchers and engineers, this paper addresses the need for robust, “plug-and-play” assistive MI-BCI technologies capable of transitioning from laboratory settings to real-world clinical environments.

	The remainder of this paper is organized as follows: Section II details our review methodology, outlining the literature review approach and research questions. Section III explores various BCI modalities, contrasting electrical, magnetic, and metabolic methods before focusing on EEG-based BCIs and their applications, specifically highlighting the advantages of motor imagery. Section IV delves into the fundamental EEG-based MI-BCI system functional blocks. Subsequently, Section V presents a comprehensive analysis of MI-BCI signal processing algorithms, covering both software-based (classic and AI algorithms) and hardware-based approaches. Section VI examines converging technologies for the next generation of MI-BCI Systems. Building upon these foundations, Section VII proposes a sample next-generation EEG-based MI-BCI system, detailing its key subsystems and an AI-based signal processing architecture. Finally, Section VIII offers our discussions, answering the research questions, identifying key opportunities and cross-disciplinary challenges, and providing concluding remarks on the future of the field. Section IX concludes the paper.

	\begin{table*} [t]
		\caption{Search strings and numbers of records retrieved from the selected databases for EEG-based BCI and related technologies. Searches were conducted in July 2026 and restricted to publications from January 1, 2014, to July 1, 2026.}
		\label{Table_Literature_Review}
		\centering
		
			
			\begin{tabular}{|>{\centering\arraybackslash}m{5cm} | >{\centering\arraybackslash}m{1.8cm} | >{\centering\arraybackslash}m{1cm} | >{\centering\arraybackslash}m{1.7cm} | >{\centering\arraybackslash}m{1cm} |
					>{\centering\arraybackslash}m{1cm} |
					>{\centering\arraybackslash}m{1cm} |
					>{\centering\arraybackslash}m{1cm} |	
				}

				\hline
				\rowcolor{lightgray} 
				\centering \rule{0pt}{9pt} \textbf{Search Strings / Search Engine} & 
				\centering \textbf{IEEE Xplore}  &
				\centering \textbf{PubMed} &
				\centering \textbf{ScienceOPEN} &
				\centering \textbf{Semantic Scholar} &
				\centering \textbf{Scopus} &
				\centering \textbf{Web of Science} &
				\textbf{Google Scholar}
				\\ \hline

				\cellcolor{cyan!10} \rule{0pt}{8pt} \tiny (EEG OR electroencephalographic OR electroencephalography) AND (BCI OR "brain-computer interface") & 
				11,751 &
				6,770 &
				3,610 &
				11,900 &
				4,916 &
				8,665 &
				22,600 
				\\ \hline

				\cellcolor{green!10} \rule{0pt}{9pt} \tiny (EEG OR electroencephalographic OR electroencephalography) AND (BCI OR "brain-computer interface") AND ("motor imagery" OR "motor-imagery" OR MI) & 
				3,267 &
				2,062 &
				1,235 &
				7,190 &
				2,301 &
				3,862 &
				17,900 
				\\ \hline

				\cellcolor{cyan!10} \rule{0pt}{9pt} \tiny (EEG OR electroencephalographic OR electroencephalography) AND (BCI OR "brain-computer interface") AND ("machine learning" OR ML OR "deep learning" OR DL OR "neural network" OR "artificial intelligence" OR AI) & 
				5,230 &
				2,961 &
				1,547 &
				4,040 & 
				1,101 &
				2,449 &
				18,100
				\\ \hline

				\cellcolor{green!10} \rule{0pt}{9pt} \tiny (EEG OR electroencephalographic OR electroencephalography) AND (BCI OR "brain-computer interface") AND (wearable OR wireless) & 
				806 &
				349 &
				287 &
				3,710 &
				1,318 &
				776 &
				17,700 
				\\ \hline

				\cellcolor{cyan!10} \rule{0pt}{9pt} \tiny (EEG OR electroencephalographic OR electroencephalography) AND (BCI OR "brain-computer interface") AND ("system-on-chip" OR "system on a chip" OR SoC OR "hardware accelerator" OR "hardware acceleration") & 
				398 &
				696 &
				234 &
				3,400 &
				44 &
				106 &
				6,190 
				\\ \hline

				\cellcolor{green!10} \rule{0pt}{9pt} \tiny (EEG OR electroencephalographic OR electroencephalography) AND (BCI OR "brain-computer interface") AND ("Internet of Things" OR IoT) & 
				284 &
				29 &
				28 &
				301 &
				376 &
				99 &
				11,200 
				\\ \hline

				\cellcolor{cyan!10} \rule{0pt}{9pt} \tiny (EEG OR electroencephalographic OR electroencephalography) AND (BCI OR "brain-computer interface") AND ("energy harvesting" OR "power harvesting" OR "energy scavenging" OR EH OR "low power" OR LP) & 
				714 &
				211 &
				179 &
				849 &
				119 &
				723 &
				2,120 
				\\ \hline

				\cellcolor{green!10} \rule{0pt}{9pt} \tiny (EEG OR electroencephalographic OR electroencephalography) AND (BCI OR "brain-computer interface") AND ("edge computing" OR "edge processing") & 
				148 &
				49 &
				35 &
				863 &
				1,306 &
				175 &
				17,200 
				\\ \hline

				\cellcolor{cyan!10} \rule{0pt}{9pt} \tiny (EEG OR electroencephalographic OR electroencephalography) AND (BCI OR "brain-computer interface") AND ("virtual reality" OR VR)
				& 
				413 &
				221 &
				186 &
				3,280 &
				940 &
				659 &
				17,300 
				\\ \hline

			\end{tabular}
		\end{table*}

		\begin{table*}[t]
			\caption{List of frequently used abbreviations in this paper.}
			\tiny  
			\label{Table_abbreviations}
			\centering
			
			\begin{tabular}{>{\centering\arraybackslash}m{1.7cm}
					>{\centering\arraybackslash}m{0.9cm}
					>{\centering\arraybackslash}m{1.7cm}
					>{\centering\arraybackslash}m{0.9cm}
					>{\centering\arraybackslash}m{1.7cm}
					>{\centering\arraybackslash}m{0.9cm}
					>{\centering\arraybackslash}m{1.7cm}
					>{\centering\arraybackslash}m{0.9cm}}		
				\hline
				\noalign{\vspace{1pt}} 
				\rowcolor{lightgray}
				
				\textbf{Full Term} & 
				\textbf{Abbreviation} &
				\textbf{Full Term} &
				\textbf{Abbreviation} & 
				\textbf{Full Term} &
				\textbf{Abbreviation} &
				\textbf{Full Term} &
				\textbf{Abbreviation} \\ 		
				\noalign{\vspace{1pt}}
				\hline 
				
				\noalign{\vspace{4pt}}

				Motor imagery-based BCI & MI-BCI & 
				Brain-computer interface & BCI &
				Artificial intelligence & AI &
				Electroencephalography & EEG
				\\
				\noalign{\vspace{4pt}}

				Machine learning & ML & 
				Artificial neural network & ANN &
				Deep learning & DL &
				Generative AI & GenAI
				\\
				\noalign{\vspace{4pt}}

				Classification accuracy & CA & 
				Convolutional neural network & CNN &
				Feedforward neural network & FFNN &
				Back propagation neural network & BPNN
				\\
				\noalign{\vspace{4pt}}

				Probabilistic neural network & PNN & 
				Euclidean Space & EUS &
				System-on-Chip & SoC &
				Band-pass filtering & BPF
				\\
				\noalign{\vspace{4pt}}

				Hybrid & HYB & 
				Deep belief network & DBN &
				Transformer & TRF &
				Common spatial pattern & CSP
				\\
				\noalign{\vspace{4pt}}

				Support vector machine & SVM & 
				Linear discriminant analysis & LDA &
				Neuro‑fuzzy system & NFS &
				Hardware accelerator & HA
				\\
				\noalign{\vspace{4pt}}

				Internet of Things  & IoT & 
				Mutual information & MIN &
				Compressive sensing & CPS &
				Energy harvesting & EH
				\\
				\noalign{\vspace{4pt}}

				Autoencoder & AEC & 
				Fuzzy integral & FZI &
				Edge processing & EP &
				Transfer learning & TL
				\\
				\noalign{\vspace{4pt}}

				Virtual reality & VR & 
				Digital signal processing & DSP &
				Naive bayes & NVB &
				long short‑term memory & LSTM
				\\
				\noalign{\vspace{4pt}}

				\hline
				
			\end{tabular}
		\end{table*}

		\section{Review Methodology}
		\label{sec:Review_Methodology}
		
		To investigate the dominant BCI applications, the signal processing techniques used for EEG-based MI-BCI systems over the past decade, and emerging technologies for future MI-BCI systems, the following research questions (RQs) are defined.
		
		\textbf{RQ1.} Which BCI application currently has the greatest practical impact and the strongest potential for future development and improving the quality of life? What is the role of EEG-based MI-BCI systems within this landscape?
		
		\textbf{RQ2.} What are the most frequently used signal processing techniques and datasets for EEG-based MI-BCI systems in the recent decade? Which processing approaches have demonstrated superior performance in recent studies and have the most potential to be used in future systems? 
		
		\textbf{RQ3.} What are the challenges of EEG-based MI-BCI systems, and what are the promising candidate solutions? 
		
		\textbf{RQ4.} How can various emerging technologies be brought together to create the next generation of EEG-based MI-BCI systems?

		To answer the above questions, this study employs a structured review methodology to minimize selection bias and ensure a reproducible, transparent synthesis of technical data, as discussed in the following sections.

		\subsection{Review Design and Search Strategy}
		
		The review was conducted in two stages. First, to address the preceding RQs, the broader EEG-based BCI literature was mapped, and a statistical analysis of article titles and abstracts was performed to identify major application areas and emerging technological trends in the BCI field. As discussed in detail in Section \ref{sec:Discussions}, this analysis indicates that EEG-based MI-BCI systems for medical and rehabilitation applications represent a major area of research within the field. Relevant studies were retrieved from IEEE Xplore, PubMed, ScienceOpen, Semantic Scholar, Scopus, Web of Science, and Google Scholar. The literature searches were conducted in July 2026, with the publication period restricted to January 1, 2014, through July 1, 2026. The complete search strings and the corresponding numbers of retrieved records are reported in Table \ref{Table_Literature_Review}. The same conceptual search strings were applied across all databases, with minor syntax adaptations according to the search interface supported by each database. Frequently used technical terms and their abbreviations are listed in Table \ref{Table_abbreviations}.

		Second, the review focused specifically on MI-BCI studies, systematically searched across IEEE Xplore, PubMed, ScienceOpen, Semantic Scholar, Scopus, Web of Science, and Google Scholar using predefined EEG, BCI, MI, and technology-specific keywords. Influential studies from the recent decade, including recent works proposing novel methodologies, were analyzed with emphasis on signal-processing algorithms, benchmark datasets, hardware implementations, and enabling technologies. The resulting findings are summarized and discussed in the following sections, while a comprehensive classification of commonly used algorithms is presented in Fig. \ref{Signal_Processing_Pipeline_Algorithms}.

		\subsection{Eligibility Criteria}
		The eligibility criteria for selecting works at both stages included studies published from January 1, 2014, to July 1, 2026, focusing on journal articles, conference proceedings, and dataset reports. Moreover, studies were eligible for the focused MI-BCI synthesis when they: (1) reported original research in a peer-reviewed journal or conference proceeding, or described a benchmark dataset relevant to MI-BCI research; (2) investigated EEG-based BCI with an explicit motor-imagery component; and (3) provided sufficient methodological or quantitative information concerning signal preprocessing, feature extraction or selection, classification, datasets, experimental validation, or hardware implementation.
		
		Studies were excluded from the focused synthesis when they: (1) investigated exclusively non-EEG modalities such as ECoG, MEG, or fMRI; (2) investigated EEG paradigms unrelated to motor imagery, such as exclusively SSVEP- or P300-based systems, unless they were used only for contextual comparison; (3) addressed EEG applications without a BCI component; (4) were reviews, editorials, news articles, commercial webpages, or other non-primary sources used only for background information; (5) duplicated another retrieved publication; or (6) did not provide sufficient methodological or outcome information for extraction. Studies outside the focused MI-BCI scope were not included in the quantitative synthesis even when they were cited elsewhere in the manuscript for general technological background.

		\subsection{Study Screening and Deduplication}
		Duplicate records were first identified using DOI, PubMed ID, and exact-title matching, followed by manual comparison of titles, authors, publication years, and journal/conference information to identify remaining duplicates. After deduplication, 1,247 unique records remained. Titles and abstracts were then screened against the eligibility criteria, resulting in the exclusion of 349 records. The full texts of the remaining 898 publications were assessed for eligibility, and 152 publications were excluded because of non-MI paradigms, non-EEG modalities, insufficient methodological information, duplicate or secondary reports, irrelevant applications, or other scope-related reasons. The remaining 746 studies constituted the relevant literature corpus. From these, 324 studies were selected for detailed analysis and inclusion in this review based on their relevance to the specific scope, research questions, and objectives of the paper.

		\subsection{Quality Assessment}
		The quality of the studies was evaluated based on: (1) adequacy of the description of participants or dataset provenance; (2) reproducibility of EEG preprocessing and feature-processing procedures; (3) completeness of model architecture and training details; (4) appropriateness and transparency of the validation protocol, including separation between training and test data and specification of subject-dependent or subject-independent evaluation; (5) completeness of performance reporting, including relevant accuracy, F1-score, Cohen's $\kappa$, variability, or other applicable metrics; and (6) comparison with appropriate baselines or benchmark methods. For hardware-oriented studies, additional consideration was given to whether the implementation platform and validation conditions were adequately described and whether relevant quantities such as power consumption, latency, memory/resource utilization, throughput, or silicon area were reported when applicable.

		\subsection{Data Extraction and Synthesis}
		A standardized data-extraction form was used for all studies included in the focused synthesis. The extracted information included publication year, EEG dataset or participant cohort, number of subjects, number of channels, sampling rate, number and type of MI classes, preprocessing procedures, feature-extraction and feature-selection methods, classification algorithm, validation protocol, subject-dependent or subject-independent evaluation setting, and reported performance metrics. For hardware implementations, the hardware platform and available implementation metrics, including resource utilization, power consumption, area, latency, inference time, memory requirements, and computational efficiency, were additionally extracted.

		The percentages reported in Figs. 30 and 32 were calculated as \(100\times n_i/N\), where \(n_i\) is the number of studies assigned to category \(i\), and \(N\) is the total number of studies included in the corresponding analysis. The denominator \(N\) for each analysis is explicitly reported in the respective figure caption. Publication trends in Figs. 31 and 34 were derived from annual Scopus search-result counts using the search strategy described in Section II. The plotted lines simply connect the annual publication counts to visualize temporal trends; no statistical regression, smoothing, or extrapolation was applied. Fig. 33 represents a qualitative synthesis of the challenges and corresponding solutions identified across the reviewed literature, with supporting references provided in the accompanying discussion.

		\section{BCI Modalities}
		\label{sec:BCI_Modalities}
		
\begin{figure*}[!t]
			\centerline{\includegraphics[scale=0.5,trim=1.2cm 0.5cm 1cm 0cm,clip=true]{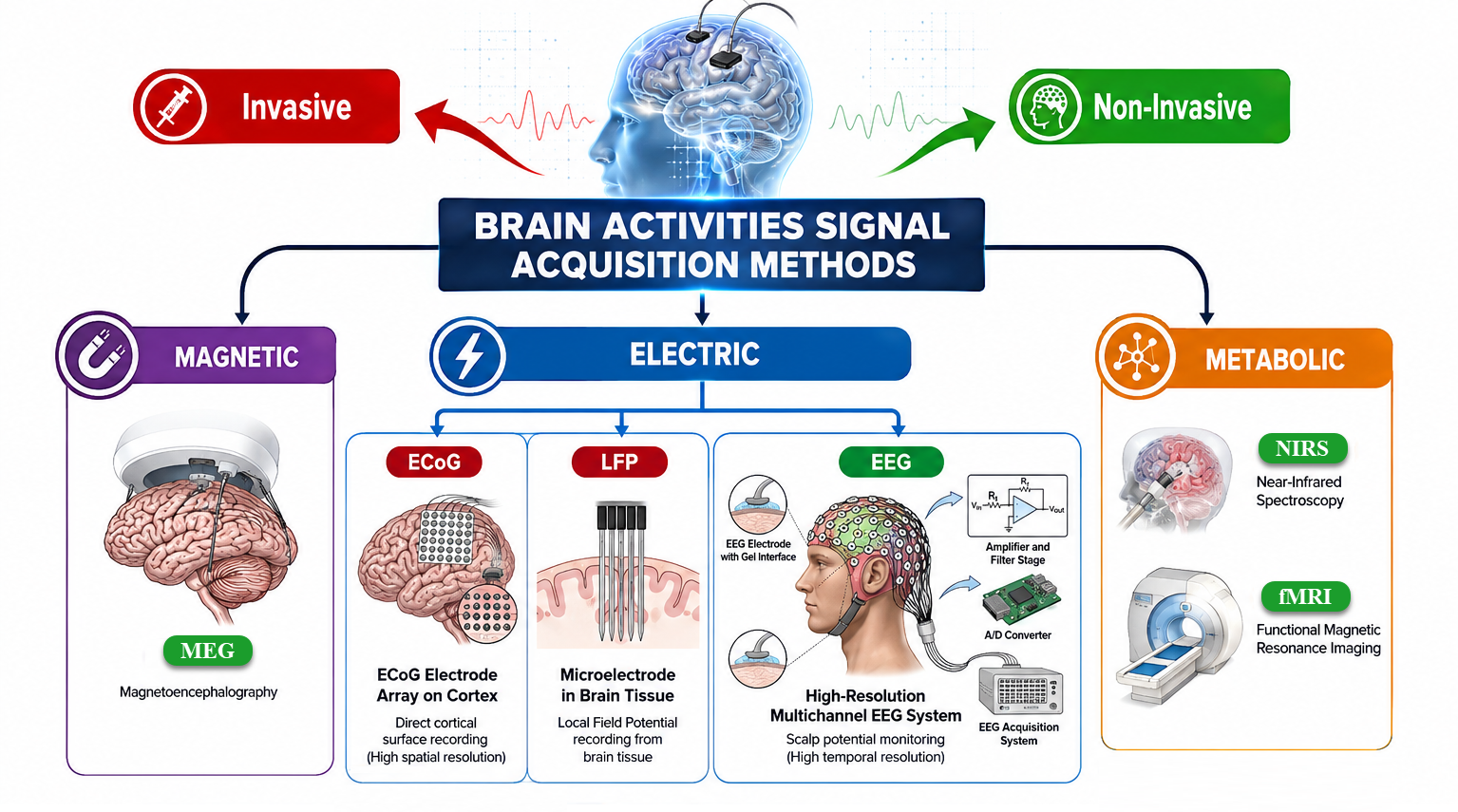}}
			\caption{Overview of brain activity signal acquisition modalities for BCI systems. The modalities are grouped into magnetic (MEG), electric (ECoG, LFP, and EEG), and metabolic (NIRS and fMRI) approaches, highlighting their invasive and non-invasive implementations and representative sensing hardware.}
			\label{Fig_BCI_Methods}
		\end{figure*}

		BCI systems are fundamentally built upon a diverse array of neuroimaging modalities, each employing different techniques to capture specific aspects of brain activity. These modalities differ considerably in their underlying physiological mechanisms for signal acquisition, leading to a spectrum of performance trade-offs in terms of resolution, invasiveness, cost, and practicality \cite{padfield2019eeg,wei2020review,hosseini2023state}. The primary categories, as depicted in Fig. \ref{Fig_BCI_Methods}, include magnetic modalities, which detect the faint magnetic fields produced by neuronal activity; electric modalities, which measure electrical potentials on the scalp or within the brain; and metabolic modalities, which track changes in blood flow or oxygenation as indicators of neural function \cite{samal2024role,he2015noninvasive}.

		\begin{figure*}[!t]
			\centerline{\includegraphics[scale=0.35,trim=5.3cm 7.3cm 4.9cm 6.6cm,clip=true]{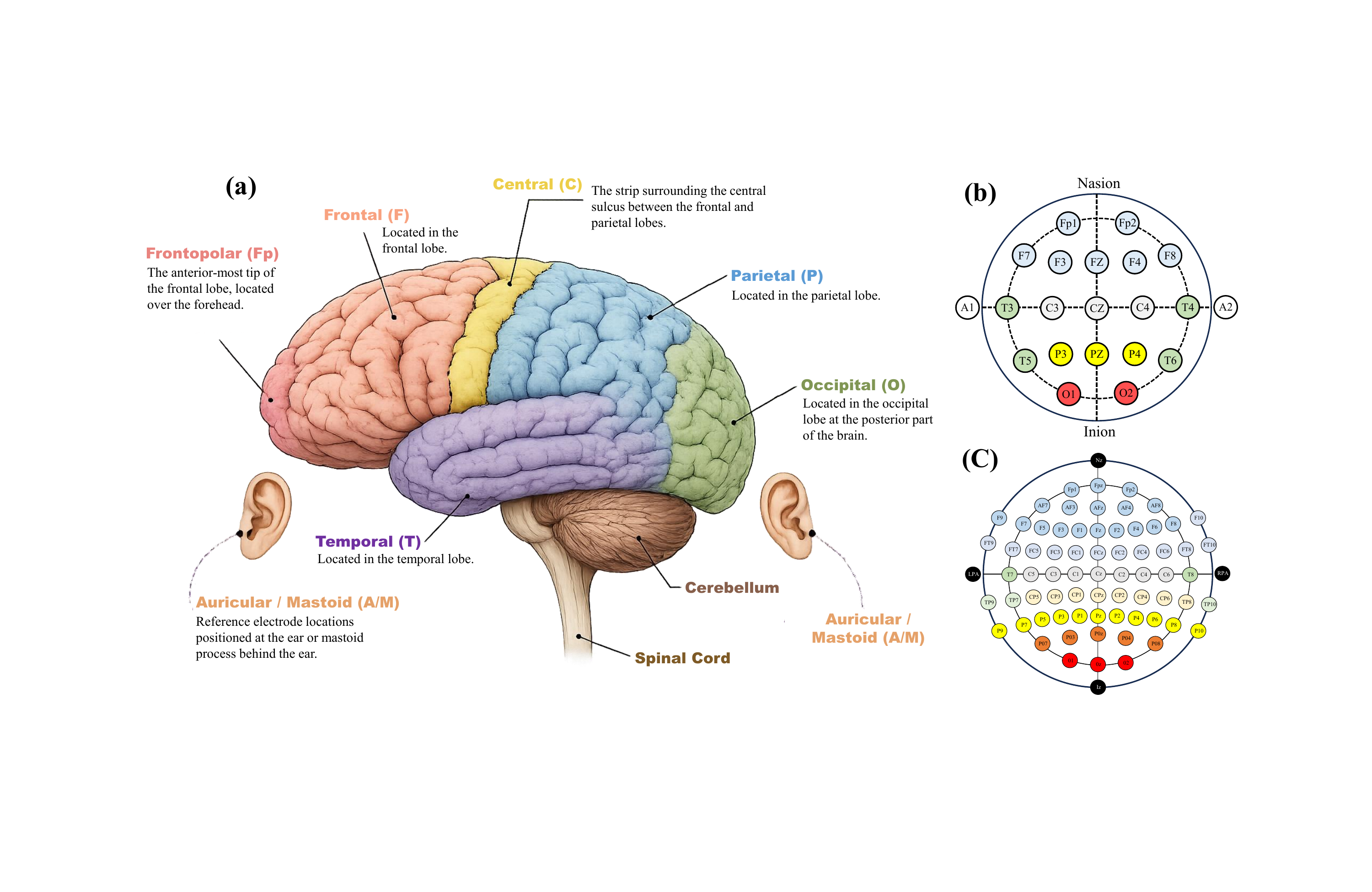}}
			\caption{Brain region labeling and the international standard EEG electrode placement systems. (a) Lateral brain view divided into the six main cortical regions referenced by electrode nomenclature: frontopolar (Fp), frontal (F), central (C), parietal (P), occipital (O), and temporal (T), with the cerebellum and spinal cord shown for anatomical context. (b) 10-20 electrode placement system (21 electrodes), with electrode labels color-coded by region and referenced to the nasion and inion. A1/A2 denote auricular (ear) reference electrodes, which are not cortical sites. (c) 10-10 electrode placement system (extended array of up to 81 electrodes), providing finer spatial resolution between the standard 10-20 sites. 
			}
			\label{Electrodes_Systems}
		\end{figure*}

		Magnetic-based approaches, such as magnetoencephalography (MEG), excel at capturing the magnetic fields generated by neuronal currents, providing exceptional temporal resolution crucial for understanding the precise timing of brain processes. This makes MEG invaluable for fundamental neuroscience research, allowing scientists to observe the rapid dynamics of neural oscillations and event-related fields \cite{tan2025brain, iivanainen2026spatial}. However, the routine application of MEG in practical settings is severely hindered by its requirement for large, prohibitively expensive, and immobile instrumentation, typically housed in specialized magnetically shielded rooms. Consequently, its use is largely confined to controlled laboratory environments rather than routine clinical or consumer applications \cite{wu2015bayesian,fred2022brief,cohen1972magnetoencephalography}.

		Electric-based BCI modalities offer a broad spectrum of approaches for capturing neural activity, ranging from highly precise invasive techniques to more accessible non-invasive methods. Invasive techniques, such as electrocorticography (ECoG) and local field potentials (LFP), involve direct access to the cortical surface or deeper brain structures, providing neural recordings with high signal-to-noise ratio (SNR) and superior spatial resolution \cite{liao2026optimal,moran2010evolution}. This makes them invaluable for applications like advanced seizure detection and prediction in epilepsy patients or for sophisticated motor control prosthetics in individuals with severe paralysis, where precise neural decoding is paramount.  In contrast, non-invasive EEG-based BCIs record electrical activity from the scalp without requiring surgical implantation \cite{wolpaw2002brain,yuan2014brain,kudale2026comprehensive}. While these methods are valued for their ease of setup, minimal risk, and affordability, enabling widespread use in clinical diagnostics and consumer-level BCI applications, they inherently sacrifice spatial resolution due to signal distortion and attenuation through the skull and tissues.

		Metabolically based BCIs, most notably near-infrared spectroscopy (NIRS) and functional magnetic resonance imaging (fMRI), infer neural activity indirectly by tracking hemodynamic responses, the localized changes in blood flow and oxygenation that occur in response to neuronal activation \cite{lim2025exploring, mirzaeian2025ultra}. By leveraging these metabolic proxies, these modalities provide valuable spatial accuracy, allowing researchers and clinicians to precisely map cognitive functions to specific cortical regions \cite{tang2026fnirnet,lim2025exploring}. This makes them well-suited for applications such as pre-surgical mapping for tumor removal, where identifying critical functional areas is vital, or in longitudinal studies of neuroplasticity and cognitive recovery. However, these benefits are frequently counterbalanced by inherent limitations: fMRI, while highly precise, requires large-scale, stationary, and expensive scanners that restrict participant mobility, while NIRS, though more portable and cost-effective, suffers from slower temporal resolution due to the inherent slow dynamics of the vascular response, which creates a significant latency compared to direct electrical measurements \cite{mirzaeian2025ultra,wang2026topology}.

		Recent technological advances have encouraged the development of hybrid (HYB) BCIs that combine multiple modalities to improve performance \cite{zhao2025novel}. For example, by integrating EEG’s high temporal resolution with NIRS’s superior spatial localization of hemodynamic responses, these HYB systems can provide more robust interfaces. In particular, a HYB BCI may enable more reliable control of a prosthetic limb by using EEG for rapid detection of motor intention and NIRS to complement the activation of specific cortical regions, thereby enhancing responsiveness \cite{ali2023correlation, kwak2022fganet}. In individuals with communication impairments, this combination can support more effective assistive technologies by enabling faster symbol selection through the complementary information (cross-validation) from electrical and metabolic brain signals, thus increasing reliability. In addition, several studies have reported the use of other sensors alongside EEG devices, including electromyography (EMG) \cite{mattia2020promotoer, li2019hybrid, peterson2020feasibility, paszkiel2020brain}, functional electrical stimulation (FES) devices \cite{gant2018eeg}, NIRS tools \cite{ khan2015motor}, and MRI devices \cite{ zich2015multimodal}.

		EEG-based BCIs have gained widespread adoption across diverse domains, including healthcare, rehabilitation, entertainment, education, and automotive systems \cite{machado2010eeg, yadav2023electroencephalogram, douibi2021toward}. Their medical applications span a broad spectrum, supporting tasks such as emotion recognition \cite{cheng2020emotion, wang2014emotional}, detection of seizure and epilepsy disorders \cite{maksimenko2017absence, liang2010closed, abiri2019comprehensive, zaghloul2019early, yang2020seizure, hosseini2017optimized, you2020unsupervised}, assessment of neurodegenerative diseases such as Parkinson’s and Alzheimer’s \cite{liberati2013development, moller2021technology, liberati2012toward}, tumor evaluation \cite{song2021evaluation}, and identification of sleep-related abnormalities \cite{luaute2015bci, handayanireal, behzad2021role}. The suitability and widespread choice of EEG for MI paradigms stem from several key advantages: its excellent temporal resolution allows for the detection of neural activity within milliseconds, crucial for capturing the real-time electrical signals generated during the mental rehearsal of movements \cite{blankertz2008berlin}; its relative non-invasiveness, portability, and cost-effectiveness compared to other neuroimaging techniques make it accessible for diverse applications; and specific patterns of brain activity, such as changes in sensorimotor rhythms, are reliably detectable and provide a clear neural correlate for decoding user intentions. These characteristics collectively make EEG a well-suited and frequently chosen modality for developing and deploying MI-BCI systems \cite{luaute2015bci, yadav2023electroencephalogram}.

		The anatomical brain regions commonly referenced in EEG electrode nomenclature are illustrated in Fig. \ref{Electrodes_Systems} (a). The cerebral cortex is divided into six major regions: frontopolar (Fp), frontal (F), central (C), parietal (P), occipital (O), and temporal (T), corresponding approximately to the cortical regions from which EEG signals primarily originate. The central region encompasses the sensorimotor cortex along the central sulcus, which plays a crucial role in MI-based BCI applications. In addition, the auricular/mastoid (A/M) locations are shown as extracerebral reference sites positioned at the ear and mastoid process, outside the cerebral cortex. This regional labeling provides a neuroanatomical framework for understanding EEG electrode placement and the cortical origins of recorded neural activity \cite{ang2016eeg,padfield2019eeg,yu2021new}.

		EEG systems measure potential differences generated primarily by synchronized postsynaptic cortical activity through electrodes placed on the scalp, typically referenced to a common electrode and recorded by an amplifier at fixed sampling rates \cite{gu2021eeg,shin2024sparse}. Two international electrode placement configurations are the 10-20 system and the 10-10 system, as depicted in Fig. \ref{Electrodes_Systems} (b) and Fig. \ref{Electrodes_Systems} (c), respectively. The 10–20 system standardizes electrode placement using proportional distances between anatomical landmarks, enabling consistent labeling of Fp, F, C, P, O, and T regions, while auricular (A) sites located near the ears serve as reference points. For denser recordings, the 10–10 configuration refines the 10–20 layout by adding more electrodes at intermediate standardized intervals, yielding finer spatial sampling across the same labeled regions. Note that compound labels (e.g., AF, FC, etc.) denote electrode positions intermediate between two adjacent regions. For example, "FC3" lies between frontal and central sites. Nz and Iz mark the nasion and inion; LPA/RPA (or A1/A2) denote left/right auricular reference points \cite{bang2021spatio,portillo2021mind}.

		For MI-BCI applications, electrode count is a trade-off between information richness and system complexity: fewer electrodes (e.g., 4–8) reduce cost, setup time, and user burden, but they may miss weaker or spatially distributed sensorimotor patterns, whereas more electrodes (e.g., 16–32) improve spatial coverage and decoding robustness, especially across inter-subject variability, at the cost of longer preparation, greater computational complexity, and increased susceptibility to noisy channels \cite{gu2021eeg,shin2024sparse}. This is particularly relevant for wearable devices and deployments in uncontrolled environments with limited computational power, where reducing data dimensionality is essential, and classification should rely on as few inputs as possible. Accordingly, beyond traditional feature reduction and selection methods, many studies focus on specific EEG channels, usually chosen from the central cortical region, which is consistent with neuroscientific evidence on MI. In practice, MI-focused montages prioritize the sensorimotor cortex, targeting the $\mu$ and $\beta$ rhythms over the central area, so electrode selections are commonly centered on C-region positions and often include nearby F and P electrodes, with practical setups frequently using around 16 electrodes, often following the international 10–20 electrode placement standard, to balance performance and usability \cite{shin2024sparse,miller2020current,varbu2022past}.

		In the next section, the theoretical fundamentals of the MI paradigm are briefly reviewed, and the functional blocks of practical EEG-based MI-BCI systems are introduced.

		\section{EEG-based MI-BCI Systems}
		\label{sec:EEG_MI-BCI_System}

		\begin{figure}[!t]
			\centerline{\includegraphics[scale=0.4,trim=2cm 0cm 0cm 0cm,clip=true]{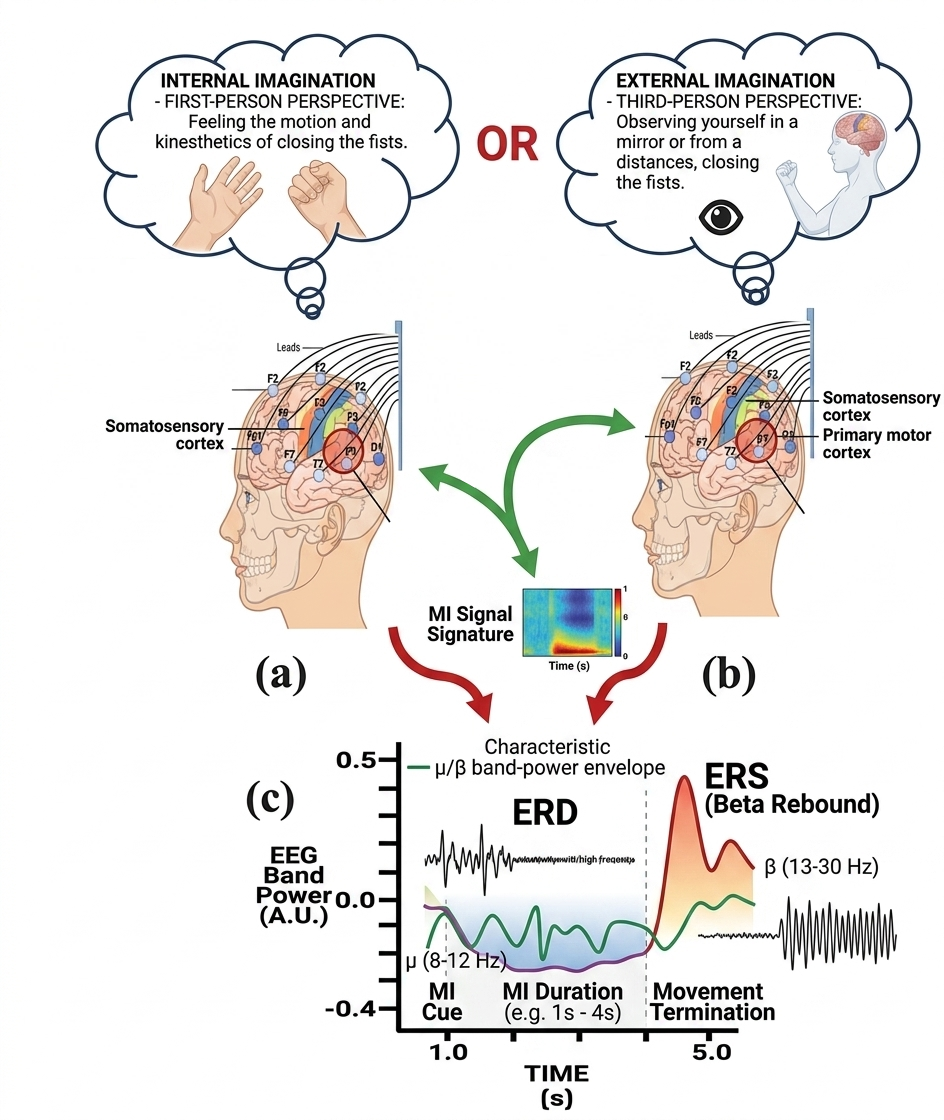}}
			\caption{The MI process in the brain. The first (internal) (a) somatosensory or third (external) (b) observational perspective imagination of opening/closing the fists activates the primary sensorimotor cortex. During imagined movement (c), the specific cortical oscillation rhythms ($\mu$, 8–12 Hz and $\beta$, 13–30 Hz) are suppressed, known as Event-Related Desynchronization (ERD). Upon movement termination, the termination of MI elicits a distinct $\beta$-band amplitude increase, known as Event-Related Synchronization (ERS) or $\beta$-rebound. 
			}
			\label{Fig_Imagination}
		\end{figure}

		\begin{figure*}[!t]
			\centerline{\includegraphics[scale=0.5,trim=0cm 0cm 0cm 0cm,clip=true]{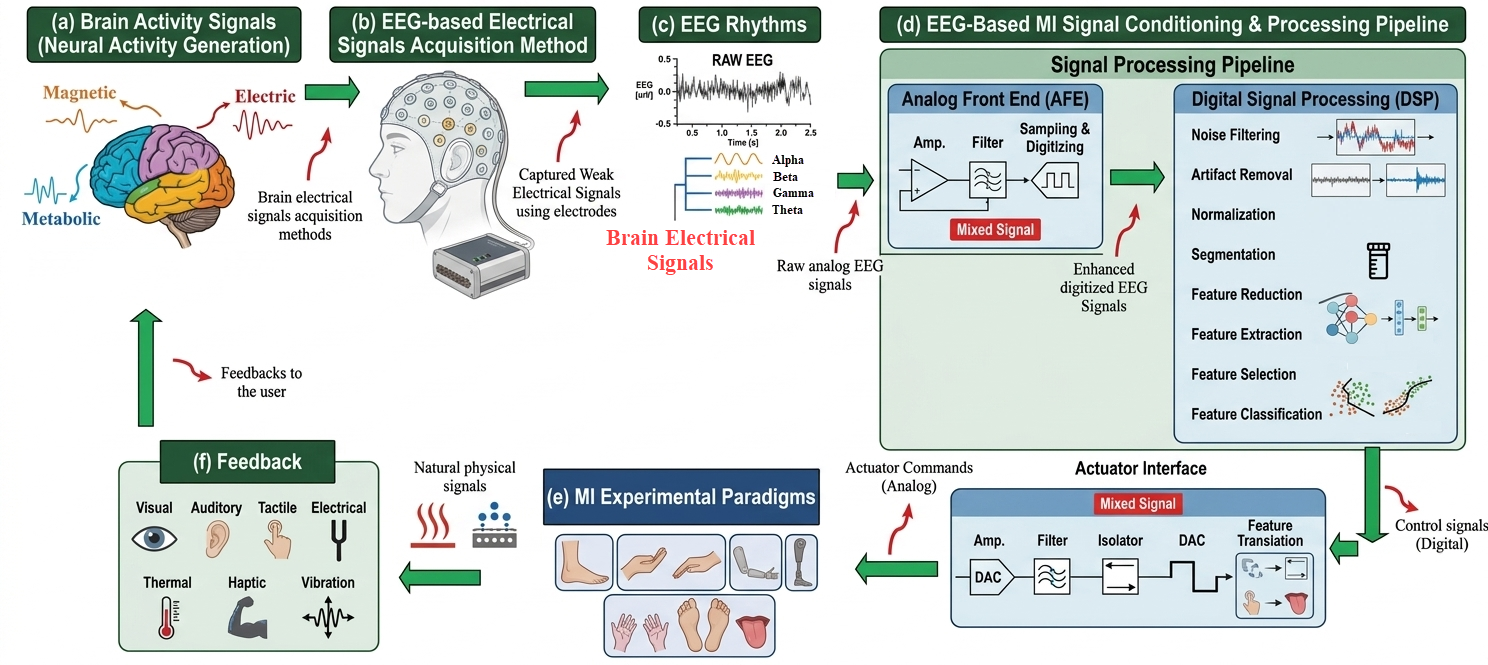}}
			\caption{The EEG-based MI-BCI life cycle overview. (a) Brain electrical, magnetic, and metabolic signals are generated by neural activity. (b) EEG-based electrical signal acquisition method. (c) Five frequency bands of EEG signals, known as EEG rhythms. (d) MI processing pipeline: analog front end (AFE), digital signal processing (DSP), and actuator interface. (e) Conventional MI experimental paradigms. (f) Actuator to brain feedback mechanisms: Vibration, Tactile, Thermal, Electrical, Haptic, Auditory, and Visual.}
			\label{Fig_BCI_Life_Cycle_MI}
		\end{figure*}

		Among the various BCI paradigms, MI has become a widely adopted paradigm for neuroprosthetics and rehabilitation due to its unique ability to engage the sensory-motor cortex and activate neural pathways similar to those utilized in actual motor execution \cite{bird2019mental,bird2018study,levine2000direct,vidal1977real,allison2007brain,wahalla2020cerebridge}. By allowing users to control external devices through imagined movements, MI-based BCIs provide a vital non-invasive means for individuals with severe disabilities to regain autonomy and interact with their environment despite severe motor impairments \cite{yuan2021bci,young2014changes,doud2011continuous,varkuti2013resting}. The primary advantage of these applications is the ability to harness neural signals without requiring residual muscle activity, which is crucial for patients with conditions such as spinal cord injuries \cite{miladinovic2020evaluation,glavas2024empowering}. However, transitioning MI-BCI from controlled laboratory settings to real-world clinical use presents significant engineering hurdles, including signal non-stationarity and subject variability. Overcoming these barriers necessitates a symbiotic optimization of both hardware acquisition layers and software classification pipelines to ensure the low latency and high accuracy required for robust, user-friendly assistive technologies \cite{mattei2024deep,wang2024towards}.

		MI refers to the mental simulation of a movement without any physical execution, and it can be carried out from either a first-person (internal) perspective, where the individual feels as if they are performing the movement, or a third-person (external) perspective, in which the individual imagines watching themselves perform the action, as illustrated in Fig. \ref{Fig_Imagination} (a) and Fig. \ref{Fig_Imagination} (b), respectively. Extensive research shows that imagined and executed movements engage functionally similar neural mechanisms and activate the same primary sensorimotor regions \cite{rao2026calibration, raza2025deep}. When a person imagines opening or closing their fists, the brain generates characteristic changes in the $\mu$ (8–12 Hz) and $\beta$ (13–30 Hz) rhythms originating from the primary sensorimotor cortex of the brain. These changes appear as event-related desynchronization (ERD), a suppression in rhythm amplitude that occurs before and during imagined movement, and event-related synchronization (ERS), an increase in $\beta$-band amplitude associated with movement termination or cortical idling, as shown in Fig. \ref{Fig_Imagination} (c). The MI-induced ERD/ERS patterns provide the fundamental neural signatures used by EEG-based MI BCIs to decode user intent \cite{kim2024meta,yu2021new,huang2022eeg}.

		The block diagram of an MI-BCI system represents a multi-stage architecture designed to decode neural signals and translate them into external commands \cite{tung2013motor,leeb2015towards}, as conceptually illustrated in Fig. \ref{Fig_BCI_Life_Cycle_MI}. Each interconnected stage plays a critical role in this complex process, from the initial detection of brain activity to the final execution of an action. These stages provide the interface between a user's thoughts and the operation of external devices, offering a powerful means of communication and control for individuals with severe motor impairments. The subsequent paragraphs will delve into the specifics of each of these crucial components.

		At the beginning of the MI process, neural activity gives rise to electrical, magnetic, and hemodynamic signals that reflect the underlying neural processes (Fig. \ref{Fig_BCI_Life_Cycle_MI} (a)). In EEG-based MI-BCI systems, the focus is on capturing the electrical signals. The EEG BCI modality employs specialized sensing electrodes to acquire these brainwave patterns from the scalp (Fig. \ref{Fig_BCI_Life_Cycle_MI} (b)). This acquisition process is the initial step, where multi-channel electrodes record raw analog EEG signals. This fundamental neuro-electronic interface captures the synchronized postsynaptic cortical activity, thereby laying the groundwork for interpreting the user's intentions \cite{dhiman2023machine,fiedler2022high}.

		\begin{table*}[t]
			\caption{The most common types of benchmark datasets used for evaluating the performance of MI algorithms.}
			\label{Table_datasets}
			\centering
			
			\begin{tabular}{
					>{\centering\arraybackslash}m{0.4cm}
					>{\centering\arraybackslash}m{3.0cm}
					>{\centering\arraybackslash}m{1.4cm}
					>{\centering\arraybackslash}m{1.2cm}
					>{\centering\arraybackslash}m{1.0cm}
					>{\centering\arraybackslash}m{0.8cm}
					>{\centering\arraybackslash}m{4.8cm}
				}
				
				\hline
				\noalign{\vspace{1pt}}
				\rowcolor{lightgray}
				
				\textbf{Cat.} &
				\textbf{Dataset} &
				\textbf{Channels} &
				\textbf{Sampling Rate} &
				\textbf{Subjects} &
				\textbf{Classes} &
				\textbf{Task}
				\\
				
				\noalign{\vspace{1pt}}
				\hline
				
				\noalign{\vspace{4pt}}
				1 & BCI Competition III dataset IIIa & 25 & 250 Hz & 9 & 4 &
				Both hands, one foot, and tongue MI \\ \hline
				
				\noalign{\vspace{4pt}}
				2 & BCI Competition III dataset IVa & 118 & 100 Hz & 5 & 3 &
				Both hands and right foot MI \\ \hline
				
				\noalign{\vspace{4pt}}
				3 & BCI Competition III dataset V & 32 & 512 Hz & 3 & 3 &
				Both hands and word generation MI \\ \hline
				
				\noalign{\vspace{4pt}}
				4 & BCI Competition IV dataset 2a & 22 + 3 EOG & 250 Hz & 9 & 4 &
				Both hands, both feet, and tongue MI \\ \hline
				
				\noalign{\vspace{4pt}}
				5 & BCI Competition IV dataset 2b & 3 EEG + 3 EOG & 250 Hz & 9 & 2 &
				Both hands MI \\ \hline
				
				\noalign{\vspace{4pt}}
				6 & Physionet EEG Motor Movement/Imagery & 64 & 160/128 Hz & 109 & 4 &
				Both fists and both feet opening/closing MI \\ \hline

				\noalign{\vspace{4pt}}
				7 & MI-OpenBCI & 15 & 125 Hz & 9 & 5 &
				Both hands, both feet, and rest MI \\ \hline
				
				\noalign{\vspace{4pt}}
				8 & Proprietary (Online) & -- & -- & -- & -- &
				-- \\ \hline
				
				\noalign{\vspace{4pt}}
				9 & High Gamma Dataset & 44 & 250 Hz & 14 & 5 &
				Both hands, both feet, and rest MI \\ \hline
				
				\noalign{\vspace{4pt}}
				10 & BCI Competition 2003 Dataset III & 3 & 128 Hz & 1 & 2 &
				Both hands MI \\ \hline
				
				\noalign{\vspace{4pt}}
				11 & BCI Competition II & 2 & 128 Hz & 3 & 2 &
				Both hands MI \\
				
				\noalign{\vspace{4pt}}
				\hline
				
			\end{tabular}
		\end{table*}

		EEG signals are characterized by distinct frequency bands, known as EEG rhythms, each associated with different brain states and cognitive processes. As illustrated in Fig. \ref{Fig_BCI_Life_Cycle_MI} (c), these critical frequency bands typically include $\delta$ (0.5-4 Hz), $\theta$ (4-8 Hz), $\alpha$ (8-12 Hz), $\beta$ (12-30 Hz), and $\gamma$ (30-100 Hz) waves. For MI, specific rhythms such as $\alpha$ and $\beta$ are particularly important for decoding the user's intentions as the power and characteristics within these bands are modulated by MI, providing the neural signatures that the processing pipeline aims to detect and interpret \cite{yuan2014brain,he2015noninvasive}.

		The core of the MI-BCI involves a processing pipeline for decoding the user's intentions that includes an analog front end (AFE), digital signal processing (DSP), and an interface to an actuator (Fig. \ref{Fig_BCI_Life_Cycle_MI} (d)). Starting from the AFE section, the AFE typically consists of mixed-signal circuitry that amplifies, filters (to remove noise and artifacts while preserving the relevant neural information), and digitizes EEG signals before digital processing \cite{cho2021neurograsp,jamil2021noninvasive}.

		Following signal conditioning, the DSP stage prepares and interprets the EEG data to discern the user's intended motor command \cite{he2015noninvasive}. This complex process encompasses several critical steps: secondary signal preprocessing, which includes noise filtering and artifact removal to clean the data, followed by normalization to standardize signal amplitudes. Subsequently, segmentation divides the continuous data stream into meaningful epochs. The pipeline then moves to feature extraction and feature reduction, where techniques like power spectral analysis (e.g., calculating power in sensorimotor $\mu$ rhythm) are employed to derive salient characteristics from the EEG signals. Feature selection further refines these features, identifying the most discriminative ones. Finally, a feature classification algorithm takes these selected features and translates them into a specific motor command \cite{dai2019eeg,wei2020review,hosseini2020review,rakhmatulin2024exploring}.

		The Actuator Interface serves as the crucial bridge between the processed neural data and the external device, translating the decoded motor intentions into physical actions. This stage begins with feature translation, where the output from the DSP stage (the classified motor command) is converted into a format understandable by the subsequent hardware \cite{yang2022enhancing,belwafi2021embedded}. When analog actuation is required, a digital-to-analog converter (DAC) converts a digital command into an analog signal. To ensure safety and protect both the user and the equipment, an isolator galvanically separates the BCI system from the actuator, preventing any electrical interference or dangerous current flow. Following isolation, a filter may be applied to the analog signal to smooth it or remove any residual noise introduced during the conversion process. Finally, an amplifier boosts the signal to the required level to effectively drive the actuator, whether it's a motor, a prosthetic limb, or another output device \cite{saeidi2021neural}.

		Experimental paradigms for MI-BCI are designed to elicit distinct and detectable neural patterns associated with the imagination of movement. As illustrated in Fig. \ref{Fig_BCI_Life_Cycle_MI} (e), these paradigms commonly include tasks like imagining left/right hand movements, shoulder flexion/extension/abduction, and execution/imagination of fist opening/closing for both left and right hands. Other paradigms involve finer movements such as finger MI, finger tapping, left/right leg MI, and even cue-based tongue movements. The specific choice of paradigm influences the brain regions activated and the resulting EEG signatures \cite{belwafi2021embedded,saibene2023eeg}.

		MI signal processing techniques can be evaluated using previously recorded EEG signals (offline) or by recording brain signals (online). To facilitate research and development, several publicly available datasets exist for MI experiments, as tabulated in Table \ref{Table_datasets}. In this table, the number of channels, sampling rate, number of subjects, number of classes, and the performed tasks are specified for each dataset. Despite these differences, most of these datasets are based on experimental paradigms involving the MI of the right hand, left hand, feet, and tongue \cite{al2021deep}. These datasets constitute essential benchmarking resources for training and validating MI-BCI systems \cite{reaves2021assessment}. A prominent example of publicly available MI datasets is the BCI Competition, which has historically provided standardized data from various MI tasks, allowing researchers worldwide to benchmark algorithms and compare results under consistent conditions \cite{tangermann2012review}.

		Based on Table \ref{Table_datasets}, the number of electrodes employed in MI‑BCI systems ranges from 1 to 64 or more. In particular, several studies reduce the number of electrodes originally provided by the acquisition devices \cite{barria2021bci,tang2020motor,mwatamotor}. While the traditional `C{3,4,z}` set located over the central cortical area is often considered to reduce data dimensionality \cite{li2019hybrid, yang2020synchronized}, electrode selection is frequently constrained by the specific device used \cite{jiang2017semiasynchronous, priyatno2022classification}. Some works focus on specific central channels like `C{3,4}` \cite{yang2020synchronized}, `Cz` \cite{ yusoff2018discrimination}, or `C{1,2}` \cite{gant2018eeg}. Researchers also frequently incorporate fronto-central and central-parietal channels, or extend the `C{3,4,z}` group with additional electrodes such as `FC{3,4}` \cite{ dehzangi2013simultaneous}, `Fpz`, and `Pz` \cite{ vourvopoulos2019eeglass}. Other studies adopt specialized configurations, including `Cz` with `CP{1,z,2}` \cite{daeglau2020challenge}, or various subsets of 8, 9, or 10 electrodes to balance dimensionality reduction and cortical coverage \cite{ vourvopoulos2019efficacy, rodriguez2018effects}. Furthermore, several datasets and authors explicitly report the sampling rates used during signal acquisition, which range from 125 Hz \cite{peterson2020feasibility}, 128 Hz \cite{tang2020motor,liu2021brain}, 250 Hz \cite{cardoso2021effect,garcia2020cnn}, 256 Hz \cite{khan2019multiclass}, and 500 Hz \cite{barria2021bci,daeglau2020challenge,lisi2018markov,yusoff2018discrimination,abdalsalam2018modulation} up to 512 Hz \cite{djamal2017brain}.

		Finally, the decoded intention is translated into feedback mechanisms delivered to the user, such as vibration, tactile, thermal, electrical, haptic, auditory, or visual cues, completing the closed-loop system \cite{chowdhury2017online,barria2021bci,du2021neurofeedback}, as illustrated in Fig. \ref{Fig_BCI_Life_Cycle_MI} (f).

		The DSP stage (shown in Fig. \ref{Fig_BCI_Life_Cycle_MI} (d)) is the computational core of the MI-BCI systems and has been the primary focus of research over the past decade. The following section provides a comprehensive review of the signal processing techniques employed at each stage of this pipeline.

		\section{MI-BCI Signal Processing Algorithms}
		\label{sec:MI-BCI_Signal_Processing_Algorithms}
		
		Signal processing plays a central role in transforming raw EEG recordings into discriminative features that can be reliably used for MI-BCI applications. EEG signals are inherently complex, non‑stationary, stochastic, and heavily contaminated by noise and artifacts, which makes the extraction of discriminative patterns a challenging task \cite{gaur2019automatic}. To overcome these limitations, multichannel EEG acquisition is often used to provide richer spatial information from the motor cortex, improving classification performance by offering a more detailed representation of neural dynamics \cite{sakhavi2015parallel}. However, this comes at the cost of increased data dimensionality and computational load, requiring sophisticated algorithms capable of efficiently handling high‑volume, high‑variance inputs. Because MI‑BCI systems must operate with minimal latency, especially in real‑time applications such as robotic control or neuroprosthetic movement, robust and optimized signal processing algorithms are essential to ensure both accuracy and responsiveness throughout the decoding pipeline \cite{she2025improved, shi2025eeg}.

		The processing of EEG MI signals typically involves several key stages, as illustrated in Fig. \ref{Fig_BCI_Life_Cycle_MI}. First, pre-processing is crucial for mitigating noise and artifacts while enhancing EEG signal quality. Typical preprocessing operations include filtering (e.g., band-pass filtering to isolate relevant frequency bands like $\alpha$ and $\beta$) and artifact removal techniques \cite{ hameed2024temporal,kwon2019subject,rao2026calibration,malekmohammadi2019efficient, gao2025effects}. Following pre-processing, feature extraction aims to extract the most informative components from the cleaned EEG data. This stage can also involve techniques for feature reduction to decrease dimensionality and computational load, and feature selection to identify the most discriminative features that most effectively represent the user's MI intent \cite{ modak2026novel, zhao2026mshanet, zhang2021adaptive,ali2023correlation,fang2022feature,bhatti2019soft}. Finally, feature classification employs classical or AI-based algorithms to map the extracted features to specific motor commands, decoding the user's intended movement into actionable control signals.

		\begin{table*}[t]
			\caption{State-of-the-art EEG-based MI-BCI signal processing algorithms reported in the literature in the recent decade.}
			\tiny  
			\label{Table_SW_Algorithms}
			\centering
			
			\begin{tabular}{|>{\centering\arraybackslash}m{0.5cm} |
					>{\centering\arraybackslash}m{0.5cm} |
					>{\centering\arraybackslash}m{1.7cm} |
					>{\centering\arraybackslash}m{1.7cm} |
					>{\centering\arraybackslash}m{1.7cm} |
					>{\centering\arraybackslash}m{1cm} |
					>{\centering\arraybackslash}m{6.4cm} |
				}
				
				\hline
				
				\textbf{Ref.} & 
				\textbf{Year} &
				\textbf{Pre-processing} &
				\textbf{Feature Extraction} &
				\textbf{Feature Classification} &
				\textbf{Dataset Category} &
				\textbf{Performance} 
				\\
				
				\hline

				\cite{modak2026novel} &
				2026 & 
				WPD, ICA, SWS &
				WVG &
				ConvNeXt &
				(4), (9) &
				Dataset (4): CA(90.20\%), Spe(95\%), Pre(87.8\%), K(0.87). Dataset (9): CA(97.73\%), Spe(98.3\%), Pre(98.57\%), K(0.94).
				\\ \hline

				\cite{raza2026maml} &
				2026 & 
				BPF &
				MAML &
				MAML-LNN &
				(4),(6) &
				Dataset (4): 4-Class CA(81.3\%), CTCA(76.3\%), 4-Class K(0.75), CT K(0.69), 4-Class macro F1 (0.8), 4-Class macro F1 (0.8), CT macro F1(0.76). Dataset (6): 4-Class CA(79.6\%), 4-Class K(0.729), 4-Class macro F1 (0.786). Eff: 0.82 M param., 0.17 G MACs, Speed: 38 fps.
				\\ \hline

				\cite{cao2026hybrid} &
				2026 & 
				ICA, FIR BPF &
				TMEF-Net &
				TMEF-Net &
				(8) &
				INACA: 89\%, INECA: 81\%
				\\ \hline

				\cite{zhao2025multi} &
				2025 & 
				ZM-STD &
				MSCFormer &
				MSCFormer, FCL &
				(4),(5) &
				Dataset (4): CA(82.95\%), K(0.7726). Dataset (5): CA(88\%), K(0.7599).
				\\ \hline

				\cite{velakanti2025improving} &
				2025 & 
				VAE &
				DAE &
				CNN-LSTM &
				(8) &
				CA: 94.36\%, Pre: 0.94, Recall: 0.947, F1-Score: 0.94
				\\ \hline

				\cite{sebait2025brain} &
				2025 & 
				BPF, CSP, ICA &
				PSD (WCM), MRCP, ERD &
				SVM, LDA, KNN, RF &
				(8) &
				CA: SVM (80\%), RF (79\%), KNN (78\%), LDA (60\%) (k-fold cross-validation)
				\\ \hline

				\cite{hameed2024temporal} &
				2024 & 
				ICA &
				TST &
				TST-FCL &
				(4),(5) &
				Dataset (4): SDCA TST-ICA(97.77\%), Pre(1), REC(0.94), Spe(1), F1-score(0.97). Dataset (5): SDCA TST-ICA(87.29\%), Pre(0.87), REC(0.84), Spe(0.88), F1-score(0.85).
				\\ \hline

				\cite{li2022motor} &
				2022 & 
				BPF, SWS &
				CNN LSTM &
				CNN LSTM FFN-FCL &
				(4) &
				CA: 94.1\%, K: 0.921
				\\ \hline

				\cite{xie2022transformer} &
				2022 & 
				ZSN &
				CNN-TRF-POE &
				TRF-FCL &
				(6) &
				CA: 83.31\% (2-class), 74.44\% (3-class), 64.22\% (4-class)
				\\ \hline

				\cite{hou2022gcns} &
				2022 & 
				PCM-GL &
				GCN &
				GCN-FCL &		
				(6), (9) &
				Dataset (6): CA(93.06\%), K(84.47\%), Pre(88.39\%), REC(88.35\%), F1-score(88.34\%). Dataset (9): CA(96.24\%).
				\\ \hline

				\cite{kwak2022fganet} &
				2022 & 
				CAR, ICA &
				FGANet &
				FGANet-WSPS &
				(8) &
				CA: 91.96\%. STD: 5.82\%
				\\ \hline

				\cite{fang2022feature} &
				2022 & 
				FBTS, BPF &
				FBRTS &
				OVR-SVM &
				(4),(5) &
				Dataset (4): CA(77.7\%), STD(13.6\%), K(0.71). Dataset (5): CA(86.9\%), STD(6.03\%), K(0.7).
				\\ \hline

				\cite{song2022eeg} &
				2022 & 
				BPF, ZSN &
				EEG Conformer &
				EEG Conformer &
				(4),(5) &
				Dataset (4): CA(78.66\%), K(0.7150). Dataset (5): CA(84.63\%), K(0.6926).
				\\ \hline

				\cite{zhang2021adaptive} &
				2021 & 
				BPF &
				CNN &
				CNN &
				(8) &
				SICA: 84.19\%, STD: 9.98\% 
				\\ \hline

				\cite{zhang2021hybrid} &
				2021 & 
				OVR-FBCSP &
				OVR-FBCSP &
				HDNN-TL &
				(4) &
				HDNN-TL: K=0.81, HDNN: K=0.78. 
				\\ \hline

				\cite{bang2021spatio} &
				2021 & 
				SWS, BPF &
				3D-CNN &
				3D-CNN &
				(4),(5),(7) &
				Dataset (4): CA(87.15\%), STD(7.31\%). Dataset (5): CA(75.85\%), STD(12.80\%). Dataset (7): CA(70.37\%), STD(17.09\%).
				\\ \hline

				\cite{autthasan2021min2net} &
				2021 & 
				BWF BPF &
				MIN2Net MTL &
				MIN2Net MTL &
				(4),(7) &
				Dataset (4): SDCA(65.23\%), SD F1-score(64.72\%), SICA(60.03\%), SI F1-score(49.09\%). Dataset (7): SDCA(61.03\%), SD F1-score(63.59\%), SICA(72.03\%), SI F1-score(72.61\%).
				\\ \hline

				\cite{yu2021new} &
				2021 & 
				MSPCA &
				EFD, IEFD &
				FFNN &
				(4),(5),(3),(11) &
				Dataset (4): SDCA(99.82\%), Dataset (5): SDCA(93.33\%), Dataset (3): SDCA(88.08\%), Dataset (11): SDCA(91.96\%), Dataset (4): SDCA(82.70\%).  
				\\ \hline

				\cite{gaur2021sliding} &
				2021 & 
				NFL BPF &
				CSP, CSP, SW-LCR, SW-MODE &
				LDA &
				(4) &
				SW-LCR: CA(86.19\%), STD(12.02\%), K(0.72). SW-MODE: CA(86.03\%), STD(12.04\%), K(0.72).
				\\ \hline

				\cite{deng2021advanced} &
				2021 & 
				NFL BPF &
				TCSGL &
				TCSGL-FCL &
				(1), (4) &
				Dataset (1): CA(88.89\%), K(0.7511). Dataset (4): CA(81.34\%), K(0.8519).
				\\ \hline

				\cite{dai2020hs} &
				2020 & 
				BPF &
				HS-CNN &
				HS-CNN &
				(4),(5) &
				Dataset (4): CA(91.57\%). Dataset (5): CA(87.6\%).
				\\ \hline

				\cite{kant2020cwt} &
				2020 & 
				BPF, WPT &
				CNN-CWT &
				CNN-CWT-TL &
				(10) &
				CA: 95.71\%, K: 0.91
				\\ \hline

				\cite{zhang2020data} &
				2020 & 
				BWF BPF , STFT &
				CNN-DCGAN &
				CNN-DCGAN &
				(4),(5) &
				Dataset (4): CA(83.2\%), STD(3.5\%), K(0.564). Dataset (5): CA(93.2\%), STD(2.8\%), K(0.677).
				\\ \hline

				\cite{zhao2020deep} &
				2020 & 
				BWF BPF &
				DRDA &
				DRDA &
				(4),(5) &
				Dataset (4): CA(74.75\%), K(0.6634). Dataset (5): CA(83.98\%),K(0.6796).
				\\ \hline

				\cite{venkatachalam2020novel} &
				2020 & 
				BPF  &
				PCA, FLDA &
				H-KELM &
				(1) &
				CA: 96.54\%
				\\ \hline

				\cite{he2019transfer} &
				2019 & 
				ESDA, FIR HMW BPF &
				CSP &
				LDA &
				(4),(5) &
				Dataset (4): CA(79.79\%). Dataset (5): CA(67.75\%).
				\\ \hline

				\cite{kwon2019subject} &
				2019 & 
				BPF, CSP, SSIG &
				CSP, MIN, CNN &
				CNN &
				(8) &
				SDCA: 71.32\%. SICA: 74.15\%.
				\\ \hline

				\cite{li2019channel} &
				2019 & 
				- &
				CP-MixedNet &
				CP-MixedNet &
				(4), (9) &
				Dataset (4): CA(74.6\%). Dataset (9): CA(93.7\%). 
				\\ \hline

				\cite{ji2019eeg} &
				2019 & 
				NFL BPF &
				DWT-EMD-SampEn &
				SVM &
				(5) &
				CA: 95.1\%
				\\ \hline

				\cite{xu2019deep} &
				2019 & 
				STFT &
				STFT &
				VGG-16-TL &
				(5) &
				CA: 74.2\%
				\\ \hline

				

				\cite{dose2018end} &
				2018 & 
				- &
				CNN &
				CNN-TL &
				(6) &
				CA: CNN(80.38\%), CNN+TL(86.49\%)
				\\ \hline

				\cite{zhang2018temporally} &
				2018 & 
				BPF, SWS &
				CSP, TSGSP &
				SVM &
				(1),(4),(5) &
				Dataset (1): CA(88.5\%), STD(12\%). Dataset (4): CA(84\%), STD(7.5\%). Dataset (5): CA(84.3\%), STD(15\%).
				\\ \hline

				\cite{luo2018exploring} &
				2018 & 
				SWCS &
				FBCSP &
				RNN-GRU-LSTM &
				(4) &
				CA: 73.56\%
				\\ \hline

				\cite{donovan2018motor} &
				2018 & 
				NFL BPF, CCA &
				DWT &
				TSK FNN &
				(6) &
				CA: 79.4\%
				\\ \hline

				\cite{sakhavi2018learning} &
				2018 & 
				BPF &
				FBCSP &
				CNN &
				(4) &
				CA:74.46\%
				\\ \hline

				\cite{xu2018wavelet} &
				2018 & 
				BPF, LAP, CWT, CAR &
				2-Layer CNN &
				2-Layer CNN &
				(4) &
				CA: 85.59\%, F1-score: 0.85, k: 0.766
				\\ \hline

				\cite{jafarifarmand2017new} &
				2017 & 
				BWF BPF &
				AR-CSP &
				SRSG-FasArt &
				(4) &
				CA: 85.49\%, STD: 11.64\%, K:0.63
				\\ \hline

				\cite{tang2017single} &
				2017 & 
				BPF &
				FFT, 5-layer CNN &
				5-layer CNN &
				(8) &
				CA: 86.41\%, STD: 0.77\%
				\\ \hline

				\cite{schirrmeister2017deep} &
				2017 & 
				BPF &
				FBCSP &
				CNN &
				(4) &
				CA: 84\%
				\\ \hline

				\cite{kevric2017comparison} &
				2017 & 
				BPF, MSPCA  &
				WPD &
				KNN &
				(2) &
				CA: 92.8\%
				\\ \hline

				\cite{tabar2017novel} &
				2017 & 
				BPF &
				CNN &
				CNN-SAE &
				(5) &
				K: 0.547
				\\ \hline

				\cite{zhang2017classification} &
				2017 & 
				BPF &
				ARM, WPD, SampEn, PCA &
				SVM &
				(8) &
				CA: 98\%
				\\ \hline

				\cite{lu2016deep} &
				2016 & 
				BWF BPF, SWS &
				FFT, WPD &
				FDBN &
				(5) &
				CA: 84\%
				\\ \hline

				\cite{wang2016detection} &
				2016 & 
				BPF &
				OAL &
				NVB &
				(5) &
				CA: 0.64\%
				\\ \hline


				\cite{zhang2015optimizing} &
				2015 & 
				BPF &
				SFBCSP &
				SVM &
				(2) &
				CA: 91.05\%, STD: 2.45\%
				\\ \hline

				\cite{yang2015use} &
				2015 & 
				BWF BPF &
				ACSP &
				CNN &
				(4) &
				CA: 69.27\%
				\\ \hline

				\cite{yu2014analysis} &
				2014 & 
				CAR, LAP, CSP &
				PCA &
				SVM &
				(4) &
				CA: 76.34\%
				\\ \hline


				\cite{ren2014convolutional} &
				2014 & 
				FFT, PCA &
				CDBN &
				CDBN &
				(1),(2),(10) &
				CA: 88.25\%, STD:5.70\%
				\\ \hline
				
			\end{tabular}
		\end{table*}

		A structured overview of impactful EEG‑based MI-BCI papers over the past decade, as well as new papers with novel ideas, is tabulated in Table \ref{Table_SW_Algorithms}. For each paper, the publication year, pre‑processing technique, feature extraction method (including any feature reduction or selection strategies), feature classification algorithm, dataset category (defined in Table \ref{Table_datasets}), and obtained performance metrics are reported. The commonly reported performance metrics in this table are the best or average obtained classification accuracy (CA), overall cross-subject classification accuracy (CSCA), overall cross-task classification accuracy (CTCA), intra-subject classification accuracy (INACA), inter-subject classification accuracy (INECA), subject-dependent classification accuracy (SDCA), subject-independent classification accuracy (SICA), F1-score, sensitivity (Sen), specificity (Spe), precision (Pre), Cohen’s kappa coefficient (K), efficiency (Eff), recall (REC), standard deviation (STD), mean difference (MEAD), and p-value. In the next section, first, a brief review of AI-based signal processing is presented, and then the most common algorithms in Table \ref{Table_SW_Algorithms} are discussed.

		\subsection{Introduction to AI-Based EEG Signal Processing}
		
\begin{figure}[!t]
			\centerline{\includegraphics[scale=0.4,trim=9.4cm 4.8cm 7cm 3.1cm,clip=true]{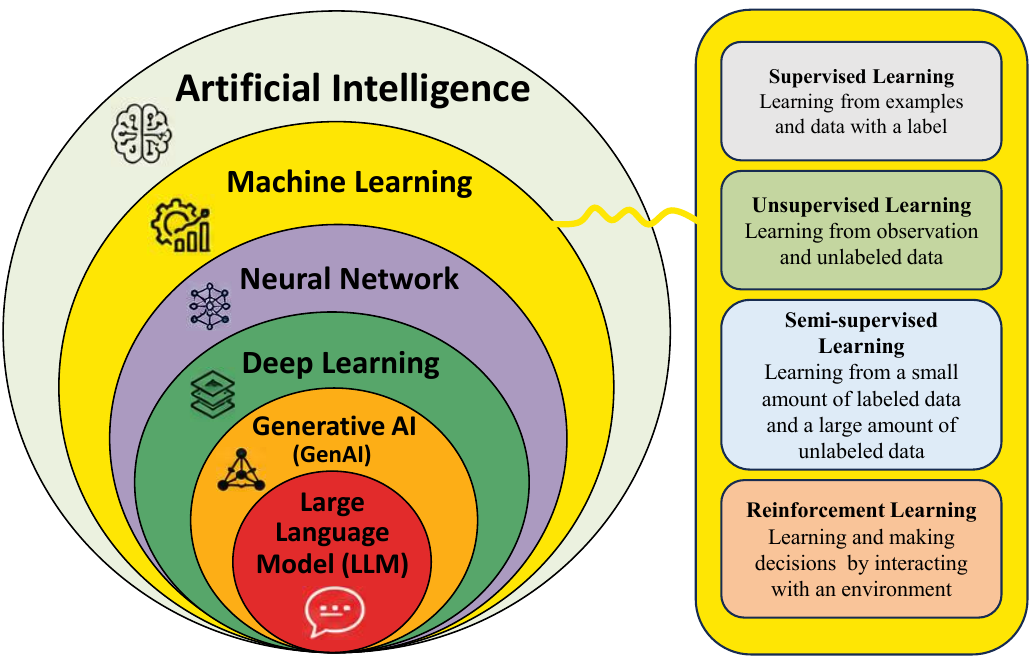}}
			\caption{The relation between various categories and sub-domains of artificial intelligence (AI).
			}
			\label{Fig_AI_Algorithms_Domains}
		\end{figure}

		Modern AI-based algorithms have dramatically enhanced MI signal processing by automatically extracting complex spatio-temporal features and improving CA and generalization. AI is the overarching field of creating intelligent systems; machine learning (ML), as illustrated in Fig \ref{Fig_AI_Algorithms_Domains}, is a subset of AI where systems learn from data without explicit programming; neural networks (NNs) are a type of ML model inspired by the human brain; deep learning (DL) uses NNs with many layers (deep architectures) for more complex feature learning \cite{cho2021neurograsp,dai2019eeg,el2024strong}; transformer models are a specific DL architecture excelling at sequential data processing, particularly useful for long-range dependencies in EEG \cite{cao2026hybrid,zhang2019convolutional}; generative AI is a type of AI focused on creating new content; and large language models (LLMs) are a type of generative AI specialized in understanding and generating human-like text. All of these algorithmic families can be tailored directly or indirectly to learn discriminative patterns in MI EEG signals for robust feature extraction and accurate motor intention decoding \cite{wang2026generative, zhao2026dg}.

		The ML algorithms are broadly classified into supervised, unsupervised, semi-supervised, and reinforcement learning, as depicted in Fig. \ref{Fig_AI_Algorithms_Domains}. Supervised learning algorithms learn from labeled data to make predictions, widely used in MI signal processing for classifying MI tasks based on pre-labeled EEG signals. Unsupervised learning algorithms discover patterns in unlabeled data, applicable to MI for clustering different EEG states or identifying artifacts. Semi-supervised learning combines labeled and unlabeled data, offering a way to improve MI models when labeled data is scarce. Reinforcement learning involves agents learning through trial and error by interacting with an environment, which can be used in MI-BCI systems for adaptive control and optimizing BCI performance over time \cite{mathiyazhagan2025motor, gupta2023machine, hosseinifard2013classifying}.

		Besides traditional signal processing algorithms, which are still being used in EEG because of their simplicity, AI-driven approaches have significantly advanced the performance and robustness of EEG-based BCI systems. Classic ML algorithms, such as support vector machines and linear Discriminant analysis, remain highly relevant and effective for MI signal processing \cite{sebait2025brain,gaur2019automatic,yu2014analysis,mccrimmon2017performance}. These algorithms excel at finding discriminative patterns within extracted EEG features, enabling accurate classification of MI intentions, particularly when combined with robust feature engineering techniques.

\begin{figure}[!t]
			\centerline{\includegraphics[scale=0.34,trim=3cm 5.8cm 4.5cm 4.5cm,clip=true]{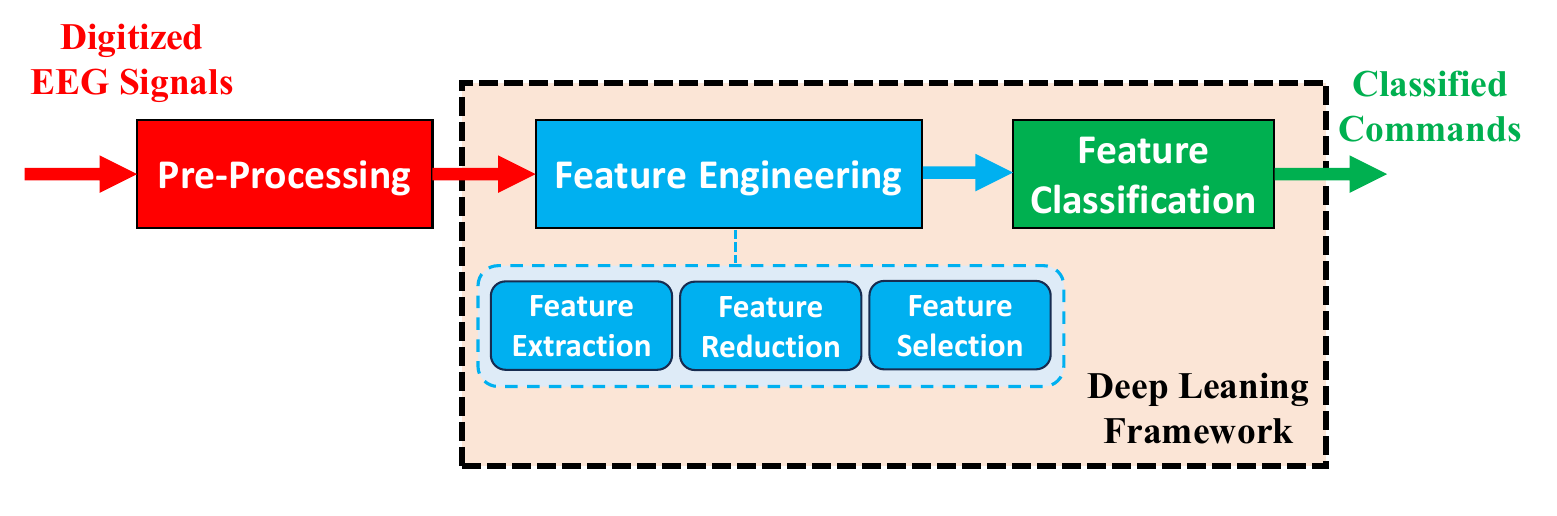}}
			\caption{Comparison of traditional processing of EEG-based MI signals and DL-based architectures.
			}
			\label{Fig_Conventional_DL_Processing}
		\end{figure}

		Specifically, as depicted in Fig. \ref{Fig_Conventional_DL_Processing}, DL-based architectures such as convolutional neural networks (CNNs) can simultaneously perform feature extraction and feature classification tasks, as opposed to traditional algorithms utilized for MI tasks \cite{li2022motor,bang2021spatio, rakhmatulin2024exploring}. Additionally, recurrent neural networks, transformers, and hybrid (HYB) models automatically learn hierarchical EEG features, reducing the need for manual preprocessing and excelling at capturing complex temporal patterns, inter-channel relationships, and user-specific variations, leading to higher CA and better generalization \cite{cao2026hybrid,xie2022transformer,sakhavi2018learning}. Also, techniques such as positional embedding (POE) incorporate temporal position information into transformer inputs, enabling attention mechanisms to preserve the sequential order of EEG signals \cite{xie2022transformer}. Moreover, while specialized models like EEGNet, ShallowConvNet, and DeepConvNet are tailored for efficiency and specific tasks like MI, newer models like EEG-TCNet, SincNet, EEG Conformer, CLTNet, and DB-BISAN integrate temporal convolutional networks, trainable sinc-filters, and self-attention mechanisms to handle temporal dynamics and long-range dependencies \cite{zhang2019convolutional,lawhern2018eegnet}. Furthermore, the MSCFormer algorithm combines multi-scale CNN branches with a transformer encoder to jointly learn local spatial patterns and long-range temporal dependencies for end-to-end MI decoding \cite{zhao2025multi}. These innovations collectively enable robust, low-latency, and power-efficient neural decoding necessary for real-time MI-BCI applications. In the following sections, a detailed discussion is provided on the most common algorithms, as listed in Table \ref{Table_SW_Algorithms}, for each signal processing block.

		\subsection{EEG Pre-Processing Algorithms}

		\begin{figure*}[!t]
			\centerline{\includegraphics[scale=0.5,trim=0.3cm 0.1cm 0.3cm 0.1cm,clip=true]{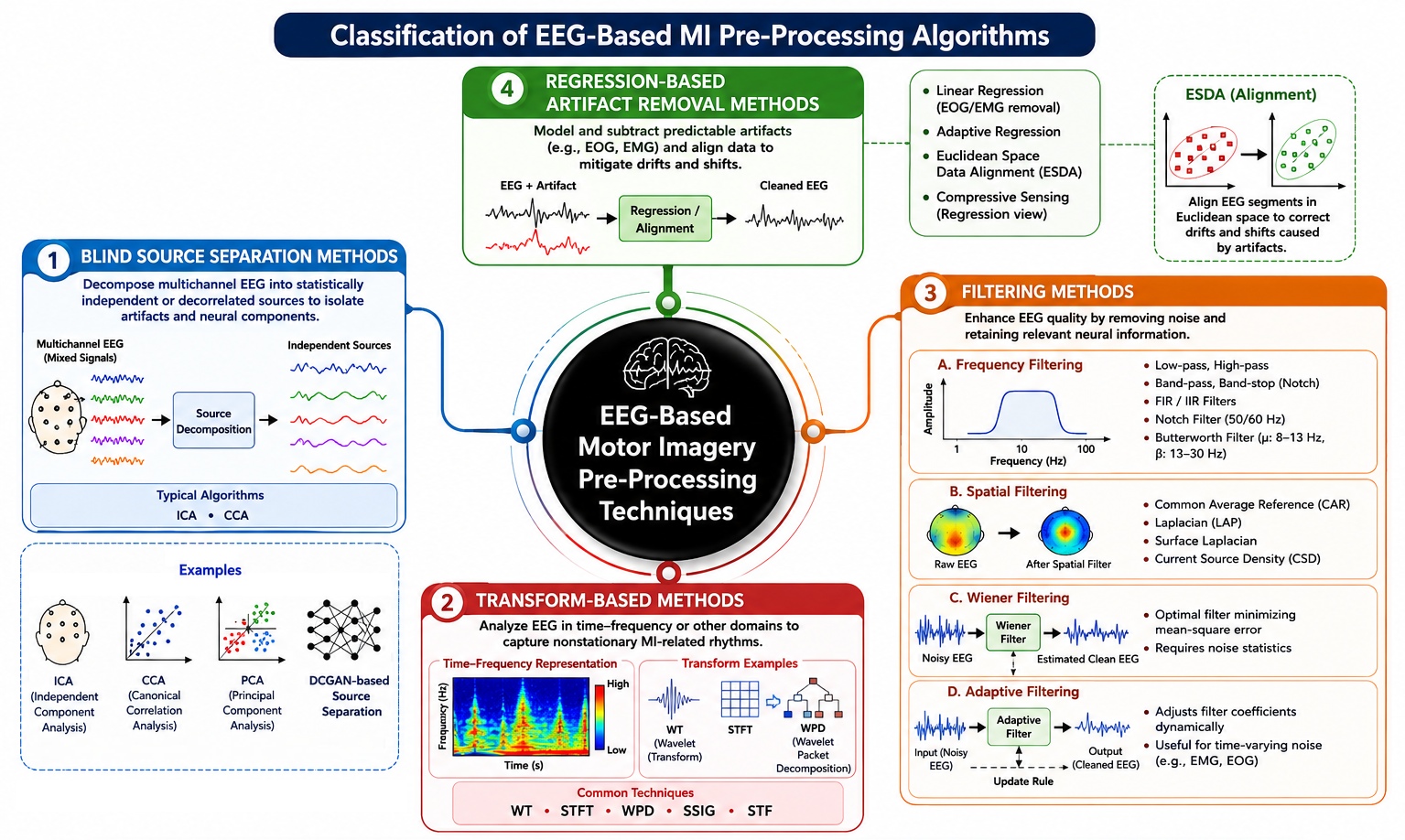}}
			\caption{Taxonomy of EEG pre-processing algorithms for MI-based BCI systems. EEG pre-processing techniques are broadly classified into blind source separation, transform-based, filtering, and regression-based artifact removal approaches. Typical algorithms and their corresponding signal processing mechanisms are illustrated, demonstrating how these methods enhance signal quality, reduce artifacts and noise, and facilitate reliable feature extraction and classification in MI-BCI applications.
			}
			\label{Fig_Preprocessing_Algorithms}
		\end{figure*}

		Signal pre-processing is a critical first step in the analysis of EEG signals, particularly for MI tasks. Its primary goal is to clean the raw EEG data by removing artifacts that can obscure the neural signals of interest. These artifacts can arise from various sources, including eye blinks, muscle movements (EMG), electrical noise from the environment, and impedance issues \cite{cho2021neurograsp}. Effective pre-processing ensures that the subsequent feature extraction and classification steps operate on cleaner, more representative neural data, thereby improving the accuracy and robustness of motor intention decoding \cite{gao2025effects}. The most common EEG signal pre-processing algorithms for MI systems can be categorized into four general groups, as shown in Fig. \ref{Fig_Preprocessing_Algorithms}.

		\textbf{1.	Blind Source Separation Methods:} The Blind source separation (BSS) methods operate by decomposing multichannel EEG into statistically independent or decorrelated source signals, allowing artifact and neural components to be isolated without prior knowledge of their origins. The independent component analysis (ICA) is the most widely used technique of this class and decomposes EEG into statistically independent components to identify and remove artifacts (e.g., eye blinks, muscle activity, cardiac noise) \cite{sebait2025brain,hameed2024temporal}. Moreover, a deep convolutional generative adversarial network (DCGAN) learns latent source representations, and as a generative model, it can separate or reconstruct independent EEG components from mixed signals \cite{you2020unsupervised,zhang2020data}. Other techniques in this group are principal component analysis (PCA), canonical correlation analysis (CCA), and some other spatial filtering methods \cite{foong2019assessment,mccrimmon2017performance}. The adaptive sparse representation (ASR) has also been employed for artifact suppression by representing EEG signals using sparse bases while preserving neural components \cite{shin2015simple}.

		\textbf{2.	Transform‑based Methods:} The transform-based methods analyze EEG signals in joint time-frequency domains. The wavelet transform (WT) is one of the most commonly used algorithms in this group, providing a flexible, multiresolution decomposition that captures transient MI-related oscillatory patterns more effectively than fixed-window transforms and is especially useful because MI-related rhythms ($\mu$ and $\beta$) are nonstationary \cite{song2025awknet, siino2025investigating}. Also, wavelet packet transform (WPT) decomposes EEG signals into multiple frequency sub-bands, such as the $\mu$ and $\beta$ rhythms, providing finer frequency resolution than conventional wavelet decomposition \cite{kant2020cwt}. The spectral-spatial input generation (SSIG) method further enhances this process by transforming raw EEG into combined spectral and spatial representations, allowing classifiers to exploit both frequency-specific activity and the spatial distribution of cortical rhythms \cite{kwon2019subject}. Furthermore, spatio-temporal formatting (STF) extends this idea by restructuring EEG into unified spatio-temporal feature maps, preserving channel relationships and temporal dynamics simultaneously, which improves compatibility with DL models that require structured, image-like inputs \cite{tang2017single}.The empirical Fourier decomposition (EFD) and its improved variant, IEFD, have recently been introduced to adaptively decompose nonstationary EEG signals into intrinsic Fourier components for improved time-frequency analysis \cite{yu2021new}. The time-frequency distribution (TFD) techniques are also widely used to characterize the temporal evolution of MI-related spectral activity \cite{al2014methods}. Finally, multilayer weighted visibility graph (WVG) converts EEG time series into weighted graph representations, preserving temporal relationships and structural characteristics that facilitate robust feature extraction from non-stationary signals \cite{modak2026novel}.

		\textbf{3.	Filtering Methods:} This is a major class and includes several subclasses:
		
		\textbf{A) Frequency Filtering:} Different types of filters can be included in this category, such as low-pass (LPF), high-pass (HPF), band-pass (BPF), and band-stop filters (BSF), such as FIR and IIR \cite{huang2026continual,olivas2019classification}. Specifically, notch filtering (NFL) is a band-stop filter designed to suppress a specific narrow frequency band, typically 50 Hz or 60 Hz power-line noise and is widely used for EEG signals \cite{rao2026calibration,olivas2019classification,li2019hybrid,malekmohammadi2019efficient,mccrimmon2017performance,leeb2015towards}. Moreover, Butterworth filtering (BWF) is a smooth band-pass/band-stop filter commonly used for BPF for MI signals to isolate $\mu$ (8--13 Hz) and $\beta$ (13--30 Hz) rhythms \cite{sebait2025brain,rao2026calibration,huang2026continual,kim2025efficacy,feng2022efficient,mahmood2021wireless,marcos2020real,wang2020accurate,schneider2020q,liu2017feature}. Sliding-window strategies, including SW-LCR and SW-Mode, are also employed to improve temporal localization and signal segmentation before feature extraction \cite{gaur2021sliding}.

		The zero-mean standardization (ZM-STD) normalizes EEG signals by centering each channel around zero mean, reducing amplitude variability and improving the robustness of subsequent learning algorithms \cite{zhao2025multi}. The Z-score normalization (ZSN) standardizes EEG features to zero mean and unit variance, reducing scale differences across channels and improving model convergence \cite{xie2022transformer}. Additionally, sliding window segmentation (SWS) \cite{modak2026novel,zhang2018temporally}, weighted overlap‑add (WOLA), and sliding window cropping strategy (SWCS) techniques are also frequently employed to improve temporal and spectral resolution in EEG preprocessing \cite{luo2018exploring}. Specifically, SWS divides continuous EEG signals into short, partially overlapping windows, ensuring local stationarity and reducing spectral smearing during subsequent analysis \cite{wang2020accurate, gaur2021sliding}. The Hamming window (HMW) is also commonly applied during segmentation, smoothing window edges to reduce spectral leakage; for example, a 3.5 s EEG segment processed with a 50\% overlapping Hamming window yields six 1 s frames for downstream analysis \cite{ma2019fpga}. The WOLA algorithm builds on similar principles, applying weighted windowing and overlap reconstruction to achieve smooth transitions between segments while preserving amplitude continuity \cite{belwafi2018embedded}. Also, the SWCS method, in turn, refines this process by cropping and aligning windowed segments to focus on the most relevant time intervals related to MI, which helps suppress noise and enhance rhythmic components in the $\mu$ and $\beta$ bands \cite{saideepthi2023sliding, zhao2023eeg}.

		\textbf{B) Spatial Filtering:} The spatial filtering methods enhance EEG signal quality by combining information across neighboring electrodes to suppress global noise and highlight localized cortical activity relevant to MI. For example, the common average reference (CAR) is a widely used spatial filtering approach that re-references each electrode to the average of all electrodes, reducing global noise and improving signal localization \cite{malekmohammadi2019efficient}. Additionally, Laplacian (LAP) is typically utilized to enhance spatial resolution by emphasizing local potentials relative to their neighbors, effectively acting like a high‑pass spatial filter \cite{yu2014analysis,rao2026calibration,kim2025efficacy,paszkiel2020brain,foong2019assessment,malekmohammadi2019efficient,leeb2015towards}. The artifact-rejected common spatial pattern (AR-CSP) extends conventional CSP by incorporating artifact rejection during spatial filtering, leading to more robust MI feature extraction \cite{jafarifarmand2017new}.

		\textbf{C) Wiener Filtering:} This technique offers an optimal approach to signal processing by minimizing the mean‑square error between the estimated signal and the desired output under known noise characteristics, making it particularly effective for enhancing the signal-to-noise ratio in EEG recordings during MI tasks, where neural signals are often obscured by significant environmental and physiological artifacts \cite{liu2026novel}.

		\textbf{D) Adaptive Filtering:} This group of filtering techniques offers a dynamic advantage by adjusting filter parameters in real-time, making them highly useful for removing time‑varying noise that is common in EEG signals during MI tasks, thereby improving the quality and reliability of the brain-computer interface \cite{wang2025smanet}.

		\textbf{4.	Regression‑based Artifact Removal Methods:} This class includes techniques that model and subtract predictable artifacts (e.g., EOG, EMG) using linear regression or adaptive regression approaches \cite{li2019hybrid,glavas2024empowering}. Among these, the Euclidean space data alignment (ESDA) methods are particularly more common by aligning EEG data segments in Euclidean space, essentially correcting for drifts and shifts that are often caused by artifacts \cite{he2019transfer}. This is formalized through the Euclidean space (EUS) representation, in which EEG covariance matrices or trial segments are aligned within a common Euclidean reference frame to reduce inter-subject and inter-session variability prior to alignment-based correction \cite{paszkiel2020brain}. This alignment implicitly models and subtracts these predictable artifactual components, helping to preserve the underlying neural activity. Furthermore, while compressive sensing (CPS) is more commonly associated with blind source separation due to its sparsity-based unmixing, it can be viewed through a regression lens when considering the reconstruction of artifact-free signals from undersampled data, effectively regressing out the artifactual components \cite{shrivastwa2018fpga}.

		\subsection{EEG Feature Engineering Algorithms}

		\begin{figure*}[!t]
			\centerline{\includegraphics[scale=0.5,trim=1.3cm 2.8cm 1.8cm 2cm,clip=true]{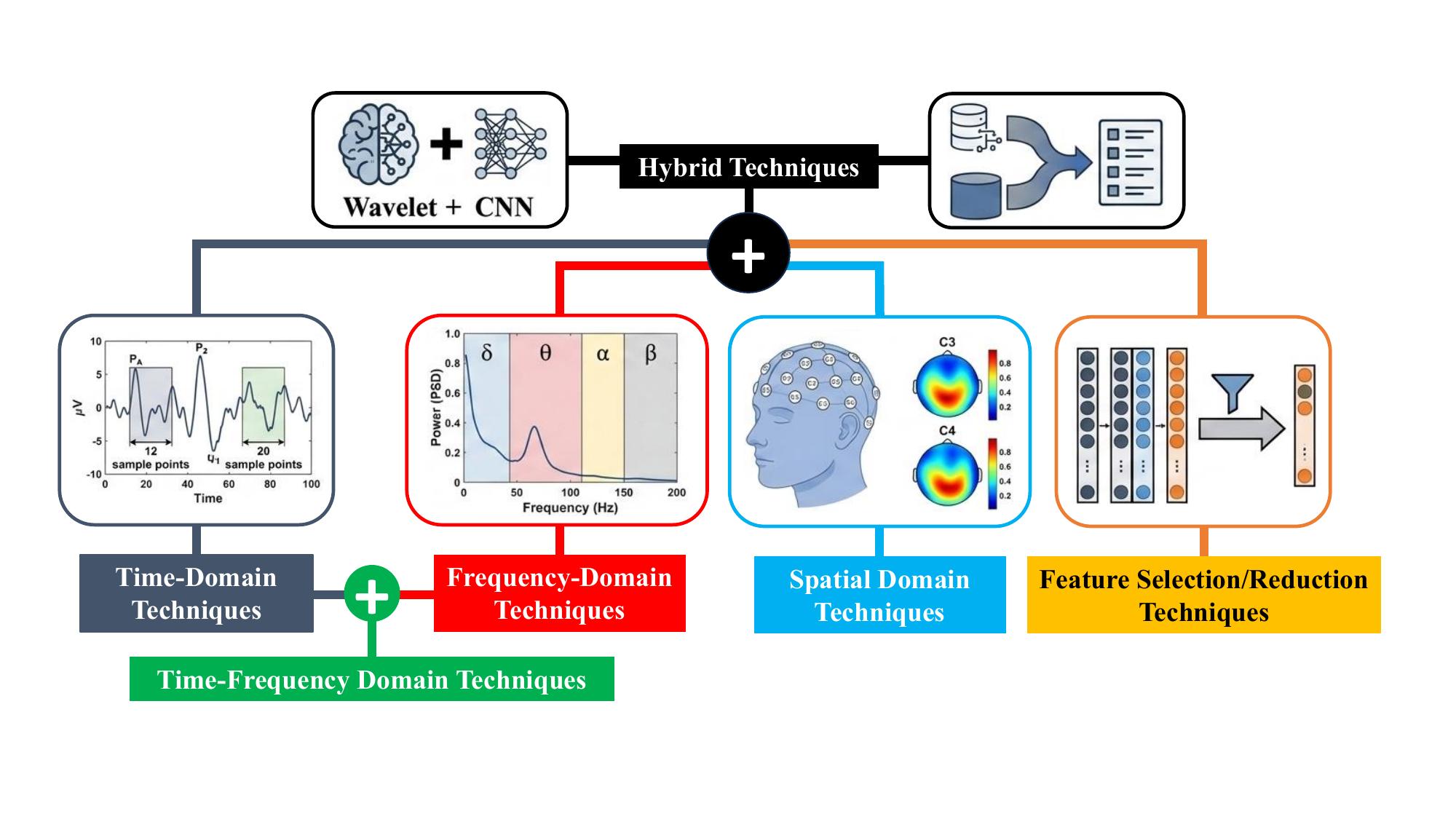}}
			\caption{EEG signals Feature Extraction algorithms classifications based on the mathematical domain of the signal representation. The classification spans five foundational domains and a combined approach: Time-Domain Techniques, Frequency-Domain Techniques, Time-Frequency Domain Techniques, Spatial Domain Techniques, Feature Selection/Reduction Techniques, and HYB Techniques.
			}
			\label{Fig_Feature_Extraction_Classes}
		\end{figure*}

		In the Feature Engineering section of an EEG-based MI processing pipeline, as depicted in Fig. \ref{Fig_Conventional_DL_Processing}, feature extraction is the initial step where raw EEG data is transformed into a set of relevant features that capture the underlying characteristics of the MI. Techniques here aim to derive meaningful information from the signals \cite{yang2018deep,yang2015use,zhang2021adaptive}. Extraction may produce a very large set of features. Feature reduction techniques are then applied to decrease the dimensionality of this feature set while retaining as much of the important information as possible. This helps in combating the curse of dimensionality and improving the efficiency of subsequent steps. Finally, feature selection techniques are used to choose a subset of the most relevant features from the reduced set (or directly from the extracted features if reduction wasn't performed or was minimal). This step focuses on identifying the features that are most discriminative for the MI task, further improving model performance and reducing overfitting \cite{hu2019eeg, rakhmatulin2024exploring}.

		The most widespread feature extraction algorithms can be broadly categorized into five groups based on the mathematical transformation applied to the raw EEG data to extract patterns, as illustrated in Fig. \ref{Fig_Feature_Extraction_Classes}. EEG feature-extraction algorithms can be classified based on the mathematical domain of the signal representation into five foundational domains and a combined approach. The time-domain techniques analyze raw data sample points over time ($P_A$, $P_2$, $Q_1$) \cite{zhang2017classification}. The frequency-domain techniques perform spectral analysis (PSD) to isolate distinct brain rhythms ($\delta$, $\theta$, $\alpha$, $\beta$) \cite{liu2017feature,mahmood2021wireless, ali2023correlation}. The time-frequency domain techniques capture joint temporal and spectral dynamics \cite{zhang2020data}. The spatial domain techniques map source separation and localized cortical activity (e.g., C3, C4 topomaps) \cite{gaur2021sliding}. The feature selection/reduction techniques optimize the feature vector via dimensionality reduction \cite{bhatti2019soft}. Moreover, the HYB techniques leverage multi-domain integration and data-driven DL architectures (e.g., Wavelet + CNN) for enhanced classification.

		\textbf{1) Time-Domain Techniques:} Time-domain techniques extract features directly from the raw EEG signal's voltage fluctuations over time, capturing characteristics like amplitude, variance, and shape, and encompass a range of techniques that analyze the signal's amplitude and statistical properties directly over time.

		Simpler methods like root-mean-square (RMS) and integrated EEG (IEEG) quantify signal power and cumulative amplitude, respectively. More complex approaches involve modeling the signal's temporal dependencies: autoregressive models (ARM) capture this by predicting future signal values based on past observations, while adaptive autoregressive (AAR) models adjust these predictions over time \cite{zhang2017classification,al2014methods}. Adaptive filtering algorithms such as recursive least-squares (RLS), least mean squares (LMS), and the Kalman filter (KF) are also employed to estimate or track signal characteristics in the time domain, often for noise reduction or state estimation. The quaternion-based signal analysis (QBSA) method offers an advanced mathematical framework for analyzing multidimensional EEG data within the time domain, leveraging quaternion algebra to capture complex signal dynamics \cite{zhao2025novel, song2025awknet}.

		In addition, statistical features such as standard deviation, skewness, kurtosis, entropy, and energy provide simple yet effective time‑domain descriptors of EEG signals. These measures quantify the variability, asymmetry, peakedness, complexity, and power of the waveform, allowing the extraction of discriminative patterns related to MI activity. Furthermore, ERD/ERS-based algorithms are crucial time-domain techniques. ERD/ERS analysis quantifies changes in the power of specific frequency bands, typically the alpha and beta bands, in the EEG signal that are time-locked to a cognitive or motor event. ERD refers to a decrease in the power of brain oscillations, often observed during motor imagery execution, while ERS indicates an increase in power, which might occur after the task or in specific conditions. By analyzing these power modulations over time, ERD/ERS provides a powerful way to capture the neural dynamics underlying motor imagery \cite{sebait2025brain,james2025hybrid,angulo2025proficiency}.

		\begin{figure}[!t]
			\centerline{\includegraphics[scale=0.15,trim=0cm 0cm 0cm 0cm,clip=true]{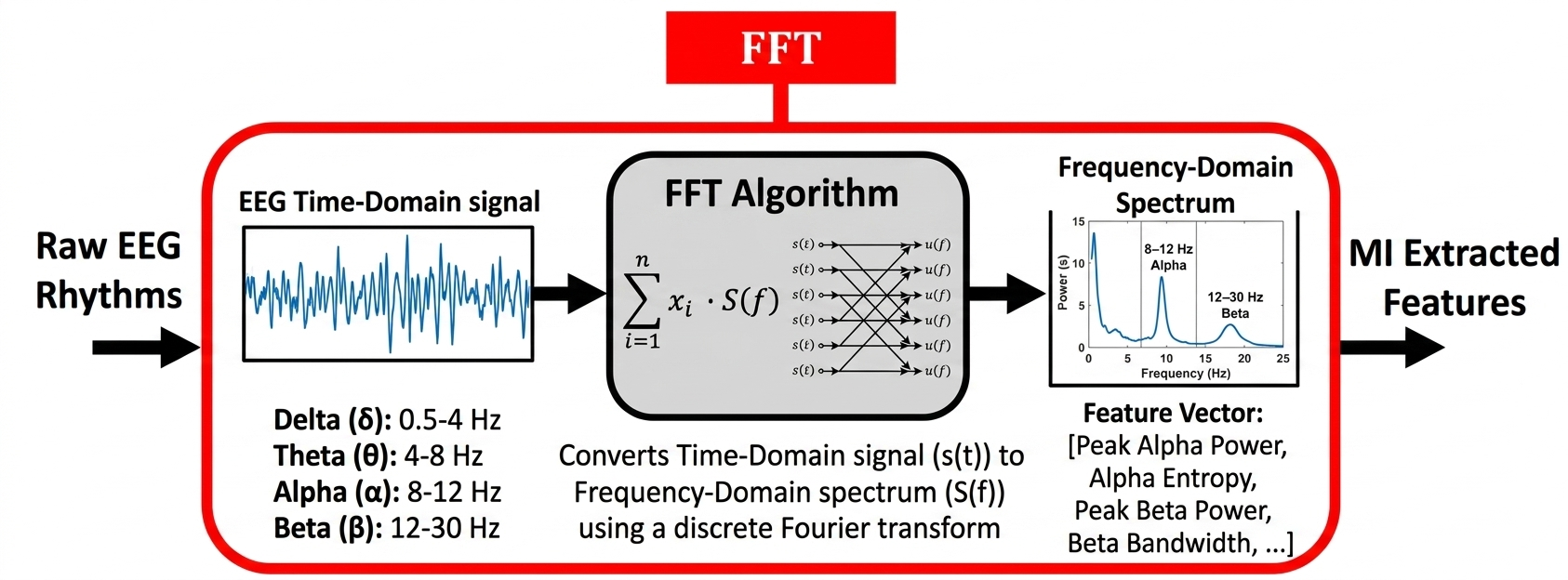}}
			\caption{Applying FFT conversion to EEG MI signals for extracting the delicate frequency features of MI oscillations. The FFT decomposes the raw, time-domain EEG rhythms into a frequency-domain spectrum, isolating power changes in target bands (e.g., alpha and beta rhythms) to construct a discriminative feature vector for classification.
			}
			\label{Fig_FFT}
		\end{figure}

		\textbf{2) Frequency-Domain Techniques:} Frequency-domain techniques extract features from EEG MI signals by analyzing how signal power is distributed across different frequency bands, rather than in the raw time waveform. Methods such as the fast Fourier transform (FFT) and Welch’s method (WCM) for power spectral estimation highlight characteristic oscillatory patterns, particularly in the $\mu$ and $\beta$ rhythms, that change during imagined movements, making them effective for identifying MI‑related neural activity \cite{lu2016deep,al2014methods,li2019hybrid}. Specifically, FFT efficiently converts a time-domain EEG signal into its constituent frequency components, as illustrated in Fig. \ref{Fig_FFT}. By analyzing the power spectral density across various frequencies, the FFT reveals the strength of oscillations in bands like alpha, beta, and gamma, which are critical for distinguishing between different MI states as these patterns shift with imagined movements \cite{paszkiel2020brain}. Specifically, Welch's method power spectral density (WCM PSD) estimation averages periodograms over overlapping windowed segments, reducing variance in the spectral estimate and yielding more stable power-band features for MI discrimination \cite{leeb2015towards}. Also, local characteristic‑scale decomposition (LCD) decomposes EEG signals into adaptive intrinsic scale components, enabling the extraction of frequency‑specific features useful for distinguishing MI patterns \cite{liu2017feature}. Additionally, mutual information (MIN) measures the statistical dependency between frequency-band features and class labels, helping select the most informative spectral components for MI discrimination \cite{kumar2017improved,foong2019assessment,malekmohammadi2019efficient}. The partial directed coherence (PDC) method further quantifies directional connectivity by estimating the direction and strength of frequency-domain interactions between EEG channels, offering insight into information flow across cortical regions during MI \cite{kim2025efficacy}.

		\textbf{3) Time-Frequency Domain Techniques:} Time–frequency domain techniques extract EEG features by capturing how signal frequency content evolves over time, making them well-suited for the non‑stationary nature of MI signals. Methods like the short‑time Fourier transform (STFT) provide a fixed time–frequency resolution by sliding a window across the signal, while wavelet‑based approaches offer more flexibility through adaptive windowing \cite{zhang2020data,xu2019deep,herman2016designing}. 
		
\begin{figure}[!t]
			\centerline{\includegraphics[scale=0.2,trim=0cm 0cm 0cm 0cm,clip=true]{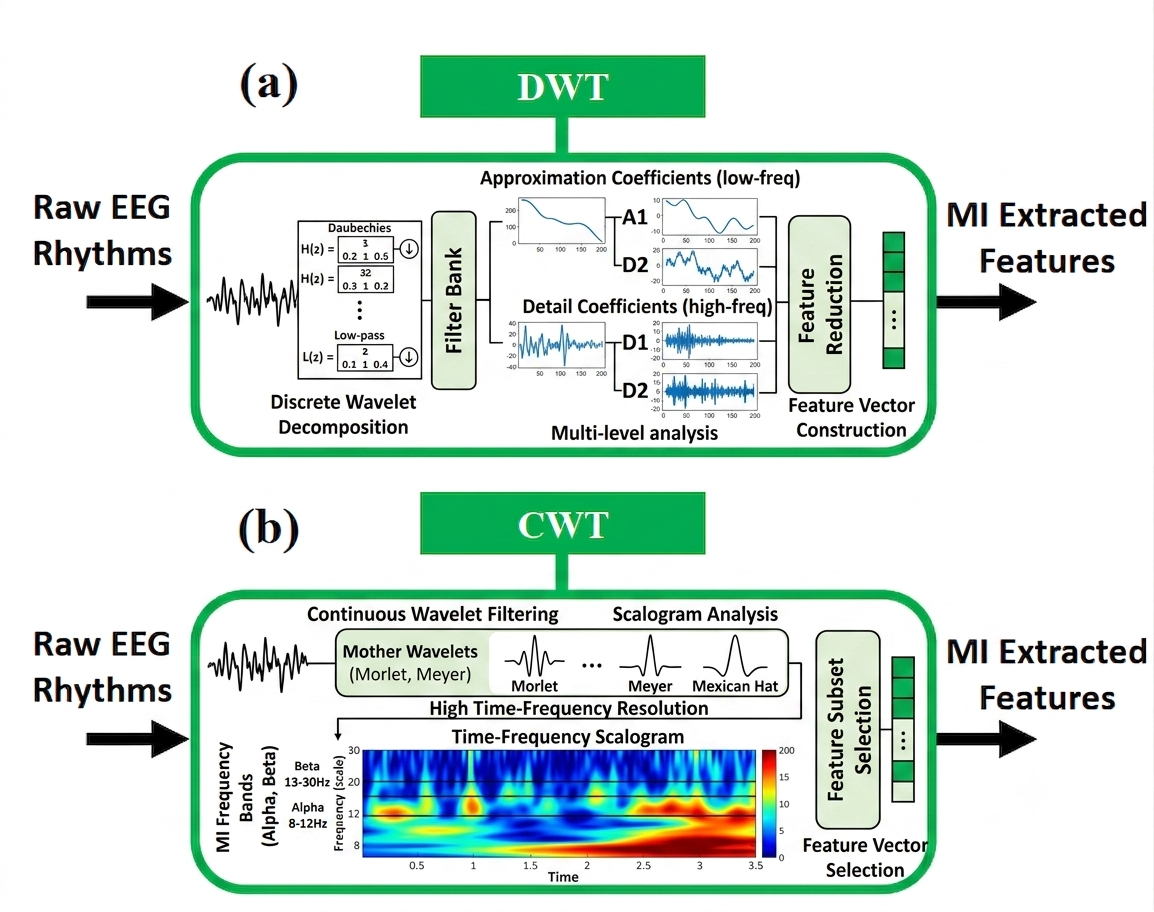}}
			\caption{EEG wavelet transform-based feature extraction methods. Applying DWT (a) and CWT (b) to MI signals. 
			}
			\label{Fig_Wavelet_Transform}
		\end{figure}

		The WT-based methods, especially the discrete wavelet transform (DWT) and continuous wavelet transform (CWT), are widely used in MI analysis because they decompose EEG into multiscale representations that localize transient oscillatory patterns linked to MI‑related $\mu$ and $\beta$ rhythms \cite{deng2021advanced,kant2020cwt,al2014methods}. The DWT pipeline passes the non-stationary EEG rhythms through successive low-pass and high-pass filter banks to yield approximation and detail coefficients, capturing macro-level trends and rapid transients across multi-level subbands before compressing them into a concise feature vector \cite{ji2019eeg,donovan2018motor,yang2018deep, doudou2023improving}. Conversely, the CWT continuously shifts and scales a selected mother wavelet across the signal to map out a high-resolution time-frequency scalogram \cite{xu2018wavelet}. Specifically, in DWT, the raw EEG rhythm is passed through a multi-level filter bank to decompose the signal into low-frequency approximation ($A_1$) and high-frequency detail ($D_1, D_2$) coefficients, which are reduced to form a structured feature vector, as illustrated in Fig. \ref{Fig_Wavelet_Transform} (a). In CWT, the signal is convolved with continuous mother wavelets (e.g., Morlet, Meyer) to generate a high-resolution time-frequency scalogram, from which energy variations within critical motor imagery bands (alpha and beta) are selected as features, as depicted in Fig. \ref{Fig_Wavelet_Transform} (b).

		Wavelet packet decomposition (WPD) extends this capability by further splitting both approximation and detail coefficients, allowing finer and more comprehensive analysis of time‑frequency components \cite{kevric2017comparison,zhang2017classification,lu2016deep,marcos2020real}. Empirical mode decomposition (EMD) offers a data-driven alternative by adaptively decomposing the EEG signal into intrinsic mode functions without requiring a predefined basis, making it well suited to the nonstationary, nonlinear nature of MI rhythms \cite{foong2019assessment}. Additionally, spatio-spectral feature representation (SSFR) integrates spatial filters with time-frequency features to enhance the extraction of MI-related patterns \cite{bang2021spatio}, while the separable common spatio spectral pattern (SCSSP) algorithm jointly optimizes spatial and spectral filters in a separable manner to improve class discrimination \cite{aghaei2015separable,malekmohammadi2019efficient}.

\begin{figure}[!t]
			\centerline{\includegraphics[scale=0.14,trim=0cm 0cm 0cm 0cm,clip=true]{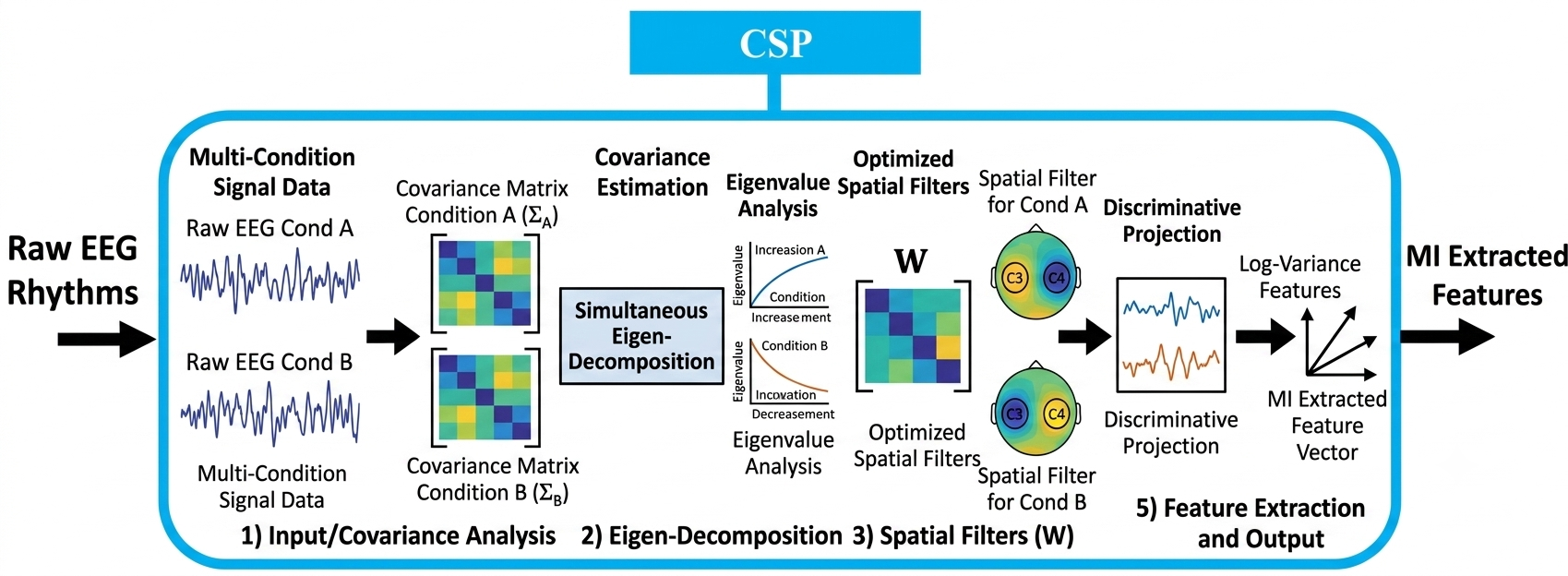}}
			\caption{Utilizing CSP-based methods for EEG MI signal feature extraction.
			}
			\label{Fig_CSP}
		\end{figure}

		\textbf{4) Spatial Domain Techniques:} Spatial‑domain feature extraction methods focus on enhancing the discriminative spatial patterns of EEG signals by exploiting the distribution of neural activity across different electrodes, with common spatial patterns (CSP) being the most widely used approach for MI \cite{gaur2021sliding,gaur2019automatic,azab2019weighted,baig2017differential,yu2014analysis,liu2017feature}. CSP learns spatial filters that maximize variance differences between MI classes, effectively highlighting the sensorimotor activation changes characteristic of left-or right‑hand imagery \cite{zhao2020deep,baig2020filtering,bhatti2019soft,ren2014convolutional}. Utilizing CSP-based methods for EEG MI signals feature extraction can be described in a few steps, as illustrated in Fig. \ref{Fig_CSP}. (1) Input and Covariance Analysis: Multi-condition EEG signal data (Condition A vs. Condition B) are used to estimate the respective spatial covariance matrices ($\sum A$ and $\sum B$). (2) Simultaneous Eigen-Decomposition: Joint diagonalization is performed via eigenvalue analysis to maximize the variance of one condition while minimizing the variance of the other. (3) Spatial Filters ($W$): An optimized spatial filter matrix ($W$) is constructed, corresponding to specific target cortical regions (e.g., C3 and C4 topomaps). (4) Discriminative Projection \& Feature Extraction: The raw EEG signals are projected onto the subband spatial filters to yield discriminative time-series waveforms, from which the log-variance is calculated to form the final optimized MI-extracted feature vector. To better preserve electrode topology in spatial-domain analysis, the azimuthal equidistant projection (AEP) method approximates the scalp as a sphere and maps electrodes from three-dimensional space onto a two-dimensional plane, facilitating image-like spatial representations for downstream CNN processing \cite{ma2019fpga}. Furthermore, the filter bank Riemannian tangent space (FBRTS) combines filter-bank processing with Riemannian geometry to extract discriminative covariance-based representations from multiband EEG signals \cite{fang2022feature}.

		Its extensions, such as the common spatio‑spectral pattern (CSSP) and common sparse spatio‑spectral pattern (CSSSP), further integrate spectral information and sparsity constraints to improve robustness \cite{kumar2017improved,zhang2018temporally}. The sub‑band CSP (SBCSP) enhances performance by applying CSP to frequency‑filtered sub‑bands, capturing finer MI‑related rhythmic variations, while methods like the temporally constrained sparse group spatial pattern (TSGSP) enforce structured sparsity to reduce noise and improve generalization \cite{zhang2018temporally,wu2016fuzzy}. Although simpler, PSD can also serve as a spatial‑domain descriptor when computed across electrodes, complementing CSP‑based techniques in MI classification \cite{mahmood2021wireless,leeb2015towards}. Additionally, the augmented CSP (ACSP) technique extends traditional CSP by incorporating auxiliary or reference data to stabilize spatial filter learning, thereby improving robustness under non-stationary MI conditions \cite{yang2015use}. Other representative feature extraction techniques include permutation entropy (PME) \cite{yang2014adaptive}, sample entropy (SampEn) \cite{yang2014adaptive}, movement-related cortical potentials (MRCP) \cite{sebait2025brain}, utilizing sensorimotor rhythms (SMR) \cite{he2015noninvasive}, Pearson correlation matrix-based graph Laplacian (PCM-based GL) \cite{hou2022gcns}, signal-space modeling (SSM) \cite{he2015noninvasive}, and eigenvector methods (EVM) \cite{al2014methods}.

		\textbf{5) Feature Selection/Reduction Techniques:} Feature selection and reduction techniques aim to eliminate irrelevant or redundant information while preserving the most discriminative patterns for MI classification \cite{bhatti2019soft}. Dimensionality-reduction algorithms such as PCA and ICA transform high-dimensional EEG features into compact representations by maximizing variance or statistical independence, respectively, thereby improving computational efficiency and robustness \cite{yu2021new,venkatachalam2020novel,yu2014analysis,herman2016designing}. The Canonical variate analysis (CVA) selects user-specific features that best reflect the MI task for each individual, which can then be used to train subject-tailored classifiers \cite{leeb2015towards}. Filter bank approaches like filter bank CSP (FBCSP) \cite{schirrmeister2017deep,zhang2017sparse,sakhavi2015parallel,foong2019assessment}, discriminant filter bank CSP (DFBCSP), and sparse filter bank CSP (SFBCSP) further refine feature spaces by decomposing EEG signals into multiple frequency sub-bands and applying spatial filters optimized for discrimination \cite{olivas2019classification,baig2020filtering,kumar2017improved,luo2018exploring}; these variants enhance class separability through discriminant weighting (DFBCSP) or sparse regularization (SFBCSP), enabling more informative and noise-resistant feature sets \cite{zhang2015optimizing, fang2022feature}. This is enabled through filter bank decomposition (FBD), in which a bank of bandpass filters splits the EEG signal into multiple sub-bands that serve as parallel inputs to subsequent spatial filtering stages such as CSP \cite{wu2016fuzzy}. The one-versus-rest FBCSP (OVR-FBCSP) extends FBCSP to multi-class MI problems by learning class-specific spatial filters for each one-versus-rest classification task \cite{zhang2021hybrid}. Moreover, an iterative algorithm (ITA) approach selects frequency bands by iteratively evaluating candidate bands and retaining the one that yields the highest CA for a given binary MI task \cite{olivas2019classification}. Furthermore, the multiscale PCA (MSPCA) extends PCA across multiple scales to simultaneously perform denoising and dimensionality reduction \cite{yu2021new}.

		A second category involves evolutionary and population-based optimization algorithms, which search for the most effective subset of features or model parameters through iterative, bio-inspired processes \cite{baig2017differential}. Methods such as particle swarm optimization (PSO), differential evolution (DE), artificial bee colony (ABC), ant colony optimization (ACO), genetic algorithm (GTA), and firefly algorithm (FFA) evaluate candidate feature subsets based on classification performance and evolve them toward optimal solutions using mechanisms like swarm behavior, mutation, selection, pheromone trails, or light-intensity attraction \cite{baig2017differential,liu2017feature}. Techniques like ACO and simulated annealing (SAN) are particularly useful for navigating large combinatorial feature spaces; ACO constructs probabilistic solution paths guided by learned pheromone levels, while SAN explores feature subsets using a temperature-controlled search strategy that helps escape local optima \cite{baig2017differential}. The learning automata (LAM) method, often combined with the firefly algorithm, have also been proposed to optimize feature selection for MI EEG by adaptively updating action probabilities based on environmental feedback \cite{liu2017feature}. The HYB approaches combine these evolutionary strategies with dimensionality-reduction or domain-specific feature extraction methods, achieving more efficient searches and often producing superior, task-adaptive feature subsets for MI-BCI systems \cite{wu2016fuzzy, chen2026multiscale}.

		Beyond population-based approaches, sparsity-driven and task-aware learning techniques provide additional mechanisms for targeted feature reduction. The joint sparse optimization (JSO) enforces shared sparsity across multiple feature vectors or sub-bands, promoting the selection of consistent, physiologically meaningful features while simultaneously suppressing noise \cite{zhang2018temporally,pan2026ctssp}. The optimal allocation (OAL) selects representative EEG trials with minimal variability, improving the quality of training data and enhancing classification robustness across subjects. Furthermore, the multitask learning (MTL) extends this idea by jointly learning feature representations across related MI tasks, sessions, or subjects, enabling knowledge transfer and improving robustness in the presence of inter-subject or temporal variability. These methods not only enhance generalization but also help identify stable neural signatures of motor imagery, making them valuable components in advanced MI-BCI pipelines \cite{shin2024sparse}.

		\textbf{6) Hybrid, Data Driven, Deep Learning Approaches:} These feature extraction methods integrate information from multiple domains to capture the complex, non‑stationary structure of EEG signals more effectively than traditional single‑domain approaches. In practice, many studies combine handcrafted features from several domains, such as time‑frequency features, spatial filters, or spectral measures, with learned representations from data‑driven models to create richer and more robust feature vectors, ultimately improving CA in MI tasks \cite{zhao2020deep}. Recent studies have proposed specialized deep architectures, including graph convolutional neural networks (GCNs) \cite{hou2022gcns}, deep autoencoders (DAEs) \cite{velakanti2025improving}, multitask learning autoencoders (MTL-AECs) \cite{autthasan2021min2net}, transformer-based multi-branch EEG fusion networks (TMEF-Net) \cite{cao2026hybrid}, and channel-projection MixedNet (CP-MixedNet) \cite{li2019channel} for learning more discriminative spatial-temporal EEG representations. Specifically, DAEs and MTL-AECs have also been employed to learn discriminative latent representations (DLRs) that improve feature compactness and class separability for MI decoding \cite{autthasan2021min2net}. The HYB EEG-fNIRS frameworks have also been introduced by converting multimodal signals into 3D tensors (EEG/fNIRS-TC) and employing fNIRS-guided attention layers (FGAL) to enhance cross-modal feature fusion \cite{kwak2022fganet}. Also, FGANet performs multimodal EEG-fNIRS fusion by using fNIRS-guided attention to emphasize informative brain regions while combining complementary neural and hemodynamic features \cite{kwak2022fganet}. The MIN2Net is a multi-task autoencoder that jointly performs latent feature learning, deep metric learning, and classification to obtain compact and discriminative EEG representations \cite{autthasan2021min2net}. Furthermore, deep representation-based domain adaptation (DRDA) performs domain adaptation by learning deep representations that minimize distribution discrepancies between source and target domains, improving cross-subject generalization \cite{zhao2020deep}.

		The HYB combinations like DWT‑AR, which apply DWT for multiresolution decomposition followed by AR modeling to characterize temporal dynamics within each sub‑band, provide a compact yet informative representation of MI‑related rhythms. Similarly, Hilbert transform (HLT)‑DWT methods use the HT to extract instantaneous amplitude or phase information from DWT‑derived components, enhancing the sensitivity to transient oscillatory changes critical for MI classification \cite{zhou2018classification}. Moreover, CNNs play a key role in this category because they can implicitly learn spatial, temporal, and frequency patterns simultaneously from raw or minimally processed EEG, even though they do not fit neatly into classical time, frequency, or spatial domain definitions \cite{zhao2025multi,dai2020hs,li2019channel}. Since CNNs also serve as classifiers, their detailed discussion is presented in the next section.

		\subsection{EEG Feature Classification Algorithms}
		
\begin{figure}[!t]
			\centerline{\includegraphics[scale=0.4,trim=6.8cm 3.1cm 6.6cm 1.9cm,clip=true]{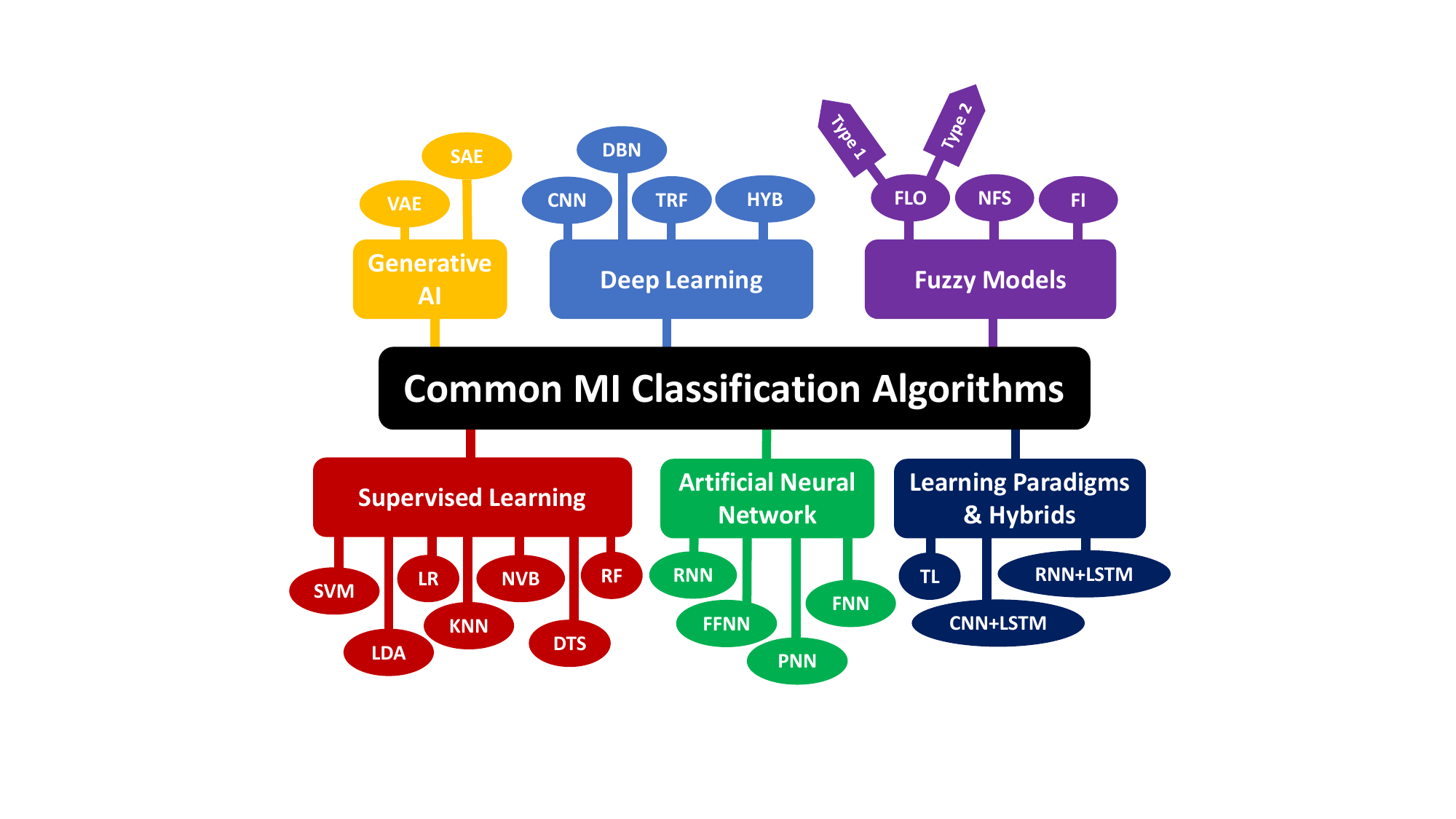}}
			\caption{Summary of prevalent AI-based MI feature classification algorithms.
			}
			\label{Fig_Feature_Classification_Classes}
		\end{figure}

		EEG‑based MI classification algorithms form the core of BCI systems designed to decode users’ imagined movements from neural activity. These algorithms process non‑invasive EEG signals to identify distinct patterns associated with different MI tasks, such as left‑ or right‑hand movement imagination. The primary goal is to translate complex, noisy, and highly individual EEG data into accurate, real‑time decisions, enabling effective interaction between the user and external devices such as prosthetics or neuro‑rehabilitation tools. In the following, the most common classification algorithms based on Table \ref{Table_SW_Algorithms} are discussed in several categories, as depicted in Fig. \ref{Fig_Feature_Classification_Classes}. Note that Some algorithms naturally fall into more than one group.

		\begin{figure*}[!t]
			\centerline{\includegraphics[scale=0.55,trim=0cm 0cm 0cm 0cm,clip=true]{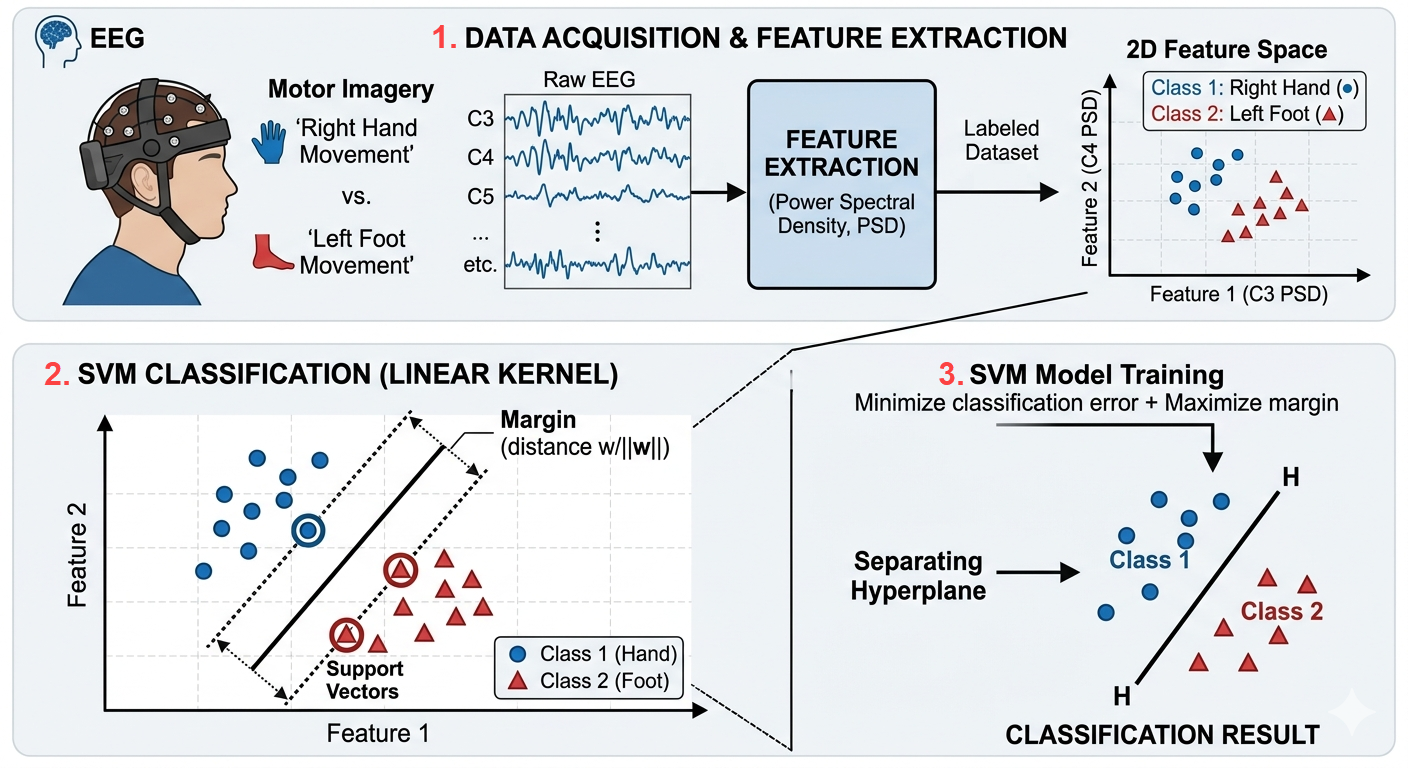}}
			\caption{Schematic workflow of an EEG-based MI-BCI classification system using the SVM algorithm. The Data Acquisition \& Feature Extraction shows the raw EEG signals are recorded from motor cortex channels (e.g., C3, C4) during imagined right-hand vs. left-foot movements, and transformed into a 2D feature space using the PSD. The SVM Classification shows that a linear kernel SVM, for example, identifies the optimal separating hyperplane ($H$) by maximizing the margin between the two MI classes, uniquely defined by the critical edge data points known as support vectors.
			}
			\label{Fig_SVM}
		\end{figure*}

		\textbf{1) Supervised Learning} \\
		The supervised learning algorithms extract spatial and temporal features from EEG signals to accurately discriminate motor intentions across subjects and sessions, where labeled EEG data, each corresponding to specific imagined movements, are used to train predictive models. These algorithms learn mappings between neural signal patterns and movement classes to enable real‑time BCI control \cite{gupta2023machine,hosseinifard2013classifying}. The most common supervised approaches utilized for MI classification can be categorized as follows:

		\textbf{A)	Linear \& Margin‑Based Classifiers}: The linear and margin‑based classifiers learn decision boundaries that separate classes using linear projections or maximum‑margin principles, making them efficient and well‑suited for high‑dimensional EEG feature spaces. Their simplicity, interpretability, and compatibility with feature‑selection or regularization strategies make them common baselines and reliable components in many MI‑BCI pipelines \cite{hosseinifard2013classifying, garcia2020cnn}.

		The support vector machines (SVMs) and their variants are among the most common and effective classifiers in this group used in MI-BCIs because they handle high‑dimensional EEG features and small training sets exceptionally well \cite{sebait2025brain,ji2019eeg,zhang2015optimizing,malekmohammadi2019efficient}. Standard SVM finds a maximum‑margin separating hyperplane, making it robust to noise and inter‑trial variability in MI signals \cite{bhatti2019soft,zhang2018temporally}. SVM classifies MI EEG signals by finding the optimal separating hyperplane that maximizes the margin between different MI classes. After extracting features, SVM maps them into a space, linear or kernel‑transformed, and determines the boundary that best separates imagined movements, allowing robust and accurate discrimination even in noisy, high‑dimensional EEG data \cite{zhang2017classification}. Moreover, hybrid-kernel extreme learning machine (H-KELM) combines multiple kernel functions to improve nonlinear decision boundaries while maintaining fast training \cite{venkatachalam2020novel}.

		The complete pipeline of an EEG-based MI-BCI system utilizing the SVM for classification is illustrated in Fig. \ref{Fig_SVM}. The process begins with data acquisition, where raw EEG signals are recorded from the motor cortex during the imagination of distinct tasks, such as right-hand versus left-foot movements. These non-stationary signals are then converted into discriminative features, typically through frequency-domain methods like PSD, mapping the data into a low-dimensional feature space. Finally, a linear SVM algorithm is applied to perform binary classification. The model iteratively trains to find the optimal separating hyperplane that maximizes the geometric margin between the two classes. This decision boundary is uniquely determined by the support vectors, which are the critical data points closest to the hyperplane, ensuring robust generalization for real-time BCI applications.

		For multi‑class decoding, the one‑versus‑rest SVM (OVR-SVM) strategy extends this framework by training one classifier per class, offering a simple and computationally efficient solution \cite{fang2022feature}. The twin SVM (TSVM) further improves scalability by learning two nonparallel hyperplanes, significantly reducing training time while maintaining strong discriminative power. Optimized TSVM variants, such as GWO‑TSVM, automatically tune critical parameters using metaheuristic search, enhancing accuracy and stability across subjects. Therefore, these SVM‑based approaches form a highly reliable and widely adopted family of classifiers for MI features \cite{sharma2025alzheimer}.

		The linear discriminant analysis (LDA) family encompasses a group of linear classifiers designed to project high‑dimensional EEG features onto a subspace that maximizes class separability \cite{sebait2025brain,gaur2019automatic,kumar2017improved,kim2025efficacy,marcos2020real,malekmohammadi2019efficient}. The LDA itself operates by finding a linear combination of features that best separates two or more MI classes (e.g., left vs. right hand). It assumes each class follows a Gaussian distribution with equal covariance, deriving a decision boundary that minimizes within‑class variance while maximizing between‑class variance \cite{gaur2021sliding,baig2017differential,he2015noninvasive,belwafi2018embedded,herman2016designing}.

\begin{figure}[!t]
			\centerline{\includegraphics[scale=0.24,trim=0cm 0cm 0cm 0cm,clip=true]{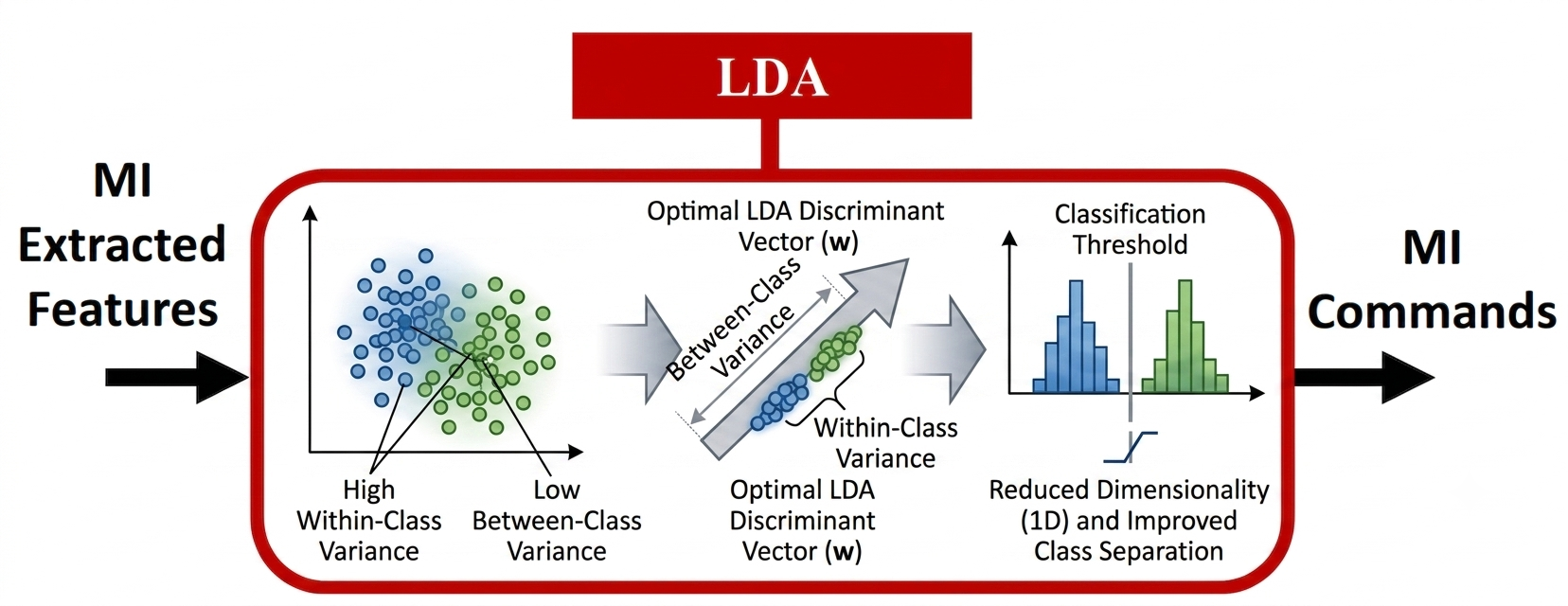}}
			\caption{Utilizing LDA-based methods for MI feature classification.
			}
			\label{Fig_LDA}
		\end{figure}

		Utilizing LDA-based methods for MI feature classification can be described in few steps, as shown in Fig. \ref{Fig_LDA}. (1) Initial Feature Space: Input MI extracted features initially exhibit high within-class variance and low between-class variance, making direct separation difficult. (2) Linear Projection: The algorithm computes an optimal discriminant vector ($w$) that maximizes the ratio of between-class variance to within-class variance, projecting the multi-dimensional features onto an optimized 1D axis. (3) Classification \& Decision: The reduced-dimensionality space yields maximum class separation, allowing a precise classification threshold to split the distributions and output final discrete MI commands.

		The fisher discriminant analysis (FDA) is conceptually similar and derived from the same statistical principle: it seeks a projection vector that maximizes the Fisher criterion, the ratio of inter‑class scatter to intra‑class scatter, yielding an optimal linear discriminant function. The linear discriminant classifier (LDC) employs the same underlying statistical foundations but emphasizes its implementation as a direct classifier, often using estimated covariance matrices to classify EEG feature vectors based on posterior probability \cite{zhang2017sparse}. Fisher's linear discriminant analysis (FLDA) is frequently paired with FBCSP-derived features, where the discriminant classifier's output accuracy directly reflects the BCI system's overall MI-detection performance \cite{foong2019assessment}. The Spectral regression discriminant analysis (SRDA) method reformulates LDA as a regression problem, substantially reducing computational cost for high-dimensional EEG feature spaces while retaining discriminative power \cite{liu2017feature}. Finally, LDA–SVM represents a HYB model in which LDA first reduces the feature dimensionality and enhances linear separability, and the SVM performs the final classification, typically leading to improved robustness and generalization in MI tasks while retaining LDA’s linear interpretability \cite{baig2017differential,kumar2017improved}.

		The logistic regression (LR) is a statistical model used for MI binary classification. It models the probability of a specific outcome (e.g., a particular MI class) using a logistic function applied to a linear combination of input features. This results in a decision boundary that is linear in the feature space, making it a simple yet effective linear classifier for differentiating between two MI states \cite{wu2025application}. Moreover, the weighted logistic regression (WLR) introduces class-dependent weighting to improve robustness under imbalanced EEG datasets \cite{azab2019weighted}.

		\textbf{B) Instance‑ and Distance‑Based Classifiers:} The instance‑ and distance‑based classifiers classify new MI samples by comparing them directly to stored training instances. Classification is determined by the proximity of a new sample to existing examples, typically using distance metrics like Euclidean or Mahalanobis distance. These methods rely on the idea that similar MI patterns occupy nearby regions in the feature space \cite{sharma2023efficient}.

		The k‑nearest neighbor (KNN) is a non‑parametric, instance‑based learning algorithm. It classifies an unknown EEG feature vector by comparing it to the closest training samples in the feature space, typically measured using Euclidean or another distance metric \cite{kevric2017comparison, rizal2022fpga}. The algorithm then assigns the class most common among these neighbors to the test sample. In MI tasks, KNN leverages spatial or spectral EEG features, relying on the assumption that samples from similar MI patterns lie close together in the feature space \cite{sebait2025brain,baig2017differential}. Additionally, the Riemannian classifiers use distance geometry principles to classify MI EEG signals by operating directly on covariance matrices viewed as points on a Riemannian manifold. Instead of relying on Euclidean distance, they measure similarity using Riemannian metrics that capture the intrinsic geometry of the data, enabling robust discrimination of MI patterns through spatial covariance structures \cite{fang2022feature, baig2017differential}.

\begin{figure}[!t]
			\centerline{\includegraphics[scale=0.24,trim=0cm 0cm 0cm 0cm,clip=true]{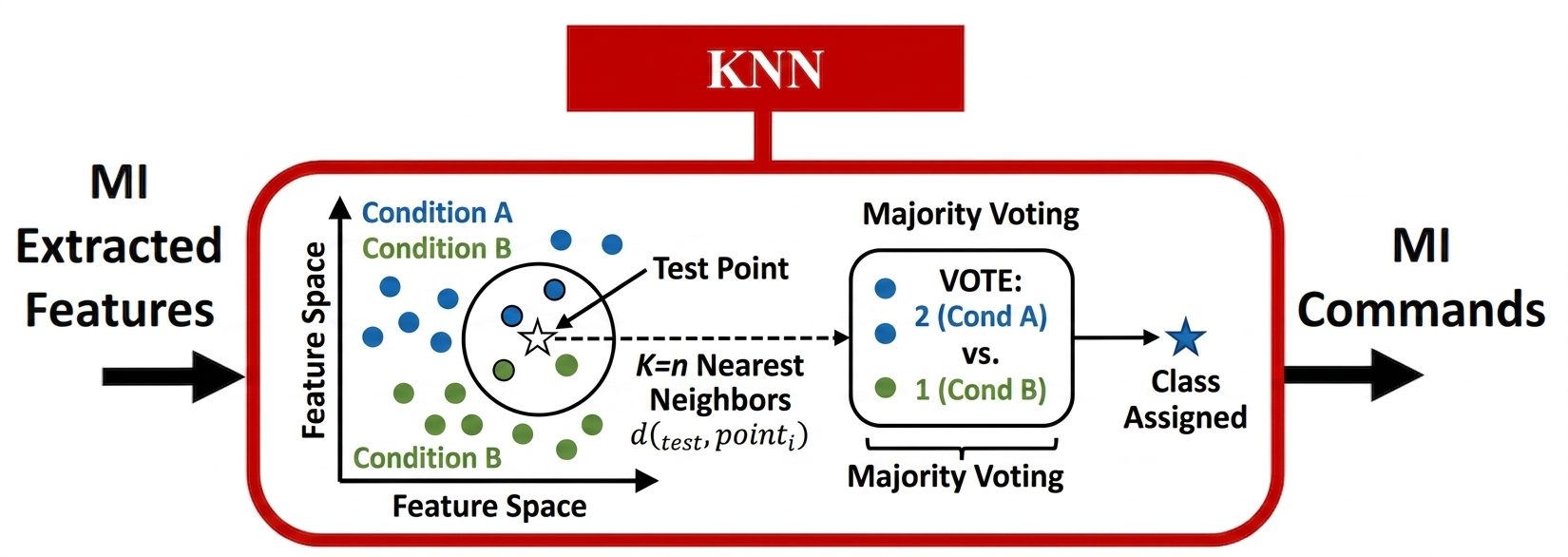}}
			\caption{Utilizing KNN-based methods for MI feature classification.
			}
			\label{Fig_KNN}
		\end{figure}

		Utilizing KNN-based methods for MI feature classification can be performed in a few steps, as depicted in Fig. \ref{Fig_KNN}. (1) Feature Space Mapping: Input MI extracted features are mapped onto a multi-dimensional feature space containing labeled reference profiles (Condition A vs. Condition B). (2) Distance Measurement: When an unknown test point (star) is introduced, the algorithm calculates the spatial distance $d(\text{test}, \text{point}_i)$ to locate its nearest neighbors ($K=n$). (3) Majority Voting \& Command Assignment: A local neighborhood decision is made in which the test point is assigned to the predominant class via majority voting (e.g., 2 votes for Condition A vs. 1 vote for Condition B), translating the feature data into a final discrete MI command.

		\textbf{C) Tree Based Methods: }The tree‑based methods capture nonlinear relationships in MI features and classify them by recursively partitioning the feature space into decision regions based on feature thresholds. Each split aims to maximize class purity, forming a hierarchical tree that predicts EEG class labels at its leaves.

\begin{figure}[!t]
			\centerline{\includegraphics[scale=0.24,trim=0cm 0cm 0cm 0cm,clip=true]{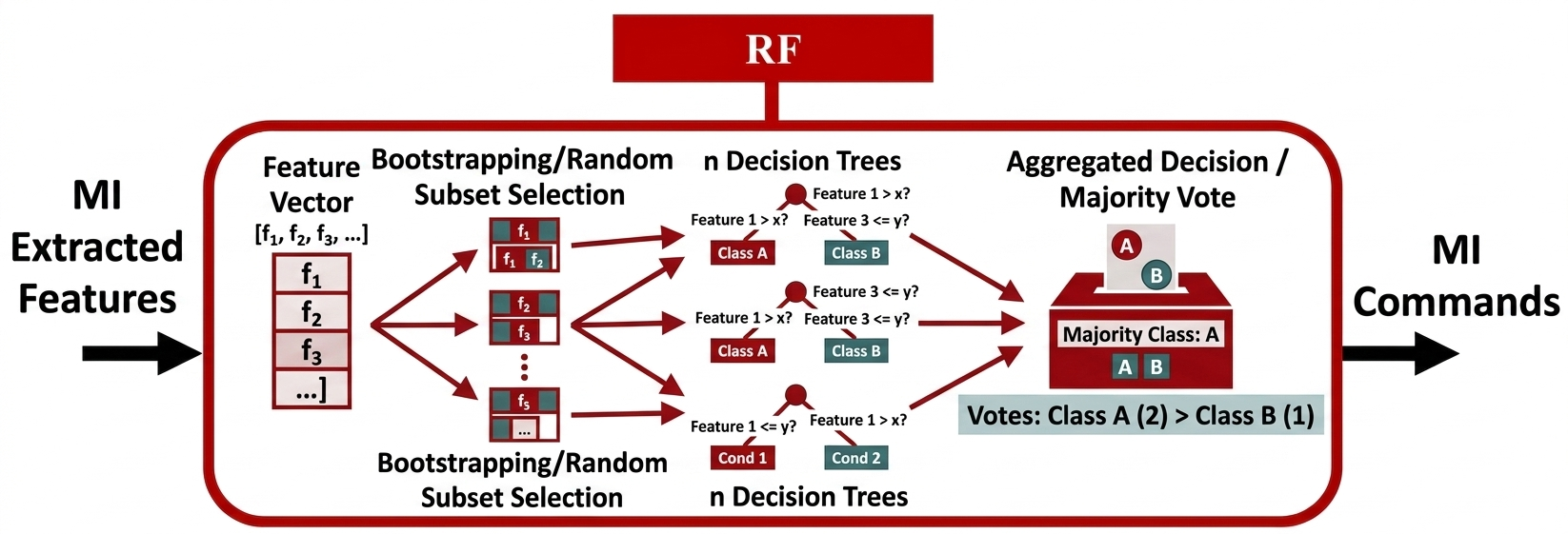}}
			\caption{Utilizing RF-based methods for MI feature classification.
			}
			\label{Fig_RF}
		\end{figure}

		Specifically, random forest (RF) is an ensemble learning method that classifies MI features by combining the outputs of many decision trees trained on different bootstrapped subsets of the data and features \cite{aydemir2014decision}. Each tree independently learns nonlinear decision rules that map MI‑related features, such as band‑power or covariance patterns, to class labels \cite{sebait2025brain}. During prediction, the forest aggregates the votes of all trees (majority voting), which reduces overfitting and improves robustness to noise, making RF well‑suited for handling the variability inherent in MI EEG recordings \cite{mohammady2026leveraging}. Utilizing RF-based methods for MI feature classification includes the following steps, as illustrated in Fig. \ref{Fig_RF}. (1) Bootstrapping and Random Subset Selection: The input multi-dimensional MI feature vector $[f_1, f_2, f_3, \dots]$ undergoes bootstrap sampling to create distinct subset combinations, ensuring diversity across the ensemble. (2) Ensemble of Decision Trees ($n$ Trees): Multiple independent decision trees are trained in parallel, where internal nodes evaluate specific threshold conditions (e.g., $Feature\,1 > x?$) to sequentially route data to candidate motor imagery classes (e.g., Class A or Class B). (3) Aggregated Decision and Majority Voting: The individual predictions from all $n$ trees are collected and aggregated via a majority voting scheme (e.g., Votes: Class A (2) > Class B (1)), combining weak estimators into a robust, unified output to generate the final target MI command.

		The decision tree structure (DTS) classifies MI data by hierarchically splitting the feature space into regions using threshold rules that maximize class purity at each node, forming an interpretable tree that maps EEG features to MI classes. Also, the regression tree (RGT) follows a similar structure but predicts continuous outputs, such as class probabilities or feature scores, by fitting simple regression models within each leaf, enabling finer modeling of MI signal relationships compared to purely categorical splits \cite{aydemir2014decision}. Furthermore, the gradient boosting models (GBMs) and RGTs have also been investigated as ensemble learning approaches for MI classification \cite{mirzaei2021eeg,baig2017differential}.

		\textbf{D) Probabilistic Models: }The probabilistic models classify MI features by estimating the likelihood of each class given the observed features, using statistical distributions (e.g., Gaussian models or Bayesian inference) to capture uncertainty and variability in brain patterns for more flexible decision making \cite{zhang2017sparse}. For example, naive bayes (NVB) is a probabilistic classifier that models MI EEG features under the assumption of conditional independence, computing class‑posterior probabilities from simple likelihood estimates, which makes it fast, lightweight, and effective even with limited training data. A Bayesian classifier (BYC) decoding approach computes the posterior probability of a given movement state directly from extracted features, providing a principled probabilistic estimate of MI intent \cite{mccrimmon2017performance}.  These CVA-derived features are often paired with a Gaussian classifier (GSC), which models each MI class as a Gaussian distribution in feature space and assigns labels based on maximum likelihood \cite{leeb2015towards}. Moreover, the Markov switching model (MSM) models MI EEG signals as transitions between latent brain states governed by a Markov process, allowing the classifier to capture temporal structure and state‑dependent feature dynamics that reflect how MI patterns evolve over time \cite{agarwal2025motor,ali2026diagonal}.

		\textbf{2)	Unsupervised Learning} \\
		Unsupervised learning algorithms, such as clustering or representation‑learning methods, attempt to group MI EEG patterns based on their intrinsic structure without using class labels. While they can reveal latent patterns or assist with feature extraction, their direct application to MI classification is relatively limited compared to supervised approaches, which typically achieve higher accuracy by leveraging labeled MI data \cite{chen2025unsupervised, raza2025deep}.

		\textbf{3)	Artificial Neural Networks}
		
\begin{figure}[!t]
			\centerline{\includegraphics[scale=0.7,trim=5.8cm 9.3cm 5.8cm 9cm,clip=true]{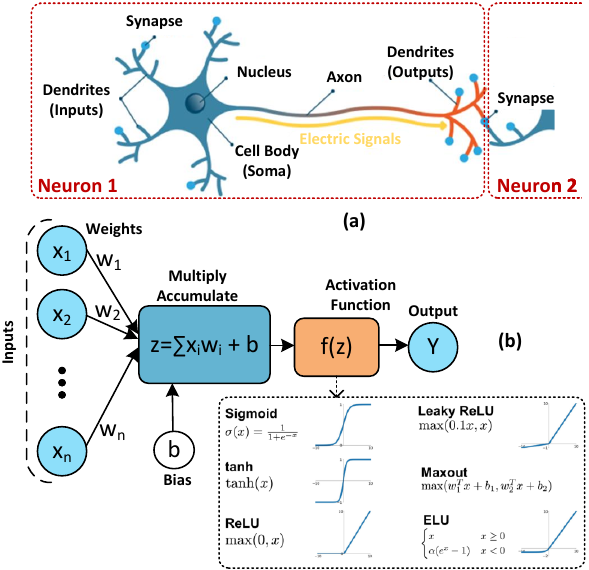}}
			\caption{(a) Natural biological neurons structures. (b) Symbolic mathematical representation of conventional artificial neurons.}
			\label{Fig_ANN}
		\end{figure}

		Artificial neural networks (ANNs) are inspired by the structure and function of biological neurons in the human brain, as depicted in Fig. \ref{Fig_ANN} (a). A natural neuron receives electrical impulses through dendrites, processes them within the cell body, and transmits responses via the axon through synapses to other neurons. Similarly, an artificial neuron receives input signals, processes them mathematically, and sends output to connected units, emulating how electrical signals propagate through biological networks to enable learning and information transfer \cite{zheng2026graph,zhang2026bayesian}. An artificial neuron, as shown in Fig. \ref{Fig_ANN} (b), consists of inputs, adjustable weights, a bias term, and an activation function. Each input is multiplied by its corresponding weight, summed up with the bias (the multiply–accumulate step), and then passed through an activation function that introduces nonlinearity. Common activation functions include Sigmoid and tanh (for smooth transitions), ReLU and Leaky ReLU (for efficient gradient propagation), Maxout, and ELU (for improved robustness), producing the final output that feeds into subsequent network layers. Typical ANN-based MI features classifiers are grouped and discussed in the following categories \cite{qamar2023artificial}.

		\textbf{A) Traditional \& Multilayer Models:} The feedforward neural networks (FFNNs) process information in a strictly forward direction, moving from input to output through one or more hidden layers without any feedback loops \cite{bhatti2019soft}. Each neuron performs a weighted sum of its inputs, adds a bias, and passes the result through a nonlinear activation function, enabling the network to learn complex decision boundaries \cite{yu2021new}. The standard neural networks and multilayer Perceptrons (MLPs) extend this structure by stacking multiple hidden layers, allowing the model to capture hierarchical feature representations \cite{shrivastwa2018fpga}. The FFNNs trained with backpropagation (BPNN) iteratively adjust weights based on error gradients to improve CA \cite{li2019hybrid, qamar2023artificial}. In MI EEG classification, these models serve as powerful discriminators by learning nonlinear mappings from extracted MI features, such as time‑frequency or spatial patterns, to class labels, offering flexible and effective pattern recognition when sufficient training data and well‑engineered features are available.

\begin{figure}[!t]
			\centerline{\includegraphics[scale=0.24,trim=0cm 0cm 0cm 0cm,clip=true]{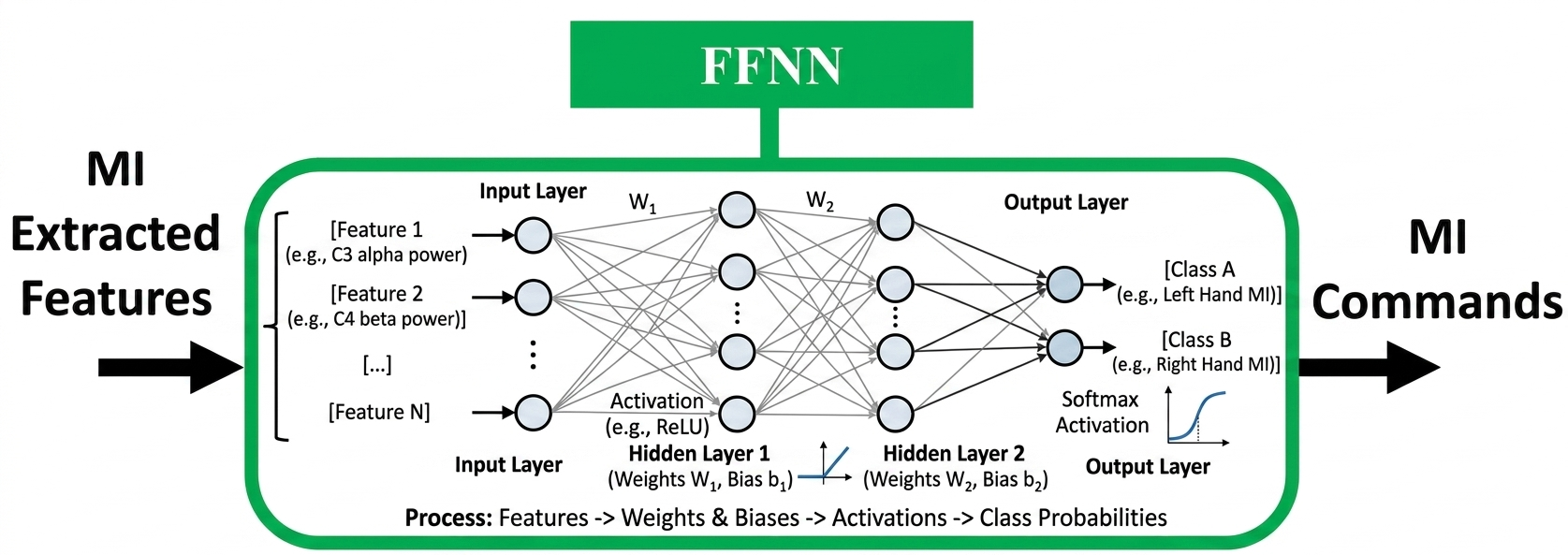}}
			\caption{FFNN-type ANNs used as MI feature classifiers.
			}
			\label{Fig_FFNN}
		\end{figure}

		FFNN-type ANNs used as the MI features classifier can be described in few steps, as shown in Fig. \ref{Fig_FFNN}. (1) Input Layer: The extracted MI feature vector—comprising elements such as C3 alpha power, C4 beta power, and other numerical signal indicators $[Feature\,1, Feature\,2, \dots, Feature\,N]$—is fed into the network. (2) Hidden Layers (Layers 1 and 2): Information moves exclusively forward through fully connected nodes, where inputs are multiplied by adaptive weights ($W_1, W_2$), combined with biases ($b_1, b_2$), and mapped via non-linear activation functions (e.g., ReLU) to resolve intricate cross-feature dependencies. (3) Output Layer: The final layer applies a Softmax activation function to compute distinct class probabilities (e.g., Left Hand vs. Right Hand MI), translating the continuous internal feature maps into a definitive, discrete MI command.

		\textbf{B) Recurrent Neural Networks:} The recurrent neural networks (RNNs) are designed to model temporal dependencies in sequential data by incorporating feedback connections that allow information from previous time steps to influence current processing \cite{zhang2019convolutional,luo2018exploring}. This structure enables them to capture dynamic patterns and context over time, an essential property for analyzing MI EEG signals, which unfold sequentially. Variants of RNNs enhance this ability: the echo state network (ESN) uses a large, fixed recurrent reservoir with trainable output weights for efficient temporal pattern learning; the long short‑term memory (LSTM) network introduces memory cells and gating mechanisms to preserve long‑range dependencies and mitigate vanishing gradients \cite{zhou2018classification,velakanti2025improving}; the gated recurrent unit (GRU) simplifies the LSTM structure while maintaining effective sequence modeling \cite{luo2018exploring}; and HYB CNN–LSTM architectures combine convolutional layers for spatial feature extraction with LSTM layers for capturing temporal evolution, achieving robust MI feature classification by learning both spatial and temporal characteristics of EEG data \cite{li2022motor,li2019channel}.

\begin{figure}[!t]
			\centerline{\includegraphics[scale=0.14,trim=0cm 0cm 0cm 0cm,clip=true]{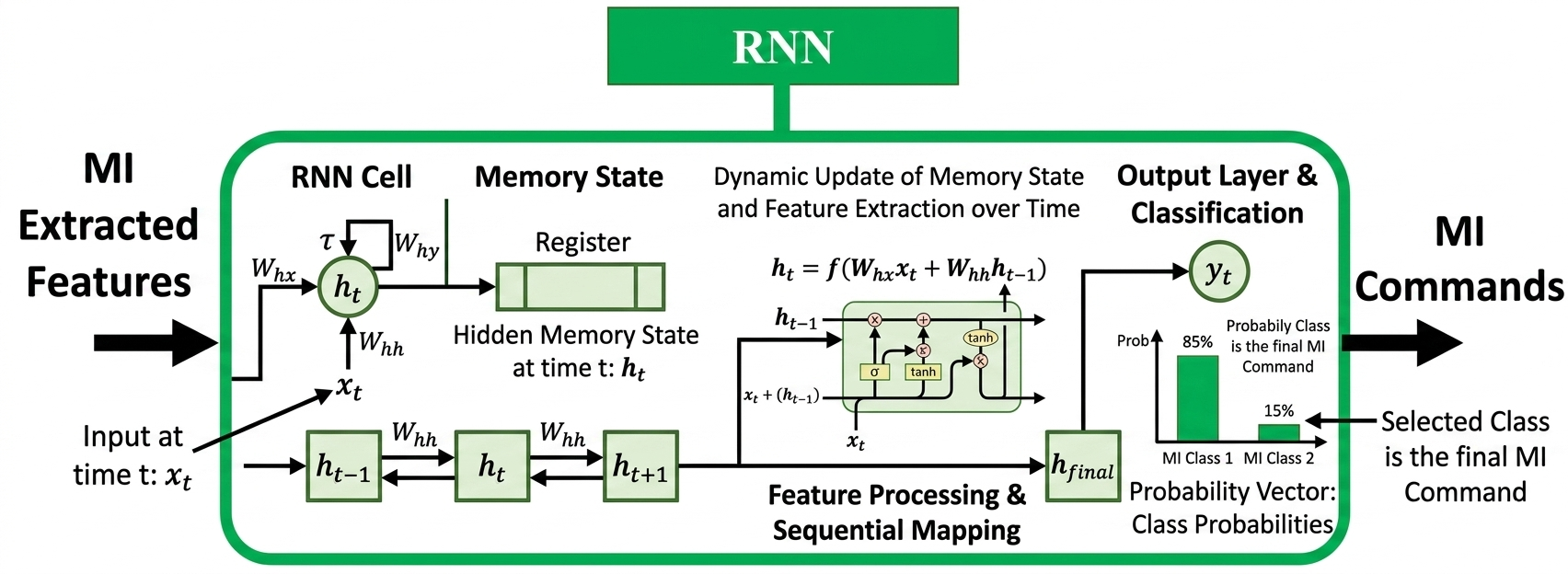}}
			\caption{The RNN-type ANNs as the MI features classifier.
			}
			\label{Fig_RNN}
		\end{figure}

		RNN-type ANNs perform MI features classification in few steps, as shown in Fig. \ref{Fig_RNN}. (1) Sequential Input Tracking: The input MI extracted features are treated as a time-series sequence, where $x_t$ represents the incoming feature slice at time step $t$. (2) Hidden Memory State Dynamics: An internal recurrence loop tracks sequential context through an adaptive hidden memory state ($h_t$). This memory state is dynamically updated over time via the mapping function $h_t = f(W_{hx}x_t + W_{hh}h_{t-1})$, effectively capturing temporal dependencies across successive motor imagery frames. (3) Output Layer and Sequential Mapping: The final cumulative sequence state ($h_{\text{final}}$) is fed into the output layer to compute a class probability vector (e.g., 85\% probability for MI Class 1 vs. 15\% for MI Class 2), routing the highest-probability index to generate the final discrete MI command.

		\begin{figure*}[!t]
			\centerline{\includegraphics[scale=0.3,trim=0cm 0cm 0cm 0cm,clip=true]{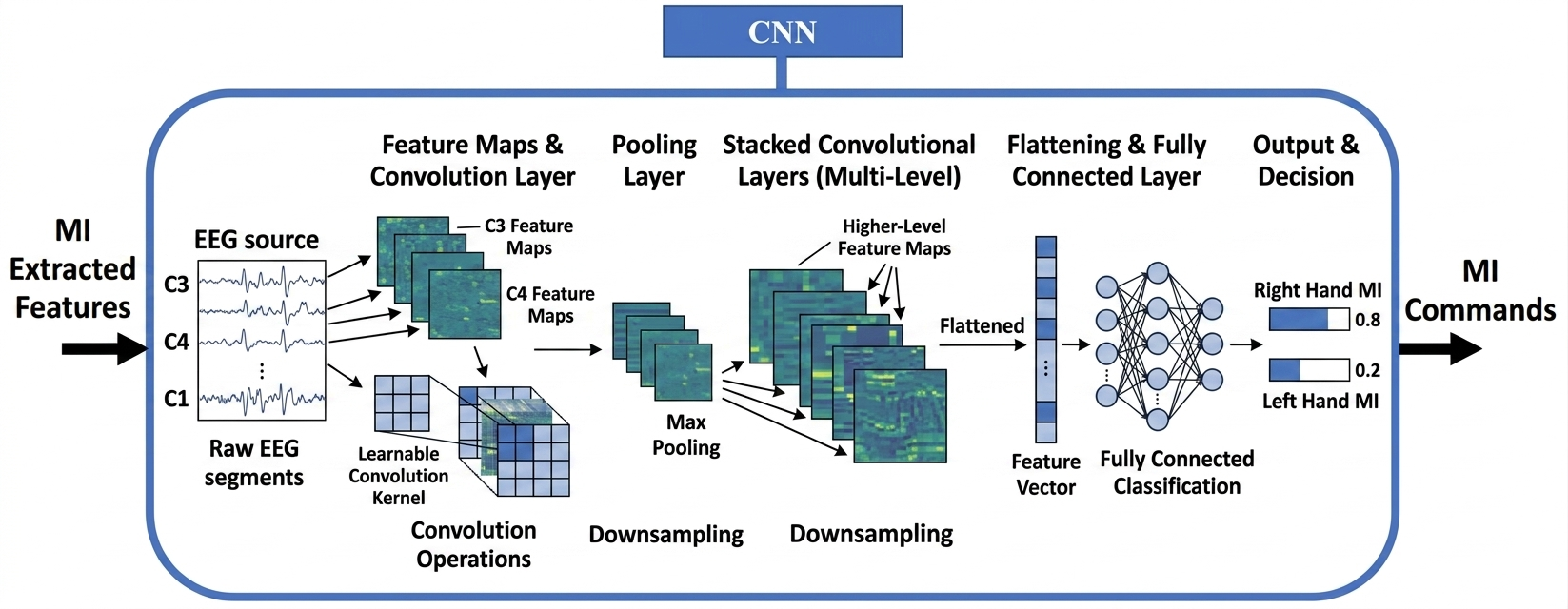}}
			\caption{Utilizing the CNN-based model for MI features classification.}
			\label{Fig_CNN}
		\end{figure*}

		\textbf{C) Fuzzy‑Integrated Neural Models:} The fuzzy neural networks (FNNs) combine the learning capability of neural networks with the reasoning mechanism of fuzzy logic to handle uncertainty and imprecision in data. They operate by first transforming input features into fuzzy membership values (fuzzification), applying fuzzy inference rules to model relationships between inputs and outputs, and then producing a final decision through defuzzification. The neural network component adjusts membership functions and rule parameters through learning algorithms, typically gradient-based optimization. In MI EEG classification, FNNs are particularly useful because EEG signals are noisy and highly variable; the fuzzy framework effectively models ambiguity in MI features (e.g., spectral or spatial patterns), while adaptive learning improves CA and robustness across subjects \cite{yang2014adaptive, li2025wavtsk}.

		\textbf{D) Probabilistic Neural Models:} The probabilistic neural models apply statistical decision principles to classify input data based on estimated probability distributions rather than deterministic boundaries. In probabilistic neural networks (PNNs), classification is achieved by computing likelihood functions for each class using kernel-based estimations, and selecting the class with the highest posterior probability; this makes them inherently robust to noise and well-suited to small training sets \cite{chen2026multiscale, yang2023motor}. Although tree‑based gradient boosting models such as XGBoost are not truly neural networks, they share a multilayer decision flow concept, aggregating many weak learners (decision trees) into a strong predictive model through iterative boosting. For MI feature classification, both PNNs and XGBoost offer complementary advantages: PNNs provide smooth probabilistic modeling of EEG patterns, while XGBoost efficiently captures nonlinear relationships and feature interactions, resulting in high accuracy and generalization across subjects and sessions \cite{xiao2025improving,zhang2026joint}.

		\textbf{4) Deep Learning}
		The DL algorithms are increasingly used for MI features classification due to their ability to automatically learn hierarchical feature representations directly from raw or minimally processed EEG signals \cite{zhao2020deep,kwon2019subject}. Unlike traditional handcrafted feature extraction methods, deep models can capture complex spatial–temporal patterns and inter-channel dependencies in EEG data \cite{dose2018end,cho2021neurograsp}. This leads to improved CA, better generalization across subjects, and reduced reliance on expert feature design, making DL a powerful and adaptive approach for modern MI systems. The most common DL-based classification algorithms can be categorized as follows.

		\textbf{A) Convolutional Neural Network:} The CNNs operate on the principle of hierarchical feature extraction through convolutional filters that learn spatial or temporal patterns from input data \cite{zhao2025multi,bang2021spatio,kwon2019subject,xu2018wavelet,schirrmeister2017deep,sakhavi2015parallel}. In MI EEG classification, 1D CNNs typically process temporal signal patterns from individual EEG channels, while 2D CNNs treat multi‑channel EEG features (e.g., time–frequency maps) as images to capture spatial correlations across brain regions \cite{zhang2021adaptive,zhang2020data,tang2017single}. The multi‑scale CNNs (MSCNNs) further enhance performance by applying filters of different sizes to extract features at multiple temporal or spatial resolutions, helping the model learn both global and local EEG dynamics. Also, VGG-16 has occasionally been adopted through transfer learning for EEG image-based classification despite its relatively high computational complexity \cite{xu2019deep}. Furthermore, hybrid-scale CNN (HS-CNN) employs convolution kernels of multiple sizes together with data augmentation to capture EEG features at different temporal scales and improve classification robustness \cite{dai2020hs}.

		Network design choices also range from monolithic CNNs, which integrate all processing stages within a single end-to-end architecture, to modular networks that decompose feature extraction and classification into separately trained, interchangeable sub-networks \cite{ olivas2019classification}. Architectural variants such as spatial CNNs, which emphasize inter-channel spatial filtering, and standard CNNs, which apply conventional convolutional stacks without specialized spatial or temporal branches, have also been benchmarked for MI decoding, often alongside classical baselines such as SVM with a cubic polynomial kernel \cite{mahmood2021wireless}. The EEGNet, a compact and efficient CNN architecture designed specifically for BCI applications, uses depthwise and separable convolutions to learn frequency‑specific activations while maintaining a lightweight design suitable for real‑time systems \cite{lawhern2018eegnet}. Additionally, pre‑trained CNNs such as AlexNet, ResNet50, and InceptionV3, originally trained on large visual datasets, can be fine‑tuned on EEG‑derived image‑like features (e.g., spectrograms), leveraging transfer learning to improve accuracy and reduce training time. The model-agnostic meta-learning (MAML) and MAML-based lightweight neural networks (MAML-LNNs) have recently been proposed to improve subject adaptation under limited calibration data by learning initialization parameters that rapidly adapt to new users \cite{raza2026maml}. Additionally, the compact temporal convolutional network (CTCNet) reduces computational complexity for online MI decoding while preserving high CA, making it well suited for real-time BCI operation \cite{kant2020cwt,wu2020transfer}. Moreover, lightweight ConvNeXt models with a Softmax classifier have also been developed to enable efficient subject-independent MI classification with reduced computational complexity \cite{modak2026novel}.

		Utilizing the CNN-based model for MI features classification includes a few steps, as depicted in Fig. \ref{Fig_CNN}. (1) Feature Maps and Convolution Layer: Input raw EEG segments from target spatial channels (e.g., C3, C4, C1) are scanned by learnable sliding convolution kernels to extract localized spatial-temporal feature maps. (2) Pooling and Multi-Level Downsampling: Max pooling layers downsample the generated matrices, reducing spatial dimensionality while retaining invariant structural features before passing them through stacked multi-level convolutional layers to form deeper high-level feature maps. (3) Flattening and Fully Connected Layer (FCL): The final multi-dimensional feature maps are flattened into a single-column 1D feature vector and routed into a fully connected classification network. (4) Output and Decision: The network processes the integrated features to compute final class probabilities (e.g., 0.8 for Right Hand MI vs. 0.2 for Left Hand MI), generating the corresponding definitive MI command.

		\textbf{B) Hybrid CNN–Recurrent Models:} The HYB CNN–recurrent models combine the strengths of convolutional and sequence‑based architectures to improve MI EEG classification by jointly learning spatial and temporal dynamics \cite{zhou2018classification}. In these models, CNN layers first extract local spatial or time–frequency features from raw EEG signals or transformed representations, capturing patterns such as sensorimotor rhythms or localized activations. The extracted feature maps are then passed to LSTM layers, which model temporal dependencies, sequential fluctuations, and long‑range correlations inherent in MI tasks \cite{li2022motor,luo2018exploring}. This HYB structure enables the system to learn both where and when discriminative patterns occur in the EEG, making CNN‑LSTM architectures particularly effective for decoding complex MI‑related brain activity \cite{zhang2021hybrid}. Furthermore, CNN-LSTM feature fusion network (FFN) extracts spatial features using CNNs and temporal features using parallel LSTM modules before fusing intermediate representations for classification \cite{li2022motor}. Additionally, hybrid deep neural network with transfer learning (HDNN-TL) combines CNN and LSTM networks with transfer learning to learn shared spatial-temporal representations while adapting efficiently to new subjects \cite{zhang2021hybrid}.

		\textbf{C) Deep Belief Network:} The deep belief networks (DBNs) are hierarchical generative models composed of stacked restricted Boltzmann machines (RBMs) that learn to capture complex, high‑order statistical dependencies within data \cite{zhuang2025self, alencar2023embedded,mewada2025low}. As MI feature classifiers, DBNs automatically extract meaningful patterns from EEG signals by learning unsupervised representations at multiple abstraction levels, reducing noise and highlighting discriminative features relevant to different MI tasks. The convolutional deep belief network (CDBN) extends this concept by introducing convolutional and pooling operations, enabling localized and translation‑invariant feature extraction in EEG time–frequency or spatial domains \cite{ren2014convolutional}. Meanwhile, the frequential deep belief network (FDBN) focuses on learning spectral features directly from EEG frequency bands, improving sensitivity to oscillatory components such as $\mu$ and $\beta$ rhythms essential for MI decoding \cite{lu2016deep}. These architectures allow robust, data‑driven feature learning that supports high CA even with limited labeled EEG data.

		\textbf{D) Transformer‑Based Models:} The DL-based transformer (TRF) architectures, the foundational models behind modern LLMs such as ChatGPT, have emerged as powerful MI feature classifiers \cite{kostas2021bendr,cao2026hybrid}. They utilize self-attention mechanisms to capture complex dependencies within EEG signals while eliminating the need for recurrent structures \cite{zhao2025multi,hameed2024temporal}. In MI decoding, TRs treat EEG time samples (or short time segments, channels, or time–frequency patches) as a sequence of tokens and use multi‑head self‑attention to learn how different time points, channels, and frequencies relate to each other \cite{kostas2021bendr,huang2026continual, kwak2022fganet}. This allows the model to emphasize the most informative neural patterns, such as task‑specific modulations of $\mu$ and $\beta$ rhythms, while down‑weighting noisy or irrelevant activity. Positional encodings preserve temporal order, enabling the TR to capture both short‑ and long‑range temporal dynamics critical for MI tasks. As a result, TR‑based models can learn rich, global spatio‑temporal representations of EEG data, often achieve competitive or superior classification performance, and offer better scalability to longer recordings and multi‑channel setups \cite{su2025transformer,qamar2026multi}. Furthermore, temporal-spatial transformer (TST) employs self-attention across both temporal and spatial EEG dimensions to capture global dependencies and improve MI classification performance \cite{hameed2024temporal}.

\begin{figure}[!t]
			\centerline{\includegraphics[scale=0.24,trim=0cm 0cm 0cm 0cm,clip=true]{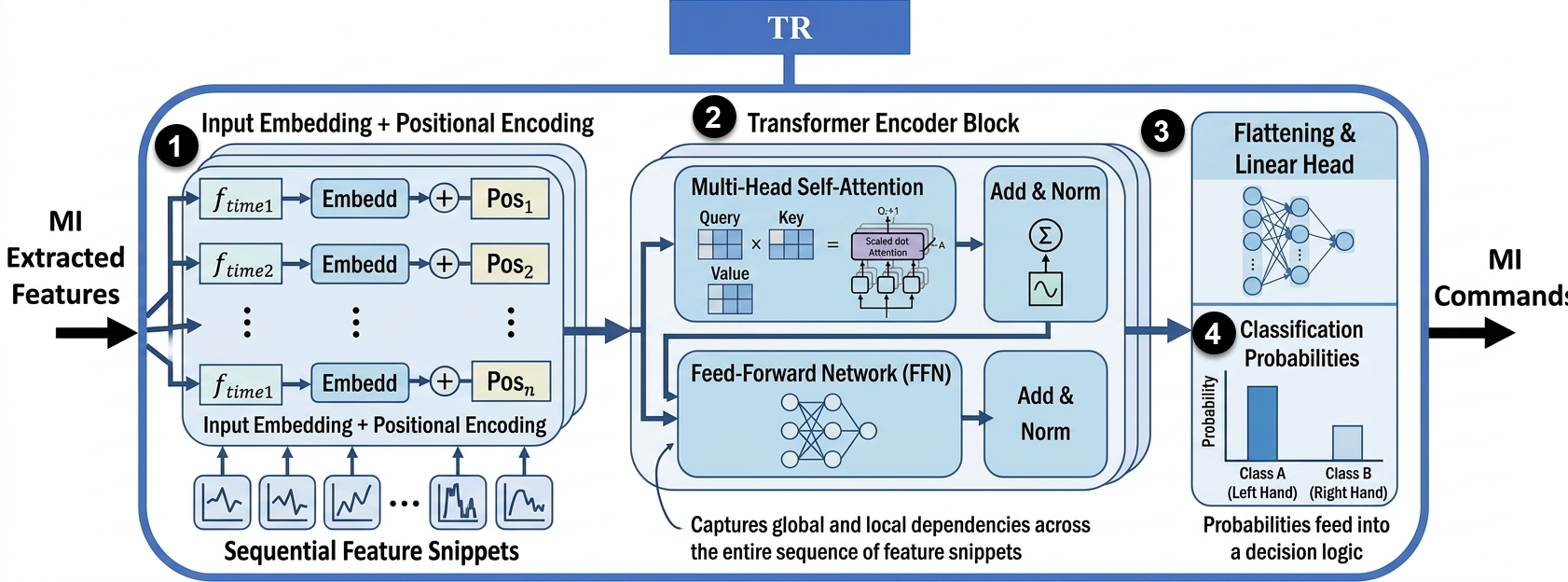}}
			\caption{The TR-based models as the MI features classifier.
			}
			\label{Fig_TR}
		\end{figure}

		TR-based models used as MI feature classifiers perform classification in a few steps, as shown in Fig. \ref{Fig_TR}. (1) Input Embedding and Positional Encoding: The incoming MI extracted features are partitioned into sequential feature snippets $[f_{\text{time}1}, f_{\text{time}2}, \dots, f_{\text{time}n}]$; these snippets are projected via an embedding layer and combined with positional encodings ($\text{Pos}_n$) to preserve critical temporal sequence information. (2) Transformer Encoder Block: The encoded sequences pass through a Multi-Head Self-Attention layer—mapping Queries ($Q$), Keys ($K$), and Values ($V$) via scaled dot-product attention to capture long-range global and local dependencies—followed by residual connections ($\Sigma$), layer normalization (Add \& Norm), and a position-wise Feed-Forward Network (FFN). (3) Flattening and Linear Head: The multi-head contextual representations are flattened into a unified classification head to extract integrated spatial-temporal profiles. (4) Classification Probabilities: The network evaluates the final vectors to yield distinct target command probabilities (e.g., Class A/Left Hand vs. Class B/Right Hand), which feed into decision logic to trigger the final discrete MI command.

		\textbf{5) Generative AI (GenAI):} The generative AI (GenAI) models are built on the principle of learning data distributions rather than just discriminative boundaries. They aim to capture and reproduce the underlying structure of the input data, enabling the generation or reconstruction of new, realistic samples. Among these, stacked autoencoders (SAEs) and variational autoencoders (VAEs) are prominent architectures \cite{mirzaei2021eeg,velakanti2025improving}. The SAEs consist of multiple layers of encoders and decoders that learn hierarchical representations by minimizing reconstruction error, allowing them to extract compact, non‑linear features from EEG data that reveal spatial and temporal patterns useful for MI classification \cite{tabar2017novel}. Moreover, the VAEs, as probabilistic extensions of autoencoders, learn latent variable distributions and enable smooth interpolation between feature states, effectively capturing the variability and complexity of EEG signals \cite{dai2019eeg}. For MI decoding, these generative frameworks can uncover deep latent representations of neural activity, improving robustness against artifacts and enhancing generalization across subjects or sessions.

\begin{figure}[!t]
			\centerline{\includegraphics[scale=0.25,trim=0cm 0cm 0cm 0cm,clip=true]{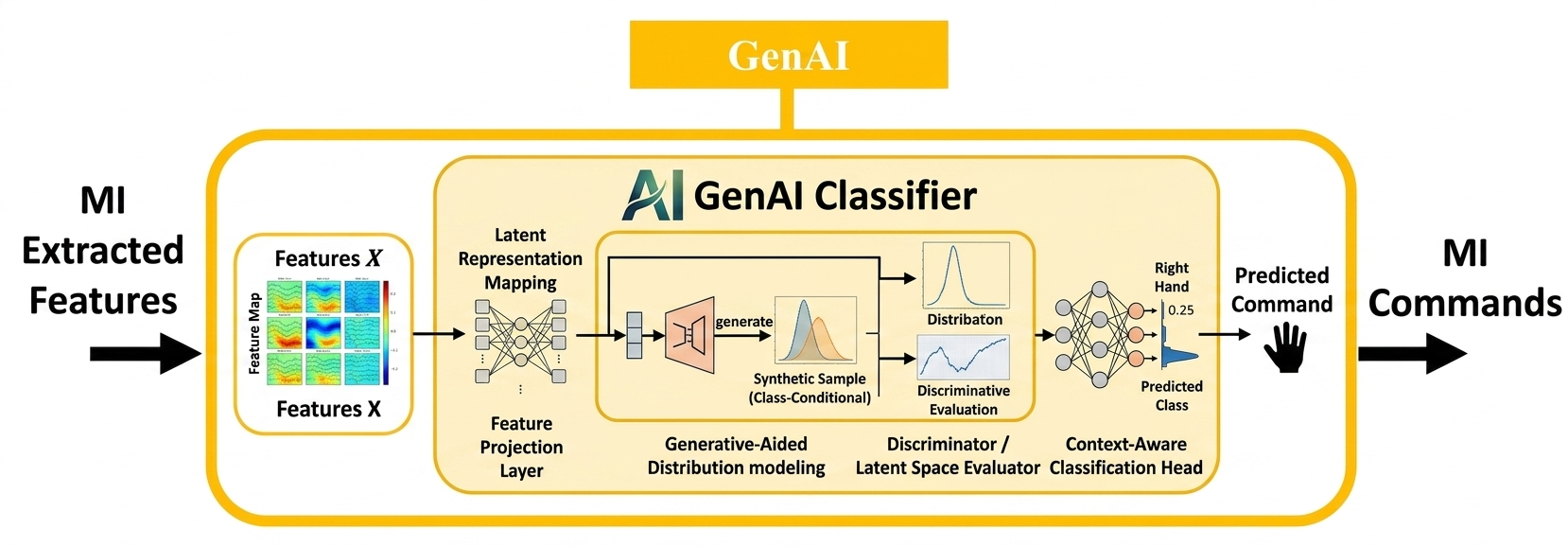}}
			\caption{The GenAI-based models as the MI features classifier.
			}
			\label{Fig_GenAI}
		\end{figure}

		The performance of GenAI-based models as the MI features classifier is shown in a few steps, as shown in Fig. \ref{Fig_GenAI}. (1) Feature Projection and Latent Mapping: The multi-dimensional input MI extracted features ($X$) are routed through a projection layer to map the signal topologies into a continuous latent representation space. (2) Generative-Aided Distribution Modeling: Class-conditional generative components (such as diffusion, VAE, or GAN architectures) model the underlying feature distributions to construct high-fidelity synthetic samples and establish robust density patterns. (3) Discriminator and Latent Space Evaluation: A discriminative evaluator assesses both raw and generated feature distributions to extract fine-grained boundary probabilities. (4) Context-Aware Classification Head: The evaluated latent features are passed to a deep classification network to compute final class metrics (e.g., Right Hand vs. Left Hand MI probabilities), yielding the predicted intent to output the final discrete MI command.

		\textbf{6) Fuzzy Models}
		Fuzzy models classify MI EEG signals by handling uncertainty through degrees of membership instead of fixed boundaries \cite{ko2019multimodal}. They use adaptive “if–then” rules and membership functions to represent nonlinear, overlapping brain patterns \cite{yang2014adaptive}. This approach captures subtle EEG variations, offering both robustness and interpretability in MI feature classification. The typical fuzzy models used for MI classification can be grouped as follows.

		\textbf{A) Type‑1 Fuzzy Logic Systems:} Type‑1 fuzzy logic–based classifiers use fixed membership functions to handle uncertainty in MI EEG signals by mapping features to classes through interpretable rules. The classical fuzzy logic (FLO) and general fuzzy rule systems define “if–then” relationships, while Takagi–Sugeno–Kang fuzzy neural network (TSK FNNs) combine fuzzy rules with neural networks for adaptive, data‑driven learning. Also, the triangular fuzzy numbers (TFNs) provide simple, efficient membership shapes, allowing these models to capture nonlinear MI patterns with clarity and low computational cost \cite{donovan2018motor, yao2026multiview}..

		\textbf{B) Type‑2 Fuzzy Logic Systems:} Type‑2 FLO systems extend classical fuzzy models by introducing uncertainty into the membership functions themselves, enabling more flexible handling of EEG variability. The interval type‑2 FLO systems (IT2FLS) model this uncertainty using upper and lower membership bounds, while IT2‑TFNs represent it with interval‑based triangular fuzzy sets. Also, SRIT2NFIS combines this higher‑order fuzzy reasoning with neural adaptation, offering enhanced robustness and generalization for complex, noisy MI feature classification \cite{herman2016designing, yao2026multiview}.

		\textbf{C) Neuro‑Fuzzy Systems:} Neuro‑fuzzy systems (NFS) combine fuzzy logic’s interpretable rules with neural networks’ learning ability to classify MI features adaptively from EEG data \cite{yang2014adaptive}. Models like FNN and adaptive neuro‑fuzzy inference system (ANFIS) learn fuzzy rules and membership parameters directly from data, while optimization‑enhanced schemes such as fuzzy system optimized using PSO (FZ‑PSO) refine fuzzy structures using evolutionary search \cite{yang2014adaptive}. Furthermore, self‑regulating interval type‑2 neuro‑fuzzy inference system (SRIT2NFIS) further extends this by integrating interval Type‑2 fuzzy sets into a neuro‑fuzzy inference architecture, improving robustness to noise and uncertainty in MI classification \cite{yang2023optimal,thamaraimanalan2025exploiting}.

		\textbf{D) Fuzzy Integral‑Based Models:} The fuzzy integral (FZI) model fuses multiple MI features or classifier outputs by modeling their interactions through fuzzy measures rather than assuming independence \cite{ko2019multimodal}. This allows the system to capture nonlinear relationships and complementary information among EEG features, leading to improved decision reliability and robustness in MI classification tasks. A related approach, multiple linear discriminant analysis with fuzzy integral fusion (MLDA-FZI), first applies MLDA to classify EEG patterns within each sub-band spectrum independently, then fuses the sub-band decisions using a fuzzy integral to produce a final, more robust MI classification \cite{wu2016fuzzy}. Other intelligent classifiers include the Self-Regulated Supervised Gaussian Fuzzy Adaptive System (SRSG-FasArt) \cite{jafarifarmand2017new}.

		\textbf{7) Learning Paradigms \& Hybrids}
		
		The HYB learning paradigms for MI EEG classification integrate complementary model types to better capture spatial, temporal, and subject‑specific variability in brain signals \cite{song2022eeg, xie2022transformer}. Architectures such as CNN–LSTM, CNN-TRF, and RNN–LSTM combine convolutional or recurrent feature extractors with LSTM layers to jointly learn spatial patterns and long‑range temporal dependencies \cite{li2022motor}. Transfer learning (TL) with CNN‑based or EEGNet‑style models reuses knowledge from related tasks or subjects, reducing training data demands and improving generalization \cite{lawhern2018eegnet,azab2019weighted,he2019transfer}. The supervised self-training (SST) strategy further reduces calibration burden by progressively fine-tuning a pre-trained subject-independent model using labeled online data, enabling calibration-free BCI operation without a dedicated offline session. The continual learning (CNL) mechanisms allow the classifier to incrementally incorporate new MI patterns while retaining previously learned knowledge, mitigating the performance degradation common in extended MI-BCI sessions. Ensemble techniques further boost robustness by aggregating diverse base classifiers, smoothing out individual model errors \cite{dai2019eeg}.

		On the other hand, the optimization‑driven HYBs enhance these pipelines by automatically tuning model structures and decision mechanisms. The ESN coupled with GTA, FI optimized via PSO, and Multi‑objective Grey Wolf optimization for twin SVMs all exemplify search‑based calibration of hyperparameters, feature weights, or fusion rules \cite{yang2018deep}. The HYB fuzzy–neural and fuzzy–optimization approaches, along with HYB CNN or RNN‑DL pipelines, exploit the interpretability of fuzzy logic, the representational power of deep networks, and the exploratory strength of metaheuristics to yield more accurate and robust MI feature classifiers under noisy and non‑stationary EEG conditions. Also, the weighted sum of prediction scores (WSPS) is a simple decision-level fusion strategy that combines outputs from multiple classifiers using weighted confidence scores \cite{kwak2022fganet}.

		\textbf{8) Other Classification Methods}
		Finally, a wide range of additional MI classification techniques falls into an “other methods” category, encompassing approaches that do not fit neatly into conventional model families. These include statistical classifiers, signal‑quality assessment frameworks, and functional‑connectivity‑based analyses that capture relationships between EEG channels rather than relying solely on localized features. Various alternative MI detection strategies, ranging from correlation‑based metrics to graph‑theoretic models and traditional pattern‑recognition schemes, also contribute valuable perspectives, offering complementary ways to improve MI decoding performance \cite{hou2022gcns}.

		\subsection{EEG Hardware-Implemented Algorithms}
		
		Practical implementation of EEG‑based MI classification systems on hardware platforms is a crucial step toward translating research algorithms into real‑world applications \cite{azghadi2020hardware,wei2020review}. Hardware deployment not only validates an algorithm’s practical feasibility but also reveals its robustness under realistic operating conditions, including noise, non‑stationary signals, and limited computational resources \cite{krishna2023sparsity}. Furthermore, implementing EEG‑based MI classifiers on physical platforms ensures that these systems can perform reliably in everyday scenarios, bridging the gap between laboratory research and functional, real‑time MI-BCI solutions. A list of the most important hardware implementations of MI systems is reported in Table \ref{Table_HW_Algorithms}. In comparison to Table \ref{Table_SW_Algorithms}, in this table, a new hardware (HW.) column is identifies the hardware platform utilized for the practical implementation. Also, the following keywords are considered in this table: resource usage (REU), power consumption (PC), area efficiency (ARE), speed-up (SPU), latency (LAT), memory reduction (MMR), inference time (INT), reconstruction accuracy (RA), and compression ratio (CR).

		\begin{table*}[t]
			\caption{State-of-the-art EEG-based MI-BCI hardware implementations reported in the literature in recent decade.}
			\tiny  
			\label{Table_HW_Algorithms}
			\centering
			
			\begin{tabular}{|>{\centering\arraybackslash}m{0.3cm} |
					>{\centering\arraybackslash}m{0.3cm} |
					>{\centering\arraybackslash}m{2.75cm} |
					>{\centering\arraybackslash}m{1cm} |
					>{\centering\arraybackslash}m{1.5cm} |
					>{\centering\arraybackslash}m{1.7cm} |
					>{\centering\arraybackslash}m{1cm} |
					>{\centering\arraybackslash}m{3.95cm} |
				}
				
				\hline
				
				\textbf{Ref.} & 
				\textbf{Year} &
				\textbf{HW.} &
				\textbf{Pre-processing} &
				\textbf{Feature Extraction} &
				\textbf{Feature Classification} &
				\textbf{Dataset Category} &
				\textbf{Performance} 
				\\ 
				
				\hline 
				

				\cite{rao2026calibration} &
				2026 & 
				Wearable EEG headband, 5 electrodes (4 wet, 1 dry) & 
				NFL, BPF, BWF, CAR &
				CTCNet &
				CTCNet, SST &
				(8) &
				SICA: 86\%
				\\ \hline

				\cite{huang2026continual} &
				2026 & 
				wireless EEG system (InMex; WellFulfill Ltd., Taoyuan, Taiwan); Bluetooth; 8 dry electrodes; Oculus VR headset (paired with EEG cap for immersive AO+MI training) & 
				BPF, BSF &
				CNN--TRF &
				CNN--TRF, CNL &
				(8) &
				CA: 47.4\% (Session 1), 66.2\% (Session 3); 52.8\% (offline cross-validation benchmark)
				\\ \hline

				\cite{kim2025efficacy} &
				2025 & 
				recoveriXPRO 16 channels stroke rehabilitation system; Estim FES device; 32-channel g.Nautilus EEG system; TMS device for motor evoked potentials (MEPs)/resting motor threshold (RMT) & 
				BPF, BWF &
				LAP, PDC &
				LDA &
				(8) &
				Improved muscle strength in the wrist extensor (MRC-WE), MEAD:0.52, p:0.036 
				\\ \hline

				\cite{feng2022efficient} &
				2022 & 
				60-channel EEG signal; Xilinx KC705 FPGA evaluation board; Artix-7 FPGA, Xilinx VC707 FPGA & 
				BPF &
				EEGNet &
				EEGNet &
				(8) &
				CA: 93.06\%; REU reduced to 2.54\% area and 3.66\% PC
				\\ \hline

				\cite{feng2022efficient_model} &
				2022 & 
				FPGA, 65 nm CMOS LP process & 
				BPF &
				MC-EEGNet &
				MC-EEGNet &
				- &
				ARE: 87.22\%, Eff: 20.77\% better than a RISC-V MCU, SPU: 3.7$\times$
				\\ \hline

				\cite{mahmood2021wireless} &
				2021 & 
				Wireless soft scalp electronics system with Samsung Gear VR; 6 flexible microneedle electrodes & 
				BPF, BWF &
				PSD, WCM, CNN &
				Spatial CNN, Standard CNN, SVM (cubic polynomial kernel) &
				(8) &
				CA: 93.22 $\pm$ 1.33\% (4-class, spatial CNN)
				\\ \hline

				\cite{marcos2020real} &
				2020 & 
				FPGA Zybo board (Zynq xc7z010-clg400) & 
				BPF &
				WPD &
				LDA &
				(10) &
				CA: 80\%; Sen: 85\%, Spe: 73.3\%; AUC $\approx$ 0.79--0.80; ITR 2.78 bits/min; LAT: 7.5 ms; PC: 0.102 W.
				\\ \hline

				\cite{wang2020accurate} &
				2020 & 
				ARM Cortex-M4F MCU and ARM Cortex-M7 & 
				BPF, SWS &
				EEGNet &
				EEGNet &
				(6) &
				CA: 82.43\% (2-class), 75.07\% (3-class), 65.07\% (4-class), LAT/energy: 101 ms \& 4.28 mJ/inference (Cortex-M4F), 44 ms \& 18.1 mJ/inference (Cortex-M7).
				\\ \hline

				\cite{schneider2020q} &
				2020 &
				Mr.Wolf PULP (Parallel Ultra-Low-Power) SoC &
				BPF &
				EEGNet CNN &
				Q-EEGNet, quantized EEGNet &
				(4) &
				SPU: 64$\times$, MMR: up to 85\%; 5.82 ms \& 0.627 mJ per inference; 256$\times$ more Eff than EEGNet.
				\\ \hline

				\cite{paszkiel2020brain} &
				2020 & 
				Emotiv EPOC Flex (32-channel wireless EEG headset); MyoWare Muscle Sensor (EMG, AD8648 amplifier) & 
				CAR, LAP, BPF &
				FFT &
				EUS &
				(8) &
				CA increased after MI training (e.g., 59--68\% $\rightarrow$ 69--79\%); overall $\sim$70\% eff in modulating 7--12 Hz (mu) activity. 
				
				\\ \hline

				\cite{olivas2019classification} &
				2019 & 
				OpenBCI device (8 dry electrodes, Ultracortex Mark IV headset) & 
				BPF, Type II IIR filter, NFL, ITA &
				DFBCSP &
				Monolithic CNN, Modular network &
				(4), (7) &
				Dataset (4): Monolithic CA:78.41\%, K=0.59. Modular: CA:80.03\%, K=0.61. Dataset (7): Monolithic: CA:73\%, K=0.64; Modular: CA:76\%, K=0.67. INT: Monolithic 0.58s, Modular 1.04s per sample (GPU).
				\\ \hline

				\cite{ma2019fpga} &
				2019 & 
				FPGA --- PYNQ-Z1 board (Xilinx Zynq-7020 chip) & 
				AEP, HMW &
				CNN &
				CNN &
				(4) &
				CA: 80.5\%, $\sim$7\% CA drop with 8-bit fixed-point; subject a: 83\%, subject b: 78\%; FPGA 8$\times$ faster than computer. PC: FPGA 0.025 W vs. computer 140 W.
				\\ \hline

				\cite{foong2019assessment} &
				2019 & 
				Neurostyle nBETTER system EEG acquisition hardware; 24 unipolar Ag/AgCl electrodes (10-20 system), 256 Hz sampling, 24-bit resolution, $\pm$300 mV range & 
				BPF, EMD, CCA, LAP &
				FBCSP, MIN &
				FLDA &
				(8) &
				CA: > 57.5\%. A significant positive correlation between relative beta power (frontal/central) and CA.
				\\ \hline

				\cite{li2019hybrid} &
				2019 & 
				NeuroScan-NuAmps 40-channel EEG; Delsys 2-channel EMG & 
				NFL, BPF, EMG processing &
				CSP, FFT &
				BPNN &
				(8) &
				CA: >80\%; MAX joint angle tracking errors <0.1 rad. 
				\\ \hline

				\cite{malekmohammadi2019efficient} &
				2019 & 
				Virtex-6 FPGA, ML605 evaluation board & 
				DC-blocker, LAP, CAR, BPF, HPF, NFL &
				SCSSP, MIN, LDA &
				LDA, SVM &
				(3),(4),(8) &
				CA: 73.54\% (2, 3 classes); 67.2\% (7, 4 classes); 80.55\% (4, first 2 classes); 81.9\% (9, 2 classes); LAT: 11.66 ms.
				\\ \hline

				\cite{belwafi2018embedded} &
				2018 & 
				Altera Stratix-IV FPGA development board & 
				WOLA &
				CSP &
				LDA &
				(1),(2),(4),(8) &
				PC: 0.7 W, LAT: 0.430 s/trial. CA: 78.85\% (4), 84.28\% (1), 67.28\% (2), 80.25\% (8).
				\\ \hline

				\cite{shrivastwa2018fpga} &
				2018 & 
				Zedboard (Xilinx Zynq, ARM Cortex-A9 + FPGA fabric); Heterogeneous ARM+FPGA architecture; FPGA on Virtex-7 (XC7VX485T-2FFG1761C) & 
				CPS &
				MLP &
				- &
				(8) &
				RA: 89.85\% at 50\% CR; CA: 78–93\%; PSNR up to 23.6 dB.
				\\ \hline

				\cite{mccrimmon2017performance} &
				2017 & 
				Custom BCI: 8-channel EEG amplifier; INA128 instrumentation amp + 2$\times$ OPA4241 op-amps. Conventional BCI (comparison): NeXus-32-channel commercial EEG amplifier. & 
				BPF, NFL &
				LDA, PCA &
				BYC-Decoding &
				(8) &
				CA: 93.6$\pm$4.3\% (custom), 96.2$\pm$1.8\% (conventional).
				\\ \hline

				\cite{liu2017feature} &
				2017 & 
				Custom real-time BCI system, 16 channels, UE-16B EEG amplifier & 
				BPF, BWF &
				CSP, LCD, FFA, LAM, GTA, PSO &
				SRDA &
				(4),(8) &
				CA: 56.91\% (4, SRDA alone), 70.2\%, K=0.6 (4, FFA-LAM-SRDA).
				\\ \hline

				\cite{herman2016designing} &
				2016 & 
				g.tec EEG amplifier with AgCl electrodes; 2 bipolar EEG channels over C3/C4 (10/20 system) & 
				BPF &
				STFT, PCA &
				IT2FLS, LDA, FLDA, SVM &
				(11),(8) &
				Offline: IT2FLS significantly outperformed LDA, FLDA, and SVM (p<0.05). Online: IT2FLS delivered statistically significantly higher CA than LDA,FLDA, and SVM.
				\\ \hline

				\cite{wu2016fuzzy} &
				2016 & 
				Custom wireless, portable EEG acquisition device; 5 dry electrodes, pre-amplifier, Bluetooth, Kinova robotic arm for real-time control output & 
				FBD, HPF &
				SBCSP &
				MLDA-FZI-PSO &
				(8) &
				Offline (AUC, 4-fold CV $\times$10): Sugeno integral improved 0.968$\pm$0.063 $\rightarrow$ 0.998$\pm$0.040 after PSO; Choquet integral improved 0.992$\pm$0.014 $\rightarrow$ 0.998$\pm$0.003 after PSO (best overall); Online real-time test CA $\approx$86\%.
				\\ \hline

				\cite{leeb2015towards} &
				2015 & 
				BCI-controlled telepresence robot; g.USBamp 16-channel active EEG electrodes amp., telepresence robot. &
				BPF, NFL, LAP, WCM PSD &
				CVA &
				GSC &
				(8) &
				CA: Online >70\%. Time ratio (BCI/manual): 109.2$\pm$11.1\% (end-users) vs. 115.1$\pm$10.3\% (healthy), not significantly different (p>0.1).
				\\ \hline
				

			\end{tabular}
		\end{table*}

		The state-of-the-art algorithms for EEG-based MI systems are evaluated primarily through rigorous benchmarking and a set of critical performance metrics, including CA, latency, robustness against noise, generalization across subjects, and computational efficiency. Since EEG signals are inherently non-stationary and noisy, software algorithms must extract discriminative temporal–spatial features while maintaining low complexity to support real-time operation. Furthermore, model compactness and decision interpretability are also essential factors for deployment in wearable or portable systems \cite{karageorgos2020hardware,feng2022efficient}.

		In contrast, hardware benchmarking for MI-BCI systems focuses on the physical implementation and operational efficiency of the platform that executes these algorithms. Hardware evaluation emphasizes metrics such as real-time processing capability, power consumption, memory utilization, throughput, latency at the system level, and silicon area or resource usage (e.g., on FPGA or ASIC) \cite{rizal2022fpga,cao2024optimized}. While software benchmarks primarily measure algorithmic performance and predictive capability, hardware benchmarks assess how efficiently those algorithms can be implemented and executed on embedded or edge devices \cite{canilang2021edge,belwafi2018embedded}. The key difference is therefore that software evaluation focuses on algorithm accuracy and robustness, whereas hardware benchmarking concentrates on implementation efficiency and system-level constraints required for practical, low-power, real-time BCI deployment.

		\begin{figure}[!t]
			\centerline{\includegraphics[scale=0.25,trim=3cm 6cm 3cm 2.7cm,clip=true]{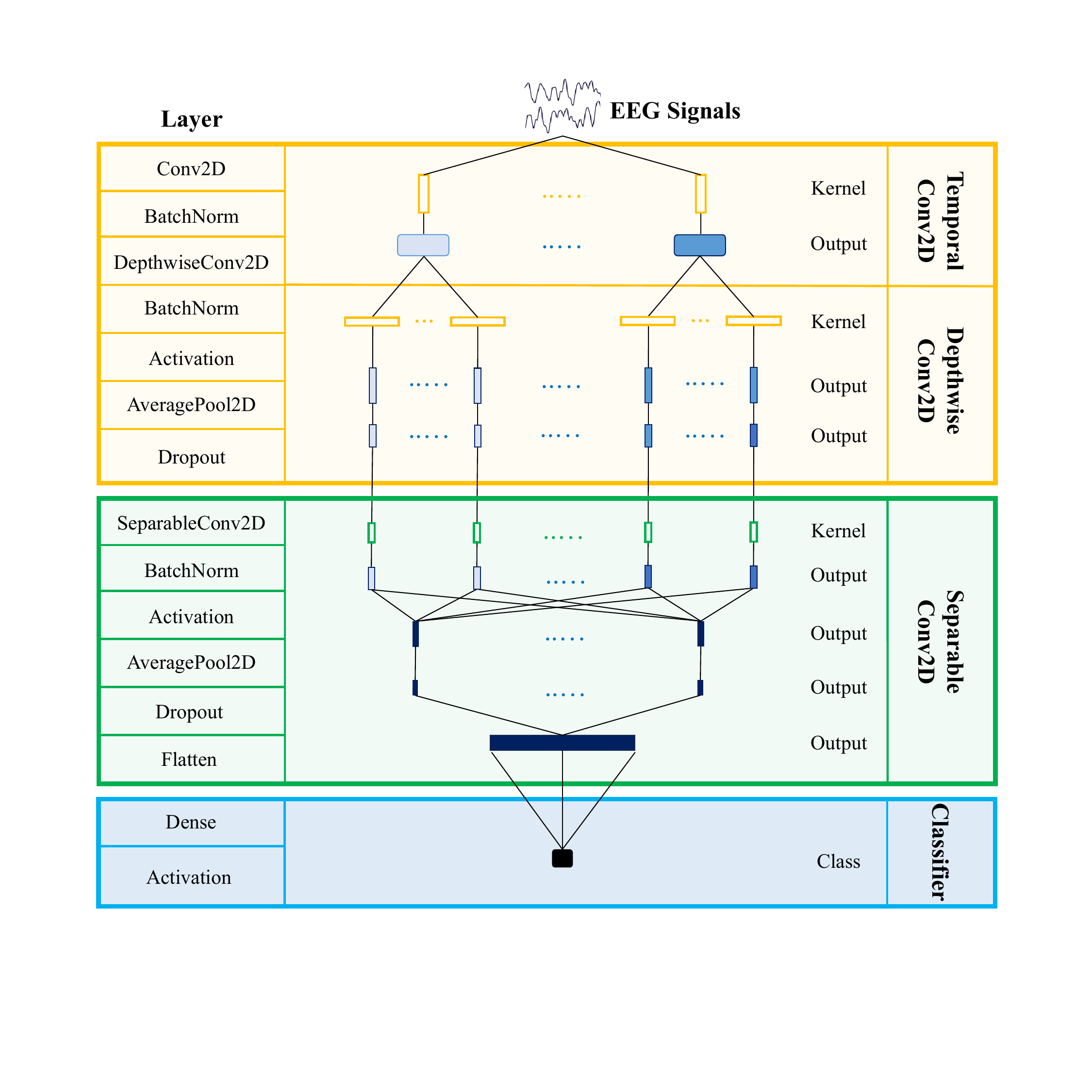}}
			\caption{The EEGNet CNN-based processing architucture for MI signals classification.}
			\label{Fig_EEGNet_V2}
		\end{figure}

		Modern DSP designs must balance high accuracy and robustness with low power consumption, real-time throughput, and hardware compactness. Achieving these optimization goals requires a holistic design approach, integrating algorithmic exploration with architectural and circuit-level innovations across both software and hardware domains. A wide range of hardware platforms, such as ASICs, FPGAs, DSP processors, and microcontroller-based systems running real-time operating systems (RTOS), play crucial roles in implementing efficient EEG-based MI-BCI systems \cite{feng2022efficient,wang2020accurate}. FPGAs offer high parallelism and reconfigurability, making them well-suited for prototyping and accelerating computationally intensive signal-processing and ML algorithms. DSP processors are optimized for real-time numerical operations such as filtering and feature extraction, making them effective for continuous EEG signal processing pipelines. RTOS-based microcontrollers enable deterministic task scheduling, low latency, and low power consumption, which are important for portable and embedded BCI devices \cite{belwafi2021embedded}. ASICs, although less flexible, provide ultra-low power consumption, compact size, and very high processing efficiency, making them suitable for final deployment in wearable or implantable BCIs \cite{cao2024optimized}. FPGAs are often preferred during research and development due to their flexibility and parallel processing capabilities, while ASICs and low-power microcontrollers become more advantageous for large-scale, power-constrained MI-BCI deployments \cite{karageorgos2020hardware}.

		On the other hand, designing hardware-compatible algorithms is essential for deploying EEG-based MI-BCI systems in real-world environments. Classical algorithms such as CSP and SVMs are well-suited for hardware implementation because of their relatively low computational complexity and predictable memory requirements \cite{zhao2020deep,li2019hybrid,belwafi2018embedded,herman2016designing}. More recently, advanced ML-based algorithms have enabled the deployment of hardware-friendly DL models, including quantized CNNs, lightweight RNNs, SNNs, and optimized architectures such as EEGNet and its low-precision variants \cite{dose2018end,huang2026continual,mahmood2021wireless,ma2019fpga}.

		EEGNet is particularly attractive for hardware realization because it employs depthwise and separable convolutions, significantly reducing the number of trainable parameters and arithmetic operations while maintaining high classification performance for MI tasks \cite{feng2022efficient,schneider2020q}, as illustrated in Fig. \ref{Fig_EEGNet_V2}. A model-compressed variant, MC-EEGNet, further reduces EEGNet's footprint through embedded channel selection, normalization merging, and product quantization, enabling more efficient on-device inference \cite{feng2022efficient_model}. Similarly, Q-EEGNet applies quantization together with algorithmic and implementation-level optimizations to execute MI-BCI inference on resource-limited edge devices, removing the need for separate network reduction steps and generalizing readily to other CNN architectures \cite{schneider2020q}.

		The architecture first applies a temporal convolution layer to learn frequency-selective representations of MI-related EEG rhythms, followed by a depthwise convolution layer that performs efficient spatial filtering across EEG channels, and a separable convolution layer that further extracts discriminative spatio-temporal features with minimal computational cost. Batch normalization, ELU activation, average pooling, and dropout layers are incorporated throughout the network to improve training stability, reduce feature dimensionality, and prevent overfitting before the extracted features are passed to a fully connected Softmax classifier for final decision making. Owing to its lightweight architecture, EEGNet is highly suitable for FPGA, ASIC, and embedded edge implementations \cite{canilang2021edge,feng2022efficient}, where techniques such as model compression, fixed-point arithmetic, quantization, and parallel pipelining further reduce hardware complexity, memory requirements, power consumption, and inference latency, enabling compact, energy-efficient, and real-time wearable BCI systems \cite{lawhern2018eegnet,feng2022efficient,wang2020accurate,schneider2020q}. Furthermore, the temporary constrained sparse group lasso (TCSGL-EEGNet) has recently been proposed to enhance EEGNet performance by enforcing structured sparsity, reducing computational complexity while maintaining CA \cite{deng2021advanced}.

		\subsection{General Overview of MI-BCI Signal Processing Pipeline Algorithms}

		\begin{figure*}[!t]
			\centerline{\includegraphics[scale=0.6,trim=7.7cm 9.6cm 7.1cm 21.5cm,clip=true]{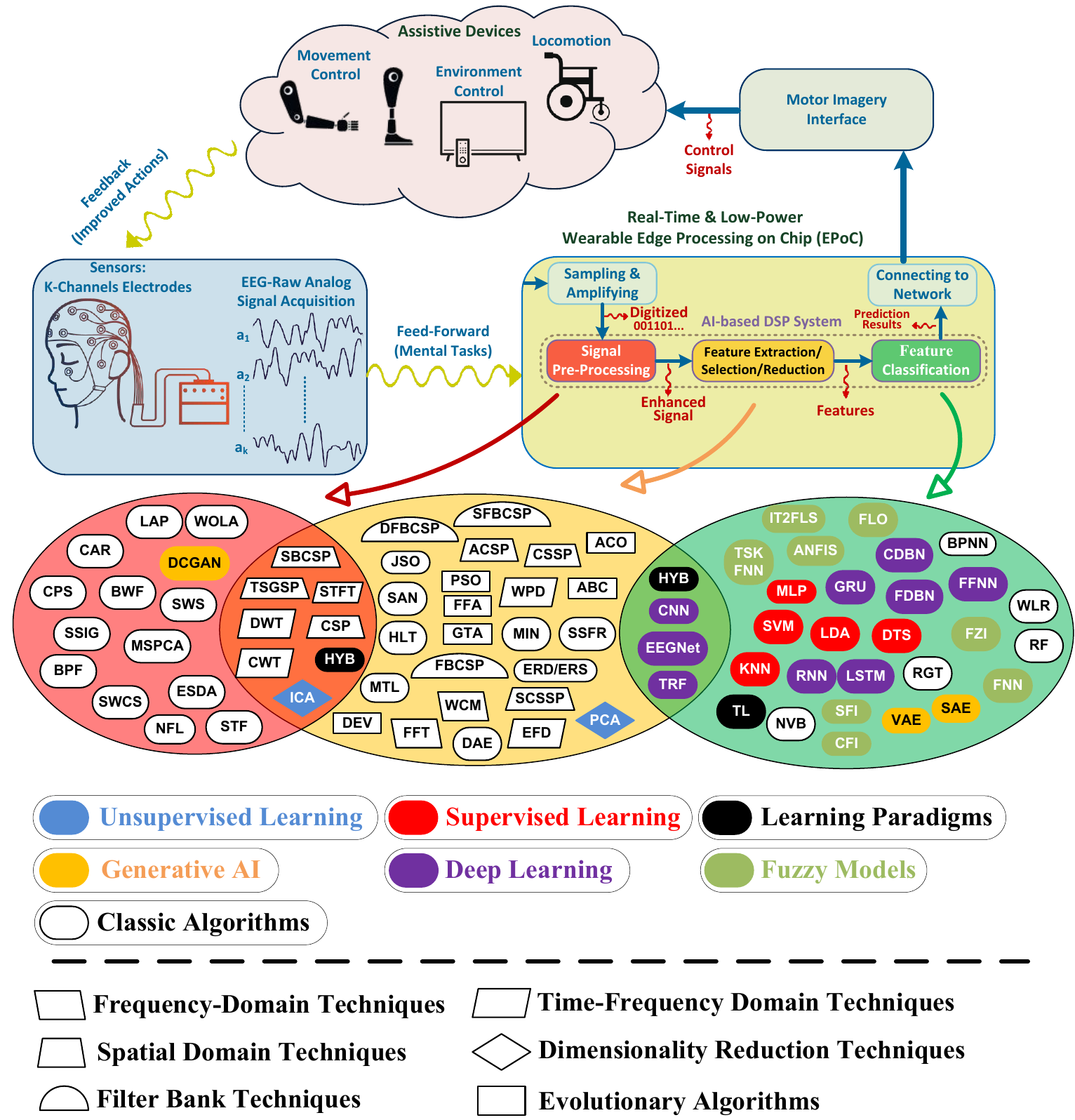}}
			\caption{State-of-the-art EEG-based MI-BCI system. The most common and most recent algorithms reported in the literature in recent decade for each blocks of the signal processing pipeline are illustrated.}
			\label{Signal_Processing_Pipeline_Algorithms}
		\end{figure*}

		As discussed in the previous sections, the signal processing pipeline for MI classification consists of several sequential stages, including preprocessing, feature extraction, feature reduction, and classification, each employing specific algorithms to transform raw neural signals into reliable control commands. In Fig. \ref{Signal_Processing_Pipeline_Algorithms}, the most common algorithms in recent literature over the past decade, as tabulated in Table \ref{Table_SW_Algorithms} and Table \ref{Table_HW_Algorithms}, for each of these signal processing blocks of EEG-based MI-BCI are illustrated and categorized. Algorithms used in these stages span multiple methodological families such as classical signal processing techniques, supervised and unsupervised learning methods, DL models, GenAI approaches, fuzzy models, and evolutionary optimization algorithms \cite{baig2017differential }. As discussed previously, the feature extraction methods are commonly categorized according to the domain in which the EEG signal is analyzed, including frequency-domain techniques, time–frequency domain approaches, and spatial-domain methods, while dimensionality reduction techniques are used to select the most discriminative features and reduce computational complexity. In addition, filter bank techniques are widely employed to capture discriminative information across multiple frequency bands \cite{fang2022feature}. This broad spectrum of algorithms reflects the ongoing effort to improve MI-BCI performance, robustness, and computational efficiency by combining advances in signal processing and ML algorithms.

		In the next section, some of the most promising technologies with the potential to revolutionize the current state of the art in MI‑BCI systems are  discussed.

		\section{Converging Technologies for MI-BCI Systems}
		\label{sec:Converging_Technologies}

		\begin{figure*}[!t]
			\centerline{\includegraphics[scale=0.3,trim=7cm 16.4cm 10cm 7cm,clip=true]{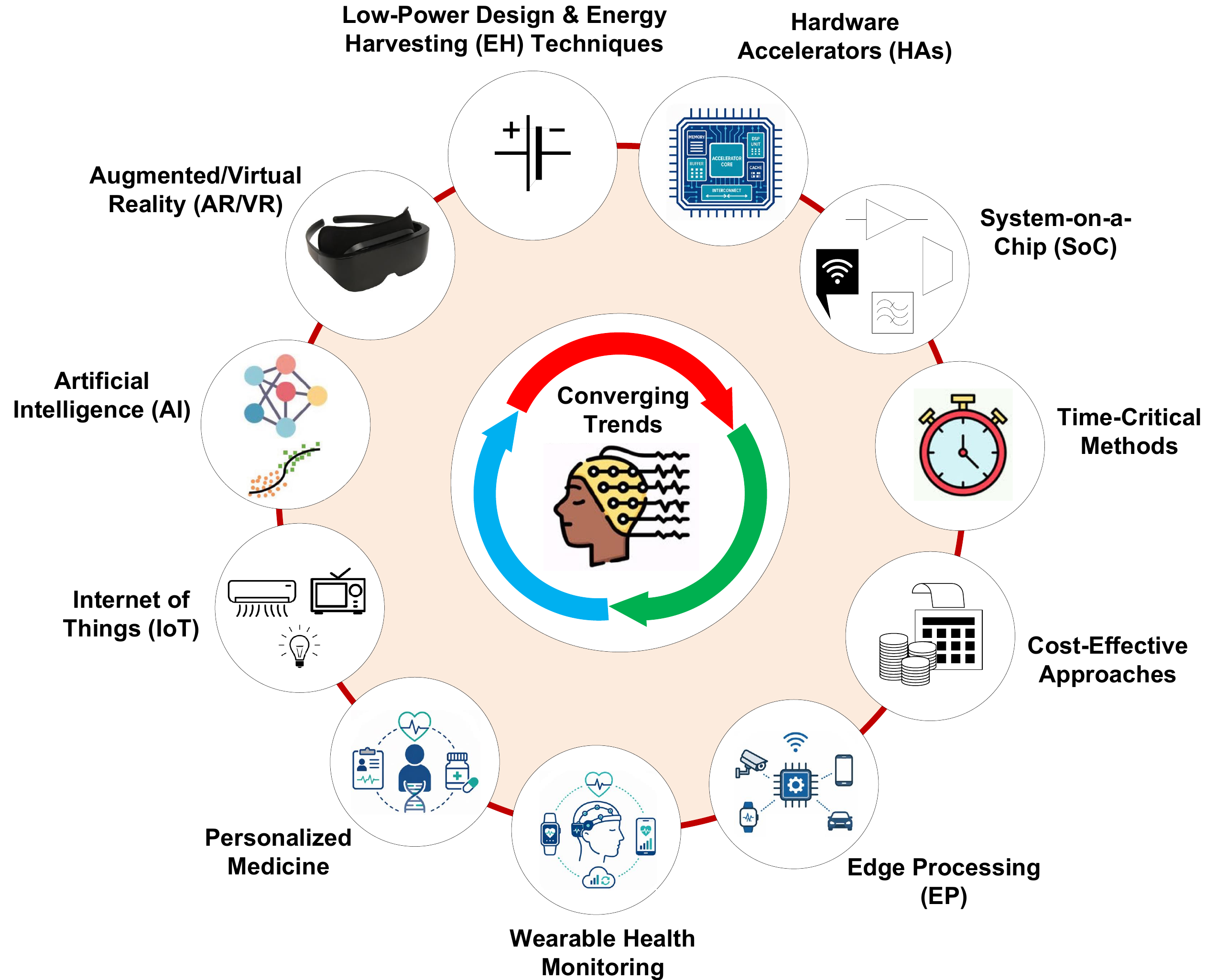}}
			\caption{Convergence of software and hardware technologies driving the commercialization of next-generation EEG-based MI-BCI applications.}
			\label{Fig_Ecosystem_Technologies}
		\end{figure*}

		As illustrated in Fig. \ref{Fig_Ecosystem_Technologies}, the convergence of emerging hardware and software technological trends is poised to revolutionize EEG‑based MI‑BCI systems by transforming them from laboratory‑bound prototypes into intelligent, wearable, and seamlessly integrated solutions. Advances in AI, wearable technology, system‑on‑chip (SoC) design, augmented and virtual reality (AR/VR), edge processing, hardware acceleration, energy harvesting, and the Internet of Things (IoT) collectively enable more adaptive, efficient, and user‑centered BCIs \cite{lu2025design}. Together, these technologies create a synergistic ecosystem that enhances real‑time processing, mobility, and long‑term usability, paving the way for next‑generation MI‑BCI systems that are both practical and pervasive in everyday life.

		At the core of any MI‑BCI system lies its signal‑processing pipeline, which is now being transformed by the integration of powerful AI‑based algorithms. Although early MI‑BCI research in the 1970s primarily explored the basic feasibility of establishing direct neural communication pathways \cite{miller2020current,varbu2022past,vidal1973toward}, recent decades have seen a major shift fueled by miniaturized, high‑fidelity EEG sensing technologies paired with sophisticated computational techniques capable of interpreting non‑stationary, low‑SNR scalp signals \cite{lebedev2017brain,boly2012brain,chatelle2012brain,edlinger2015many,gibson2014multiple,yuan2010negative,hohne2014motor}. The performance of modern MI‑BCIs is increasingly shaped by the combined evolution of high‑precision software methods and improved hardware designs, where ML, DL, and HYB models have dramatically advanced the decoding of neural activity by enabling more reliable recognition of complex patterns, substantially boosting decoding performance and CA.

		Wearable MI‑BCI technologies should fulfill key design criteria such as portability, low‑power consumption, wireless connectivity, ease of use for non‑expert users, minimal setup time, high signal quality, robustness to motion artifacts, comfort during long‑term wear, and cost‑effectiveness \cite{casson2019wearable,mihajlovic2014wearable,saibene2023eeg}. Meeting these criteria ensures that EEG‑based wearable systems can move beyond laboratory settings into everyday environments, enabling practical and continuous brain‑computer interaction in real‑life scenarios \cite{fiedler2022high,casson2010wearable}.

		EEG electrodes can be categorized by technology as dry, wet, and semi‑dry, each offering distinct advantages and limitations. The dry electrodes provide fast setup and comfort without gels but often suffer from higher impedance and reduced signal quality. The wet electrodes, whether gel‑based or saline‑based, offer superior conductivity and lower movement artifacts but require time‑consuming preparation, careful cleanup, and can cause user discomfort \cite{soufineyestani2020electroencephalography,hu2019eeg}. The semi‑dry electrodes strike a balance by using minimal electrolyte to improve signal conduction while maintaining an easier setup \cite{li2020review}. For instance, innovative flexible inkjet-printed electrodes \cite{xie2012heterogeneous} use silver nanoparticles to enable cost-effective, gel-free integration with mixed-signal SoC, demonstrating success in personalized healthcare applications. Moreover, emerging tattoo‑like electrodes further enhance comfort and stability through ultra‑thin, skin‑conforming designs \cite{wang2022robust}. As these technologies advance, wearable EEG platforms significantly improve MI‑BCI systems' usability by reducing cable‑related artifacts, increasing mobility, guiding non‑expert users in electrode placement, and enabling more portable, robust, and user‑friendly systems suitable for real‑life operation.

\begin{figure}[!t]
			\centerline{\includegraphics[scale=0.27,trim=0cm 0cm 0cm 0cm,clip=true]{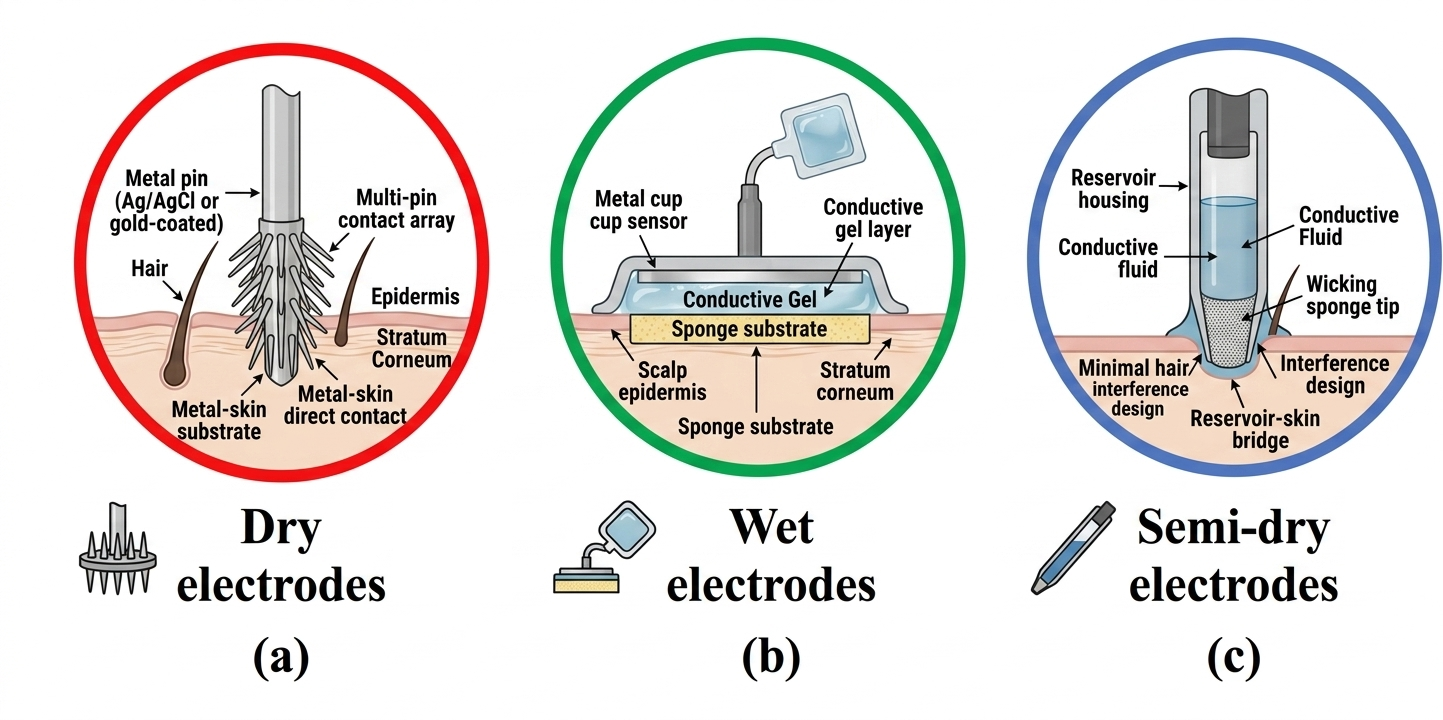}}
			\caption{Different types of EEG electrodes. (a) Dry electrodes. (b) Wet electrodes. (c) Semi-dry electrodes.
			}
			\label{Fig_Electrodes_Wet_dry_semi-dry}
		\end{figure}

		Some examples of different types of EEG electrodes are illustrated in Fig. \ref{Fig_Electrodes_Wet_dry_semi-dry}. Dry electrodes utilize a multi-pin array designed to mechanically bypass the hair layer and establish direct metal-to-skin contact with the stratum corneum without the need for conductive gel. Wet electrodes feature a traditional metal cup sensor paired with a sponge substrate or electrolyte chamber that contains a specialized conductive gel layer, lowering skin-electrode impedance by establishing a secure bridge to the scalp epidermis. The semi-dry electrodes employ a HYB reservoir housing that slowly releases a minimal amount of conductive fluid through a wicking sponge tip, combining a minimal hair-interference design with a continuous reservoir-to-skin liquid bridge to optimize preparation time and long-term signal stability \cite{wang2026eeg,razaghi2026systematic}.

		In EEG-based MI-BCI systems, the captured scalp signals can be transmitted to the processing unit either through traditional wired connections or via wireless technologies such as Bluetooth, proprietary low-power radio, or Wi‑Fi \cite{dou2026wireless,do2014wireless}. While wired EEG systems typically offer higher bandwidth and more stable data transmission, they require bulky helmets and numerous cables that prolong setup time and restrict mobility, making them unsuitable for everyday use. In contrast, modern wearable EEG devices integrate microchip electrodes with on-board amplifiers, quantizers, and wireless transmitters, enabling cable-free operation that is more comfortable, portable, and less prone to cable‑movement artifacts. Bluetooth and similar low-power wireless protocols thus play a key role in realizing small, head‑mounted systems that can stream MI‑BCI data reliably over extended periods in real-life environments \cite{saibene2023eeg,heim2025real,li2025wireless}.

		The integration of AR/VR into EEG-based MI‑BCI systems has introduced powerful new avenues for neuro‑rehabilitation and user interaction \cite{khan2020review,pfurtscheller2010neurofeedback}. By immersing users in realistic, closed‑loop environments, AR/VR enhances engagement, motivation, and sensory feedback while enabling motor priming, an essential process for restoring or strengthening neural pathways \cite{vourvopoulos2019efficacy,pfurtscheller2010neurofeedback}. These immersive interfaces allow individuals to visualize and control virtual limbs or objects through thought‑driven commands, thereby bridging the gap between neural intention and physical response \cite{carrino2012self}. As a result, AR/VR‑enabled MI‑BCI systems not only improve training outcomes and motor recovery but also create more intuitive and interactive experiences for both clinical rehabilitation and everyday neuroadaptive applications \cite{alanis2020assessment}.

		SoC designs, HAs, and edge processing (EP) are transforming EEG-based MI-BCI systems by tackling key challenges such as latency, size, and power consumption \cite{canilang2021edge,fang2019development,krishna2023sparsity}. Integrating signal acquisition, preprocessing, and classification into compact SoC platforms enables miniaturized and less bulky wearable devices capable of recording EEG signals continuously for extended periods (days or even weeks). However, moving from high-power laboratory setups to real-world, resource-constrained environments remains a critical bottleneck. In this context, EP plays a central role by shifting computation closer to the sensors, thereby reducing latency, enhancing real-time responsiveness, and improving data privacy \cite{covi2021adaptive,jin2022survey}. To further support computationally intensive AI algorithms, dedicated HAs are increasingly employed to efficiently execute feature extraction and classification tasks while maintaining low power consumption.

		Despite these advances, battery life continues to limit the widespread adoption of wearable MI-BCI systems outside controlled settings. EH techniques offer a promising solution by enabling autonomous and maintenance-free operation through the capture of ambient energy sources such as body heat or mechanical vibrations. When combined with low-power SoC architectures and optimized hardware accelerators, EH can significantly extend device lifetime and reduce reliance on frequent battery recharging \cite{kanemoto2026battery,tatistscheff2025ultra}. Overall, the convergence of SoC integration, edge computing, hardware acceleration, and sustainable energy solutions is paving the way toward fully wearable, intelligent, and long-term EEG-based MI-BCI systems suitable for real-world applications \cite{fang2019development,du2026machine}.

		The integration of IoT technologies into EEG-based MI systems extends their functionality beyond standalone operation toward interconnected, intelligent environments. By embedding BCI modules into IoT-enabled devices, these systems can continuously monitor neural signals and infer cognitive states such as mental fatigue, stress, attention, and frustration, enabling adaptive and context-aware responses in real time \cite{myrden2017passive,li2025enhancing}. This connectivity allows seamless data exchange with cloud or edge platforms for further analysis, remote monitoring, and personalized feedback, while also facilitating applications in smart healthcare, rehabilitation, and assistive living \cite{jiang2017semiasynchronous,foong2019assessment}. Consequently, IoT-enhanced MI-BCI systems can act as intelligent assistants within smart environments, improving user experience and enabling more responsive and scalable neurotechnology solutions.

		Emerging directions in EEG-based MI-BCI systems include the exploration of quantum computing, ultra-low-latency processing methods, and cost-effective design and fabrication strategies to enable broader adoption \cite{huang2022survey}. Although still in its infancy, the integration of quantum computing with BCI has introduced novel paradigms such as HYB classical–quantum frameworks and quantum brain networks, where neural signals can potentially interact with quantum systems (e.g., controlling or measuring qubits), opening new interdisciplinary research opportunities. At the same time, time-critical methods such as optimized signal pipelines, edge-based inference, and lightweight algorithms are essential to minimize latency and ensure real-time responsiveness, which is crucial for practical MI-BCI applications \cite{miranda2022quantum,miranda2022approach}. Finally, for commercialization and widespread use, reducing system cost through scalable hardware, efficient architectures, and affordable wearable components remains a key priority, ensuring that these advanced neurotechnologies become accessible beyond specialized research environments.

\begin{figure}[!t]
			\centerline{\includegraphics[scale=0.3,trim=0cm 1.8cm 0cm 2.2cm,clip=true]{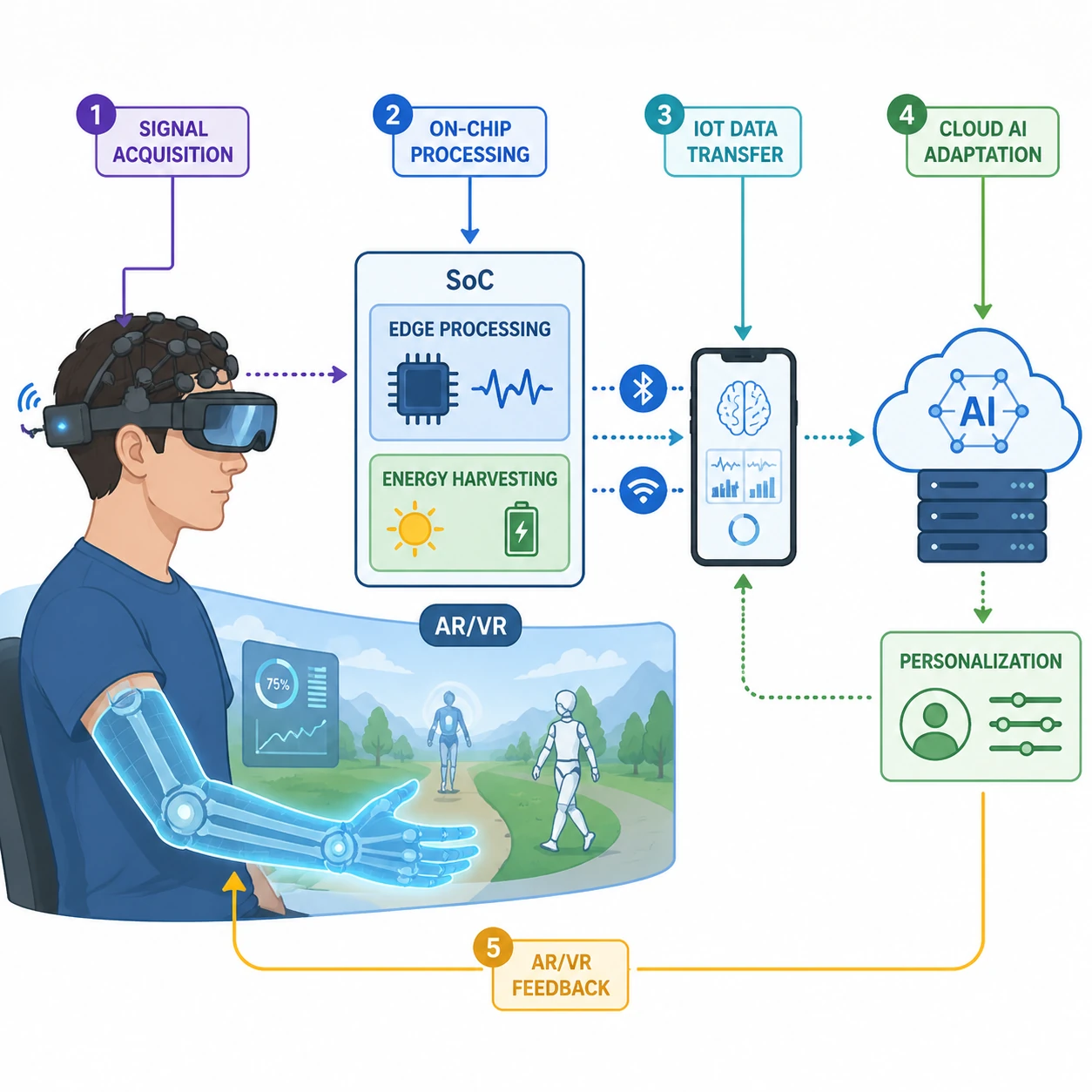}}
			\caption{Intelligent integrated MI-BCI ecosystem for personalized neuro interaction.
			}
			\label{Fig_Ecosystem_Technologies_2}
		\end{figure}

		The utilization of several emerging technologies in an MI-BCI ecosystem is illustrated in Fig. \ref{Fig_Ecosystem_Technologies_2} through an example of a system-level architecture that combines wearable sensing, EP, and cloud intelligence to enable mobile, adaptive BCI interaction. The user wears a lightweight wireless EEG headset equipped with flexible electrodes and an SoC that integrates amplification, ADC, and an EP core for real‑time signal decoding. The SoC manages power through EH modules that capture body heat or motion, allowing low‑power operation. Processed neural features are transmitted via Bluetooth or Wi‑Fi, connecting the headset to nearby devices and the cloud in an IoT framework.

		In the cloud–AI layer, accumulated data refines personalized neural models for improved movement decoding and adaptive performance. These personalized classifiers enable applications such as AR/VR control, neurorehabilitation, and cognitive training, where users interact with responsive virtual environments through MI. The figure illustrates how sensing, EP, wireless connectivity, and intelligent personalization converge into an energy‑efficient, user‑centered MI‑BCI ecosystem, highlighting the synergy between HA, on‑chip integration, and sustainable design principles.

		In the next section, a modern AI‑based EP SoC architecture is proposed for next‑generation EEG‑based MI‑BCI systems. This architecture is designed to tightly integrate low‑noise EEG acquisition, energy‑efficient signal processing, and intelligent inference within a unified hardware platform, enabling real‑time operation under strict power and latency constraints.

		\section{CONCEPTUAL HARDWARE ARCHITECTURE FOR NEXT-GENERATION MI-BCI SYSTEMS}
		\label{sec:Hardware_Architectural_Recommendations}

		\begin{figure*}[!t]
			\centerline{\includegraphics[scale=0.6,trim=6cm 11.1cm 6cm 3cm,clip=true]{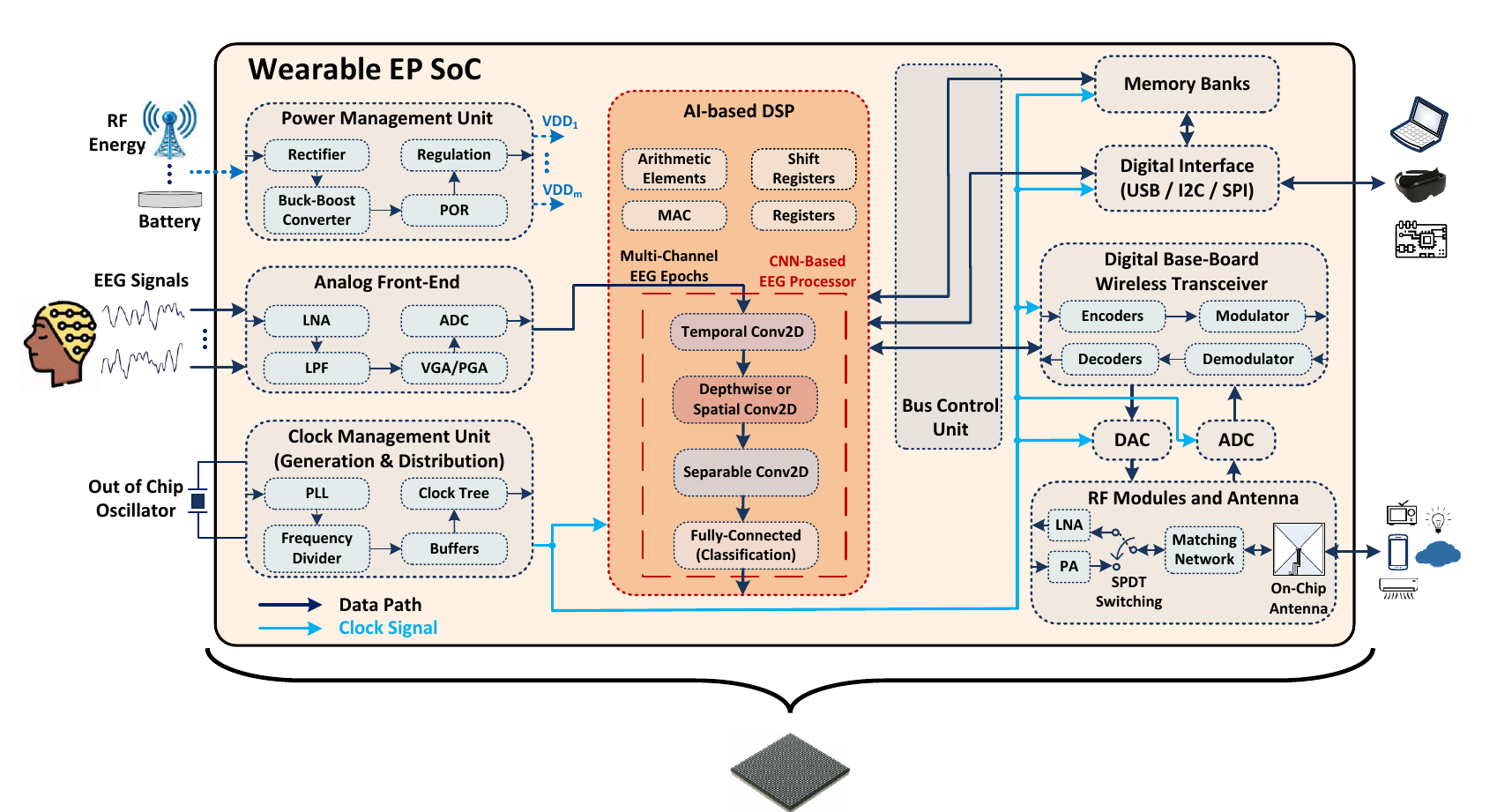}}
			\caption{Conceptual block diagram and major subsystems of a next-generation wearable edge-processing EEG-based MI-BCI SoC.}
			\label{Fig_edge_processing_BCI_chips}
		\end{figure*}

		Based on the technologies and hardware implementations reviewed in the preceding sections, this section presents a conceptual SoC architecture for next-generation EEG-based MI-BCI systems. The proposed architecture is intended to illustrate how low-power EEG acquisition, on-chip signal conditioning, AI-based signal processing, wireless communication, and power-management functions could be integrated within a unified platform. It should be emphasized that the proposed architecture is a conceptual system-level design and has not been physically implemented or fabricated.

		The proposed block diagram of an AI‑powered EEG-based EP MI-BCI chip is illustrated in Fig. \ref{Fig_edge_processing_BCI_chips}. The conceptual SoC is intended to illustrate how efficient power management and real-time processing could be supported in future MI-BCI implementations. The primary features of this SoC include EH techniques to eliminate or reduce the need for a battery, integration with the IoT ecosystem through AI-enabled EP, and an optimized transceiver circuit for seamless wireless communication. These architectural features could potentially support a range of applications, such as medical signal recording for health monitoring and emergency alert systems. The proposed SoC comprises multiple interconnected sub‑systems that collaboratively enable efficient and reliable processing of EEG signals for MI tasks.

		At the core of this SoC is the AI-based DSP unit, designed to enhance EEG signal interpretation and classification using deep learning algorithms. This unit forms the computational backbone of the system and operates in a structured pipeline including pre‑processing, feature extraction, classification, and post‑processing. A CNN-based classification engine could be employed to support real-time MI prediction with potentially reduced processing latency, while a post-processing module could apply smoothing and multi-channel decision fusion to improve robustness.

		The DSP architecture is inspired by EEGNet, a lightweight CNN model optimized for EEG‑based BCIs. EEGNet employs temporal convolutions, depthwise spatial convolutions, and separable convolutions to efficiently capture both temporal dynamics and spatial correlations across channels while significantly reducing parameter count and power consumption. This compact yet powerful architecture has demonstrated strong performance on public datasets, outperforming traditional shallow and deep ConvNet models. Furthermore, its hardware‑friendly structure makes it well‑suited for real‑time deployment on edge processors and ASIC‑based EP platforms, where energy efficiency, speed, and classification reliability are critical.
		
		The other sections of the proposed EP SoC are as follows:

		A power-management unit (PMU), potentially integrated with energy-harvesting interfaces, could support low-power operation suitable for wearable applications. The PMU is intended to manage energy from multiple inputs, such as RF energy harvesters, micro-batteries, or other ambient sources, and to provide the voltage levels required by different system components. The rectifier first converts incoming RF or AC energy into DC voltage. The buck‑boost converter then elevates and stabilizes this voltage to a usable level, buffering energy to handle load fluctuations. A power‑on reset (POR) circuit ensures safe system startup by holding the SoC in reset until the supply reaches a defined threshold. Finally, the regulation stage provides stable, low‑noise output voltages for the AFE and DSP blocks, ensuring consistent and efficient operation of the MI‑BCI system.

		The AFE unit filters, amplifies, and conditions the weak EEG signals before they reach the digital domain, ensuring that meaningful neural information is preserved with minimal distortion. It typically includes low‑noise amplifiers (LNAs) to boost microvolt‑level EEG signals while adding as little noise as possible, LPFs to remove high‑frequency noise and interference, and variable or programmable gain amplifiers (VGA/PGA) to adapt the amplification level based on signal amplitude and recording conditions. Together, these blocks maximize signal integrity and dynamic range, enabling the downstream ADC and processing engines to operate with higher accuracy and robustness.

		The clock management unit is responsible for generating and distributing clock signals across the SoC. This sub-system includes phase-locked loops (PLL), frequency dividers, clock trees, and buffers to ensure precise timing and synchronization for efficient processing.

		The memory-bank unit provides temporary storage for EEG signal processing, intermediate computations, and algorithm execution. These memory units support high-speed data access and real-time performance.

		The digital interface unit enables communication with external components via universal serial bus (USB), inter‑integrated circuit (I2C), and serial peripheral interface (SPI) interfaces, ensuring flexible integration with external devices and controllers. This interface also facilitates seamless connectivity with AR/VR platforms, allowing real‑time transmission of decoded MI‑BCI commands to head‑mounted displays or immersive rehabilitation systems, thereby enabling interactive neurofeedback, virtual training environments, and low‑latency brain‑controlled user experiences.

		The digital base‑board wireless transceiver unit enables reliable wireless communication between the EEG‑based MI‑BCI system and external devices. In this section, encoders prepare outgoing data for transmission, the modulator maps it onto a carrier signal, the demodulator recovers the received signal, and the decoders reconstruct the original data, ensuring efficient and accurate bidirectional communication. Additionally, by integrating IoT technology, this transceiver supports seamless connectivity with cloud platforms and smart devices, enabling remote monitoring, data sharing, and adaptive learning for continuous improvement of user performance and device intelligence in real‑world applications.

		The wireless transceiver and RF modules handle digital signal transmission and reception, while RF modules and antennas manage radio frequency communication. The on-chip antenna, LNA, power amplifier (PA), and RF matching networks facilitate reliable wireless data exchange.

		Finally, the bus control unit coordinates data flow between internal sub-systems, ensuring efficient and organized communication within the SoC. It also manages ADC and digital-to-analog converter (DAC) conversions, enabling seamless transitions between analog and digital signals.

		In conclusion, the proposed wearable edge-processing EEG-based MI-BCI SoC represents a conceptual framework integrating AI-based DSP, power management, wireless connectivity, and on-chip computation within a unified architecture. Such integration could potentially reduce external data transmission, support low-latency processing, and enable more compact and energy-efficient future MI-BCI implementations. However, the proposed architecture has not been implemented or fabricated, and its power consumption, latency, memory requirements, silicon area, and system-level performance remain to be quantitatively evaluated in future work.

		\section{Discussions}
		\label{sec:Discussions}

		\begin{figure*}[!t]
			\centerline{\includegraphics[scale=0.35,trim=42cm 33cm 47cm 70cm,clip=true]{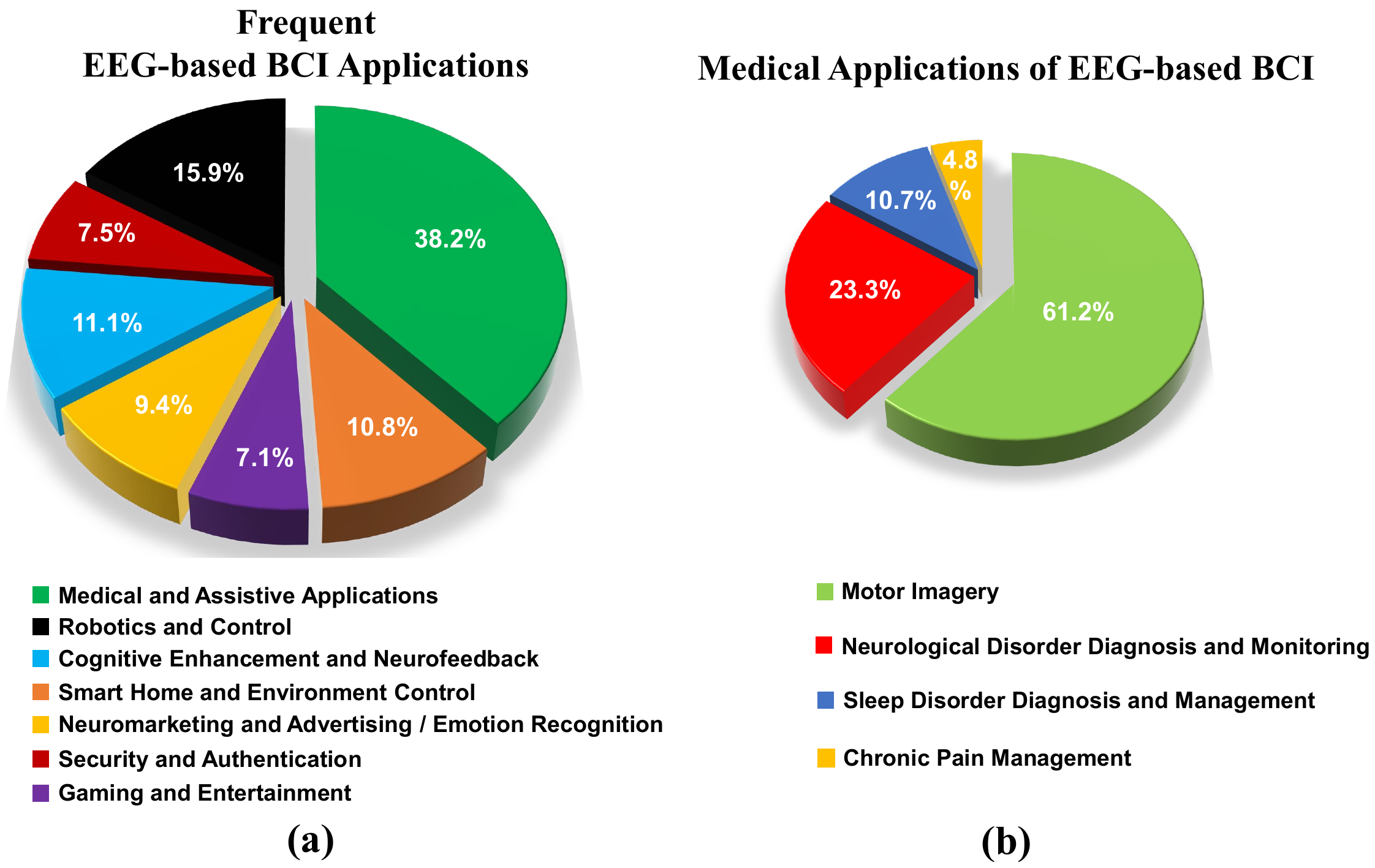}}
			\caption{Prevalence percentages of medical and non-medical applications of EEG-based BCI. (a) Distribution of major BCI application categories (\(N=\) 318 studies). (b) Distribution of medical EEG-based BCI applications (\(N=\) 122 studies). Percentages were calculated based on the studies included in the literature analysis described in Section II.}
			\label{Fig_EEG-based BCI Applications}
		\end{figure*}

		\begin{figure}[!t]
			\centerline{\includegraphics[scale=0.5,trim=7.5cm 4.6cm 7cm 4cm,clip=true]{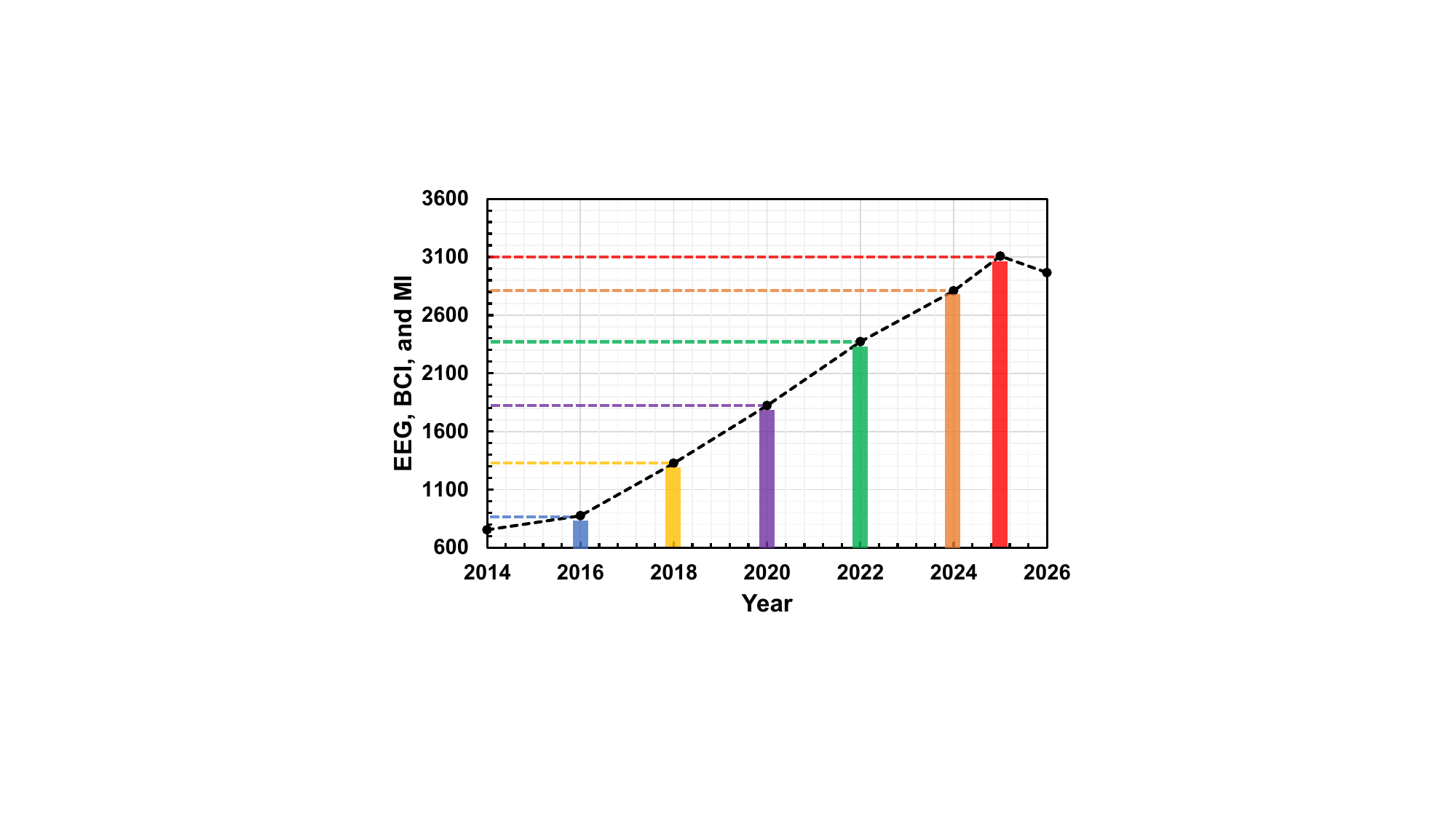}}
			\caption{Annual number of MI-BCI publications from January 2014 to July 2026, retrieved from Scopus using the search strategy described in Section II. The connecting line represents the yearly publication counts and does not indicate statistical regression, smoothing, or extrapolation.}
			\label{Fig_BCI_MI_Growth}
		\end{figure}

		After conducting an extensive and systematic exploration of the literature in the field of EEG-based MI-BCI systems, this section synthesizes the key findings and technical insights gathered from prior research. Building upon this foundation, we now proceed to address the research question outlined in Section~\ref{sec:Review_Methodology}, connecting the surveyed evidence to the core objectives and motivations of this study.

		\textbf{RQ1. Which BCI application currently has the greatest practical impact and the strongest potential for future development and improving the quality of life? What is the role of EEG-based MI-BCI systems within this landscape?}

		To answer this question, a literature analysis was performed with a detailed review of titles, keywords, abstracts, and data description sections to ensure comprehensive coverage. Hence, the most popular and widely explored applications of EEG-based BCI technology are extracted and illustrated in Fig. \ref{Fig_EEG-based BCI Applications}(a). The figure presents both medical and non-medical applications, offering insights into the prevalence of each category. The percentages reflect the current research focus and potential areas for future advancements in the BCI domain. The most dominant application in EEG-based BCI is medical and assistive applications, accounting for 38.2\%, highlighting its critical role in healthcare and rehabilitation. Other notable applications include: robotics and control (15.9\%), cognitive enhancement and neurofeedback (11.1\%), smart home and environment control (10.8\%), neuromarketing and emotion recognition (9.4\%), security and authentication (7.5\%), and gaming and entertainment (7.1\%).

		The Fig. \ref{Fig_EEG-based BCI Applications}(b) provides a closer look at the prevalence of different medical applications of EEG-based BCI technology. These applications are crucial in assisting individuals with neurological conditions and improving healthcare solutions \cite{jamil2021noninvasive}. The primary medical applications include: MI (61.2\%), neurological disorder diagnosis and monitoring (23.3\%), sleep disorder diagnosis and management (10.7\%), and chronic pain management (4.8\%). Based on the literature, while assistive technology and rehabilitation remain the most widespread applications, the range of medical uses for EEG-based BCI has expanded significantly in recent years to include diagnosis and monitoring of various neurological and mental health conditions, as well as chronic pain management. However, some emerging applications like neurostimulation for depression and personalized interventions for autism are still in the early research stages.

		Among the various medical applications of EEG-based BCIs, MI stands out as a crucial area of focus due to its significance in rehabilitation capability and its positive impact on improving human lives, enabling individuals with severe disabilities to control assistive devices such as wheelchairs, prosthetic limbs, and communication aids using brain signals. MI involves the mental simulation of movement without actual physical execution, allowing users to engage with BCI systems effectively. Case studies have highlighted the successful use of MI in enhancing recovery outcomes for individuals post-stroke or with spinal cord injuries \cite{mattia2020promotoer}. For instance, patients have demonstrated improved motor function by practicing imagined movements while connected to EEG systems that provide real-time feedback. This approach not only aids in physical rehabilitation but also empowers users by restoring a sense of agency over their movements. Fig. \ref{Fig_BCI_MI_Growth} illustrates the increasing number of published works on MI-BCI applications in recent years. As research continues to advance in this field, the integration of AI algorithms and emerging technologies will further enhance the capabilities of MI-BCI systems, paving the way for innovative solutions in both clinical and everyday applications.

		\textbf{RQ2. What are the most frequently used signal processing techniques and datasets for EEG-based MI-BCI systems in the recent decade? Which processing approaches have demonstrated superior performance in recent studies and have the most potential to be used in future systems?}

		\begin{figure*}[!t]
			\centerline{\includegraphics[scale=0.32,trim=7.9cm 9cm 9cm 3.8cm,clip=true]{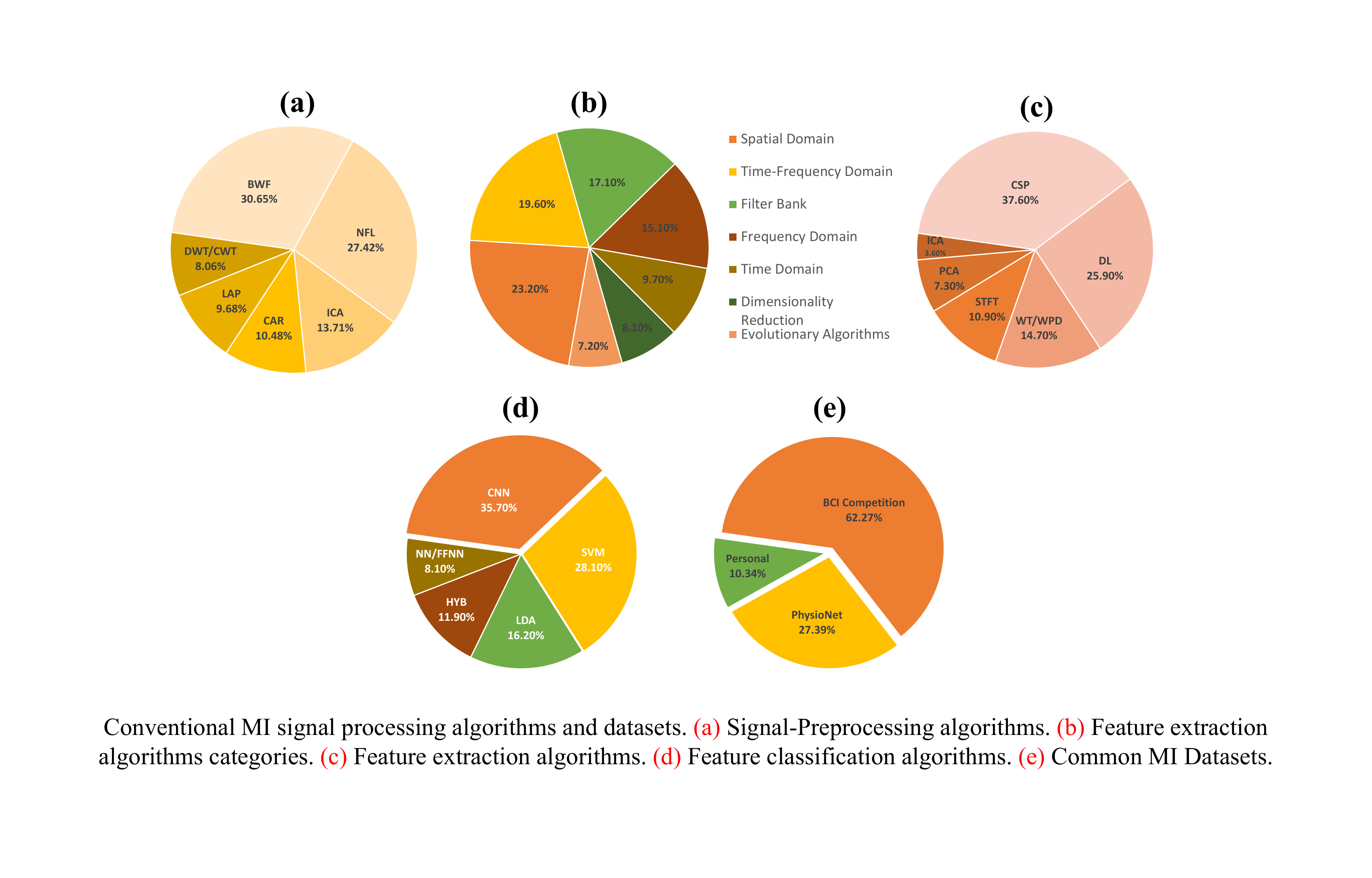}}
			\caption{Prevalence percentages of MI signal-processing algorithms and datasets based on the subsets of studies providing sufficient information for each analysis: (a) signal preprocessing algorithms (\(N=124\) studies); (b) feature-extraction algorithm categories (\(N=168\) studies); (c) feature-extraction algorithms (\(N=286\) studies); (d) feature-classification algorithms (\(N=297\) studies); and (e) common MI datasets (\(N=301\) studies).}
			\label{Fig_Algorithms_Comparisons}
		\end{figure*}

		Based on the papers reviewed in Section~\ref{sec:MI-BCI_Signal_Processing_Algorithms}, several important observations can be made regarding the most frequently used signal processing techniques and datasets in EEG‑based MI‑BCI systems over the past decade. The analysis reveals the dominant algorithms employed in each major signal‑processing block, including preprocessing, feature extraction, feature selection, and classification. It also highlights the datasets that have been most widely adopted for benchmarking and performance evaluation in recent studies. By comparing these trends with reported classification accuracies and implementation feasibility, it becomes possible to identify which processing approaches have demonstrated greater success in recent research and which ones show the strongest potential for integration into future EEG‑based MI‑BCI systems.

		Among the signal preprocessing methods used in recent EEG based MI BCI studies, BWF is the most common approach, appearing in 30.71\% of the reviewed works, as illustrated in Fig. \ref{Fig_Algorithms_Comparisons} (a). It is followed by the NFL with 27.07\%, which is primarily used for its effectiveness in suppressing power line interference. ICA accounts for 13.43\% of the studies, while CAR follows with 10.81\%, serving to improve spatial signal quality by reducing common noise across electrodes. The LAP filtering approach is utilized in 9.70\% of cases, reflecting its role in enhancing local cortical activity relevant to MI tasks. Finally, DWT/CWT techniques appear in 8.28\% of the studies, showing a more limited but still meaningful role in time–frequency based preprocessing.

		Regarding feature extraction algorithm categories in recent EEG‑based MI‑BCI studies, spatial domain methods are the most frequently used, accounting for 23.20\% of the reviewed approaches, as shown in Fig. \ref{Fig_Algorithms_Comparisons} (b). They are followed by time–frequency domain methods with 19.60\%, highlighting the importance of capturing both temporal and spectral characteristics of EEG signals. Filter bank methods represent 17.10\% of the studies, while frequency domain techniques account for 15.10\%, reflecting their strong relevance for analyzing rhythm‑based MI patterns. Time domain methods appear in 9.70\% of the studies, showing a more limited but still meaningful role in direct signal analysis. Dimensionality reduction approaches account for 8.10\%, indicating their usefulness in simplifying feature sets and reducing complexity. Finally, evolutionary algorithms are the least frequently used at 7.20\%, suggesting a more specialized role in optimizing feature extraction processes.

		Based on Fig. \ref{Fig_Algorithms_Comparisons} (c), among the specific feature extraction algorithms used in EEG‑based MI‑BCI systems, CSP is the most dominant technique, accounting for 37.60\% of the reviewed studies, mainly because of its strong ability to enhance discriminative spatial patterns between MI classes. LDA follows with 25.90\%, reflecting its effectiveness in improving class separability and supporting robust feature discrimination. WT/WPD represents 14.70\% of the studies, highlighting its usefulness in capturing the nonstationary time–frequency characteristics of EEG signals. STFT accounts for 10.90\%, showing its continued relevance for analyzing spectral variations over time. PCA is used in 7.30\% of the studies, indicating its role in reducing feature dimensionality while preserving important signal information. Finally, ICA appears in 3.60\% of the reviewed works, suggesting a more limited but still meaningful application in artifact removal and feature enhancement.

		Among the commonly used classification algorithms in EEG‑based MI‑BCI systems, CNN is the most prevalent, appearing in 35.70\% of the reviewed studies, as depicted in Fig. \ref{Fig_Algorithms_Comparisons} (d). This reflects the growing adoption of deep learning approaches capable of automatically learning spatial–temporal patterns from EEG data. SVM follows with 28.10\%, due to its strong performance on high‑dimensional and nonlinearly separable EEG features. Traditional classifiers remain relevant as well, with LDA used in 16.20\% of studies, appreciated for its simplicity and fast computation. The hybrid (HYB) methods account for 11.90\%, indicating increasing interest in combining multiple techniques to improve classification performance. Finally, NN/FFNN appears in 8.10\% of the studies, showing that neural‑based models continue to play a useful but less dominant role in MI‑BCI classification.

		Based on Fig. \ref{Fig_Algorithms_Comparisons} (e), commonly used MI datasets in research include the BCI Competition datasets (62.27\%), which are popular because they provide standardized benchmarks and a wide range of MI tasks; PhysioNet datasets (27.39\%), which are appreciated for their open availability and detailed physiological recordings; and personal or experimental datasets (10.34\%), which are usually gathered by individual research groups to support specific experimental designs or new protocols. Even when the same benchmark dataset is used, the classification setting can still differ considerably, ranging from a binary one-vs.-one task, such as left versus right hand, to one vs. rest tasks, such as imagined right-hand movement versus resting state, as well as multiclass problems involving three to five classes.

		Direct comparison or averaging of classification accuracies across studies was not performed because the reported results were obtained using different datasets, numbers of classes, subject-dependent or subject-independent evaluation settings, and validation protocols. Accordingly, performance results are reported individually in Table \ref{Table_SW_Algorithms} and Table \ref{Table_HW_Algorithms}, together with their corresponding experimental conditions.

		\textbf{RQ3. What are the challenges of EEG-based MI-BCI systems, and what are the promising candidate solutions?}

\begin{figure}[!t]
			\centerline{\includegraphics[scale=0.6,trim=5cm 7cm 4cm 8cm,clip=true]{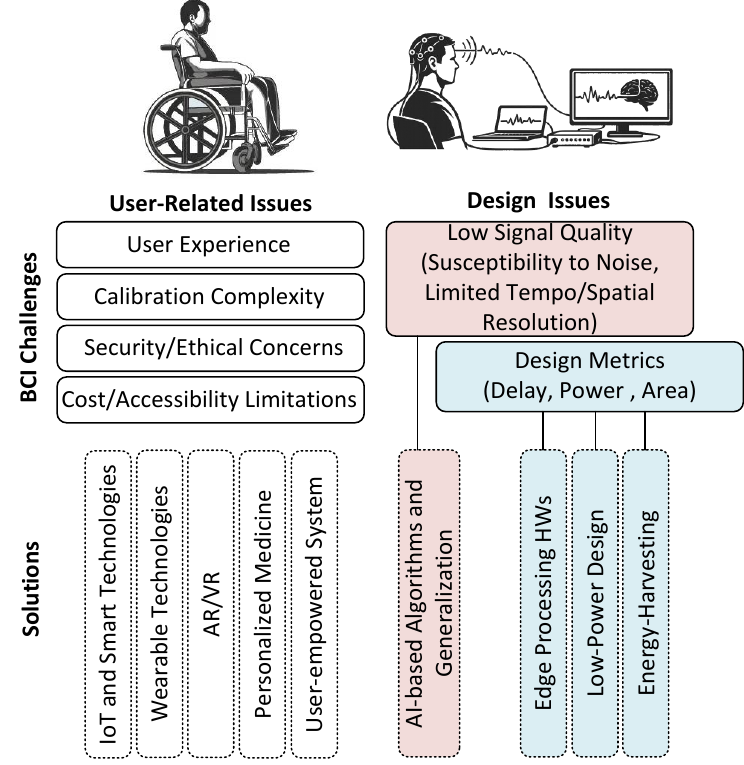}}
			\caption{Current MI-BCI systems challenges and solutions. 
			}
			\label{Challenges_Solutions_Ghamkhari}
		\end{figure}

		The challenges and solutions associated with MI-BCI technologies can be grouped into two categories, user-oriented and design-oriented, as illustrated in Fig.~\ref{Challenges_Solutions_Ghamkhari}. This classification provides a structured view of the diverse obstacles that hinder the practical deployment of MI-BCI systems and highlights the corresponding strategies proposed in the literature. Together, these categories help contextualize the multilayered nature of MI-BCI limitations and offer a unified framework to guide ongoing research and system development. The challenges and corresponding solutions summarized in Fig.~\ref{Challenges_Solutions_Ghamkhari} were synthesized from the reviewed literature and the studies discussed throughout Sections V and VI.

		User-oriented challenges encompass those that directly affect usability and comfort, including discomfort from wet electrodes, bulky setups restricting mobility, and cognitive fatigue induced by sustained MI tasks. Long setup times and repeated calibrations further reduce accessibility, hindering adoption among non-expert users and patients \cite{rao2026calibration}. Economic and ethical factors, such as high equipment costs and data privacy concerns, compound these barriers. Emerging solutions leverage IoT-enabled and wearable EEG systems, AR/VR integration \cite{li2024combining}, and personalized medicine to enhance user comfort and adaptability. Advances in affordable, low-power SoC architectures \cite{gawali2016mixed,bhattacharyya2018design,tu2017body,modak202165nm,xia20140,xie2012heterogeneous,bhattacharyya2015design,bhamra201524,bhattacharyya2014rfid,tekeste2013survey}, combined with cost efficient electrode materials such as dry or flexible sensors, significantly reduce production costs while improving comfort and usability \cite{saibene2023eeg}. Open-source ecosystems like OpenBCI, EEGNet, and BrainFlow facilitate lower software overhead, while cloud and edge computing employ lightweight ML and federated learning for real-time decoding on low-cost hardware. These user centric advancements, strengthened by IoT connectivity and smart environments, promote seamless interaction and everyday usability \cite{talpur2024illuminating,shaown2019iot,rawat2024health,shwetha2024iot,massot2009wearable,carlino2024sensor,ruby2023review,beach2021edge,dogan2012multi,shajari2023emergence,ma2022recent,covi2021adaptive,gupta2023machine,jebelli2018eeg,kobayashi2025parameter}.

		Design-oriented challenges relate to limitations in signal quality, system architecture, and real-time computational performance. EEG's weak and noise-prone signals, low spatial resolution, artifacts, and nonstationary inter-subject variability hinder accurate MI decoding, motivating advanced approaches such as DL, TL, Riemannian methods, and domain adaptation to learn more transferable representations and improve robustness across subjects and sessions \cite{xue2020multifrequency,dos2023residual,lawhern2018eegnet,congedo2017riemannian,schirrmeister2017deep,hosseinifard2013classifying,wu2015bayesian,wang2020accurate,chen2022transfer,zhao2020deep,autthasan2021min2net,lu2016deep,sakhavi2015parallel}. Implementation constraints such as latency, limited energy resources, and bulkiness are being addressed through EPoC, miniaturized wearables, hardware accelerators, and energy-efficient integrated-circuit techniques, including advanced-node FinFET technologies and low-power standard-cell architectures \cite{clark2016asap7}. Additional improvements stem from advanced shielding against EMI/ESD \cite{li2023design}, AI-enabled SoCs \cite{eichler2021mastermind}, ultra-low-power implants \cite{chatterjee20221}, and open-source SoC platforms like ESP4ML \cite{giri2020esp4ml}. Future directions highlight neuromorphic computing, EH, sub-millisecond decoding, and quantum-inspired algorithms as key enablers of next-generation, real-time MI-BCI systems.

		\begin{figure*}[!t]
			\centerline{\includegraphics[scale=0.36,trim=11.8cm 3.6cm 15.4cm 1cm,clip=true]{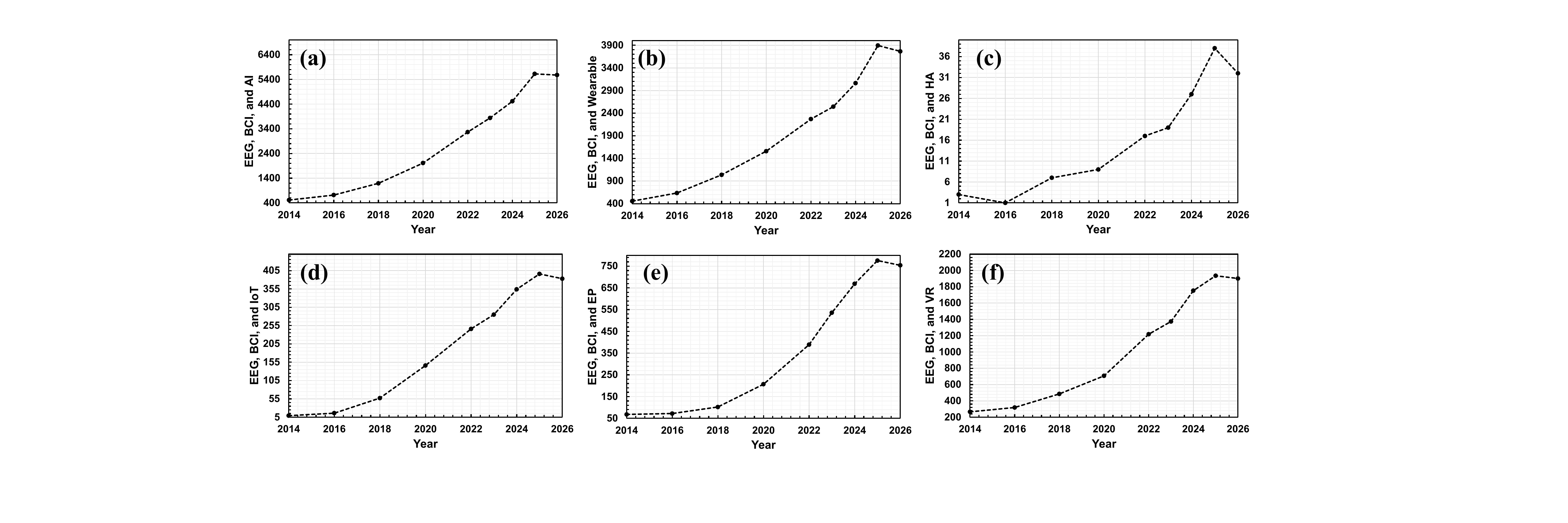}}
			\caption{Trends of converging technologies for EEG-based BCI from January 2014 to July 2026 based on annual publication counts retrieved from Scopus using the search strategy described in Section II: (a) EEG-based BCI and AI. (b) EEG-based BCI and wearable technology. (c) EEG-based BCI and HA technology. (d) EEG-based BCI and IoT. (e) EEG-based BCI and EP. (f) EEG-based BCI and VR. Scopus is used as the reference (see Section \ref{sec:Review_Methodology} for more details). The connecting lines represent yearly publication counts and do not indicate statistical regression, smoothing, or extrapolation.}
			\label{Fig_Trends_technologies_EEG_BCI}
		\end{figure*}

		\textbf{RQ4. How can various emerging technologies be brought together to create the next generation of EEG-based MI-BCI systems?}

		The trends in converging technologies for EEG-based BCIs from January 2014 to July 2026 are illustrated in Fig. \ref{Fig_Trends_technologies_EEG_BCI} based on annual publication counts (the sources are listed in Section \ref{sec:Review_Methodology}), highlighting key developments in the field. The integration of AI (Fig. \ref{Fig_Trends_technologies_EEG_BCI}(a)) has enhanced feature extraction and classification, improving the accuracy of brain signal interpretation. The adoption of wearable technology (Fig. \ref{Fig_Trends_technologies_EEG_BCI}(b)) has made EEG-based BCIs more portable and comfortable, facilitating their use in daily activities and long-term monitoring. Hardware acceleration (Fig. \ref{Fig_Trends_technologies_EEG_BCI}(c)) has enabled power-efficient and high-performance processing, supporting real-time signal computation even in resource-constrained environments. The IoT (Fig. \ref{Fig_Trends_technologies_EEG_BCI}(d)) has enhanced BCI connectivity, enabling data exchange between devices for cloud-based analysis, remote healthcare, and adaptive neurofeedback applications. Edge processing (Fig. \ref{Fig_Trends_technologies_EEG_BCI}(e)) enables local EEG signal processing and MI decoding, potentially reducing communication latency, cloud dependence, and data transmission requirements while supporting privacy-aware and energy-efficient implementations. Finally, the convergence of EEG-based BCI and VR (Fig. \ref{Fig_Trends_technologies_EEG_BCI}(f)) has facilitated immersive neuroadaptive environments for neurorehabilitation, cognitive training, and human-computer interaction.

		In the near future, the integration of emerging technologies will greatly improve the performance and accessibility of EEG‑based MI‑BCI systems. Advances in ultra‑low‑power processors, flexible bioelectronic interfaces, federated learning, and multimodal sensing will enable more adaptive and personalized BCIs. Enhanced wireless connectivity, cloud and edge AI, and smarter learning algorithms will deliver faster, more accurate, and intuitive brain‑device interactions. As BCIs become increasingly efficient and seamlessly integrated into digital ecosystems, they will expand applications across healthcare, rehabilitation, assistive devices, and consumer neurotechnology, bringing human‑machine synergy closer than ever before.

		\section{Conclusion}
		
		This paper has provided a comprehensive review of the technological evolution of EEG‑based MI‑BCI systems, synthesizing findings from a decade of research across software algorithms, hardware platforms, datasets, and application domains. By examining both statistical trends in the broader MI‑BCI literature and detailed analyses of influential papers, the study highlights MI‑BCI as the dominant and most impactful modality within non‑invasive BCI research. The review demonstrates how advances in signal processing, feature extraction, and classification, especially those driven by modern AI models, have significantly enhanced the reliability and performance of MI‑BCI systems for medical, assistive, and neurorehabilitation applications.
		
		Despite substantial progress, EEG-based MI‑BCIs continue to face challenges related to low SNR, limited spatial resolution, non-stationarity of neural signals, and the need for user‑specific calibration. These limitations have historically restricted their accuracy, latency, and practicality in real‑world environments. However, the field’s trajectory shows that recent developments in deep learning, time‑critical computation, and adaptive decoding algorithms are reshaping the landscape. These emerging methods are increasingly capable of handling complex EEG dynamics, reducing training time, and improving robustness across user sessions and environmental conditions.
		
		On the hardware side, the convergence of SoC‑based embedded processing, low‑power accelerators, wearable EEG technologies, and IoT‑enabled connectivity marks a significant shift toward portable and scalable MI‑BCI architectures. Energy‑efficient designs, wireless communication, and cloud/edge computing pipelines now permit real-time decoding and seamless integration with smart environments. Moreover, immersive VR ecosystems and neurofeedback-driven rehabilitation frameworks provide new avenues for enhancing user engagement, learning efficiency, and neuroplasticity, thereby widening the clinical and consumer applicability of MI‑BCIs.
		
		Looking forward, future research should focus on cost‑effective and standardized MI‑BCI platforms that unify AI‑driven personalization, edge‑level computation, and interoperable hardware architectures. The democratization of neurotechnology, supported by open-source toolchains, accessible datasets, and unified benchmarking protocols, will be essential for ensuring reproducibility and accelerating innovation. Ultimately, the continued convergence of AI, SoC design, IoT ecosystems, and immersive training technologies will shape the next generation of intelligent, adaptive, and practical EEG‑based MI‑BCI systems, enabling their transition from laboratory prototypes to widespread clinical, assistive, and consumer deployment.

		\section*{Acknowledgment}
		The authors would like to thank Diana Babaei and Mahyar Mehri for their valuable support and insightful discussions during the course of this work.

		
		\bibliographystyle{ieeetr}
		\bibliography{Refs}

@article{cohen2017does,
	title={Where does EEG come from and what does it mean?},
	author={Cohen, Michael X},
	journal={Trends in neurosciences},
	volume={40},
	number={4},
	pages={208--218},
	year={2017},
	publisher={Elsevier}
}

@article{michel2012towards,
	title={Towards the utilization of EEG as a brain imaging tool},
	author={Michel, Christoph M and Murray, Micah M},
	journal={Neuroimage},
	volume={61},
	number={2},
	pages={371--385},
	year={2012},
	publisher={Elsevier}
}

@article{nicolas2012brain,
	title={Brain computer interfaces, a review},
	author={Nicolas-Alonso, Luis Fernando and Gomez-Gil, Jaime},
	journal={sensors},
	volume={12},
	number={2},
	pages={1211--1279},
	year={2012},
	publisher={Molecular Diversity Preservation International (MDPI)}
}

@article{shin2015simple,
	title={Simple adaptive sparse representation based classification schemes for EEG based brain--computer interface applications},
	author={Shin, Younghak and Lee, Seungchan and Ahn, Minkyu and Cho, Hohyun and Jun, Sung Chan and Lee, Heung-No},
	journal={Computers in biology and medicine},
	volume={66},
	pages={29--38},
	year={2015},
	publisher={Elsevier}
}

@inproceedings{modak202165nm,
	title={A 65nm resonant electro-quasistatic 5-240uw human whole-body powering and 2.19 uw communication soc with automatic maximum resonant power tracking},
	author={Modak, Nirmoy and Das, Debayan and Nath, Mayukh and Chatterjee, Baibhab and Kumar, K Gaurav and Maity, Shovan and Sen, Shreyas},
	booktitle={2021 IEEE Custom Integrated Circuits Conference (CICC)},
	pages={1--2},
	year={2021},
	organization={IEEE}
}

@article{canilang2021edge,
	title={Edge EEG: Edge AI Device-based EEG Signal Processing for Emotion Recognition},
	author={Canilang, Henar Mike O and Caliwag, Ej Miguel Francisco C and Nyechinyere, Judith Njoku and Caliwag, Angela C and Lim, Wansu},
	journal={Proceedings of the Korean Institute of Communications and Information Sciences Conference
	},
	pages={260--262},
	year={2021}
}

@article{gong2025multi,
	title={Multi-source Discriminant Dynamic Domain Adaptation for Cross-subject Motor Imagery EEG Recognition},
	author={Gong, Y. and Shi, K. and Niu, X. and Yang, L. and Yang, X. and Zheng, C.},
	journal={IEEE Journal of Biomedical and Health Informatics},
	year={2025},
	month={Sep},
	publisher={IEEE}
}

@article{li2025enhancing,
	title={Enhancing eeg-based authentication with transformer in internet of things},
	author={Li, Chunxue and Meng, Weizhi and Li, Wenjuan},
	journal={IEEE Transactions on Information Forensics and Security},
	year={2025},
	publisher={IEEE}
}

@article{du2026machine,
	title={Machine Learning in Vibration Control, Vibration Energy Harvesting, and Structural Health Monitoring: A Review},
	author={Du, Houfan and Duan, Bohao and Huang, Dongmei and Wang, Lu and Litak, Grzegorz and Xu, Haitao and Zhou, Shengxi and Jiang, Zhuangde},
	journal={IEEE Sensors Journal},
	year={2026},
	publisher={IEEE}
}

@inproceedings{tatistscheff2025ultra,
	title={Ultra-Low Power Millimeter-Sized Subcutaneous EEG Implant with NFC Power Harvesting},
	author={Tatistscheff, Caterina Azais and Rapeaux, Adrien and Constandinou, Timothy G},
	booktitle={2025 IEEE Biomedical Circuits and Systems Conference (BioCAS)},
	pages={409--412},
	year={2025},
	organization={IEEE}
}

@inproceedings{kanemoto2026battery,
	title={A battery-free wireless EEG transmission system using compressed sensing and powered by body--ambient temperature difference: Outdoor demonstration at expo 2025},
	author={Kanemoto, Daisuke and Yoshimoto, Kazane and Motomochi, Shodai and Hirose, Tetsuya},
	booktitle={2026 IEEE International Conference on Consumer Electronics (ICCE)},
	pages={1--6},
	year={2026},
	organization={IEEE}
}

@article{dou2026wireless,
	title={A Wireless EEG Acquisition System Capable of Real-time BCI Applications},
	author={Dou, Chenlong and Luo, Jinhong and Lv, Jidong and Wang, Dong and Zhao, Bochuang and Zou, Ling},
	journal={IEEE Sensors Journal},
	year={2026},
	publisher={IEEE}
}

@article{li2025wireless,
	title={A Wireless Neural Signal Acquisition System for Interventional Brain--Computer Interface},
	author={Li, Sining and Kuang, Wenchuan and Zhang, Yiwen and Liu, Gan and Wang, Wenzhi and Jia, Hao and Ma, Yudong and Duan, Feng},
	journal={IEEE Sensors Journal},
	volume={26},
	number={1},
	pages={1147--1155},
	year={2025},
	publisher={IEEE}
}

@article{heim2025real,
	title={Real-time eeg-based bci for self-paced motor imagery and motor execution using functional neural networks},
	author={Heim, Mavin and Heinrichs, Florian and Hueppe, Michael and Nunez, Fran and Szameitat, Alexander and Reuter, Muriel and Goetz, Stefan M and Weber, Corinna},
	journal={IEEE Access},
	year={2025},
	publisher={IEEE}
}

@article{razaghi2026systematic,
	title={A systematic evaluation of EEG electrode geometry for enhanced signals: an experimental approach},
	author={Razaghi, Zahra and Faraji, Mehrbod and Mohammadpour, Raheleh and Ebrahimpour, Reza and Zad, Azam Iraji},
	journal={Scientific Reports},
	year={2026},
	publisher={Nature Publishing Group UK London}
}

@article{wang2026eeg,
	title={EEG-Based Driving Fatigue Detection in Urban Road Conditions Using a Novel Semi-Dry Electrode},
	author={Wang, Fuwang and Guo, Qi and Luo, Anni and Zhang, Xiaolei and Fu, Rongrong},
	journal={IEEE Transactions on Instrumentation and Measurement},
	year={2026},
	publisher={IEEE}
}

@article{thamaraimanalan2025exploiting,
	title={Exploiting adaptive neuro-fuzzy inference systems for cognitive patterns in multimodal brain signal analysis},
	author={Thamaraimanalan, T and Gopal, Dhanalakshmi and Vignesh, S and Kishore Kumar, K},
	journal={Scientific Reports},
	volume={15},
	number={1},
	pages={9029},
	year={2025},
	publisher={Nature Publishing Group UK London}
}

@article{yang2023optimal,
	title={Optimal fuzzy logic enabled EEG motor imagery classification for brain computer interface},
	author={Yang, Eunmok and Shankar, K and Perumal, Eswaran and Seo, Changho},
	journal={IEEE Access},
	volume={12},
	pages={46002--46011},
	year={2023},
	publisher={IEEE}
}

@article{yao2026multiview,
	title={Multiview Transfer Fuzzy Classification with Soft Variable Embedded and Discriminative Structure Preservation on Motor Imagery Electroencephalogram},
	author={Yao, Jian and Qian, Pengjiang and Gu, Xiaoqing and Sun, Jing and Wang, Liang and Yang, Guisong and Wang, Shitong},
	journal={IEEE Transactions on Fuzzy Systems},
	year={2026},
	publisher={IEEE}
}

@article{qamar2026multi,
	title={Multi-scale EEG feature decoding with Swin Transformers for subject independent motor imagery BCIs},
	author={Qamar, Wasi Ur Rehman and Abibullaev, Berdakh},
	journal={Scientific Reports},
	volume={16},
	number={1},
	pages={2503},
	year={2026},
	publisher={Nature Publishing Group UK London}
}

@article{su2025transformer,
	title={Transformer-based multiscale 3-D convolutional network for motor imagery classification},
	author={Su, Jingyu and An, Shan and Wang, Guoxin and Sun, Xinlin and Hao, Yushi and Li, Haoyu and Gao, Zhongke},
	journal={IEEE Sensors Journal},
	volume={25},
	number={5},
	pages={8621--8630},
	year={2025},
	publisher={IEEE}
}

@article{mewada2025low,
	title={A Low Computational EEG-Based Hand Movements Classification Using a Restricted Boltzmann Machine for Brain-Computer Interface Applications},
	author={Mewada, Hiren and Desai, Miral and Pires, Ivan Miguel},
	journal={Procedia Computer Science},
	volume={265},
	pages={722--727},
	year={2025},
	publisher={Elsevier}
}

@article{alencar2023embedded,
	title={Embedded restricted Boltzmann Machine Approach for adjustments of repetitive physical activities using IMU Data},
	author={Alencar, M{\'a}rcio and Barreto, Raimundo and Oliveira, Hor{\'a}cio and Souto, Eduardo},
	journal={IEEE Embedded Systems Letters},
	volume={16},
	number={2},
	pages={102--105},
	year={2023},
	publisher={IEEE}
}

@article{zhuang2025self,
	title={A Self-Optimizing Deep Belief Network With Adaptive-Active Learning: Dynamic Optimization for Neural Network},
	author={Zhuang, Kaize and Luan, Xinwei and Jiang, Xiantao and Wang, Gongming and Wang, Zi-Peng},
	journal={IEEE Systems, Man, and Cybernetics Magazine},
	volume={11},
	number={2},
	pages={75--83},
	year={2025},
	publisher={IEEE}
}

@article{zhang2026joint,
	title={Joint-Level Intent Recognition of Upper Limb based on XGBoost and Viterbi Decoding},
	author={Zhang, Qi and Tao, Yong and Xiao, Zikang and Wang, Xingsong},
	journal={IEEE Sensors Journal},
	year={2026},
	publisher={IEEE}
}

@article{xiao2025improving,
	title={Improving brain-computer interface performance with optimized frequency interaction and enhancement techniques: CFC-PSO-XGBoost (CPX)},
	author={Xiao, Xiao and Li, Haoyue},
	journal={Medical Engineering \& Physics},
	volume={143},
	pages={104392},
	year={2025},
	publisher={Elsevier}
}

@inproceedings{yang2023motor,
	title={Motor Imagery Recognition Based on an Optimized Probabilistic Neural Network with the Particle Swarm Optimization},
	author={Yang, Yufei and Li, Mingai},
	booktitle={2023 35th Chinese Control and Decision Conference (CCDC)},
	pages={4743--4748},
	year={2023},
	organization={IEEE}
}

@article{chen2026multiscale,
	title={Multiscale Pooling Spatial--Temporal Attention Network: Elevating Cross Session and Small Sample Decoding in Motor Imagery Brain--Computer Interfaces},
	author={Chen, Weijie and Daly, Ian and Chen, Yixin and Wu, Xiao and He, Xinjie and Wang, Xingyu and Cichocki, Andrzej and Jin, Jing},
	journal={IEEE Transactions on Systems, Man, and Cybernetics: Systems},
	year={2026},
	publisher={IEEE}
}

@article{li2025wavtsk,
	title={WavTSK: An Interpretable Fuzzy Network with Learnable Wavelet-Based Feature Extraction for Motor Imagery EEG Decoding},
	author={Li, Zhiying and Chen, Yi-Feng and Yu, Jingwan and Zhang, Mingming},
	journal={IEEE Transactions on Fuzzy Systems},
	year={2025},
	publisher={IEEE}
}

@article{zhang2026bayesian,
	title={Bayesian Physics-Informed Neural Networks With MIQPSO-Backstepping Control for Vibration Suppression in Nonuniform Quay Cranes},
	author={Zhang, Huapeng and Yu, Gan and Duan, Kairong and Zhang, Weidong and Sun, Ning and Xie, Wei},
	journal={IEEE Transactions on Cybernetics},
	year={2026},
	publisher={IEEE}
}

@article{qamar2023artificial,
	title={Artificial neural networks: An overview},
	author={Qamar, Roheen and Zardari, Baqar Ali},
	journal={Mesopotamian Journal of Computer Science},
	volume={2023},
	pages={124--133},
	year={2023}
}

@article{zheng2026graph,
	title={Graph neural networks for graphs with heterophily: A survey},
	author={Zheng, Xin and Wang, Yi and Liu, Yixin and Li, Ming and Zhang, Miao and Jin, Di and Yu, Philip S and Pan, Shirui},
	journal={IEEE Transactions on Knowledge and Data Engineering},
	year={2026},
	publisher={IEEE}
}

@article{chen2025unsupervised,
	title={Unsupervised domain adaptation with synchronized self-training for cross-domain motor imagery recognition},
	author={Chen, Peiyin and Liu, Xiaofeng and Ma, Chao and Wang, He and Yang, Xiong and Grebogi, Celso and Gu, Xiao and Gao, Zhongke},
	journal={IEEE journal of biomedical and health informatics},
	volume={29},
	number={5},
	pages={3664--3677},
	year={2025},
	publisher={IEEE}
}

@article{ali2026diagonal,
	title={Diagonal loading common spatial patterns with Pearson correlation coefficient based feature selection for efficient motor imagery classification},
	author={Ali, Hanaa S and Ismail, Asmaa I and El-Rabaie, El-Sayed M and Abd El-Samie, Fathi E},
	journal={Computer Methods in Biomechanics and Biomedical Engineering},
	volume={29},
	number={8},
	pages={1752--1766},
	year={2026},
	publisher={Taylor \& Francis}
}

@article{agarwal2025motor,
	title={Motor imagery-based neural networks for assisting tetraplegic patients},
	author={Agarwal, Prabhakar and Kumar, Sandeep and Singh, Rishav},
	journal={Medical \& Biological Engineering \& Computing},
	volume={63},
	number={12},
	pages={3793--3807},
	year={2025},
	publisher={Springer}
}

@article{mohammady2026leveraging,
  title={Leveraging Segmentation and Visibility Graph Analysis To Enhance Motor Imagery Classification in EEG Signals},
  author={Mohammady, Fatemeh and Asadi Amiri, Sekineh and Mohammadpoory, Zeynab},
  journal={Cognitive Computation},
  volume={18},
  number={1},
  pages={8},
  year={2026},
  publisher={Springer}
}

@article{sharma2023efficient,
  title={An efficient approach for recognition of motor imagery eeg signals using the fourier decomposition method},
  author={Sharma, Neha and Sharma, Manoj and Singhal, Amit and Vyas, Ritesh and Malik, Hasmat and Hossaini, Mohammad Asef and Afthanorhan, Asyraf},
  journal={IEEE Access},
  volume={11},
  pages={122782--122791},
  year={2023},
  publisher={IEEE}
}

@article{wu2025application,
  title={The Application of Entropy in Motor Imagery Paradigms of Brain--Computer Interfaces},
  author={Wu, Chengzhen and Yao, Bo and Zhang, Xin and Li, Ting and Wang, Jinhai and Pu, Jiangbo},
  journal={Brain Sciences},
  volume={15},
  number={2},
  pages={168},
  year={2025},
  publisher={MDPI}
}

@article{sharma2025alzheimer,
  title={Alzheimer's disease diagnosis using ensemble of random weighted features and fuzzy least square twin support vector machine},
  author={Sharma, Rahul and Goel, Tripti and Tanveer, M and Al-Dhaifallah, Mujahed},
  journal={IEEE Transactions on Emerging Topics in Computational Intelligence},
  volume={9},
  number={2},
  pages={1281--1291},
  year={2025},
  publisher={IEEE}
}

@article{pan2026ctssp,
  title={CTSSP: A temporal--spectral-spatial joint optimization algorithm for motor imagery EEG decoding},
  author={Pan, Lincong and Wang, Kun and Yi, Weibo and Zhang, Yang and Xu, Minpeng and Ming, Dong},
  journal={Journal of Neural Engineering},
  volume={23},
  number={1},
  pages={016009},
  year={2026},
  publisher={IOP Publishing}
}

@article{angulo2025proficiency,
  title={Proficiency in motor imagery is linked to the lateralization of focused ERD patterns and beta PDC},
  author={Angulo-Sherman, Irma Nayeli and Le{\'o}n-Dom{\'\i}nguez, Umberto and Martinez-Torteya, Antonio and Fragoso-Gonz{\'a}lez, Gilberto Andr{\'e}s and Mart{\'\i}nez-P{\'e}rez, Mayt{\'e} Ver{\'o}nica},
  journal={Journal of NeuroEngineering and Rehabilitation},
  volume={22},
  number={1},
  pages={30},
  year={2025},
  publisher={Springer}
}

@article{james2025hybrid,
  title={Hybrid ERD/ERS and Time--Frequency Analysis Methods for Motor Prediction},
  author={James, Micheal},
  year={2025}
}

@article{zhao2025novel,
  title={A novel hybrid brain—computer interface for EEG and EMG based on an ensemble learning approach for motor imagery classification tasks},
  author={Zhao, Zhenxi and Cao, Yingyu and Yu, Hongbin and Yu, Huixian and Huang, Junfen},
  journal={IEEE Transactions on Instrumentation and Measurement},
  year={2025},
  publisher={IEEE}
}

@article{wang2025smanet,
  title={SMANet: a model combining SincNet, multi-branch spatial-temporal CNN and attention mechanism for motor imagery BCI},
  author={Wang, Danjie and Wei, Qingguo},
  journal={IEEE Transactions on Neural Systems and Rehabilitation Engineering},
  year={2025},
  publisher={IEEE}
}

@article{liu2026novel,
  title={A Novel Drone RF Signal Enhancement Method Using Adaptive Phase-Compensated Wiener Filter for Improved Detection},
  author={Liu, Peizhou and Wang, Hailiang and Li, Zan and Ji, Yu and Liu, Ningxi and Yang, Bo and Wang, Chao and Si, Jiangbo},
  journal={IEEE Signal Processing Letters},
  year={2026},
  publisher={IEEE}
}

@inproceedings{zhao2023eeg,
  title={Eeg motor imagery classification based on sliding window and attention},
  author={Zhao, Jinke and Liu, Mingliang},
  booktitle={2023 IEEE International Conference on Mechatronics and Automation (ICMA)},
  pages={1638--1643},
  year={2023},
  organization={IEEE}
}

@article{saideepthi2023sliding,
  title={Sliding window along with EEGNet-based prediction of EEG motor imagery},
  author={Saideepthi, Pabba and Chowdhury, Anirban and Gaur, Pramod and Pachori, Ram Bilas},
  journal={IEEE Sensors Journal},
  volume={23},
  number={15},
  pages={17703--17713},
  year={2023},
  publisher={IEEE}
}

@article{siino2025investigating,
  title={Investigating the impact of rational dilated wavelet transform on motor imagery EEG decoding with deep learning models},
  author={Siino, Marco and Bonomo, Giuseppe and Sorbello, Rosario and Tinnirello, Ilenia},
  journal={IEEE Access},
  volume={13},
  pages={214223--214235},
  year={2025},
  publisher={IEEE}
}

@article{song2025awknet,
  title={AWKNet: A lightweight neural network for motor imagery electroencephalogram classification based on adaptive wavelet transform Kolmogorov--Arnold},
  author={Song, Yu and Zhang, Hang and Man, Janzhi and Jin, Xiaoqian and Li, Qi},
  journal={IEEE Transactions on Consumer Electronics},
  year={2025},
  publisher={IEEE}
}

@article{mathiyazhagan2025motor,
  title={Motor imagery EEG signal classification using novel deep learning algorithm},
  author={Mathiyazhagan, Sathish and Devasena, MS Geetha},
  journal={Scientific Reports},
  volume={15},
  number={1},
  pages={24539},
  year={2025},
  publisher={Nature Publishing Group UK London}
}

@article{zhao2026dg,
  title={DG-MTLNet: A Domain Generalization Motor Imagery Classification Model With a Novel Temporal Generation Component},
  author={Zhao, Zhenxi and Cao, Yingyu and Yu, Hongbin and Yu, Huixian and Huang, Junfen},
  journal={IEEE Transactions on Industrial Informatics},
  year={2026},
  publisher={IEEE}
}

@article{wang2026generative,
  title={Generative AI Empowers Brain-Computer Interfaces: A Review-Perspective on Technical Realities and Future Visions},
  author={Wang, Shuqiang and Guo, Yi and Dong, Yihang and Shen, Yanyan and Zhang, Zhiguo and Cheung, Albert C and Pu, Jiangbo and Zhong, Sheng-hua and Tong, Raymond Kai-Yu and Li, Ye and others},
  journal={IEEE Transactions on Consumer Electronics},
  year={2026},
  publisher={IEEE}
}

@article{zhao2026mshanet,
  title={MSHANet: A Multiscale Hybrid Attention Network for Motor Imagery EEG Decoding},
  author={Zhao, Yanlong and Cao, Dianguo and Yu, Haoyang and Liang, Guangjin and Chen, Zhicheng},
  journal={IEEE Transactions on Biomedical Engineering},
  year={2026},
  publisher={IEEE}
}

@article{modak2026novel,
  title={A Novel Multi-Layer Functional Brain Connectivity-Based Motor Imagery Classification Model Using EEG Sensor Data},
  author={Modak, Sudip and Halder, Suman and Chatterjee, Soumya},
  journal={IEEE Sensors Letters},
  year={2026},
  publisher={IEEE}
}

@article{she2025improved,
  title={Improved few-shot learning based on triplet metric for motor imagery eeg classification},
  author={She, Qingshan and Li, Chengjun and Tan, Tongcai and Fang, Feng and Zhang, Yingchun},
  journal={IEEE Transactions on Cognitive and Developmental Systems},
  volume={17},
  number={4},
  pages={987--999},
  year={2025},
  publisher={IEEE}
}

@article{shi2025eeg,
  title={EEG-based motor imagery classification with tuned heuristic fusion graph convolutional network for rehabilitation training},
  author={Shi, Kecheng and Huang, Rui and Lyu, Jianzhi and Li, Zhe and Mu, Fengjun and Peng, Zhinan and Zou, Chaobin and Cheng, Hong and Zhang, Jianwei and Ghosh, Bijoy Kumar},
  journal={IEEE Transactions on Automation Science and Engineering},
  year={2025},
  publisher={IEEE}
}

@article{gao2025effects,
  title={Effects of different preprocessing pipelines on motor imagery-based brain-computer interfaces},
  author={Gao, Xin and Gui, Kai and Wu, Xiaolong and Metcalfe, Benjamin and Zhang, Dingguo},
  journal={IEEE Journal of Biomedical and Health Informatics},
  volume={29},
  number={5},
  pages={3343--3355},
  year={2025},
  publisher={IEEE}
}

@article{lim2025exploring,
  title={Exploring Functional Connectivity in Attention Deficit/Hyperactivity Disorder: A Functional Near-infrared Spectroscopy Study with Machine Learning Analysis.},
  author={Lim, S and Dong, S--Y and McIntyre, RS and Chiang, SK and Ho, R},
  journal={IEEE Journal of Biomedical and Health Informatics},
  year={2025},
  publisher={IEEE}
}

@inproceedings{mirzaeian2025ultra,
  title={Ultra-High Order Independent Component Analysis for Intrinsic Connectivity Networks in Resting-State Functional Magnetic Resonance Imaging Data},
  author={Mirzaeian, Shiva and Jensen, Kyle M and Ballem, Adithya Ram and Calhoun, Vince D and Iraji, Armin},
  booktitle={2025 IEEE 22nd International Symposium on Biomedical Imaging (ISBI)},
  pages={1--4},
  year={2025},
  organization={IEEE}
}

@article{iivanainen2026spatial,
  title={Spatial-jitter model for magnetoencephalography sensor arrays},
  author={Iivanainen, Joonas},
  journal={IEEE Transactions on Medical Imaging},
  year={2026},
  publisher={IEEE}
}

@article{tan2025brain,
  title={A brain-computer interface system based on magnetoencephalography with optically pumped magnetometers},
  author={Tan, Gaobo and Gai, Jinming and Li, Fei and Huang, Qiying and Yu, Wengyue and Huang, Yuxiang and Zhang, GuiYing and Zhao, Xiaohu and Lin, Qiang and Hu, Zhenghui},
  journal={IEEE Transactions on Instrumentation and Measurement},
  year={2025},
  publisher={IEEE}
}

@article{raza2025deep,
  title={Deep learning approaches for eeg-motor imagery based bcis: Current models, generalization challenges, and emerging trends},
  author={Raza, Aaqib and Yusoff, Mohd Zuki},
  journal={IEEE Access},
  year={2025},
  publisher={IEEE}
}

@article{tangermann2012review,
	title={Review of the BCI competition IV},
	author={Tangermann, Michael and M{\"u}ller, Klaus-Robert and Aertsen, Ad and Birbaumer, Niels and Braun, Christoph and Brunner, Clemens and Leeb, Robert and Mehring, Carsten and Miller, Kai J and M{\"u}ller-Putz, Gernot R and others},
	journal={Frontiers in neuroscience},
	volume={6},
	pages={55},
	year={2012},
	publisher={Frontiers Research Foundation}
}

@article{wang2026topology,
	title={Topology-Learnable Static-Dynamic Graph Convolutional Network for Brain Disorder Detection with Functional MRI},
	author={Wang, Mingliang and Li, Xizhi and Sun, Qiyu and Li, Wenyang and Wang, Yulin and Huang, Jiashuang and Sun, Liang and Zhang, Daoqiang and Liu, Mingxia},
	journal={IEEE Transactions on Neural Systems and Rehabilitation Engineering},
	year={2026},
	publisher={IEEE}
}

@article{tang2026fnirnet,
	title={FNIRNet: A Spatiotemporal Statistical Feature-Based Convolutional Neural Network for Low-Channel fNIRS Brain Function Classification},
	author={Tang, Lin and Zhou, Shuang and Zhou, Nan and Gong, Liang and Wang, Wenhao and Yu, Yue and Soh, YengChai and Pedrycz, Witold},
	journal={IEEE Transactions on Cognitive and Developmental Systems},
	year={2026},
	publisher={IEEE}
}

@article{wolpaw2002brain,
	title={Brain--computer interfaces for communication and control},
	author={Wolpaw, Jonathan R and Birbaumer, Niels and McFarland, Dennis J and Pfurtscheller, Gert and Vaughan, Theresa M},
	journal={Clinical neurophysiology},
	volume={113},
	number={6},
	pages={767--791},
	year={2002},
	publisher={Elsevier}
}

@article{kudale2026comprehensive,
	title={A comprehensive review for seizure prediction based on eeg dataset: From handcrafted features to deep neural models},
	author={Kudale, Trupti and Pagare, Reena},
	journal={IEEE Access},
	year={2026},
	publisher={IEEE}
}

@article{moran2010evolution,
	title={Evolution of brain--computer interface: action potentials, local field potentials and electrocorticograms},
	author={Moran, Daniel},
	journal={Current opinion in neurobiology},
	volume={20},
	number={6},
	pages={741--745},
	year={2010},
	publisher={Elsevier}
}

@article{liao2026optimal,
	title={Optimal Electrode Design and Four-Port Equivalent-Circuit Matching for Backscatter Communication in ECoG Systems},
	author={Liao, Wei and Meng, Ziyi},
	journal={IEEE Sensors Letters},
	year={2026},
	publisher={IEEE}
}

@article{fred2022brief,
	title={A brief introduction to magnetoencephalography (MEG) and its clinical applications},
	author={Fred, Alfred Lenin and Kumar, Subbiahpillai Neelakantapillai and Kumar Haridhas, Ajay and Ghosh, Sayantan and Purushothaman Bhuvana, Harishita and Sim, Wei Khang Jeremy and Vimalan, Vijayaragavan and Givo, Fredin Arun Sedly and Jousm{\"a}ki, Veikko and Padmanabhan, Parasuraman and others},
	journal={Brain sciences},
	volume={12},
	number={6},
	pages={788},
	year={2022},
	publisher={MDPI}
}

@article{cohen1972magnetoencephalography,
	title={Magnetoencephalography: detection of the brain's electrical activity with a superconducting magnetometer},
	author={Cohen, David},
	journal={Science},
	volume={175},
	number={4022},
	pages={664--666},
	year={1972},
	publisher={American Association for the Advancement of Science}
}

@article{al2021deep,
	title={Deep learning for motor imagery EEG-based classification: A review},
	author={Al-Saegh, Ali and Dawwd, Shefa A and Abdul-Jabbar, Jassim M},
	journal={Biomedical Signal Processing and Control},
	volume={63},
	pages={102172},
	year={2021},
	publisher={Elsevier}
}

@article{jamil2021noninvasive,
	title={Noninvasive  equipment for assistive, adaptive, and rehabilitative brain--computer interfaces: a systematic literature review},
	author={Jamil, Nuraini and Belkacem, Abdelkader Nasreddine and Ouhbi, Sofia and Lakas, Abderrahmane},
	journal={Sensors},
	volume={21},
	number={14},
	pages={4754},
	year={2021},
	publisher={MDPI}
}

@article{reaves2021assessment,
	title={Assessment and application of EEG: A literature review},
	author={Reaves, J and Flavin, T and Mitra, B and Mahantesh, K and Nagaraju, V},
	journal={Journal of Applied Bioinformatics \& Computational Biology},
	volume={10},
	number={7},
	year={2021}
}

@article{barria2021bci,
	title={BCI-based control for ankle exoskeleton T-FLEX: comparison of visual and haptic stimuli with stroke survivors},
	author={Barria, Patricio and Pino, Angie and Tovar, Nicol{\'a}s and Gomez-Vargas, Daniel and Baleta, Karim and D{\'\i}az, Camilo AR and M{\'u}nera, Marcela and Cifuentes, Carlos A},
	journal={Sensors},
	volume={21},
	number={19},
	pages={6431},
	year={2021},
	publisher={MDPI}
}

@article{tang2020motor,
	title={Motor imagery EEG recognition based on conditional optimization empirical mode decomposition and multi-scale convolutional neural network},
	author={Tang, Xianlun and Li, Wei and Li, Xingchen and Ma, Weichang and Dang, Xiaoyuan},
	journal={Expert Systems with Applications},
	volume={149},
	pages={113285},
	year={2020},
	publisher={Elsevier}
}

@misc{mwatamotor,
	title={Motor imagery classification based on a recurrent-convolutional architecture to control a hexapod robot, Mathematics 9 (6)},
	author={Mwata-Velu, T and Ruiz-Pinales, J and Rostro-Gonzalez, H and Ibarra-Manzano, M and Cruz-Duarte, J and Avina-Cervantes, J}
}

@article{yang2020synchronized,
	title={A Synchronized Hybrid Brain-Computer Interface System for Simultaneous Detection and Classification of Fusion EEG Signals},
	author={Yang, Dalin and Nguyen, Trung-Hau and Chung, Wan-Young},
	journal={Complexity},
	volume={2020},
	number={1},
	pages={4137283},
	year={2020},
	publisher={Wiley Online Library}
}

@article{gant2018eeg,
	title={EEG-controlled functional electrical stimulation for hand opening and closing in chronic complete cervical spinal cord injury},
	author={Gant, Katie and Guerra, Santiago and Zimmerman, Lauren and Parks, Brandon A and Prins, Noeline W and Prasad, Abhishek},
	journal={Biomedical physics \& engineering express},
	volume={4},
	number={6},
	pages={065005},
	year={2018},
	publisher={IOP Publishing}
}

@article{peterson2020feasibility,
	title={A feasibility study of a complete low-cost consumer-grade brain-computer interface system},
	author={Peterson, Victoria and Galv{\'a}n, Catalina and Hern{\'a}ndez, Hugo and Spies, Ruben},
	journal={Heliyon},
	volume={6},
	number={3},
	year={2020},
	publisher={Elsevier}
}

@article{liu2021brain,
	title={Brain-computer interface for hands-free teleoperation of construction robots},
	author={Liu, Yizhi and Habibnezhad, Mahmoud and Jebelli, Houtan},
	journal={Automation in Construction},
	volume={123},
	pages={103523},
	year={2021},
	publisher={Elsevier}
}

@article{cardoso2021effect,
	title={Effect of a brain--computer interface based on pedaling motor imagery on cortical excitability and connectivity},
	author={Cardoso, Vivianne Fl{\'a}via and Delisle-Rodriguez, Denis and Romero-Laiseca, Maria Alejandra and Loterio, Fl{\'a}via A and Gurve, Dharmendra and Floriano, Alan and Valad{\~a}o, Carlos and Silva, Leticia and Krishnan, Sridhar and Frizera-Neto, Anselmo and others},
	journal={Sensors},
	volume={21},
	number={6},
	pages={2020},
	year={2021},
	publisher={MDPI}
}

@inproceedings{garcia2020cnn,
	title={A CNN-LSTM deep learning classifier for motor imagery EEG detection using a low-invasive and low-cost BCI headband},
	author={Garcia-Moreno, Francisco M and Bermudez-Edo, Maria and Rodr{\'\i}guez-F{\'o}rtiz, Mar{\'\i}a Jos{\'e} and Garrido, Jos{\'e} Luis},
	booktitle={2020 16th international conference on intelligent environments (IE)},
	pages={84--91},
	year={2020},
	organization={IEEE}
}

@article{khan2019multiclass,
	title={Multiclass EEG motor-imagery classification with sub-band common spatial patterns},
	author={Khan, Javeria and Bhatti, Muhammad Hamza and Khan, Usman Ghani and Iqbal, Razi},
	journal={EURASIP Journal on Wireless Communications and Networking},
	volume={2019},
	number={1},
	pages={174},
	year={2019},
	publisher={Springer}
}

@article{daeglau2020challenge,
	title={Challenge accepted? Individual performance gains for motor imagery practice with humanoid robotic EEG neurofeedback},
	author={Daeglau, Mareike and Wallhoff, Frank and Debener, Stefan and Condro, Ignatius Sapto and Kranczioch, Cornelia and Zich, Catharina},
	journal={Sensors},
	volume={20},
	number={6},
	pages={1620},
	year={2020},
	publisher={MDPI}
}

@article{lisi2018markov,
	title={Markov switching model for quick detection of event related desynchronization in EEG},
	author={Lisi, Giuseppe and Rivela, Diletta and Takai, Asuka and Morimoto, Jun},
	journal={Frontiers in neuroscience},
	volume={12},
	pages={24},
	year={2018},
	publisher={Frontiers Media SA}
}

@article{yusoff2018discrimination,
	title={Discrimination of four class simple limb motor imagery movements for brain--computer interface},
	author={Yusoff, Mohd Zuki and Mahmoud, Dalia and Malik, Aamir Saeed and Bahloul, Mohammad Rida and others},
	journal={Biomedical Signal Processing and Control},
	volume={44},
	pages={181--190},
	year={2018},
	publisher={Elsevier}
}

@article{abdalsalam2018modulation,
	title={Modulation of sensorimotor rhythms for brain-computer interface using motor imagery with online feedback},
	author={Abdalsalam, Eltaf and Yusoff, Mohd Zuki and Malik, Aamir and Kamel, Nidal S and Mahmoud, Dalia},
	journal={Signal, Image and Video Processing},
	volume={12},
	number={3},
	pages={557--564},
	year={2018},
	publisher={Springer}
}

@article{djamal2017brain,
	title={Brain computer interface game controlling using fast fourier transform and learning vector quantization},
	author={Djamal, Esmeralda C and Abdullah, Maulana Y and Renaldi, Faiza},
	journal={Journal of Telecommunication, Electronic and Computer Engineering (JTEC)},
	volume={9},
	number={2-5},
	pages={71--74},
	year={2017}
}

@article{mattia2020promotoer,
	title={The Promotoer, a brain-computer interface-assisted intervention to promote upper limb functional motor recovery after stroke: a study protocol for a randomized controlled trial to test early and long-term efficacy and to identify determinants of response},
	author={Mattia, Donatella and Pichiorri, Floriana and Colamarino, Emma and Masciullo, Marcella and Morone, Giovanni and Toppi, Jlenia and Pisotta, Iolanda and Tamburella, Federica and Lorusso, Matteo and Paolucci, Stefano and others},
	journal={BMC neurology},
	volume={20},
	number={1},
	pages={254},
	year={2020},
	publisher={Springer}
}

@article{li2019hybrid,
	title={Hybrid brain/muscle signals powered wearable walking exoskeleton enhancing motor ability in climbing stairs activity},
	author={Li, Zhijun and Yuan, Yuxia and Luo, Ling and Su, Wenbin and Zhao, Kuankuan and Xu, Cuichao and Huang, Junliang and Pi, Ming},
	journal={IEEE Transactions on Medical Robotics and Bionics},
	volume={1},
	number={4},
	pages={218--227},
	year={2019},
	publisher={IEEE}
}

@article{paszkiel2020brain,
	title={Brain--computer technology-based training system in the field of motor imagery},
	author={Paszkiel, Szczepan and Dobrakowski, Pawe{\l}},
	journal={IET Science, Measurement \& Technology},
	volume={14},
	number={10},
	pages={1014--1018},
	year={2020},
	publisher={Wiley Online Library}
}

@inproceedings{khan2015motor,
	title={Motor imagery performance evaluation using hybrid EEG-NIRS for BCI},
	author={Khan, M Jawad and Hong, Keum-Shik and Naseer, Noman and Bhutta, M Raheel},
	booktitle={2015 54th Annual Conference of the Society of Instrument and Control Engineers of Japan (SICE)},
	pages={1150--1155},
	year={2015},
	organization={IEEE}
}

@inproceedings{zich2015multimodal,
	title={Multimodal evaluation of motor imagery training supported by mobile EEG at home: A case report},
	author={Zich, Catharina and Schweinitz, Clara and Debener, Stefan and Kranczioch, Cornelia},
	booktitle={2015 IEEE International Conference on Systems, Man, and Cybernetics},
	pages={3181--3186},
	year={2015},
	organization={IEEE}
}

@article{priyatno2022classification,
	title={Classification of motor imagery brain wave for bionic hand movement using multilayer perceptron},
	author={Priyatno, Sapto Budi and Prakoso, Teguh and Riyadi, Munawar Agus},
	journal={Sinergi},
	volume={26},
	number={1},
	pages={57--64},
	year={2022},
	publisher={Universitas Mercu Buana}
}

@inproceedings{dehzangi2013simultaneous,
	title={Simultaneous classification of motor imagery and SSVEP EEG signals},
	author={Dehzangi, Omid and Zou, Yuan and Jafari, Roozbeh},
	booktitle={2013 6th International IEEE/EMBS Conference on Neural Engineering (NER)},
	pages={1303--1306},
	year={2013},
	organization={IEEE}
}

@inproceedings{vourvopoulos2019eeglass,
	title={EEGlass: An EEG-eyeware prototype for ubiquitous brain-computer interaction},
	author={Vourvopoulos, Athanasios and Niforatos, Evangelos and Giannakos, Michail},
	booktitle={Adjunct proceedings of the 2019 ACM international joint conference on pervasive and ubiquitous computing and proceedings of the 2019 ACM international symposium on wearable computers},
	pages={647--652},
	year={2019}
}

@article{rodriguez2018effects,
	title={Effects of tDCS on real-time BCI detection of pedaling motor imagery},
	author={Rodriguez-Ugarte, Maria de la Soledad and I{\'a}{\~n}ez, Eduardo and Ortiz-Garcia, Mario and Azor{\'\i}n, Jos{\'e} M},
	journal={Sensors},
	volume={18},
	number={4},
	pages={1136},
	year={2018},
	publisher={MDPI}
}

@article{rao2026calibration,
	title={Calibration-Free Online Detection in Wearable Motor Imagery Brain-Computer Interfaces},
	author={Rao, Zuguang and Lu, Zilin and Xiao, Jing and Li, Kendi and Zhu, Mengnan and Guan, Zijin and Chen, Yantao and Cao, Hui and Li, Yuanqing},
	journal={IEEE Transactions on Neural Systems and Rehabilitation Engineering},
	year={2026},
	publisher={IEEE}
}

@inproceedings{blankertz2008berlin,
	title={The Berlin brain-computer interface},
	author={Blankertz, Benjamin and Tangermann, Michael and Popescu, Florin and Krauledat, Matthias and Fazli, Siamac and D{\'o}naczy, M{\'a}rton and Curio, Gabriel and M{\"u}ller, Klaus-Robert},
	booktitle={IEEE World Congress on Computational Intelligence},
	pages={79--101},
	year={2008},
	organization={Springer}
}

@article{jiang2017semiasynchronous,
	title={Semiasynchronous BCI using wearable two-channel EEG},
	author={Jiang, Yubing and Hau, Nguyen Trung and Chung, Wan-Young},
	journal={IEEE Transactions on Cognitive and Developmental Systems},
	volume={10},
	number={3},
	pages={681--686},
	year={2017},
	publisher={IEEE}
}

@article{chowdhury2017online,
	title={Online covariate shift detection-based adaptive brain--computer interface to trigger hand exoskeleton feedback for neuro-rehabilitation},
	author={Chowdhury, Anirban and Raza, Haider and Meena, Yogesh Kumar and Dutta, Ashish and Prasad, Girijesh},
	journal={IEEE Transactions on Cognitive and Developmental Systems},
	volume={10},
	number={4},
	pages={1070--1080},
	year={2017},
	publisher={IEEE}
}

@article{du2021neurofeedback,
	title={Neurofeedback with low-cost, wearable electroencephalography (EEG) reduces symptoms in chronic Post-Traumatic Stress Disorder},
	author={Du Bois, Naomi and Bigirimana, Alain Desire and Korik, Attila and K{\'e}thina, L Gaju and Rutembesa, Eug{\`e}ne and Mutabaruka, Jean and Mutesa, Leon and Prasad, Girijesh and Jansen, Stefan and Coyle, DH},
	journal={Journal of affective disorders},
	volume={295},
	pages={1319--1334},
	year={2021},
	publisher={Elsevier}
}

@inproceedings{lu2025design,
	title={Design of A Real-Time Motor Imagery BCI-VR System with Sample Segmentation and Soft Voting},
	author={Lu, Annan and Huang, Mengjie and Liao, Kai-Lun and Sun, Yue and Yang, Rui},
	booktitle={2025 8th International Conference on Big Data and Artificial Intelligence (BDAI)},
	pages={47--52},
	year={2025},
	organization={IEEE}
}

@article{huang2026continual,
	title={Continual-Learning-Enhanced CNN--Transformer Framework for Real-Time Motor-Imagery BCI in Virtual Environments},
	author={Huang, Chao-Jen and Cao, Cheng-Fu and Shyu, Kuo Kai and Lee, Te-Min and Lee, Po-Lei},
	journal={Bioengineering},
	volume={13},
	number={5},
	pages={536},
	year={2026},
	publisher={MDPI}
}

@article{raza2026maml,
	title={A MAML-Based Lightweight Neural Network with Domain Adaptation for Cross-Subject and Generalized EEG-Based Motor Imagery Classification},
	author={Raza, Aaqib and Yusoff, Mohd Zuki},
	journal={IEEE Sensors Letters},
	year={2026},
	publisher={IEEE}
}

@article{cao2026hybrid,
	title={A Hybrid Covert Attention--Augmented Motor Imagery Paradigm for Brain--Computer Interfaces},
	author={Cao, Beining and Guo, Yixin and Tian, Yutong and Cao, Fuqun and Chen, Ying and Lin, Chin-Teng},
	journal={IEEE Transactions on Neural Systems and Rehabilitation Engineering},
	year={2026},
	publisher={IEEE}
}

@article{zhao2025multi,
	title={Multi-scale convolutional transformer network for motor imagery brain-computer interface},
	author={Zhao, Wei and Zhang, Baocan and Zhou, Haifeng and Wei, Dezhi and Huang, Chenxi and Lan, Quan},
	journal={Scientific Reports},
	volume={15},
	number={1},
	pages={12935},
	year={2025},
	publisher={Nature Publishing Group UK London}
}

@inproceedings{velakanti2025improving,
	title={Improving Motor Imagery EEG Signal Classification Using Deep Learning Techniques},
	author={Velakanti, Gouthami and Mahveen, Safa and Rao, A Abhinav and Hanith, M Sai and Priya, G Krishna and Reddy, K Sushrith},
	booktitle={2025 IEEE International Conference on Compute, Control, Network \& Photonics (ICCCNP)},
	pages={1--6},
	year={2025},
	organization={IEEE}
}

@inproceedings{sebait2025brain,
	title={Brain-Computer Interfaces for Motor Activity and Rehabilitation},
	author={Sebait, Surab and Rane, Rashmi and Sathe, Neha and Khadkatkar, Arnav and Chidrupee, Marasanapalle Kalle and Hunchalkar, Abubakar Siddiq AbdulKhadar},
	booktitle={2025 9th International Conference on Computing, Communication, Control and Automation (ICCCBEA)},
	pages={1--7},
	year={2025},
	organization={IEEE}
}

@article{hameed2024temporal,
	title={Temporal--spatial transformer based motor imagery classification for BCI using independent component analysis},
	author={Hameed, Adel and Fourati, Rahma and Ammar, Boudour and Ksibi, Amel and Alluhaidan, Ala Saleh and Ayed, Mounir Ben and Khleaf, Hussain Kareem},
	journal={Biomedical Signal Processing and Control},
	volume={87},
	pages={105359},
	year={2024},
	publisher={Elsevier}
}

@article{li2022motor,
	title={Motor imagery EEG classification algorithm based on CNN-LSTM feature fusion network},
	author={Li, Hongli and Ding, Man and Zhang, Ronghua and Xiu, Chunbo},
	journal={Biomedical signal processing and control},
	volume={72},
	pages={103342},
	year={2022},
	publisher={Elsevier}
}

@article{xie2022transformer,
	title={A transformer-based approach combining deep learning network and spatial-temporal information for raw EEG classification},
	author={Xie, Jin and Zhang, Jie and Sun, Jiayao and Ma, Zheng and Qin, Liuni and Li, Guanglin and Zhou, Huihui and Zhan, Yang},
	journal={IEEE Transactions on Neural Systems and Rehabilitation Engineering},
	volume={30},
	pages={2126--2136},
	year={2022},
	publisher={IEEE}
}

@article{hou2022gcns,
	title={GCNs-net: a graph convolutional neural network approach for decoding time-resolved eeg motor imagery signals},
	author={Hou, Yimin and Jia, Shuyue and Lun, Xiangmin and Hao, Ziqian and Shi, Yan and Li, Yang and Zeng, Rui and Lv, Jinglei},
	journal={IEEE Transactions on Neural Networks and Learning Systems},
	volume={35},
	number={6},
	pages={7312--7323},
	year={2022},
	publisher={IEEE}
}

@article{kwak2022fganet,
	title={FGANet: fNIRS-guided attention network for hybrid EEG-fNIRS brain-computer interfaces},
	author={Kwak, Youngchul and Song, Woo-Jin and Kim, Seong-Eun},
	journal={IEEE Transactions on Neural Systems and Rehabilitation Engineering},
	volume={30},
	pages={329--339},
	year={2022},
	publisher={IEEE}
}

@article{fang2022feature,
	title={Feature extraction method based on filter banks and Riemannian tangent space in motor-imagery BCI},
	author={Fang, Hua and Jin, Jing and Daly, Ian and Wang, Xingyu},
	journal={IEEE journal of biomedical and health informatics},
	volume={26},
	number={6},
	pages={2504--2514},
	year={2022},
	publisher={IEEE}
}

@article{song2022eeg,
	title={EEG conformer: Convolutional transformer for EEG decoding and visualization},
	author={Song, Yonghao and Zheng, Qingqing and Liu, Bingchuan and Gao, Xiaorong},
	journal={IEEE Transactions on Neural Systems and Rehabilitation Engineering},
	volume={31},
	pages={710--719},
	year={2022},
	publisher={IEEE}
}

@article{zhang2021adaptive,
	title={Adaptive transfer learning for EEG motor imagery classification with deep convolutional neural network},
	author={Zhang, Kaishuo and Robinson, Neethu and Lee, Seong-Whan and Guan, Cuntai},
	journal={Neural Networks},
	volume={136},
	pages={1--10},
	year={2021},
	publisher={Elsevier}
}

@article{zhang2021hybrid,
	title={Hybrid deep neural network using transfer learning for EEG motor imagery decoding},
	author={Zhang, Ruilong and Zong, Qun and Dou, Liqian and Zhao, Xinyi and Tang, Yifan and Li, Zhiyu},
	journal={Biomedical Signal Processing and Control},
	volume={63},
	pages={102144},
	year={2021},
	publisher={Elsevier}
}

@article{bang2021spatio,
	title={Spatio-spectral feature representation for motor imagery classification using convolutional neural networks},
	author={Bang, Ji-Seon and Lee, Min-Ho and Fazli, Siamac and Guan, Cuntai and Lee, Seong-Whan},
	journal={IEEE Transactions on Neural Networks and Learning Systems},
	volume={33},
	number={7},
	pages={3038--3049},
	year={2021},
	publisher={IEEE}
}

@article{autthasan2021min2net,
	title={MIN2Net: End-to-end multi-task learning for subject-independent motor imagery EEG classification},
	author={Autthasan, Phairot and Chaisaen, Rattanaphon and Sudhawiyangkul, Thapanun and Rangpong, Phurin and Kiatthaveephong, Suktipol and Dilokthanakul, Nat and Bhakdisongkhram, Gun and Phan, Huy and Guan, Cuntai and Wilaiprasitporn, Theerawit},
	journal={IEEE Transactions on Biomedical Engineering},
	volume={69},
	number={6},
	pages={2105--2118},
	year={2021},
	publisher={IEEE}
}

@article{yu2021new,
	title={A new framework for automatic detection of motor and mental imagery EEG signals for robust BCI systems},
	author={Yu, Xiaojun and Aziz, Muhammad Zulkifal and Sadiq, Muhammad Tariq and Fan, Zeming and Xiao, Gaoxi},
	journal={IEEE Transactions on Instrumentation and Measurement},
	volume={70},
	pages={1--12},
	year={2021},
	publisher={IEEE}
}

@article{gaur2021sliding,
	title={A sliding window common spatial pattern for enhancing motor imagery classification in EEG-BCI},
	author={Gaur, Pramod and Gupta, Harsh and Chowdhury, Anirban and McCreadie, Karl and Pachori, Ram Bilas and Wang, Hui},
	journal={IEEE Transactions on Instrumentation and Measurement},
	volume={70},
	pages={1--9},
	year={2021},
	publisher={IEEE}
}

@article{cho2021neurograsp,
	title={NeuroGrasp: Real-time EEG classification of high-level motor imagery tasks using a dual-stage deep learning framework},
	author={Cho, Jeong-Hyun and Jeong, Ji-Hoon and Lee, Seong-Whan},
	journal={IEEE Transactions on Cybernetics},
	volume={52},
	number={12},
	pages={13279--13292},
	year={2021},
	publisher={IEEE}
}

@article{kostas2021bendr,
	title={BENDR: Using transformers and a contrastive self-supervised learning task to learn from massive amounts of EEG data},
	author={Kostas, Demetres and Aroca-Ouellette, Stephane and Rudzicz, Frank},
	journal={Frontiers in Human Neuroscience},
	volume={15},
	pages={653659},
	year={2021},
	publisher={Frontiers Media SA}
}

@article{deng2021advanced,
	title={Advanced TSGL-EEGNet for motor imagery EEG-based brain-computer interfaces},
	author={Deng, Xin and Zhang, Boxian and Yu, Nian and Liu, Ke and Sun, Kaiwei},
	journal={IEEE access},
	volume={9},
	pages={25118--25130},
	year={2021},
	publisher={IEEE}
}

@article{mirzaei2021eeg,
	title={EEG motor imagery classification using dynamic connectivity patterns and convolutional autoencoder},
	author={Mirzaei, Sayeh and Ghasemi, Parisa},
	journal={Biomedical Signal Processing and Control},
	volume={68},
	pages={102584},
	year={2021},
	publisher={Elsevier}
}

@article{dai2020hs,
	title={HS-CNN: a CNN with hybrid convolution scale for EEG motor imagery classification},
	author={Dai, Guanghai and Zhou, Jun and Huang, Jiahui and Wang, Ning},
	journal={Journal of neural engineering},
	volume={17},
	number={1},
	pages={016025},
	year={2020},
	publisher={IOP Publishing}
}

@article{kant2020cwt,
	title={CWT based transfer learning for motor imagery classification for brain computer interfaces},
	author={Kant, Piyush and Laskar, Shahedul Haque and Hazarika, Jupitara and Mahamune, Rupesh},
	journal={Journal of Neuroscience Methods},
	volume={345},
	pages={108886},
	year={2020},
	publisher={Elsevier}
}

@article{zhang2020data,
	title={Data augmentation for motor imagery signal classification based on a hybrid neural network},
	author={Zhang, Kai and Xu, Guanghua and Han, Zezhen and Ma, Kaiquan and Zheng, Xiaowei and Chen, Longting and Duan, Nan and Zhang, Sicong},
	journal={Sensors},
	volume={20},
	number={16},
	pages={4485},
	year={2020},
	publisher={MDPI}
}

@article{zhao2020deep,
	title={Deep representation-based domain adaptation for nonstationary EEG classification},
	author={Zhao, He and Zheng, Qingqing and Ma, Kai and Li, Huiqi and Zheng, Yefeng},
	journal={IEEE Transactions on Neural Networks and Learning Systems},
	volume={32},
	number={2},
	pages={535--545},
	year={2020},
	publisher={IEEE}
}

@article{baig2020filtering,
	title={Filtering techniques for channel selection in motor imagery EEG applications: a survey},
	author={Baig, Muhammad Zeeshan and Aslam, Nauman and Shum, Hubert PH},
	journal={Artificial intelligence review},
	volume={53},
	number={2},
	pages={1207--1232},
	year={2020},
	publisher={Springer}
}

@article{venkatachalam2020novel,
	title={A Novel Method of motor imagery classification using eeg signal},
	author={Venkatachalam, K and Devipriya, A and Maniraj, J and Sivaram, M and Ambikapathy, A and others},
	journal={Artificial intelligence in medicine},
	volume={103},
	pages={101787},
	year={2020},
	publisher={Elsevier}
}

@article{gaur2019automatic,
	title={An automatic subject specific intrinsic mode function selection for enhancing two-class EEG-based motor imagery-brain computer interface},
	author={Gaur, Pramod and Pachori, Ram Bilas and Wang, Hui and Prasad, Girijesh},
	journal={IEEE Sensors Journal},
	volume={19},
	number={16},
	pages={6938--6947},
	year={2019},
	publisher={IEEE}
}

@article{he2019transfer,
	title={Transfer learning for brain--computer interfaces: A Euclidean space data alignment approach},
	author={He, He and Wu, Dongrui},
	journal={IEEE Transactions on Biomedical Engineering},
	volume={67},
	number={2},
	pages={399--410},
	year={2019},
	publisher={IEEE}
}

@article{kwon2019subject,
	title={Subject-independent brain--computer interfaces based on deep convolutional neural networks},
	author={Kwon, O-Yeon and Lee, Min-Ho and Guan, Cuntai and Lee, Seong-Whan},
	journal={IEEE transactions on neural networks and learning systems},
	volume={31},
	number={10},
	pages={3839--3852},
	year={2019},
	publisher={IEEE}
}

@article{ko2019multimodal,
	title={Multimodal fuzzy fusion for enhancing the motor-imagery-based brain computer interface},
	author={Ko, Li-Wei and Lu, Yi-Chen and Bustince, Humberto and Chang, Yu-Cheng and Chang, Yang and Ferandez, Javier and Wang, Yu-Kai and Sanz, Jose Antonio and Dimuro, Gracaliz Pereira and Lin, Chin-Teng},
	journal={IEEE Computational Intelligence Magazine},
	volume={14},
	number={1},
	pages={96--106},
	year={2019},
	publisher={IEEE}
}

@article{li2019channel,
	title={A channel-projection mixed-scale convolutional neural network for motor imagery EEG decoding},
	author={Li, Yang and Zhang, Xian-Rui and Zhang, Bin and Lei, Meng-Ying and Cui, Wei-Gang and Guo, Yu-Zhu},
	journal={IEEE Transactions on Neural Systems and Rehabilitation Engineering},
	volume={27},
	number={6},
	pages={1170--1180},
	year={2019},
	publisher={IEEE}
}

@article{azab2019weighted,
	title={Weighted transfer learning for improving motor imagery-based brain--computer interface},
	author={Azab, Ahmed M and Mihaylova, Lyudmila and Ang, Kai Keng and Arvaneh, Mahnaz},
	journal={IEEE Transactions on Neural Systems and Rehabilitation Engineering},
	volume={27},
	number={7},
	pages={1352--1359},
	year={2019},
	publisher={IEEE}
}

@article{ji2019eeg,
	title={EEG signals feature extraction based on DWT and EMD combined with approximate entropy},
	author={Ji, Na and Ma, Liang and Dong, Hui and Zhang, Xuejun},
	journal={Brain sciences},
	volume={9},
	number={8},
	pages={201},
	year={2019},
	publisher={MDPI}
}

@article{xu2019deep,
	title={A deep transfer convolutional neural network framework for EEG signal classification},
	author={Xu, Gaowei and Shen, Xiaoang and Chen, Sirui and Zong, Yongshuo and Zhang, Canyang and Yue, Hongyang and Liu, Min and Chen, Fei and Che, Wenliang},
	journal={IEEe Access},
	volume={7},
	pages={112767--112776},
	year={2019},
	publisher={IEEE}
}

@article{dai2019eeg,
	title={EEG classification of motor imagery using a novel deep learning framework},
	author={Dai, Mengxi and Zheng, Dezhi and Na, Rui and Wang, Shuai and Zhang, Shuailei},
	journal={Sensors},
	volume={19},
	number={3},
	pages={551},
	year={2019},
	publisher={MDPI}
}

@article{zhang2019convolutional,
	title={A convolutional recurrent attention model for subject-independent EEG signal analysis},
	author={Zhang, Dalin and Yao, Lina and Chen, Kaixuan and Monaghan, Jessica},
	journal={IEEE signal processing letters},
	volume={26},
	number={5},
	pages={715--719},
	year={2019},
	publisher={IEEE}
}

@article{lawhern2018eegnet,
	title={EEGNet: a compact convolutional neural network for EEG-based brain--computer interfaces},
	author={Lawhern, Vernon J and Solon, Amelia J and Waytowich, Nicholas R and Gordon, Stephen M and Hung, Chou P and Lance, Brent J},
	journal={Journal of neural engineering},
	volume={15},
	number={5},
	pages={056013},
	year={2018},
	publisher={iOP Publishing}
}

@article{dose2018end,
	title={An end-to-end deep learning approach to MI-EEG signal classification for BCIs},
	author={Dose, Hauke and M{\o}ller, Jakob S and Iversen, Helle K and Puthusserypady, Sadasivan},
	journal={Expert Systems with Applications},
	volume={114},
	pages={532--542},
	year={2018},
	publisher={Elsevier}
}

@article{zhang2018temporally,
	title={Temporally constrained sparse group spatial patterns for motor imagery BCI},
	author={Zhang, Yu and Nam, Chang S and Zhou, Guoxu and Jin, Jing and Wang, Xingyu and Cichocki, Andrzej},
	journal={IEEE transactions on cybernetics},
	volume={49},
	number={9},
	pages={3322--3332},
	year={2018},
	publisher={IEEE}
}

@article{luo2018exploring,
	title={Exploring spatial-frequency-sequential relationships for motor imagery classification with recurrent neural network},
	author={Luo, Tian-jian and Zhou, Chang-le and Chao, Fei},
	journal={BMC bioinformatics},
	volume={19},
	number={1},
	pages={344},
	year={2018},
	publisher={Springer}
}

@inproceedings{donovan2018motor,
	title={Motor imagery classification using TSK fuzzy inference neural networks},
	author={Donovan, Rory and Yu, Xiao-Hua},
	booktitle={2018 international joint conference on neural networks (IJCNN)},
	pages={1--6},
	year={2018},
	organization={IEEE}
}

@article{jafarifarmand2017new,
	title={A new self-regulated neuro-fuzzy framework for classification of EEG signals in motor imagery BCI},
	author={Jafarifarmand, Aysa and Badamchizadeh, Mohammad Ali and Khanmohammadi, Sohrab and Nazari, Mohammad Ali and Tazehkand, Behzad Mozaffari},
	journal={IEEE transactions on fuzzy systems},
	volume={26},
	number={3},
	pages={1485--1497},
	year={2017},
	publisher={IEEE}
}

@article{sakhavi2018learning,
	title={Learning temporal information for brain-computer interface using convolutional neural networks},
	author={Sakhavi, Siavash and Guan, Cuntai and Yan, Shuicheng},
	journal={IEEE transactions on neural networks and learning systems},
	volume={29},
	number={11},
	pages={5619--5629},
	year={2018},
	publisher={IEEE}
}

@article{xu2018wavelet,
	title={Wavelet transform time-frequency image and convolutional network-based motor imagery EEG classification},
	author={Xu, Baoguo and Zhang, Linlin and Song, Aiguo and Wu, Changcheng and Li, Wenlong and Zhang, Dalin and Xu, Guozheng and Li, Huijun and Zeng, Hong},
	journal={Ieee Access},
	volume={7},
	pages={6084--6093},
	year={2018},
	publisher={IEEE}
}

@article{yang2018deep,
	title={Deep fusion feature learning network for MI-EEG classification},
	author={Yang, Jun and Yao, Shaowen and Wang, Jin},
	journal={Ieee Access},
	volume={6},
	pages={79050--79059},
	year={2018},
	publisher={IEEE}
}

@inproceedings{zhou2018classification,
	title={Classification of motor imagery EEG using wavelet envelope analysis and LSTM networks},
	author={Zhou, Jie and Meng, Ming and Gao, Yunyuan and Ma, Yuliang and Zhang, Qizhong},
	booktitle={2018 Chinese Control And Decision Conference (CCDC)},
	pages={5600--5605},
	year={2018},
	organization={IEEE}
}

@article{tang2017single,
	title={Single-trial EEG classification of motor imagery using deep convolutional neural networks},
	author={Tang, Zhichuan and Li, Chao and Sun, Shouqian},
	journal={Optik},
	volume={130},
	pages={11--18},
	year={2017},
	publisher={Elsevier}
}

@article{schirrmeister2017deep,
	title={Deep learning with convolutional neural networks for EEG decoding and visualization},
	author={Schirrmeister, Robin Tibor and Springenberg, Jost Tobias and Fiederer, Lukas Dominique Josef and Glasstetter, Martin and Eggensperger, Katharina and Tangermann, Michael and Hutter, Frank and Burgard, Wolfram and Ball, Tonio},
	journal={Human brain mapping},
	volume={38},
	number={11},
	pages={5391--5420},
	year={2017},
	publisher={Wiley Online Library}
}

@article{baig2017differential,
	title={Differential evolution algorithm as a tool for optimal feature subset selection in motor imagery EEG},
	author={Baig, Muhammad Zeeshan and Aslam, Nauman and Shum, Hubert PH and Zhang, Li},
	journal={Expert Systems with Applications},
	volume={90},
	pages={184--195},
	year={2017},
	publisher={Elsevier}
}

@article{kumar2017improved,
	title={An improved discriminative filter bank selection approach for motor imagery EEG signal classification using mutual information},
	author={Kumar, Shiu and Sharma, Alok and Tsunoda, Tatsuhiko},
	journal={BMC bioinformatics},
	volume={18},
	number={Suppl 16},
	pages={545},
	year={2017},
	publisher={Springer}
}

@article{kevric2017comparison,
	title={Comparison of signal decomposition methods in classification of EEG signals for motor-imagery BCI system},
	author={Kevric, Jasmin and Subasi, Abdulhamit},
	journal={Biomedical Signal Processing and Control},
	volume={31},
	pages={398--406},
	year={2017},
	publisher={Elsevier}
}

@article{zhang2017sparse,
	title={Sparse Bayesian learning for obtaining sparsity of EEG frequency bands based feature vectors in motor imagery classification},
	author={Zhang, Yu and Wang, Yu and Jin, Jing and Wang, Xingyu},
	journal={International journal of neural systems},
	volume={27},
	number={02},
	pages={1650032},
	year={2017},
	publisher={World Scientific}
}

@article{tabar2017novel,
	title={A novel deep learning approach for classification of EEG motor imagery signals},
	author={Tabar, Yousef Rezaei and Halici, Ugur},
	journal={Journal of neural engineering},
	volume={14},
	number={1},
	pages={016003},
	year={2017},
	publisher={IOP Publishing}
}

@article{zhang2017classification,
	title={Classification of EEG signals based on autoregressive model and wavelet packet decomposition},
	author={Zhang, Yong and Liu, Bo and Ji, Xiaomin and Huang, Dan},
	journal={Neural Processing Letters},
	volume={45},
	number={2},
	pages={365--378},
	year={2017},
	publisher={Springer}
}

@article{lu2016deep,
	title={A deep learning scheme for motor imagery classification based on restricted Boltzmann machines},
	author={Lu, Na and Li, Tengfei and Ren, Xiaodong and Miao, Hongyu},
	journal={IEEE transactions on neural systems and rehabilitation engineering},
	volume={25},
	number={6},
	pages={566--576},
	year={2016},
	publisher={IEEE}
}

@article{bhatti2019soft,
	title={Soft computing-based EEG classification by optimal feature selection and neural networks},
	author={Bhatti, Muhammad Hamza and Khan, Javeria and Khan, Muhammad Usman Ghani and Iqbal, Razi and Aloqaily, Moayad and Jararweh, Yaser and Gupta, Brij},
	journal={IEEE Transactions on Industrial Informatics},
	volume={15},
	number={10},
	pages={5747--5754},
	year={2019},
	publisher={IEEE}
}

@article{wang2016detection,
	title={Detection of motor imagery EEG signals employing Na{\"\i}ve Bayes based learning process},
	author={Wang, Hua and Zhang, Yanchun and others},
	journal={Measurement},
	volume={86},
	pages={148--158},
	year={2016},
	publisher={Elsevier}
}

@article{aghaei2015separable,
	title={Separable common spatio-spectral patterns for motor imagery BCI systems},
	author={Aghaei, Amirhossein S and Mahanta, Mohammad Shahin and Plataniotis, Konstantinos N},
	journal={IEEE Transactions on Biomedical Engineering},
	volume={63},
	number={1},
	pages={15--29},
	year={2015},
	publisher={IEEE}
}

@article{he2015noninvasive,
	title={Noninvasive brain-computer interfaces based on sensorimotor rhythms},
	author={He, Bin and Baxter, Bryan and Edelman, Bradley J and Cline, Christopher C and Ye, Wenjing W},
	journal={Proceedings of the IEEE},
	volume={103},
	number={6},
	pages={907--925},
	year={2015},
	publisher={IEEE}
}

@article{zhang2015optimizing,
	title={Optimizing spatial patterns with sparse filter bands for motor-imagery based brain--computer interface},
	author={Zhang, Yu and Zhou, Guoxu and Jin, Jing and Wang, Xingyu and Cichocki, Andrzej},
	journal={Journal of neuroscience methods},
	volume={255},
	pages={85--91},
	year={2015},
	publisher={Elsevier}
}

@inproceedings{yang2015use,
	title={On the use of convolutional neural networks and augmented CSP features for multi-class motor imagery of EEG signals classification},
	author={Yang, Huijuan and Sakhavi, Siavash and Ang, Kai Keng and Guan, Cuntai},
	booktitle={2015 37th annual international conference of the IEEE engineering in medicine and biology society (EMBC)},
	pages={2620--2623},
	year={2015},
	organization={IEEE}
}

@inproceedings{sakhavi2015parallel,
	title={Parallel convolutional-linear neural network for motor imagery classification},
	author={Sakhavi, Siavash and Guan, Cuntai and Yan, Shuicheng},
	booktitle={2015 23rd European signal processing conference (EUSIPCO)},
	pages={2736--2740},
	year={2015},
	organization={IEEE}
}

@article{yu2014analysis,
	title={Analysis the effect of PCA for feature reduction in non-stationary EEG based motor imagery of BCI system},
	author={Yu, Xinyang and Chum, Pharino and Sim, Kwee-Bo},
	journal={Optik},
	volume={125},
	number={3},
	pages={1498--1502},
	year={2014},
	publisher={Elsevier}
}

@article{aydemir2014decision,
	title={Decision tree structure based classification of EEG signals recorded during two dimensional cursor movement imagery},
	author={Aydemir, Onder and Kayikcioglu, Temel},
	journal={Journal of neuroscience methods},
	volume={229},
	pages={68--75},
	year={2014},
	publisher={Elsevier}
}

@misc{al2014methods,
	title={Methods of EEG signal features extraction using linear analysis in frequency and time-frequency domains. ISRN Neuroscience, 2014, 730218},
	author={Al-Fahoum, AS and Al-Fraihat, AA},
	year={2014}
}

@inproceedings{ren2014convolutional,
	title={Convolutional deep belief networks for feature extraction of EEG signal},
	author={Ren, Yuanfang and Wu, Yan},
	booktitle={2014 International joint conference on neural networks (IJCNN)},
	pages={2850--2853},
	year={2014},
	organization={IEEE}
}

@article{yang2014adaptive,
	title={Adaptive Neuro-Fuzzy Inference System for Classification of Background EEG Signals from ESES Patients and Controls},
	author={Yang, Zhixian and Wang, Yinghua and Ouyang, Gaoxiang},
	journal={The Scientific World Journal},
	volume={2014},
	number={1},
	pages={140863},
	year={2014},
	publisher={Wiley Online Library}
}

@article{kim2025efficacy,
	title={Efficacy of brain-computer interface training with motor imagery-contingent feedback in improving upper limb function and neuroplasticity among persons with chronic stroke: a double-blinded, parallel-group, randomized controlled trial},
	author={Kim, Myeong Sun and Park, Hyunju and Kwon, Ilho and An, Kwang-Ok and Kim, Hayeon and Park, Gyulee and Hyung, Wooseok and Im, Chang-Hwan and Shin, Joon-Ho},
	journal={Journal of NeuroEngineering and Rehabilitation},
	volume={22},
	number={1},
	pages={1},
	year={2025},
	publisher={Springer}
}

@article{feng2022efficient,
	title={An efficient EEGNet processor design for portable EEG-Based BCIs},
	author={Feng, Lichen and Yang, Liying and Liu, Shubin and Han, Chenxi and Zhang, Yueqi and Zhu, Zhangming},
	journal={Microelectronics Journal},
	volume={120},
	pages={105356},
	year={2022},
	publisher={Elsevier}
}

@article{mahmood2021wireless,
	title={Wireless Soft Scalp Electronics and Virtual Reality System for Motor Imagery-Based Brain--Machine Interfaces},
	author={Mahmood, Musa and Kwon, Shinjae and Kim, Hojoong and Kim, Yun-Soung and Siriaraya, Panote and Choi, Jeongmoon and Otkhmezuri, Boris and Kang, Kyowon and Yu, Ki Jun and Jang, Young C and others},
	journal={Advanced Science},
	volume={8},
	number={19},
	pages={2101129},
	year={2021},
	publisher={Wiley Online Library}
}

@article{marcos2020real,
	title={Real time identification of motor imagery actions on EEG signals.},
	author={Marcos Rojas, Carlos Luis and Chailloux Peguero, Juan David and Alba Blanco, Emiliano},
	journal={Ingenier{\'\i}a Electr{\'o}nica, Autom{\'a}tica y Comunicaciones},
	volume={41},
	number={1},
	pages={101--117},
	year={2020},
	publisher={Facultad de El{\'e}ctrica, Instituto Superior Polit{\'e}cnico Jos{\'e} Antonio~…}
}

@inproceedings{schneider2020q,
	title={Q-EEGNet: An energy-efficient 8-bit quantized parallel EEGNet implementation for edge motor-imagery brain-machine interfaces},
	author={Schneider, Tibor and Wang, Xiaying and Hersche, Michael and Cavigelli, Lukas and Benini, Luca},
	booktitle={2020 IEEE International Conference on Smart Computing (SMARTCOMP)},
	pages={284--289},
	year={2020},
	organization={IEEE}
}

@article{olivas2019classification,
	title={Classification of multiple motor imagery using deep convolutional neural networks and spatial filters},
	author={Olivas-Padilla, Brenda E and Chacon-Murguia, Mario I},
	journal={Applied Soft Computing},
	volume={75},
	pages={461--472},
	year={2019},
	publisher={Elsevier}
}

@inproceedings{ma2019fpga,
	title={Fpga-based rapid electroencephalography signal classification system},
	author={Ma, Xunguang and Zheng, Wenkai and Peng, Zujian and Yang, Jimin},
	booktitle={2019 IEEE 11th International Conference on Advanced Infocomm Technology (ICAIT)},
	pages={223--227},
	year={2019},
	organization={IEEE}
}

@article{foong2019assessment,
	title={Assessment of the efficacy of EEG-based MI-BCI with visual feedback and EEG correlates of mental fatigue for upper-limb stroke rehabilitation},
	author={Foong, Ruyi and Ang, Kai Keng and Quek, Chai and Guan, Cuntai and Phua, Kok Soon and Kuah, Christopher Wee Keong and Deshmukh, Vishwanath Arun and Yam, Lester Hon Lum and Rajeswaran, Deshan Kumar and Tang, Ning and others},
	journal={IEEE Transactions on Biomedical Engineering},
	volume={67},
	number={3},
	pages={786--795},
	year={2019},
	publisher={IEEE}
}

@article{belwafi2018embedded,
	title={An embedded implementation based on adaptive filter bank for brain--computer interface systems},
	author={Belwafi, Kais and Romain, Olivier and Gannouni, Sofien and Ghaffari, Fakhreddine and Djemal, Ridha and Ouni, Bouraoui},
	journal={Journal of neuroscience methods},
	volume={305},
	pages={1--16},
	year={2018},
	publisher={Elsevier}
}

@article{malekmohammadi2019efficient,
	title={An efficient hardware implementation for a motor imagery brain computer interface system},
	author={Malekmohammadi, Alireza and Mohammadzade, Hoda and Chamanzar, Alireza and Shabany, Mahdi and Ghojogh, Benyamin},
	journal={Scientia Iranica},
	volume={26},
	number={Special Issue on: Socio-Cognitive Engineering},
	pages={72--94},
	year={2019},
	publisher={Sharif University of Technology}
}

@inproceedings{shrivastwa2018fpga,
	title={An FPGA-based brain computer interfacing using compressive sensing and machine learning},
	author={Shrivastwa, Ritu Ranjan and Pudi, Vikramkumar and Chattopadhyay, Anupam},
	booktitle={2018 IEEE computer society annual symposium on VLSI (ISVLSI)},
	pages={726--731},
	year={2018},
	organization={IEEE}
}

@article{wu2016fuzzy,
	title={Fuzzy integral with particle swarm optimization for a motor-imagery-based brain--computer interface},
	author={Wu, Shang-Lin and Liu, Yu-Ting and Hsieh, Tsung-Yu and Lin, Yang-Yin and Chen, Chih-Yu and Chuang, Chun-Hsiang and Lin, Chin-Teng},
	journal={IEEE Transactions on Fuzzy Systems},
	volume={25},
	number={1},
	pages={21--28},
	year={2016},
	publisher={IEEE}
}

@article{mccrimmon2017performance,
	title={Performance assessment of a custom, portable, and low-cost brain--computer interface platform},
	author={McCrimmon, Colin M and Fu, Jonathan Lee and Wang, Ming and Lopes, Lucas Silva and Wang, Po T and Karimi-Bidhendi, Alireza and Liu, Charles Y and Heydari, Payam and Nenadic, Zoran and Do, An Hong},
	journal={IEEE Transactions on Biomedical Engineering},
	volume={64},
	number={10},
	pages={2313--2320},
	year={2017},
	publisher={IEEE}
}

@article{liu2017feature,
	title={Feature selection for motor imagery EEG classification based on firefly algorithm and learning automata},
	author={Liu, Aiming and Chen, Kun and Liu, Quan and Ai, Qingsong and Xie, Yi and Chen, Anqi},
	journal={Sensors},
	volume={17},
	number={11},
	pages={2576},
	year={2017},
	publisher={MDPI}
}

@article{herman2016designing,
	title={Designing an interval type-2 fuzzy logic system for handling uncertainty effects in brain--computer interface classification of motor imagery induced EEG patterns},
	author={Herman, Pawel Andrzej and Prasad, Girijesh and McGinnity, Thomas Martin},
	journal={IEEE transactions on fuzzy systems},
	volume={25},
	number={1},
	pages={29--42},
	year={2016},
	publisher={IEEE}
}

@article{leeb2015towards,
	title={Towards independence: a BCI telepresence robot for people with severe motor disabilities},
	author={Leeb, Robert and Tonin, Luca and Rohm, Martin and Desideri, Lorenzo and Carlson, Tom and Millan, Jose del R},
	journal={Proceedings of the IEEE},
	volume={103},
	number={6},
	pages={969--982},
	year={2015},
	publisher={IEEE}
}

@article{saibene2023eeg,
	title={EEG-based BCIs on motor imagery paradigm using wearable technologies: a systematic review},
	author={Saibene, Aurora and Caglioni, Mirko and Corchs, Silvia and Gasparini, Francesca},
	journal={Sensors},
	volume={23},
	number={5},
	pages={2798},
	year={2023},
	publisher={MDPI}
}

@article{fiedler2022high,
	title={A high-density 256-channel cap for dry electroencephalography},
	author={Fiedler, Patrique and Fonseca, Carlos and Supriyanto, Eko and Zanow, Frank and Haueisen, Jens},
	journal={Human brain mapping},
	volume={43},
	number={4},
	pages={1295--1308},
	year={2022},
	publisher={Wiley Online Library}
}

@article{casson2010wearable,
	title={Wearable electroencephalography},
	author={Casson, Alexander J and Yates, David C and Smith, Shelagh JM and Duncan, John S and Rodriguez-Villegas, Esther},
	journal={IEEE engineering in medicine and biology magazine},
	volume={29},
	number={3},
	pages={44--56},
	year={2010},
	publisher={IEEE}
}

@article{soufineyestani2020electroencephalography,
	title={Electroencephalography (EEG) technology applications and available devices},
	author={Soufineyestani, Mahsa and Dowling, Dale and Khan, Arshia},
	journal={Applied Sciences},
	volume={10},
	number={21},
	pages={7453},
	year={2020},
	publisher={MDPI}
}

@book{hu2019eeg,
	title={EEG signal processing and feature extraction},
	author={Hu, Li and Zhang, Zhiguo},
	year={2019},
	publisher={Springer}
}

@article{li2020review,
	title={Review of semi-dry electrodes for EEG recording},
	author={Li, Guang-Li and Wu, Jing-Tao and Xia, Yong-Hui and He, Quan-Guo and Jin, Hong-Guang},
	journal={Journal of Neural Engineering},
	volume={17},
	number={5},
	pages={051004},
	year={2020},
	publisher={IOP Publishing}
}

@article{wang2022robust,
	title={Robust tattoo electrode prepared by paper-assisted water transfer printing for wearable health monitoring},
	author={Wang, Hongfei and Wang, Jingjing and Chen, Da and Ge, Song and Liu, Yijian and Wang, Zhengjie and Zhang, Xiaojun and Guo, Qiuquan and Yang, Jun},
	journal={IEEE Sensors Journal},
	volume={22},
	number={5},
	pages={3817--3827},
	year={2022},
	publisher={IEEE}
}

@article{casson2019wearable,
	title={Wearable EEG and beyond},
	author={Casson, Alexander J},
	journal={Biomedical engineering letters},
	volume={9},
	number={1},
	pages={53--71},
	year={2019},
	publisher={Springer}
}

@article{mihajlovic2014wearable,
	title={Wearable, wireless EEG solutions in daily life applications: what are we missing?},
	author={Mihajlovi{\'c}, Vojkan and Grundlehner, Bernard and Vullers, Ruud and Penders, Julien},
	journal={IEEE journal of biomedical and health informatics},
	volume={19},
	number={1},
	pages={6--21},
	year={2014},
	publisher={IEEE}
}

@inproceedings{do2014wireless,
	title={Wireless behind-the-ear EEG recording device with wireless interface to a mobile device (iPhone/iPod touch)},
	author={Do Valle, Bruno G and Cash, Sydney S and Sodini, Charlie G},
	booktitle={2014 36th Annual International Conference of the IEEE Engineering in Medicine and Biology Society},
	pages={5952--5955},
	year={2014},
	organization={IEEE}
}

@article{khan2020review,
	title={Review on motor imagery based BCI systems for upper limb post-stroke neurorehabilitation: From designing to application},
	author={Khan, Muhammad Ahmed and Das, Rig and Iversen, Helle K and Puthusserypady, Sadasivan},
	journal={Computers in biology and medicine},
	volume={123},
	pages={103843},
	year={2020},
	publisher={Elsevier}
}

@article{pfurtscheller2010neurofeedback,
	title={Neurofeedback Training for BCI Control, Brain--Computer Interfaces},
	author={Pfurtscheller, G and Neuper, C},
	journal={The Frontiers Collection},
	pages={65},
	year={2010}
}

@article{vourvopoulos2019efficacy,
	title={Efficacy and brain imaging correlates of an immersive motor imagery BCI-driven VR system for upper limb motor rehabilitation: A clinical case report},
	author={Vourvopoulos, Athanasios and Jorge, Carolina and Abreu, Rodolfo and Figueiredo, Patr{\'\i}cia and Fernandes, Jean-Claude and Bermudez i Badia, Sergi},
	journal={Frontiers in human neuroscience},
	volume={13},
	pages={244},
	year={2019},
	publisher={Frontiers Media SA}
}

@inproceedings{carrino2012self,
	title={A self-paced BCI system to control an electric wheelchair: Evaluation of a commercial, low-cost EEG device},
	author={Carrino, Francesco and Dumoulin, Joel and Mugellini, Elena and Abou Khaled, Omar and Ingold, Rolf},
	booktitle={2012 ISSNIP biosignals and biorobotics conference: biosignals and robotics for better and safer living (BRC)},
	pages={1--6},
	year={2012},
	organization={IEEE}
}

@article{alanis2020assessment,
	title={On the assessment of functional connectivity in an immersive brain-computer interface during motor imagery},
	author={Alanis-Espinosa, Myriam and Guti{\'e}rrez, David},
	journal={Frontiers in Psychology},
	volume={11},
	pages={1301},
	year={2020},
	publisher={Frontiers Media SA}
}

@article{covi2021adaptive,
	title={Adaptive extreme edge computing for wearable devices},
	author={Covi, Erika and Donati, Elisa and Liang, Xiangpeng and Kappel, David and Heidari, Hadi and Payvand, Melika and Wang, Wei},
	journal={Frontiers in Neuroscience},
	volume={15},
	pages={611300},
	year={2021},
	publisher={Frontiers Media SA}
}

@article{jin2022survey,
	title={A survey on edge computing for wearable technology},
	author={Jin, Xinqi and Li, Lingkun and Dang, Fan and Chen, Xinlei and Liu, Yunhao},
	journal={Digital Signal Processing},
	volume={125},
	pages={103146},
	year={2022},
	publisher={Elsevier}
}

@article{myrden2017passive,
	title={A passive EEG-BCI for single-trial detection of changes in mental state},
	author={Myrden, Andrew and Chau, Tom},
	journal={IEEE Transactions on neural systems and rehabilitation Engineering},
	volume={25},
	number={4},
	pages={345--356},
	year={2017},
	publisher={IEEE}
}

@article{huang2022survey,
	title={A survey of quantum computing hybrid applications with brain-computer interface},
	author={Huang, Dandan and Wang, Mei and Wang, Jianping and Yan, Jiaxin},
	journal={Cognitive Robotics},
	volume={2},
	pages={164--176},
	year={2022},
	publisher={Elsevier}
}

@article{miranda2022quantum,
	title={Quantum brain networks: A perspective},
	author={Miranda, Eduardo R and Mart{\'\i}n-Guerrero, Jos{\'e} D and Venkatesh, Satvik and Hernani-Morales, Carlos and Lamata, Lucas and Solano, Enrique},
	journal={Electronics},
	volume={11},
	number={10},
	pages={1528},
	year={2022},
	publisher={MDPI}
}

@article{miranda2022approach,
	title={An approach to interfacing the brain with quantum computers: practical steps and caveats},
	author={Miranda, Eduardo Reck and Venkatesh, Satvik and Mart{\i}n-Guerrero, Jos{\'e} D and Hernani-Morales, Carlos and Lamata, Lucas and Solano, Enrique},
	journal={arXiv preprint arXiv:2201.00817},
	year={2022}
}

@article{miller2020current,
	title={The current state of electrocorticography-based brain--computer interfaces},
	author={Miller, Kai J and Hermes, Dora and Staff, Nathan P},
	journal={Neurosurgical focus},
	volume={49},
	number={1},
	pages={E2},
	year={2020},
	publisher={American Association of Neurological Surgeons}
}

@article{varbu2022past,
	title={Past, present, and future of EEG-based BCI applications},
	author={V{\"a}rbu, Kaido and Muhammad, Naveed and Muhammad, Yar},
	journal={Sensors},
	volume={22},
	number={9},
	pages={3331},
	year={2022},
	publisher={MDPI}
}

@article{kawala2021summary,
	title={Summary of over fifty years with brain-computer interfaces—a review},
	author={Kawala-Sterniuk, Aleksandra and Browarska, Natalia and Al-Bakri, Amir and Pelc, Mariusz and Zygarlicki, Jaroslaw and Sidikova, Michaela and Martinek, Radek and Gorzelanczyk, Edward Jacek},
	journal={Brain sciences},
	volume={11},
	number={1},
	pages={43},
	year={2021},
	publisher={MDPI}
}

@article{vidal1973toward,
	title={Toward direct brain-computer communication},
	author={Vidal, Jacques J},
	journal={Annual review of Biophysics and Bioengineering},
	volume={2},
	number={1},
	pages={157--180},
	year={1973},
	publisher={Annual Reviews 4139 El Camino Way, PO Box 10139, Palo Alto, CA 94303-0139, USA}
}

@article{carmena2003learning,
	title={Learning to control a brain--machine interface for reaching and grasping by primates},
	author={Carmena, Jose M and Lebedev, Mikhail A and Crist, Roy E and O'Doherty, Joseph E and Santucci, David M and Dimitrov, Dragan F and Patil, Parag G and Henriquez, Craig S and Nicolelis, Miguel A L},
	journal={PLoS biology},
	volume={1},
	number={2},
	pages={e42},
	year={2003},
	publisher={Public Library of Science San Francisco, USA}
}

@article{ifft2013brain,
	title={A brain-machine interface enables bimanual arm movements in monkeys},
	author={Ifft, Peter J and Shokur, Solaiman and Li, Zheng and Lebedev, Mikhail A and Nicolelis, Miguel AL},
	journal={Science translational medicine},
	volume={5},
	number={210},
	pages={210ra154--210ra154},
	year={2013},
	publisher={American Association for the Advancement of Science}
}

@article{kim2006continuous,
	title={Continuous shared control for stabilizing reaching and grasping with brain-machine interfaces},
	author={Kim, Hyun K and Biggs, J and Schloerb, W and Carmena, M and Lebedev, Mikhail A and Nicolelis, Miguel AL and Srinivasan, Mandayam A},
	journal={IEEE Transactions on Biomedical Engineering},
	volume={53},
	number={6},
	pages={1164--1173},
	year={2006},
	publisher={IEEE}
}

@inproceedings{bird2019mental,
	title={Mental emotional sentiment classification with an eeg-based brain-machine interface},
	author={Bird, Jordan J and Ekart, Aniko and Buckingham, Christopher D and Faria, Diego R},
	booktitle={Proceedings of theInternational Conference on Digital Image and Signal Processing (DISP’19)},
	year={2019}
}

@inproceedings{bird2018study,
	title={A study on mental state classification using eeg-based brain-machine interface},
	author={Bird, Jordan J and Manso, Luis J and Ribeiro, Eduardo P and Ekart, Aniko and Faria, Diego R},
	booktitle={2018 international conference on intelligent systems (IS)},
	pages={795--800},
	year={2018},
	organization={IEEE}
}

@article{levine2000direct,
	title={A direct brain interface based on event-related potentials},
	author={Levine, Simon P and Huggins, Jane E and BeMent, Spencer L and Kushwaha, Ramesh K and Schuh, Lori A and Rohde, Mitchell M and Passaro, Erasmo A and Ross, Donald A and Elisevich, Kost V and Smith, Brien J},
	journal={IEEE Transactions on Rehabilitation Engineering},
	volume={8},
	number={2},
	pages={180--185},
	year={2000},
	publisher={IEEE}
}

@article{vidal1977real,
	title={Real-time detection of brain events in EEG},
	author={Vidal, Jacques J},
	journal={Proceedings of the IEEE},
	volume={65},
	number={5},
	pages={633--641},
	year={1977},
	publisher={IEEE}
}

@article{allison2007brain,
	title={Brain--computer interface systems: progress and prospects},
	author={Allison, Brendan Z and Wolpaw, Elizabeth Winter and Wolpaw, Jonathan R},
	journal={Expert review of medical devices},
	volume={4},
	number={4},
	pages={463--474},
	year={2007},
	publisher={Taylor \& Francis}
}

@article{lebedev2017brain,
	title={Brain-machine interfaces: from basic science to neuroprostheses and neurorehabilitation},
	author={Lebedev, Mikhail A and Nicolelis, Miguel AL},
	journal={Physiological reviews},
	year={2017},
	publisher={American Physiological Society Bethesda, MD}
}

@inproceedings{wahalla2020cerebridge,
	title={CereBridge: An Efficient, FPGA-based Real-Time Processing Platform for True Mobile Brain-Computer Interfaces},
	author={Wahalla, Marc-Nils and Vaya, Guillermo Paya and Blume, Holger},
	booktitle={2020 42nd Annual International Conference of the IEEE Engineering in Medicine \& Biology Society (EMBC)},
	pages={4046--4050},
	year={2020},
	organization={IEEE}
}

@article{boly2012brain,
	title={Brain connectivity in disorders of consciousness},
	author={Boly, M{\'e}lanie and Massimini, Marcello and Garrido, Marta Isabel and Gosseries, Olivia and Noirhomme, Quentin and Laureys, Steven and Soddu, Andrea},
	journal={Brain connectivity},
	volume={2},
	number={1},
	pages={1--10},
	year={2012},
	publisher={Mary Ann Liebert, Inc. 140 Huguenot Street, 3rd Floor New Rochelle, NY 10801 USA}
}

@article{chatelle2012brain,
	title={Brain--computer interfacing in disorders of consciousness},
	author={Chatelle, Camille and Chennu, Srivas and Noirhomme, Quentin and Cruse, Damian and Owen, Adrian M and Laureys, Steven},
	journal={Brain injury},
	volume={26},
	number={12},
	pages={1510--1522},
	year={2012},
	publisher={Taylor \& Francis}
}

@article{edlinger2015many,
	title={How many people can use a BCI system?},
	author={Edlinger, G{\"u}nter and Allison, Brendan Z and Guger, Christoph},
	journal={Clinical systems neuroscience},
	pages={33--66},
	year={2015},
	publisher={Springer}
}

@article{gibson2014multiple,
	title={Multiple tasks and neuroimaging modalities increase the likelihood of detecting covert awareness in patients with disorders of consciousness},
	author={Gibson, Raechelle M and Fern{\'a}ndez-Espejo, Davinia and Gonzalez-Lara, Laura E and Kwan, Benjamin Y and Lee, Donald H and Owen, Adrian M and Cruse, Damian},
	journal={Frontiers in Human Neuroscience},
	volume={8},
	pages={950},
	year={2014},
	publisher={Frontiers Media SA}
}

@article{machado2010eeg,
	title={EEG-based brain-computer interfaces: an overview of basic concepts and clinical applications in neurorehabilitation},
	author={Machado, Sergio and Ara{\'u}jo, Fernanda and Paes, Fl{\'a}via and Velasques, Bruna and Cunha, Mario and Budde, Henning and Basile, Luis F and Anghinah, Renato and Arias-Carri{\'o}n, Oscar and Cagy, Mauricio and others},
	journal={Reviews in the Neurosciences},
	volume={21},
	number={6},
	pages={451--468},
	year={2010},
	publisher={De Gruyter}
}

@article{yadav2023electroencephalogram,
	title={Electroencephalogram based brain-computer interface: Applications, challenges, and opportunities},
	author={Yadav, Hitesh and Maini, Surita},
	journal={Multimedia Tools and Applications},
	volume={82},
	number={30},
	pages={47003--47047},
	year={2023},
	publisher={Springer}
}

@article{douibi2021toward,
	title={Toward EEG-based BCI applications for industry 4.0: Challenges and possible applications},
	author={Douibi, Khalida and Le Bars, Sol{\`e}ne and Lemontey, Alice and Nag, Lipsa and Balp, Rodrigo and Breda, Gabri{\`e}le},
	journal={Frontiers in Human Neuroscience},
	volume={15},
	pages={705064},
	year={2021},
	publisher={Frontiers Media SA}
}

@article{cheng2020emotion,
	title={Emotion recognition from multi-channel EEG via deep forest},
	author={Cheng, Juan and Chen, Meiyao and Li, Chang and Liu, Yu and Song, Rencheng and Liu, Aiping and Chen, Xun},
	journal={IEEE Journal of Biomedical and Health Informatics},
	volume={25},
	number={2},
	pages={453--464},
	year={2020},
	publisher={IEEE}
}

@article{wang2014emotional,
	title={Emotional state classification from EEG data using machine learning approach},
	author={Wang, Xiao-Wei and Nie, Dan and Lu, Bao-Liang},
	journal={Neurocomputing},
	volume={129},
	pages={94--106},
	year={2014},
	publisher={Elsevier}
}

@article{maksimenko2017absence,
	title={Absence seizure control by a brain computer interface},
	author={Maksimenko, Vladimir A and Van Heukelum, Sabrina and Makarov, Vladimir V and Kelderhuis, Janita and L{\"u}ttjohann, Annika and Koronovskii, Alexey A and Hramov, Alexander E and Van Luijtelaar, Gilles},
	journal={Scientific Reports},
	volume={7},
	number={1},
	pages={2487},
	year={2017},
	publisher={Nature Publishing Group UK London}
}

@inproceedings{liang2010closed,
	title={A closed-loop brain computer interface for real-time seizure detection and control},
	author={Liang, Sheng-Fu and Shaw, Fu-Zen and Young, Chung-Ping and Chang, Da-Wei and Liao, Yi-Cheng},
	booktitle={2010 annual international conference of the IEEE engineering in medicine and biology},
	pages={4950--4953},
	year={2010},
	organization={IEEE}
}

@article{abiri2019comprehensive,
	title={A comprehensive review of EEG-based brain--computer interface paradigms},
	author={Abiri, Reza and Borhani, Soheil and Sellers, Eric W and Jiang, Yang and Zhao, Xiaopeng},
	journal={Journal of neural engineering},
	volume={16},
	number={1},
	pages={011001},
	year={2019},
	publisher={IOP Publishing}
}

@article{zaghloul2019early,
	title={Early prediction of epilepsy seizures vlsi bci system},
	author={Zaghloul, Zaghloul Saad and Bayoumi, Magdy},
	journal={arXiv preprint arXiv:1906.02894},
	year={2019}
}

@article{yang2020seizure,
	title={From seizure detection to smart and fully embedded seizure prediction engine: A review},
	author={Yang, Jie and Sawan, Mohamad},
	journal={IEEE Transactions on Biomedical Circuits and Systems},
	volume={14},
	number={5},
	pages={1008--1023},
	year={2020},
	publisher={IEEE}
}

@article{hosseini2017optimized,
	title={Optimized deep learning for EEG big data and seizure prediction BCI via internet of things},
	author={Hosseini, Mohammad-Parsa and Pompili, Dario and Elisevich, Kost and Soltanian-Zadeh, Hamid},
	journal={IEEE Transactions on Big Data},
	volume={3},
	number={4},
	pages={392--404},
	year={2017},
	publisher={IEEE}
}

@article{you2020unsupervised,
	title={Unsupervised automatic seizure detection for focal-onset seizures recorded with behind-the-ear EEG using an anomaly-detecting generative adversarial network},
	author={You, Sungmin and Cho, Baek Hwan and Yook, Soonhyun and Kim, Joo Young and Shon, Young-Min and Seo, Dae-Won and Kim, In Young},
	journal={Computer Methods and Programs in Biomedicine},
	volume={193},
	pages={105472},
	year={2020},
	publisher={Elsevier}
}

@inproceedings{liberati2013development,
	title={Development of a binary fMRI-BCI for Alzheimer patients: a semantic conditioning paradigm using affective unconditioned stimuli},
	author={Liberati, Giulia and Veit, Ralf and Kim, Sunjung and Birbaumer, Niels and Von Arnim, Christine and Jenner, Anne and Lul{\'e}, Doroth{\'e}e and Ludolph, Albert Christian and Raffone, Antonino and Belardinelli, Marta Olivetti and others},
	booktitle={2013 Humaine Association Conference on Affective Computing and Intelligent Interaction},
	pages={838--842},
	year={2013},
	organization={IEEE}
}

@article{moller2021technology,
	title={Technology-based neurorehabilitation in Parkinson’s disease—A narrative review},
	author={M{\"o}ller, Jens Carsten and Zutter, Daniel and Riener, Robert},
	journal={Clinical and Translational Neuroscience},
	volume={5},
	number={3},
	pages={23},
	year={2021},
	publisher={MDPI}
}

@article{liberati2012toward,
	title={Toward a brain-computer interface for Alzheimer's disease patients by combining classical conditioning and brain state classification},
	author={Liberati, Giulia and Da Rocha, Josu{\'e} Luiz Dalboni and Van der Heiden, Linda and Raffone, Antonino and Birbaumer, Niels and Olivetti Belardinelli, Marta and Sitaram, Ranganatha},
	journal={Journal of Alzheimer’s disease},
	volume={31},
	number={s3},
	pages={S211--S220},
	year={2012},
	publisher={SAGE Publications Sage UK: London, England}
}

@inproceedings{song2021evaluation,
	title={Evaluation and Diagnosis of Brain Diseases based on Non-invasive BCI},
	author={Song, Zuoting and Fang, Tao and Ma, Jing and Zhang, Yuan and Le, Song and Zhan, Gege and Zhang, Xueze and Wang, Shouyan and Li, Hui and Lin, Yifang and others},
	booktitle={2021 9th International Winter Conference on Brain-Computer Interface (BCI)},
	pages={1--6},
	year={2021},
	organization={IEEE}
}

@article{luaute2015bci,
	title={BCI in patients with disorders of consciousness: clinical perspectives},
	author={Luaut{\'e}, Jacques and Morlet, Dominique and Mattout, J{\'e}r{\'e}mie},
	journal={Annals of Physical and Rehabilitation Medicine},
	volume={58},
	number={1},
	pages={29--34},
	year={2015},
	publisher={Elsevier}
}

@article{handayanireal,
	title={A Real-Time Brain-Computer Interface (BCI) Framework for Sleep State Stimulation Using a Deep-Learning Technique: Proposal},
	author={Handayani, Dini and Yaacob, Hamwira and Attarbashi, Zainab and Osmani, Noor Mohammad and Jamaludin, Mohammad Aizat and Altaleb, Abdulazeez E}
}

@article{behzad2021role,
	title={The role of EEG in the diagnosis and management of patients with sleep disorders},
	author={Behzad, Ramina and Behzad, Aida},
	journal={Journal of Behavioral and Brain Science},
	volume={11},
	number={10},
	pages={257--266},
	year={2021},
	publisher={Scientific Research Publishing}
}

@article{yuan2021bci,
	title={BCI training effects on chronic stroke correlate with functional reorganization in motor-related regions: a concurrent EEG and fMRI study},
	author={Yuan, Kai and Chen, Cheng and Wang, Xin and Chu, Winnie Chiu-wing and Tong, Raymond Kai-yu},
	journal={Brain sciences},
	volume={11},
	number={1},
	pages={56},
	year={2021},
	publisher={MDPI}
}

@article{young2014changes,
	title={Changes in functional connectivity correlate with behavioral gains in stroke patients after therapy using a brain-computer interface device},
	author={Young, Brittany Mei and Nigogosyan, Zack and Remsik, Alexander and Walton, L{\'e}o M and Song, Jie and Nair, Veena A and Grogan, Scott W and Tyler, Mitchell E and Edwards, Dorothy Farrar and Caldera, Kristin and others},
	journal={Frontiers in neuroengineering},
	volume={7},
	pages={25},
	year={2014},
	publisher={Frontiers Media SA}
}

@article{doud2011continuous,
	title={Continuous three-dimensional control of a virtual helicopter using a motor imagery based brain-computer interface},
	author={Doud, Alexander J and Lucas, John P and Pisansky, Marc T and He, Bin},
	journal={PloS one},
	volume={6},
	number={10},
	pages={e26322},
	year={2011},
	publisher={Public Library of Science San Francisco, USA}
}

@article{varkuti2013resting,
	title={Resting state changes in functional connectivity correlate with movement recovery for BCI and robot-assisted upper-extremity training after stroke},
	author={Varkuti, Balint and Guan, Cuntai and Pan, Yaozhang and Phua, Kok Soon and Ang, Kai Keng and Kuah, Christopher Wee Keong and Chua, Karen and Ang, Beng Ti and Birbaumer, Niels and Sitaram, Ranganathan},
	journal={Neurorehabilitation and neural repair},
	volume={27},
	number={1},
	pages={53--62},
	year={2013},
	publisher={SAGE Publications Sage CA: Los Angeles, CA}
}

@inproceedings{miladinovic2020evaluation,
	title={Evaluation of Motor Imagery-Based BCI methods in neurorehabilitation of Parkinson’s Disease patients},
	author={Miladinovi{\'c}, Aleksandar and Aj{\v{c}}evi{\'c}, Milo{\v{s}} and Busan, Pierpaolo and Jarmolowska, Joanna and Silveri, Giulia and Deodato, Manuela and Mezzarobba, Susanna and Battaglini, Piero Paolo and Accardo, Agostino},
	booktitle={2020 42nd Annual International Conference of the IEEE Engineering in Medicine \& Biology Society (EMBC)},
	pages={3058--3061},
	year={2020},
	organization={IEEE}
}

@article{qin2004motor,
	title={Motor imagery classification by means of source analysis for brain--computer interface applications},
	author={Qin, Lei and Ding, Lei and He, Bin},
	journal={Journal of neural engineering},
	volume={1},
	number={3},
	pages={135},
	year={2004},
	publisher={IOP Publishing}
}

@inproceedings{tung2013motor,
	title={Motor imagery BCI for upper limb stroke rehabilitation: An evaluation of the EEG recordings using coherence analysis},
	author={Tung, Sau Wai and Guan, Cuntai and Ang, Kai Keng and Phua, Kok Soon and Wang, Chuanchu and Zhao, Ling and Teo, Wei Peng and Chew, Effie},
	booktitle={2013 35th Annual International Conference of the IEEE Engineering in Medicine and Biology Society (EMBC)},
	pages={261--264},
	year={2013},
	organization={IEEE}
}

@article{yuan2010negative,
	title={Negative covariation between task-related responses in alpha/beta-band activity and BOLD in human sensorimotor cortex: an EEG and fMRI study of motor imagery and movements},
	author={Yuan, Han and Liu, Tao and Szarkowski, Rebecca and Rios, Cristina and Ashe, James and He, Bin},
	journal={Neuroimage},
	volume={49},
	number={3},
	pages={2596--2606},
	year={2010},
	publisher={Elsevier}
}

@article{hohne2014motor,
	title={Motor imagery for severely motor-impaired patients: evidence for brain-computer interfacing as superior control solution},
	author={H{\"o}hne, Johannes and Holz, Elisa and Staiger-S{\"a}lzer, Pit and M{\"u}ller, Klaus-Robert and K{\"u}bler, Andrea and Tangermann, Michael},
	journal={PloS one},
	volume={9},
	number={8},
	pages={e104854},
	year={2014},
	publisher={Public Library of Science San Francisco, USA}
}

@article{xie2012heterogeneous,
	title={Heterogeneous integration of bio-sensing system-on-chip and printed electronics},
	author={Xie, Li and Yang, Geng and Mantysalo, Matti and Xu, Lin-Lin and Jonsson, Fredrik and Zheng, Li-Rong},
	journal={IEEE Journal on Emerging and Selected Topics in Circuits and Systems},
	volume={2},
	number={4},
	pages={672--682},
	year={2012},
	publisher={IEEE}
}

@article{padfield2019eeg,
	title={EEG-based brain-computer interfaces using motor-imagery: Techniques and challenges},
	author={Padfield, Natasha and Zabalza, Jaime and Zhao, Huimin and Masero, Valentin and Ren, Jinchang},
	journal={Sensors},
	volume={19},
	number={6},
	pages={1423},
	year={2019},
	publisher={MDPI}
}

@article{samal2024role,
	title={Role of machine learning and deep learning techniques in EEG-based BCI emotion recognition system: a review},
	author={Samal, Priyadarsini and Hashmi, Mohammad Farukh},
	journal={Artificial Intelligence Review},
	volume={57},
	number={3},
	pages={50},
	year={2024},
	publisher={Springer}
}

@article{azghadi2020hardware,
	title={Hardware implementation of deep network accelerators towards healthcare and biomedical applications},
	author={Azghadi, Mostafa Rahimi and Lammie, Corey and Eshraghian, Jason K and Payvand, Melika and Donati, Elisa and Linares-Barranco, Bernabe and Indiveri, Giacomo},
	journal={IEEE Transactions on Biomedical Circuits and Systems},
	volume={14},
	number={6},
	pages={1138--1159},
	year={2020},
	publisher={IEEE}
}

@article{rizal2022fpga,
	title={FPGA-based implementation for real-time epileptic EEG classification using Hjorth descriptor and KNN},
	author={Rizal, Achmad and Hadiyoso, Sugondo and Ramdani, Ahmad Zaky},
	journal={Electronics},
	volume={11},
	number={19},
	pages={3026},
	year={2022},
	publisher={MDPI}
}

@article{rakhmatulin2024exploring,
	title={Exploring convolutional neural network architectures for EEG feature extraction},
	author={Rakhmatulin, Ildar and Dao, Minh-Son and Nassibi, Amir and Mandic, Danilo},
	journal={Sensors},
	volume={24},
	number={3},
	pages={877},
	year={2024},
	publisher={MDPI}
}

@article{fang2019development,
	title={Development and validation of an EEG-based real-time emotion recognition system using edge AI computing platform with convolutional neural network system-on-chip design},
	author={Fang, Wai-Chi and Wang, Kai-Yen and Fahier, Nicolas and Ho, Yun-Lung and Huang, Yu-De},
	journal={IEEE Journal on Emerging and Selected Topics in Circuits and Systems},
	volume={9},
	number={4},
	pages={645--657},
	year={2019},
	publisher={IEEE}
}

@article{wei2020review,
	title={A review of algorithm \& hardware design for AI-based biomedical applications},
	author={Wei, Ying and Zhou, Jun and Wang, Yin and Liu, Yinggang and Liu, Qingsong and Luo, Jiansheng and Wang, Chao and Ren, Fengbo and Huang, Li},
	journal={IEEE transactions on biomedical circuits and systems},
	volume={14},
	number={2},
	pages={145--163},
	year={2020},
	publisher={IEEE}
}

@inproceedings{karageorgos2020hardware,
	title={Hardware-software co-design for brain-computer interfaces},
	author={Karageorgos, Ioannis and Sriram, Karthik and Vesel{\`y}, J{\'a}n and Wu, Michael and Powell, Marc and Borton, David and Manohar, Rajit and Bhattacharjee, Abhishek},
	booktitle={2020 ACM/IEEE 47th Annual International Symposium on Computer Architecture (ISCA)},
	pages={391--404},
	year={2020},
	organization={IEEE}
}

@inproceedings{eichler2021mastermind,
	title={MasterMind: Many-accelerator SoC architecture for real-time brain-computer interfaces},
	author={Eichler, Guy and Piccolboni, Luca and Giri, Davide and Carloni, Luca P},
	booktitle={2021 IEEE 39th International Conference on Computer Design (ICCD)},
	pages={101--108},
	year={2021},
	organization={IEEE}
}

@inproceedings{krishna2023sparsity,
	title={A Sparsity-driven tinyML Accelerator for Decoding Hand Kinematics in Brain-Computer Interfaces},
	author={Krishna, Adithya and Ramanathan, Vignesh and Yadav, Satyapreet Singh and Shah, Sahil and van Schaik, Andr{\'e} and Mehendale, Mahesh and Thakur, Chetan Singh},
	booktitle={2023 IEEE Biomedical Circuits and Systems Conference (BioCAS)},
	pages={1--5},
	year={2023},
	organization={IEEE}
}

@article{cao2024optimized,
	title={An optimized EEGNet processor for low-power and real-time EEG classification in wearable brain--computer interfaces},
	author={Cao, Jiacheng and Xiong, Wei and Lu, Jie and Chen, Peilin and Wang, Jian and Lai, Jinmei and Huang, Miaoqing},
	journal={Microelectronics Journal},
	volume={145},
	pages={106134},
	year={2024},
	publisher={Elsevier}
}

@inproceedings{wang2020accurate,
	title={An accurate eegnet-based motor-imagery brain--computer interface for low-power edge computing},
	author={Wang, Xiaying and Hersche, Michael and T{\"o}mekce, Batuhan and Kaya, Burak and Magno, Michele and Benini, Luca},
	booktitle={2020 IEEE international symposium on medical measurements and applications (MeMeA)},
	pages={1--6},
	year={2020},
	organization={IEEE}
}

@article{el2024strong,
	title={A Strong and Simple Deep Learning Baseline for BCI Motor Imagery decoding},
	author={El Ouahidi, Yassine and Gripon, Vincent and Pasdeloup, Bastien and Bouallegue, Ghaith and Farrugia, Nicolas and Lioi, Giulia},
	journal={IEEE Transactions on Neural Systems and Rehabilitation Engineering},
	year={2024},
	publisher={IEEE}
}

@inproceedings{mattei2024deep,
	title={Deep Learning Architecture analysis for EEG-Based BCI Classification under Motor Execution},
	author={Mattei, Enrico and Lozzi, Daniele and Di Matteo, Alessandro and Polsinelli, Matteo and Manes, Costanzo and Mignosi, Filippo and Placidi, Giuseppe},
	booktitle={2024 IEEE 37th International Symposium on Computer-Based Medical Systems (CBMS)},
	pages={549--555},
	year={2024},
	organization={IEEE}
}

@article{glavas2024empowering,
	title={Empowering Individuals With Disabilities: A 4-DoF BCI Wheelchair Using MI and EOG Signals},
	author={Glavas, Kosmas and Tzimourta, Katerina D and Tzallas, Alexandros T and Giannakeas, Nikolaos and Tsipouras, Markos G},
	journal={IEEE Access},
	year={2024},
	publisher={IEEE}
}

@article{wang2024towards,
	title={Towards a Fast and Robust MI-BCI: Online Adaptation of Stimulus Paradigm and Classification Model},
	author={Wang, Jiaxing and Wang, Weiqun and Su, Jianqiang and Wang, Yihan and Hou, Zeng-Guang},
	journal={IEEE Transactions on Instrumentation and Measurement},
	year={2024},
	publisher={IEEE}
}

@inproceedings{kim2024meta,
	title={Meta-Learning-based Cross-Dataset Motor Imagery Brain-Computer Interface},
	author={Kim, Jun-Mo and Bak, Soyeon and Nam, Hyeonyeong and Choi, WooHyeok and Kam, Tae-Eui},
	booktitle={2024 12th International Winter Conference on Brain-Computer Interface (BCI)},
	pages={1--4},
	year={2024},
	organization={IEEE}
}

@inproceedings{shin2024sparse,
	title={Sparse Multitask Learning for Efficient Neural Representation of Motor Imagery and Execution},
	author={Shin, Hye-Bin and Yin, Kang and Lee, Seong-Whan},
	booktitle={2024 12th International Winter Conference on Brain-Computer Interface (BCI)},
	pages={1--4},
	year={2024},
	organization={IEEE}
}

@article{ali2023correlation,
	title={Correlation-filter-based channel and feature selection framework for hybrid EEG-fNIRS BCI applications},
	author={Ali, Muhammad Umair and Zafar, Amad and Kallu, Karam Dad and Masood, Haris and Mannan, Malik Muhammad Naeem and Ibrahim, Malik Muhammad and Kim, Sangil and Khan, Muhammad Attique},
	journal={IEEE Journal of Biomedical and Health Informatics},
	volume={28},
	number={6},
	pages={3361--3370},
	year={2023},
	publisher={IEEE}
}

@article{dhiman2023machine,
	title={Machine learning techniques for electroencephalogram based brain-computer interface: A systematic literature review},
	author={Dhiman, Rohtash and others},
	journal={Measurement: Sensors},
	volume={28},
	pages={100823},
	year={2023},
	publisher={Elsevier}
}

@inproceedings{doudou2023improving,
	title={Improving classification accuracy for motor imagery BCI using DWT based Features},
	author={Doudou, Abdel Fateh and Reffad, Aicha and Mebarkia, Kamel},
	booktitle={2023 International Conference on Electrical Engineering and Advanced Technology (ICEEAT)},
	volume={1},
	pages={1--5},
	year={2023},
	organization={IEEE}
}

@article{hosseini2023state,
	title={State-based decoding of continuous hand movements using EEG signals},
	author={Hosseini, Seyyed Moosa and Shalchyan, Vahid},
	journal={IEEE Access},
	volume={11},
	pages={42764--42778},
	year={2023},
	publisher={IEEE}
}

@article{feng2022efficient_model,
	title={An efficient model-compressed EEGNet accelerator for generalized brain-computer interfaces with near sensor intelligence},
	author={Feng, Lichen and Shan, Hongwei and Zhang, Yueqi and Zhu, Zhangming},
	journal={IEEE Transactions on Biomedical Circuits and Systems},
	volume={16},
	number={6},
	pages={1239--1249},
	year={2022},
	publisher={IEEE}
}

@article{huang2022eeg,
	title={EEG-based vibrotactile evoked brain-computer interfaces system: A systematic review},
	author={Huang, Xiuyu and Liang, Shuang and Li, Zengguang and Lai, Cynthia Yuen Yi and Choi, Kup-Sze},
	journal={PLoS One},
	volume={17},
	number={6},
	pages={e0269001},
	year={2022},
	publisher={Public Library of Science San Francisco, CA USA}
}

@inproceedings{yang2022enhancing,
	title={Enhancing EEG motor imagery decoding performance via deep temporal-domain information extraction},
	author={Yang, Qihong and Yang, Mingzhao and Liu, Ke and Deng, Xin},
	booktitle={2022 IEEE 11th Data Driven Control and Learning Systems Conference (DDCLS)},
	pages={420--424},
	year={2022},
	organization={IEEE}
}

@article{gu2021eeg,
	title={EEG-based brain-computer interfaces (BCIs): A survey of recent studies on signal sensing technologies and computational intelligence approaches and their applications},
	author={Gu, Xiaotong and Cao, Zehong and Jolfaei, Alireza and Xu, Peng and Wu, Dongrui and Jung, Tzyy-Ping and Lin, Chin-Teng},
	journal={IEEE/ACM transactions on computational biology and bioinformatics},
	volume={18},
	number={5},
	pages={1645--1666},
	year={2021},
	publisher={IEEE}
}

@article{belwafi2021embedded,
	title={Embedded brain computer interface: State-of-the-art in research},
	author={Belwafi, Kais and Gannouni, Sofien and Aboalsamh, Hatim},
	journal={Sensors},
	volume={21},
	number={13},
	pages={4293},
	year={2021},
	publisher={MDPI}
}

@article{portillo2021mind,
	title={Mind the gap: State-of-the-art technologies and applications for EEG-based brain--computer interfaces},
	author={Portillo-Lara, Roberto and Tahirbegi, Bogachan and Chapman, Christopher AR and Goding, Josef A and Green, Rylie A},
	journal={APL bioengineering},
	volume={5},
	number={3},
	year={2021},
	publisher={AIP Publishing}
}

@article{saeidi2021neural,
	title={Neural decoding of EEG signals with machine learning: a systematic review},
	author={Saeidi, Maham and Karwowski, Waldemar and Farahani, Farzad V and Fiok, Krzysztof and Taiar, Redha and Hancock, Peter A and Al-Juaid, Awad},
	journal={Brain sciences},
	volume={11},
	number={11},
	pages={1525},
	year={2021},
	publisher={MDPI}
}

@article{wu2020transfer,
	title={Transfer learning for EEG-based brain--computer interfaces: A review of progress made since 2016},
	author={Wu, Dongrui and Xu, Yifan and Lu, Bao-Liang},
	journal={IEEE Transactions on Cognitive and Developmental Systems},
	volume={14},
	number={1},
	pages={4--19},
	year={2020},
	publisher={IEEE}
}

@article{hosseini2020review,
	title={A review on machine learning for EEG signal processing in bioengineering},
	author={Hosseini, Mohammad-Parsa and Hosseini, Amin and Ahi, Kiarash},
	journal={IEEE reviews in biomedical engineering},
	volume={14},
	pages={204--218},
	year={2020},
	publisher={IEEE}
}

@article{lazarou2018eeg,
	title={EEG-based brain--computer interfaces for communication and rehabilitation of people with motor impairment: a novel approach of the 21 st Century},
	author={Lazarou, Ioulietta and Nikolopoulos, Spiros and Petrantonakis, Panagiotis C and Kompatsiaris, Ioannis and Tsolaki, Magda},
	journal={Frontiers in human neuroscience},
	volume={12},
	pages={14},
	year={2018},
	publisher={Frontiers Media SA}
}

@article{ang2016eeg,
	title={EEG-based strategies to detect motor imagery for control and rehabilitation},
	author={Ang, Kai Keng and Guan, Cuntai},
	journal={IEEE Transactions on Neural Systems and Rehabilitation Engineering},
	volume={25},
	number={4},
	pages={392--401},
	year={2016},
	publisher={IEEE}
}

@article{yuan2014brain,
	title={Brain--computer interfaces using sensorimotor rhythms: current state and future perspectives},
	author={Yuan, Han and He, Bin},
	journal={IEEE Transactions on Biomedical Engineering},
	volume={61},
	number={5},
	pages={1425--1435},
	year={2014},
	publisher={IEEE}
}

@article{chen2022transfer,
	title={Transfer learning with optimal transportation and frequency mixup for EEG-based motor imagery recognition},
	author={Chen, Peiyin and Wang, He and Sun, Xinlin and Li, Haoyu and Grebogi, Celso and Gao, Zhongke},
	journal={IEEE transactions on neural systems and rehabilitation engineering},
	volume={30},
	pages={2866--2875},
	year={2022},
	publisher={IEEE}
}

@article{xue2020multifrequency,
	title={A Multifrequency Brain Network-Based Deep Learning Framework for Motor Imagery Decoding},
	author={Xue, Juntao and Ren, Feiyue and Sun, Xinlin and Yin, Miaomiao and Wu, Jialing and Ma, Chao and Gao, Zhongke},
	journal={Neural Plasticity},
	volume={2020},
	number={1},
	pages={8863223},
	year={2020},
	publisher={Wiley Online Library}
}

@inproceedings{dos2023residual,
	title={Residual Attention Module on EEGN et for Brain-Computer Interface},
	author={Dos Santos, Davi Esteves and De Souza, Gabriel Henrique and Bernardino, Heder and Vieira, Alex Borges and Motta, Luciana Paix{\~a}o},
	booktitle={2023 IEEE Symposium Series on Computational Intelligence (SSCI)},
	pages={58--63},
	year={2023},
	organization={IEEE}
}

@article{congedo2017riemannian,
	title={Riemannian geometry for EEG-based brain-computer interfaces; a primer and a review},
	author={Congedo, Marco and Barachant, Alexandre and Bhatia, Rajendra},
	journal={Brain-Computer Interfaces},
	volume={4},
	number={3},
	pages={155--174},
	year={2017},
	publisher={Taylor \& Francis}
}

@article{hosseinifard2013classifying,
	title={Classifying depression patients and normal subjects using machine learning techniques and nonlinear features from EEG signal},
	author={Hosseinifard, Behshad and Moradi, Mohammad Hassan and Rostami, Reza},
	journal={Computer methods and programs in biomedicine},
	volume={109},
	number={3},
	pages={339--345},
	year={2013},
	publisher={Elsevier}
}

@article{wu2015bayesian,
	title={Bayesian Machine Learning: EEG$\backslash$/MEG signal processing measurements},
	author={Wu, Wei and Nagarajan, Srikantan and Chen, Zhe},
	journal={IEEE Signal Processing Magazine},
	volume={33},
	number={1},
	pages={14--36},
	year={2015},
	publisher={IEEE}
}

@article{li2024combining,
	title={Combining VR with electroencephalography as a frontier of brain-computer interfaces},
	author={Li, Hongbian and Shin, Hyonyoung and Sentis, Luis and Siu, Ka-Chun and Mill{\'a}n, Jos{\'e} del R and Lu, Nanshu},
	journal={Device},
	volume={2},
	number={6},
	year={2024},
	publisher={Elsevier}
}

@inproceedings{gawali2016mixed,
	title={Mixed signal SoC based Bio-Sensor Node for long term health monitoring},
	author={Gawali, Dhanashri H and Wadhai, Vijay M},
	booktitle={2016 IEEE International WIE Conference on Electrical and Computer Engineering (WIECON-ECE)},
	pages={194--198},
	year={2016},
	organization={IEEE}
}

		
		\begin{IEEEbiography}[{\includegraphics[width=1in,height=1.25in,clip,keepaspectratio]{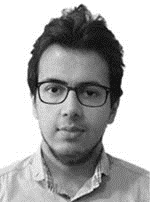}}]
			{Mohammad Hossein Koohi Ghamsari} received his M.S. degree in electrical engineering with a focus on communications and optics from Sharif University of Technology, Tehran, Iran, in 2022. 
			
			Following his master's studies, he began collaborating with the Electrical Engineering Department at Sharif University of Technology as a Researcher. He is also an active lecturer, delivering university courses and conducting workshops at conferences. His research interests include electronic circuits, biosensors, communication systems, applied electromagnetics, Metasurfaces, optics, and applied machine learning.
			
			Mr. Koohi Ghamsari has received several honors and awards, including the Best National Student Book of the Year Award (University of Tehran) and an M.S. Thesis Fellowship. In 2018, he ranked 1st in the National Technical \& Innovation Competition among more than 14,500 participants at Sharif University of Technology. He has been actively involved in various professional societies, serving as a student member of the IEEE from 2017 to 2018, the Iran Microelectronic Association from 2020 to 2021, and the Iranian Mathematical Society (IMS) from 2022 to 2023.
		\end{IEEEbiography}

		\begin{IEEEbiography}[{\includegraphics[width=1in,height=1.25in,clip,keepaspectratio]{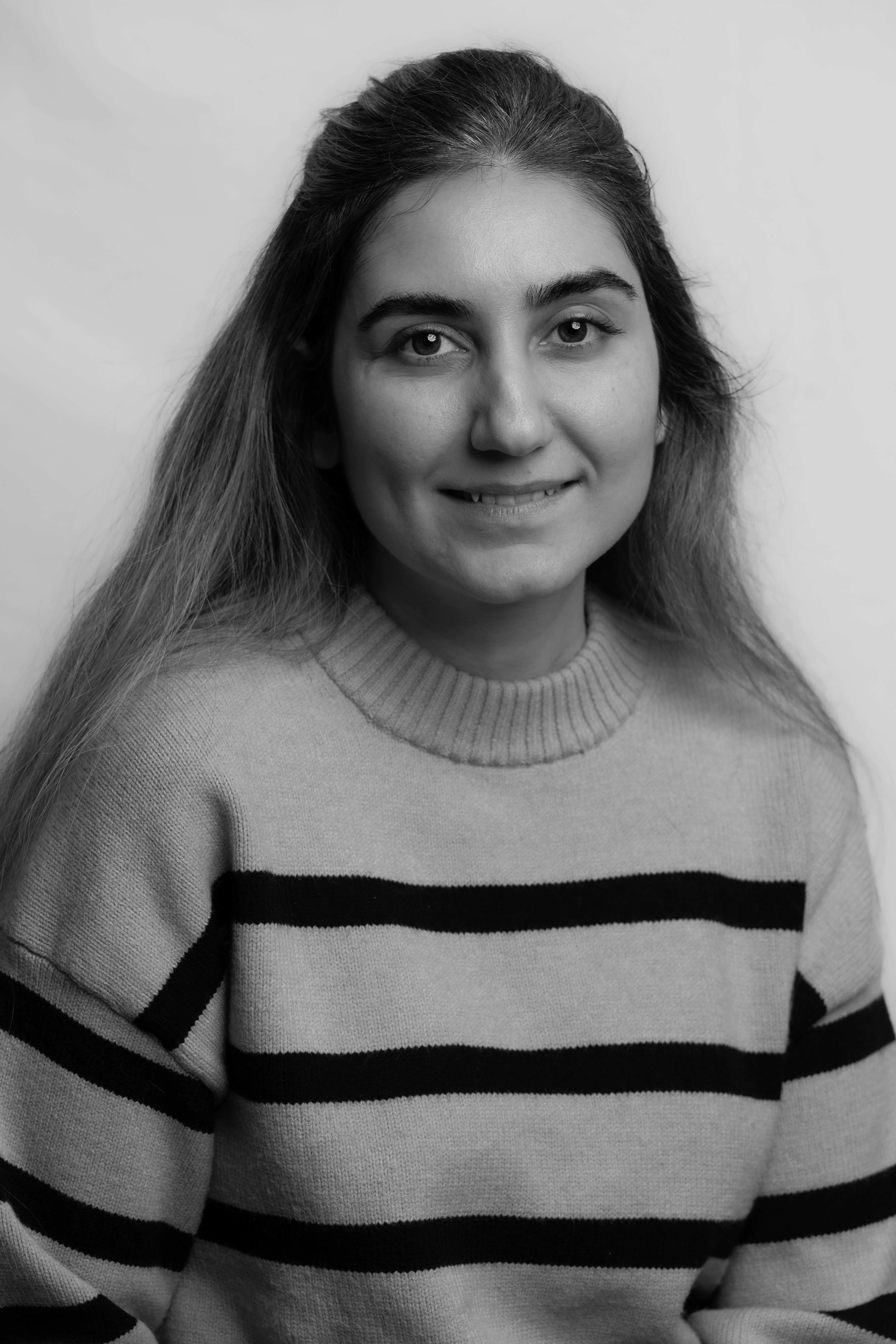}}]{Seyede Fatemeh Ghamkhari}
			received the Ph.D. degree in electrical engineering from Shahed University, Tehran, Iran, in 2021. From 2022 to 2024, she was a Postdoctoral Researcher at Sharif University of Technology, where she focused on energy-efficient designs of convolutional filters for IoT edge processing applications. Since 2025, she has been a Postdoctoral Researcher at KTH Royal Institute of Technology, Stockholm, Sweden, focusing on hardware accelerators for signal processing applications. Her research interests include digital signal processing, hardware accelerators, low-power integrated circuits, and IoT edge processing. She has published several research papers and delivered presentations in both academic and industrial settings.
		\end{IEEEbiography}

		\begin{IEEEbiography}[{\includegraphics[width=1in,height=1.25in,clip,keepaspectratio]{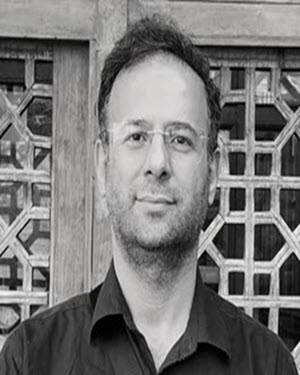}}]{Siavash Bayat} (S’10–M’13) received a Ph.D. degree in electrical engineering from the University of Sydney, Sydney, NSW, Australia in 2013.
			
			He was then a Post-Doctoral Research Fellow with the University of Sydney, prior to joining the Sharif University of Technology, Tehran, Iran, in 2014, where he is currently an Assistant Professor and the Director of the Electronics Research Institute (ERI). He is also the Founder and Head of the IoT Research Center. His current research interests include wireless communications, signal processing, wireless resource management, cognitive radio, game theory, and physical layer security.
			
			Prof. Bayat was recognized as an outstanding researcher by the Sharif University Research Department in 2017 and 2022 for his exceptional contributions and innovation to the field of telecommunications, being selected as the outstanding researcher among all research units at Sharif University of Technology. In 2018, he was honored as the Outstanding Technology Director by the same department for his leadership and achievements in directing a multidisciplinary research center. Under his guidance, the center successfully carried out multiple industrial research projects in the areas of IoT and communication systems and he was awarded the title of distinguished technologist at Sharif University of Technology in 2023. 
			
		\end{IEEEbiography}

		\begin{IEEEbiography}[{\includegraphics[width=1in,height=1.25in,clip,keepaspectratio]{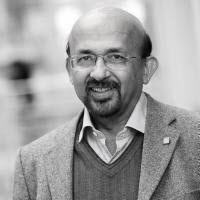}}]{Ahmed Hemani} received the Ph.D. degree in electrical engineering from the KTH Royal Institute of Technology (KTH), Stockholm, Sweden, in 1993.
			
			He is currently a Professor of Electronic Systems Design with the School of Electrical Engineering and Computer Science (EECS), KTH. Prior to joining academia, he held research and development positions at several leading semiconductor and telecommunications companies, including National Semiconductor, ABB, Ericsson, NewLogic, and Philips Semiconductors (now NXP), where he contributed to the design of embedded systems, wireless communication ASICs, and low-power multicore processors. His current research interests include electronic system design, design automation, networks-on-chip, clocking and power management, latency-insensitive systems, and reconfigurable VLSI architectures.
			
			Prof. Hemani pioneered the concept of Networks-on-Chip (NoC) architectures and has made significant contributions to high-level synthesis, Globally Asynchronous Locally Synchronous (GALS) systems, the Globally Ratiochronous Locally Synchronous (GRLS) design methodology, and the SiLago (Silicon Lego) design framework. He is a Senior Member of the IEEE and has founded several technology start-ups based on his research innovations.
		\end{IEEEbiography}
	\end{document}